\documentclass[twoside,12pt]{PhDthesisPSnPDF}

\usepackage{lineno}
\usepackage{amsbsy}
\usepackage{amssymb,amsmath}
\usepackage{xspace}
\usepackage{wtmmPkg}
\usepackage{natbib}
\usepackage{multirow}
\usepackage{paralist}
\usepackage{fancyhdr} % for better header layout
\usepackage{subfigure}
\usepackage{rotating}
\usepackage{mathrsfs}

\DeclareRobustCommand{\ion}[2]{%
\relax\ifmmode
\ifx\testbx\f@series
{\mathbf{#1\,\mathsc{#2}}}\else
{\mathrm{#1\,\mathsc{#2}}}\fi
\else\textup{#1\,{\mdseries\textsc{#2}}}%
\fi}

\newcommand{\aap}{    {\it Astronomy \& Astrophysics}}

\ifpdf  
    \pdfcatalog { /PageMode (/UseOutlines)
                  /OpenAction (fitbh)  }
\fi

\title{The Kinematics and Morphology of Solar Coronal Mass Ejections}

\ifpdf
  \author{\href{mailto:jbyrne6@tcd.ie}{Jason P. Byrne, B.A.(Mod.)}}
  \collegeordept{\href{http://www.physics.tcd.ie}{School of Physics}}
  \university{\href{http://www.tcd.ie}{University of Dublin, Trinity College}}

  \crest{
  \vspace{1cm}
  \includegraphics[width=0.3\textwidth]{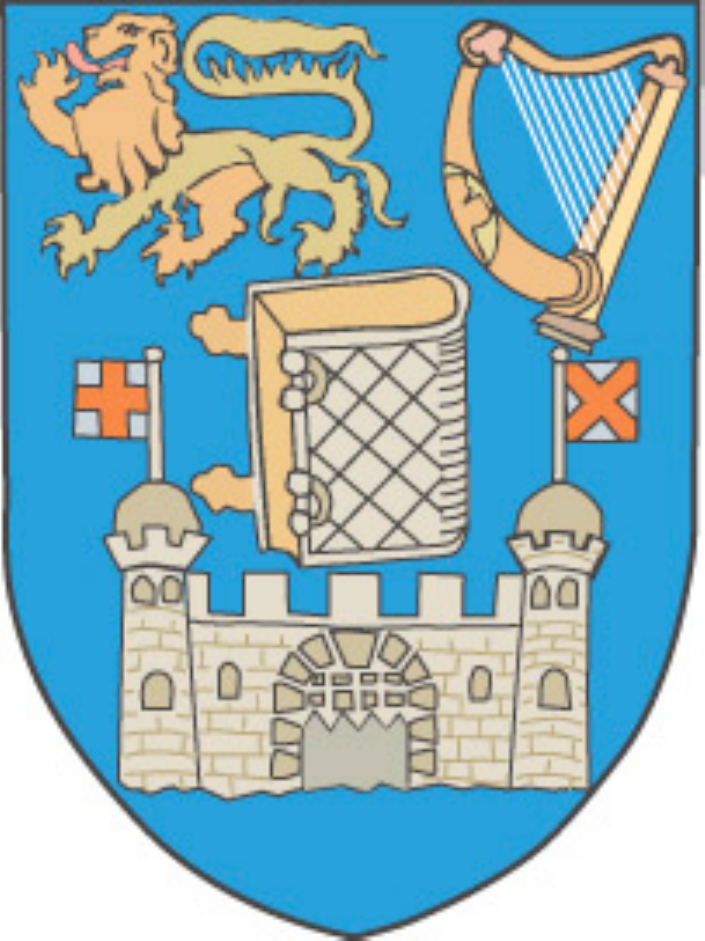}
   \vspace{1cm}
    }
  
\else
  \author{\\ Jason P. Byrne}
  \collegeordept{School of Physics}
  \university{University of Dublin, Trinity College}
  \crest{\includegraphics[width=6cm]{TCD_Crest.pdf}}
\fi
\degree{Philosophi\ae Doctor (PhD)}
\degreedate{September 2010}

\usepackage{setspace}
\begin{document}

%\language{english}

% sets line spacing
\renewcommand\baselinestretch{1.2}
\baselineskip=18pt plus1pt

%: ----------------------- generate cover page ------------------------

\maketitle  % command to print the title page with above variables

%
%%: ----------------------- cover page back side ------------------------
%% Your research institution may require reviewer names, etc.
%% This cover back side is required by Dresden Med Fac; uncomment if needed.

%\newpage
%\vspace{10mm}
%1. Reviewer: Name

%\vspace{10mm}
%2. Reviewer: 

%\vspace{20mm}
%Day of the defense:

%\vspace{20mm}
%\hspace{70mm}Signature from head of PhD committee:

%: ----------------------- abstract ------------------------

% Your institution may have specific regulations if you need an abstract and where it is to be placed in the document. The default here is just after title.

% The original template provides and abstractseparate environment, if your institution requires them to be separate. I think it's easier to print the abstract from the complete thesis by restricting printing to the relevant page.
% \begin{abstractseparate}
%   \input{Abstract/abstract}
% \end{abstractseparate}

%: ----------------------- tie in front matter ------------------------B

\frontmatter
%: Declaration of originality
%\include{backmatter/declaration}

% Thesis Abstract -----------------------------------------------------

%\begin{abstractslong}    %uncommenting this line, gives a different abstract heading
\begin{abstracts}        %this creates the heading for the abstract page

Solar coronal mass ejections (CMEs) are large-scale eruptions of plasma and magnetic field from the Sun into the corona and interplanetary space. They are the most significant drivers of adverse space weather at Earth and other locations in the heliosphere, so it is important to understand the physics governing their eruption and propagation. However the diffuse morphology and transient nature of CMEs makes them difficult to identify and track using traditional image processing techniques. Furthermore, the true three-dimensional geometry of CMEs has remained elusive due to the limitations of coronagraph plane-of-sky images with restricted fields-of-view. For these reasons the Solar Terrestrial Relations Observatory (STEREO) was launched as a twin-spacecraft mission to fly in orbits ahead and behind the Earth in order to triangulate independent observations of CME structure. It is the first time CMEs have been observed from vantage points off the Sun-Earth line and each spacecraft carries an instrument suite designed to image from the low solar corona out to the orbit of Earth in order to observe and study CME propagation towards Earth.

In this thesis the implementation of multiscale image processing techniques to identify and track the CME front through coronagraph images is detailed. An ellipse characterisation of the CME front is used to determine the CME kinematics and morphology with increased precision as compared to techniques used in current CME catalogues, and efforts are underway to automate this procedure for applying to a large number of CME observations for future analysis. It was found that CMEs do not simply undergo constant acceleration, but rather tend to show a higher acceleration early in their propagation. The angular width of CMEs was also found to change as they propagate, normally increasing with height from the Sun. However these results were derived from plane-of-sky measurements with no correction for how the true CME geometry and direction affect the kinematics and morphology observed.

With the advent of the unique dual perspectives of the STEREO spacecraft, the multiscale methods were extended to an elliptical tie-pointing technique in order reconstruct the front of a CME in three-dimensions. Applying this technique to the Earth-directed CME of 12 December 2008 allowed an accurate determination of its true kinematics and morphology, and the CME was found to undergo early acceleration, non-radial motion, angular width expansion, and aerodynamic drag in the solar wind as it propagated towards Earth. This study and its conclusions are of vital importance to the fields of space weather monitoring and forecasting.
\newline
\newline
\newline
\newline
\newline
\newline
*This is the online version of this thesis which has reduced quality images due to file size concerns.

%\textbf{\underline{Keywords:}}

%Sun: activity -- Sun: coronal mass ejections (CMEs) -- Techniques: image processing -- Methods: data analysis

\end{abstracts}
%\end{abstractslong}

% ---------------------------------------------------------------------- 

% Thesis Dedictation ---------------------------------------------------

\begin{dedication} %this creates the heading for the dedication page

For My Parents.

\end{dedication}

% ----------------------------------------------------------------------
% Thesis Acknowledgements ------------------------------------------------

%\begin{acknowledgementslong} %uncommenting this line, gives a different acknowledgements heading
\begin{acknowledgements}      %this creates the heading for the acknowlegments

I am hugely grateful to my supervisor Peter Gallagher. You have been a constant source of inspiration, and it has been an absolute pleasure to work with you.

A big thank you goes to Alex Young for your guidance and support during my time at Goddard. Working in NASA was everything I could have hoped it would be. Thank you also to James McAteer, Jack Ireland, Angelos Vourlidas, everyone in the SOHO and STEREO teams and extended Goddard network. Your guidance and helpful discussion have been invaluable.

A collective thank you goes to Shane Maloney, James McAteer, Jose Refojo and Peter Gallagher for the huge amount of time and effort you contributed to achieving the final stages of work in this thesis.
 
A massive thank you goes to all the staff and students in the Astrophysics Research Group in Trinity College, including those who have since left. You made my time here as enjoyable as possible, and I'm sorry it has to come to an end.

And finally, a heartfelt thank you to my friends and family. You mean more to me than you can ever know.

\end{acknowledgements}
%\end{acknowledgmentslong}

% ------------------------------------------------------------------------

%: ----------------------- contents ------------------------

\setcounter{secnumdepth}{3} % organisational level that receives a numbers
\setcounter{tocdepth}{3}    % print table of contents for level 3
\tableofcontents            % print the table of contents
% levels are: 0 - chapter, 1 - section, 2 - subsection, 3 - subsection

%: ----------------------- list of figures/tables ------------------------

\chapter{List of Publications}% \& Presentations}
%\label{chapter:publications}

\textbf{\Large{Refereed}}

\begin{enumerate}

\item \textbf{Byrne, J. P.}, Maloney, S. A., McAteer, R. T. J., Refojo, J. M. \& Gallagher, P. T.\\
``Propagation of an Earth-directed coronal mass ejection in three-dimensions"\\
{\it Nature Communications}, {\bf 1}:74 doi:\,10.1038/ncomms1077 (2010).
 
 \item Gallagher, P. T., Young, C. A., \textbf{Byrne, J. P.} \& McAteer, R. T. J.\\
``Coronal mass ejection detection using wavelets, curvelets and ridgelets: Applications for space weather monitoring"\\
{\it Advances in Space Research}, doi:\,10.1016/j.asr.2010.03.028 (2010).

\item Mierla, M., Inhester, B., Antunes, A., Boursier, Y., \textbf{Byrne, J. P.}, Colaninno, R., Davila, J., de Koning, C.A., Gallagher, P.T., Gissot, S., Howard, R.A., Howard, T.A., Kramar, M., Lamy, P., Liewer, P.C., Maloney, S., Marqu{\'e}, C., McAteer, R.T.J., Moran, T., Rodriguez, L., Srivastava, N., St. Cyr, O.C., Stenborg, G., Temmer, M., Thernisien, A., Vourlidas, A., West, M.J., Wood, B.E. \& Zhukov, A.N.\\
``On the 3-D reconstruction of coronal mass ejections using coronagraph data"\\
{\it Annales Geophysicae}, {\bf 28,} 203--215 (2010).

\item \textbf{Byrne, J. P.}, Gallagher, P. T., McAteer, R. T. J. \& Young, C. A.\\
 ``The kinematics of coronal mass ejections using multiscale methods''\\
 \aap, {\bf 495,} 325--334 (2009).
 
 \item P{\'e}rez-Su{\'a}rez, D., Higgins, P. A., Bloomfield, D. S., McAteer, R. T. J., Krista, L. D., \textbf{Byrne, J. P.} \& Gallagher, P. T.\\
``Automated Solar Feature Detection for Space Weather Applications"\\
in Qahwaji, R., Green, R. \& Hines, E. [Eds] {\it Applied Signal and Image Processing: Multidisciplinary Advancements}, IGI, in press.
 
\end{enumerate}

\vspace{10mm}

\textbf{\Large{Contributed}}

\begin{enumerate}

\item \textbf{Byrne, J. P.}, Young, C. A., Gallagher, P. T. \& McAteer, R. T. J.\\
``Multiscale Characterization of Eruptive Events" \\
In S. A. Matthews, J. M. Davis \& L. K. Harra, eds., {\it First Results from Hinode}, vol. 397 of {\it Astronomical Society of the Pacific Conference Series}, 162--163 (2008). 

\end{enumerate}

\listoffigures	% print list of figures

\listoftables  % print list of tables

%: ----------------------- glossary ------------------------

% Tie in external source file for definitions: /0_frontmatter/glossary.tex
% Glossary entries can also be defined in the main text. See glossary.tex
%\include{0_frontmatter/glossary} 

%\begin{multicols}{2} % \begin{multicols}{#columns}[header text][space]
%\begin{footnotesize} % scriptsize(7) < footnotesize(8) < small (9) < normal (10)

%\printnomenclature[1.5cm] % [] = distance between entry and description
%\label{nom} % target name for links to glossary

%\end{footnotesize}
%\end{multicols}

%: --------------------------------------------------------------
%:                  MAIN DOCUMENT SECTION
% --------------------------------------------------------------

% the main text starts here with the introduction, 1st chapter,...
\mainmatter

%\renewcommand{\chaptername}{} % uncomment to print only "1" not "Chapter 1"

%: ----------------------- subdocuments ------------------------

% Parts of the thesis are included below. Rename the files as required.
% But take care that the paths match. You can also change the order of appearance by moving the include commands.
\doublespacing
\pagestyle{fancy}

\chapter{Introduction}
\markright{Introduction}
\label{chapter:intro}

% So max of 10 pages
% Introduction
% Sun in general

The Sun, as provider of light and heat to all life on Earth, has been a constant source of mystery and wonder to humankind. History recounts numerous tales inspired by our connection with the Sun: from its worship as a deity in the earliest civilisations, to the appreciation of its seasonal influence marked by structures like Newgrange, and the eventual observance of its complex behaviour with the development of telescopes and scientific intrigue. As our nearest star, astronomers have increasingly taken interest in the complexities of the Sun, and now in the modern age of space exploration numerous observatories have been built specifically to monitor solar activity and further our understanding of its dynamic behaviour.

%The main theory on the origin of the Sun is by the collapse of a molecular cloud in the interstellar medium. While other theories of star formation exist (e.g., the coalescence of two or more lower mass stars) they all depend on the gravitational collapse of matter. In a molecular cloud or stellar nursery the gravitational field of the mass exceeds the gas pressure and by the process of accretion the mass condenses to form a ball of plasma, conserving angular momentum which leads to rotation of the protostar. For a solar-mass star the formation can take about 100,000 years but various different mass stars exist through the universe, our Sun being a typical low to medium mass star. The governing description of their lifetimes is conveyed with the Hertzsprung-Russell diagram which relates absolute magnitude and effective temperature of stars as they evolve. Our Sun is thought to be about 4.57 billion years old. Its fate is unclear but it does not have enough mass to explode as a supernova. It is currently in a state of virial equilibrium with forces of expansion and contraction balancing out: %($\nabla P  = - \rho g$).
%\begin{equation}
%\nabla P \; = \; - \rho g
%\int^{R_\odot}_0 3kTdN = \int^{M_\odot}_0 \frac{GM_\odot(r)}{r}dM_\odot(r)
%\end{equation}
%When this equilibrium is lost, it may enter a red giant phase as it approaches 10 billion years of age, expanding out past 1 AU before throwing off its outer layers to leave a hot stellar core which slowly cools and fades over billions of years as a white dwarf.

\newpage
\section{The Solar Interior}

The Sun is a G2V main sequence star of luminosity $L_{\odot}=3.85\times10^{26}$~W, mass $M_{\odot}=1.99\times10^{30}$~kg and radius $R_{\odot}=6.96\times10^{8}$~m \citep{2000itss.book.....P}. It was born from the gravitational collapse of a molecular cloud approximately 4.6$\times$10$^9$~years ago, is currently in a state of hydrostatic equilibrium ($\nabla P = - \rho g$), and predicted to enter a red giant phase in another $\sim$\,5 billion years before ending its life as a white dwarf \citep{1999phst.book.....P}. Since we cannot directly observe the interior of the Sun, its structure and evolution are fundamentally realised with the use of the `standard solar model'  \citep[SSM;][]{1989neas.book.....B}, which is a mathematical treatment of stellar structure described by several differential equations derived from basic physical principles. The SSM is constrained by the well-determined boundary conditions of the Sun's luminosity, radius, age and composition, and thus provides a basis for understanding the mechanisms of energy transport in the solar interior. It assumes hydrostatic equilibrium, with energy generated by nuclear fusion, although small effects of contraction or expansion are included, and any abundance changes are caused solely by the nuclear reactions. The SSM is the end product of an iterative process that converges on an optimum description of the internal energy generation and transport, and overall evolution of the Sun.%Thermal and gravitational diffusion are estimated to be small over the lifetime of the Sun and are thus neglected.
\begin{figure}[!p]
\centerline{\includegraphics[width = \linewidth]{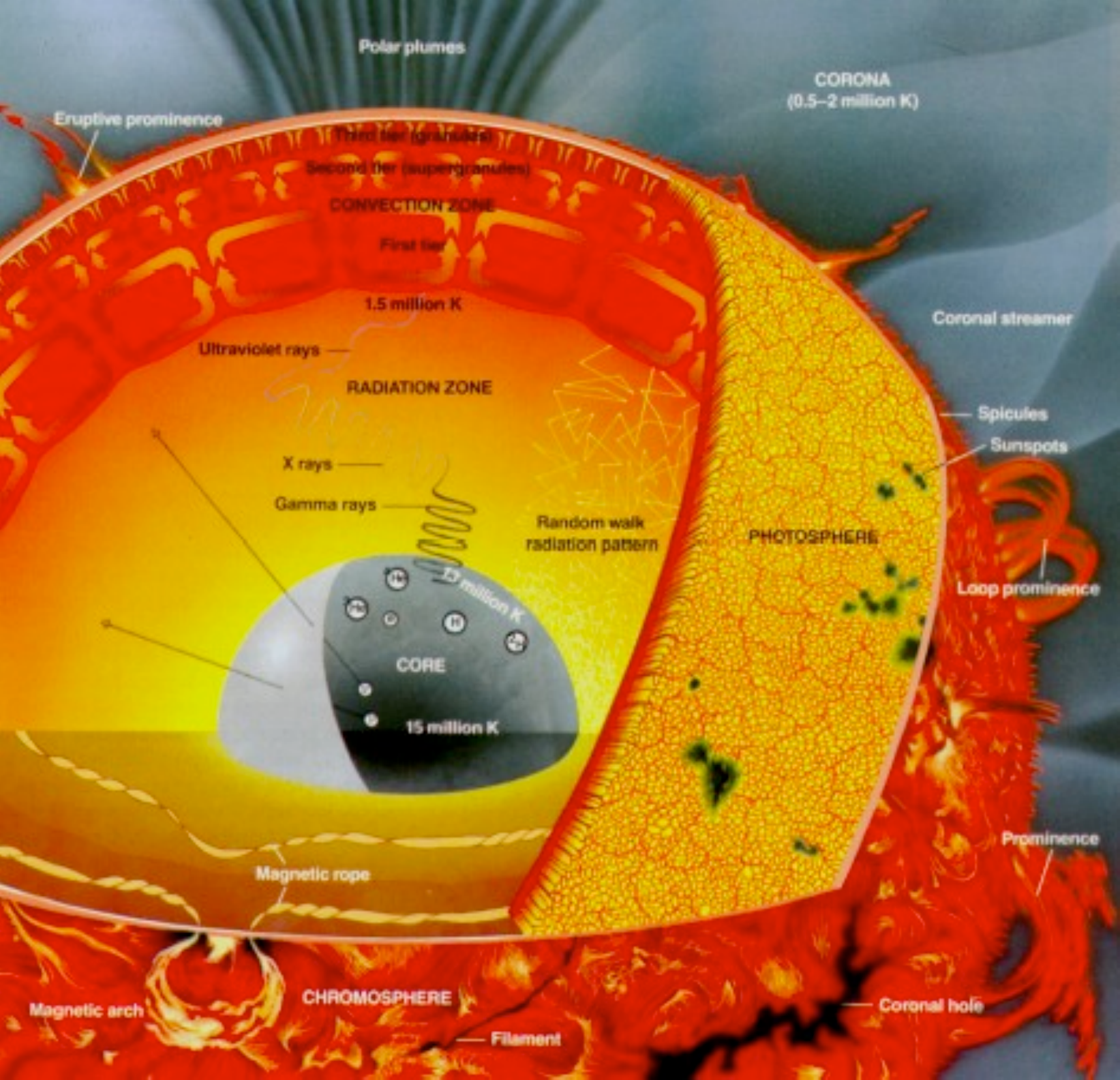}}
\caption{Illustration of the structure of the Sun. The core is the source of energy, where fusion heats the plasma to $\sim$\,15~MK. Energy is transported from the core by radiative processes in the radiation zone. The convection zone is heated from the base at the tachocline, allowing convective currents to flow to the photosphere. Locations of strong magnetic fields inhibit convection and appear as dark sunspots on the photosphere. These strong magnetic fields extend into the upper atmosphere of the Sun, responsible for coronal loops, prominences and streamers. 
\newline
\emph{Image credit: eu.spaceref.com}.}
\label{sun_core}
\end{figure}

\begin{figure}[!t]
\centerline{\includegraphics[scale=0.6]{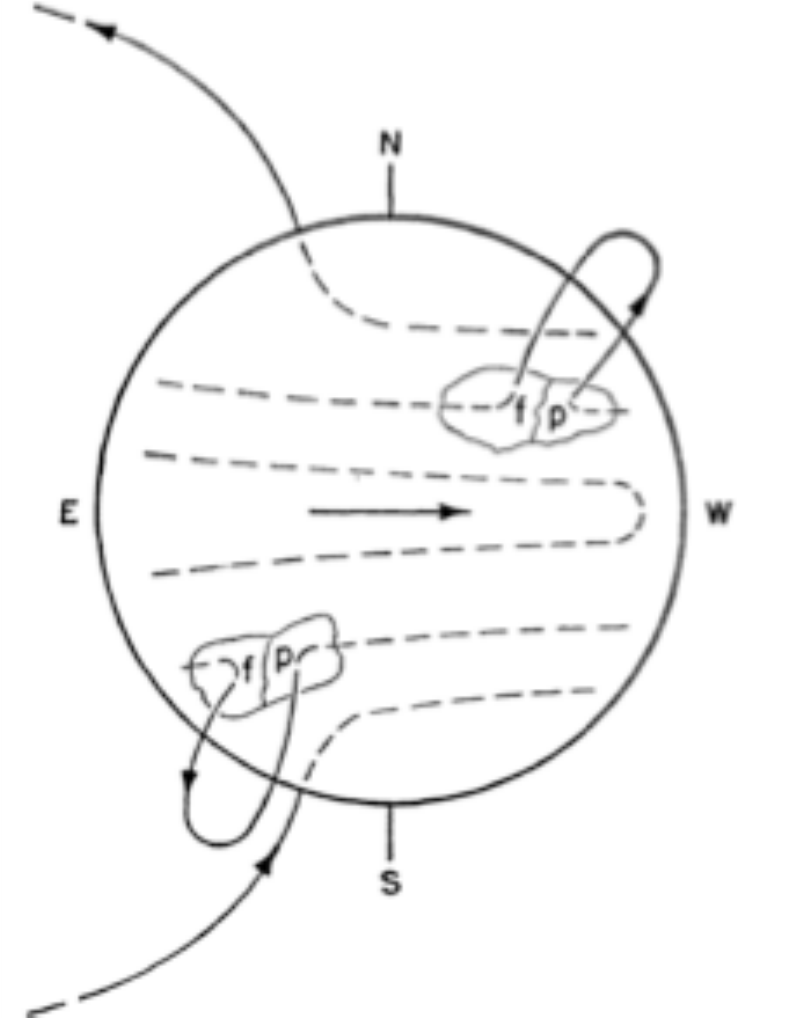}}
\caption{Illustration of the $\alpha \Omega$ effect of winding-up magnetic field due to the differential rotation of the Sun, reproduced from \citet{1961ApJ...133..572B}. Sunspots visible on the disk are as a result of protruding field with positive $p$ and negative $f$ polarity as shown.}
\label{babcock}
\end{figure}

\newpage
\indent The fundamental energy process driving the Sun is nuclear fusion in the core, through the proton-proton chain at temperatures of $\sim$\,15~MK:
\begin{eqnarray}
\rm{_1^1H} + \rm{_1^1H}	& \to & \rm{_1^2H} + \rm{e^+} + \nu_e \\ %+ 1.44~MeV\\
\rm{_1^2H} + \rm{_1^1H} & \to & \rm{_2^3He} + \gamma \\ %+ 5.49~MeV\\
\rm{_2^3He} + \rm{_2^3He} & \to & \rm{_2^4He} + 2\, \rm{_1^1H} %+ 12.86~MeV, 
\end{eqnarray}
where $\rm{_1^1H}$ is a proton, $\rm{_1^2H}$ is the deuteron isotope of hydrogen, $\rm{_2^3He}$ and $\rm{_2^4He}$ are helium isotopes with 1 and 2 neutrons respectively, $\rm{e^+}$ a positron, $\nu_e$ an electron-neutrino and $\gamma$ a gamma ray. The resulting energy release for one complete reaction chain is approximately $4.3\times10^{-12}\rm~J$ \citep{1995gusu.book.....P}. The core extends from the centre out to $\sim$\,0.25~R$_{\odot}$, followed by the radiation zone out to $\sim$\,0.75~R$_{\odot}$, then the convection zone out to the solar surface at 1~R$_{\odot}$ (Figure~\ref{sun_core}). The temperature across the radiation zone drops to $\sim$\,5~MK with radiation being the most efficient method of energy transport. This radiation field is closely approximated by a black body, for which the spectral radiance is described by the Planck equation:
\begin{equation}
B_{\lambda}(T) = \frac{2hc^{2}\mu^{2}}{\lambda^{5}\left[\exp\left(hc/\lambda kT\right)-1\right]}
\end{equation}
where $B_{\lambda}(T)$ is the intensity of radiation per unit wavelength interval (at temperature $T$), $h$ is the Plank constant, $c$ is the speed of light, $\mu$ is the refractive index of the medium, and $k$ is the Boltzmann constant. By Wien's law $\lambda_{max}T=2.8979\times10^{-3}$~m~K we determine that the radiation is in the form of X-rays, and these high-energy photons undergo random walks in the plasma, escaping into the convection zone on time-scales of 10$^6$~years. The optically thick convection zone then transports energy by fluid motion across the temperature gradient between its base ($\sim$\,1\,--\,2~MK) and the solar surface ($\sim$\,5,800~K). Plasma elements move sufficiently rapidly for the energy interchange with their surroundings to be negligible, i.e., they change adiabatically. A useful measure of when convection is likely to occur is given by the Schwarzschild criterion:
\begin{equation}
\frac{d\log T}{d\log P}\Biggl\vert_{star} > \frac{\gamma-1}{\gamma}
 \end{equation}
where $\gamma=C_P/C_V$ is the ratio of specific heats, equal to $5/3$ for a perfect monatomic gas. Essentially convection occurs once the absolute magnitude of the radiative gradient becomes larger than the absolute magnitude of the adiabatic gradient, so that rising elements of plasma remain buoyant and move towards the surface before they can lose heat to their surroundings. The rising and falling parcels of plasma create the granulation effects observed on the surface, with granules ranging in size from hundreds to thousands of kilometres and dissipating over tens of minutes. (Details of the above radiative and convective processes are found in, e.g., \citet{1987snim.book.....K, 1998assu.book.....Z}).
\newline
\indent Between the radiation and convection zones is a relatively thin interface called the tachocline, where the solid body rotation of the radiative interior meets the differentially rotating outer convection zone. It thus has a very large shear profile which could account for the formation of large scale magnetic fields in the solar dynamo. The magnetic field of the Sun has an overall dipolar configuration, with opposite polarities dominant at each pole. The differential rotation of the Sun's convection zone causes a large-scale winding up of the magnetic field, named the $\Omega$-effect, while the effects of the coriolis force and smaller scale motions of the plasma can give twist and writhe to the field, named the $\alpha$-effect (Figure~\ref{babcock}). Buoyancy effects cause the magnetic field to rise up through the convection zone and protrude through the surface of the Sun, observed as sunspots on-disk marking the footpoints of over-arching concentrations of magnetic flux extending up through the solar atmosphere. In a given hemisphere one magnetic polarity leads the sunspot group and the opposite follows, while in the opposite hemisphere the situation is reversed (Hale's law). The tilt angle between leading and trailing sunspots has an average value of 5.6$^{\circ}$ relative to the E-W line (Joy's law). Furthermore, sunspots are observed to migrate from high latitudes towards the equator over an 11 year cycle due to the continual build-up of field by the $\alpha\Omega$-effect (Sp\"{o}rer's law). These combined effects lead to an increase of oppositely oriented poloidal field at each of the poles, neutralising the field there and resulting in the magnetic dipole flipping every 11 years. This 22 year periodicity is known as the solar cycle, and gives rise to periods of increased and decreased solar activity manifested by the frequency of phenomena such as active regions, flares and transients in the solar atmosphere \citep{2000ssma.book.....S}.

\newpage
\section{The Solar Atmosphere}

\begin{figure}[!t]
\centerline{\includegraphics[scale=0.8, trim=0 410 0 80]{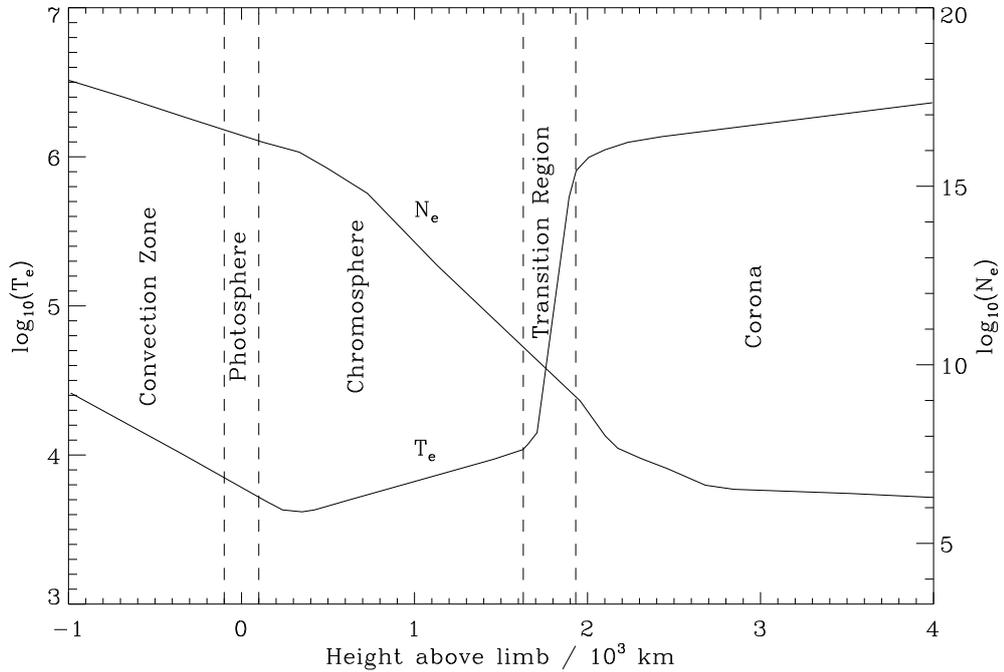}}
\caption{A 1D static model of electron density $N_e~[\rm cm^{-3}]$ and temperature $T_e~[\rm K]$ profiles in the solar atmosphere, reproduced from \citet{1982aacp....1..345G}. In the chromosphere, the plasma is only partially ionised. The plasma becomes fully ionized at the sharp transition from chromospheric to coronal temperatures.}
\label{Profile}
\end{figure}

\begin{figure}[!t]
\centerline{\includegraphics[scale=0.8, trim=0 220 20 240]{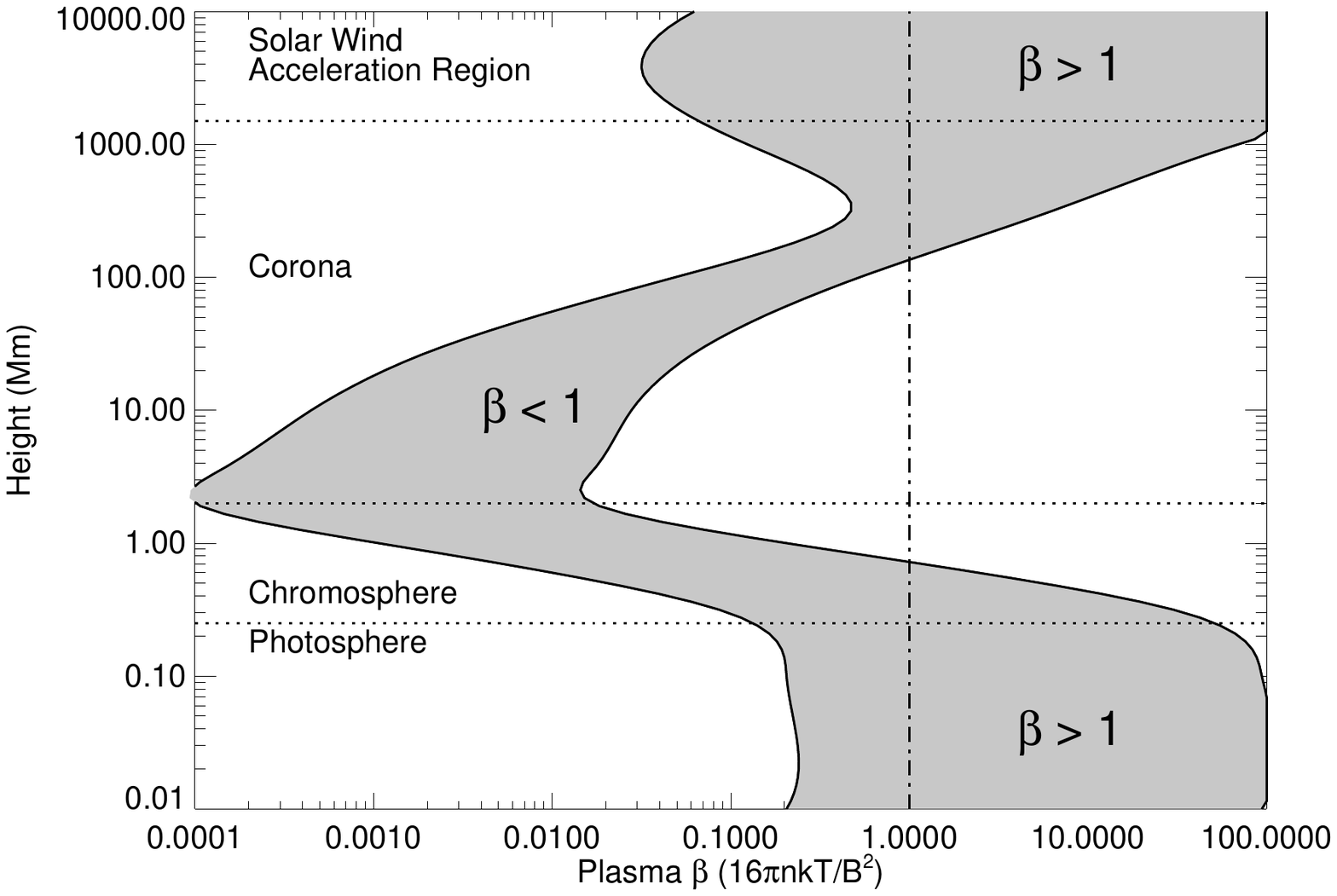}}
\caption{Plasma $\beta$ as a function of height for a regime of magnetic field strengths between $\sim$\,100 and $\sim$\,2,500~G, reproduced from \citet{2001SoPh..203...71G}. The dotted lines segregate the layers of photosphere ($\beta>1$), chromosphere and corona ($\beta<1$), and the solar wind ($\beta>1$).}
\label{plasmabeta}
\end{figure}

The Sun's atmosphere is composed of all regions extending from the photosphere out into the heliosphere. It may be separated into distinct regimes dependent on the density and temperature profiles. These are plotted in Figure~\ref{Profile} for a 1D static model of the solar atmosphere. The layers are generally stratified into photosphere, chromosphere, transition region, and corona; having a decreasing density with increasing height from the photosphere, but from the chromosphere up the temperature increases with a dramatic jump in the transition region giving rise to the so-called `coronal heating problem'. However this stratification is a simplified view and the solar atmosphere in reality is an inhomogenous mix of photospheric, chromospheric and coronal zones due to complex dynamic processes such as heated upflows, cooling downflows, intermittent heating, nonthermal electron beams, field line motions and reconnections, emission from hot plasma, absorption and scattering in cool plasma, acoustic waves, and shocks \citep{2005psci.book.....A}. The interplay between the magnetic and gas pressure represents an important determining factor in the behaviour of structures throughout the solar atmosphere, quantified by the plasma-$\beta$ term:
\begin{equation}
\beta \;=\; \frac{p_{gas}}{p_{mag}} \;=\; \frac{nkT}{\left( B^2 / 8\pi \right)}
\end{equation}
This is illustrated in Figure~\ref{plasmabeta} for the different layers of the atmosphere. At the photospheric level the plasma-$\beta$ is large, and plasma motions dominate over the magnetic field forces. Through the chromosphere and corona the plasma-$\beta$ decreases to low values where the magnetic field structures are seen to suspend plasma in loops and filaments. Finally in the extended upper atmosphere the plasma-$\beta$ rises again, and the magnetic field is advected out with the solar wind plasma flow to ultimately form the Parker spiral.

\subsection{Photosphere}

The surface of the Sun is the photosphere defined as the point where the optical depth equals 2/3 for wavelengths of visible light, centred on 5,000~\AA~($\tau_{5000}\sim2/3$ for $I/I_0=e^{-\tau}$). The spectrum of light emitted has a profile like that of a black body with an effective temperature of 5,800~K interspersed with the Fraunhofer absorption lines due to the tenuous layers above the photosphere. It has a particle density of $\sim$\,10$^{23}$~m$^{-3}$ and a thickness of less than 500~km. Cooler regions called sunspots have temperatures of 4,000\,--\,4,500~K and are due to intense magnetic field activity that acts to suppress convective plasma motion. Granulation of the photosphere is observed as the manifestation of plasma motion in the convection zone below, with typical cell sizes on the order of 1,000~km in diameter. They occur when hot plasma rises to the surface and is transported along it to the granule edges, which appear darker as the plasma cools and descends. The gas pressure dominates the magnetic pressure ($\beta>1$), and the magnetic field is effectively coupled to the plasma motion which sweeps it into the inter-granular network.

\subsection{Chromosphere}

Above the photosphere lies the chromosphere where the temperature initially drops to a minimum of $\sim$\,4,500~K before increasing to $\sim$\,20,000~K with increasing height from the Sun (Figure~\ref{Profile}). It is approximately 2,000~km thick and the density falls by a factor of almost a million from bottom to top, so the magnetic field begins to dominate the chromospheric structure ($\beta < 1$). The second law of thermodynamics does not permit heating of the chromosphere with the thermal energy of the cooler photosphere below. \citet{1948ZA.....25..161B}, \citet{1948ApJ...107....1S} and \citet{1949AnAp...12..203S} put forward ideas on the acoustic wave heating of the chromosphere as a result of the convective plasma motions in the photosphere and convection zone beneath. Referred to as the BSS model, the hypothesis is that acoustic waves transport energy upward with little dissipation once the velocity is below the sound speed. As the density drops and the velocity reaches the sound speed, the waves steepen into shocks and rapidly dissipate the energy, consequently heating the material \citep{1998assu.book.....Z}. However, acoustic wave heating does not apply in regions of strong magnetic field where motions, and therefore heating, are suppressed. This has led to work on Alfv{\'e}n wave heating theories, first introduced by \citet{1961ApJ...134..347O}. An Alfv{\'e}n wave is a type of magnetohydrodynamic wave that propagates in the direction of the magnetic field with the magnetic tension providing the restoring force and the ion mass density providing the inertia. In Alfv{\'e}n wave heating theories the magnetic field itself is thus responsible for depositing energy from the subsurface into the chromosphere and above. These theories better sit with observations of vigourous heating above plages and emerging flux regions, since they imply the amount of heating is proportional to the rate of magnetic change.
\newline
\indent While the brightness of the photosphere overwhelms that of the chromosphere in the optical continuum, the hotter chromospheric temperatures lead to the hydrogen being ionised, resulting in strong H$\alpha$ emission. Filaments are observed as dark channels on-disk in H$\alpha$ images (called prominences when seen on the limb). Numerous plasma columns called spicules are also observed on the limb, that typically reach heights of $\sim$\,3,000\,--\,10,000~km above the Sun's surface and are very short-lived (rising and falling over $\sim$\,5\,--\,15~minutes).
\newline
\indent Between the chromosphere and corona lies the transition region where the temperature jumps rapidly to over 1~MK. It is only about 100~km thick and it marks the point where magnetic forces dominate completely over gravity, gas pressure and fluid motion ($\beta \ll 1$). The extreme temperatures result in prominent UV and EUV emission from carbon, oxygen and silicon ions \citep{1992str..book.....M}.

\subsection{Corona}

\begin{figure}[!p]
\centerline{\includegraphics[width=\linewidth]{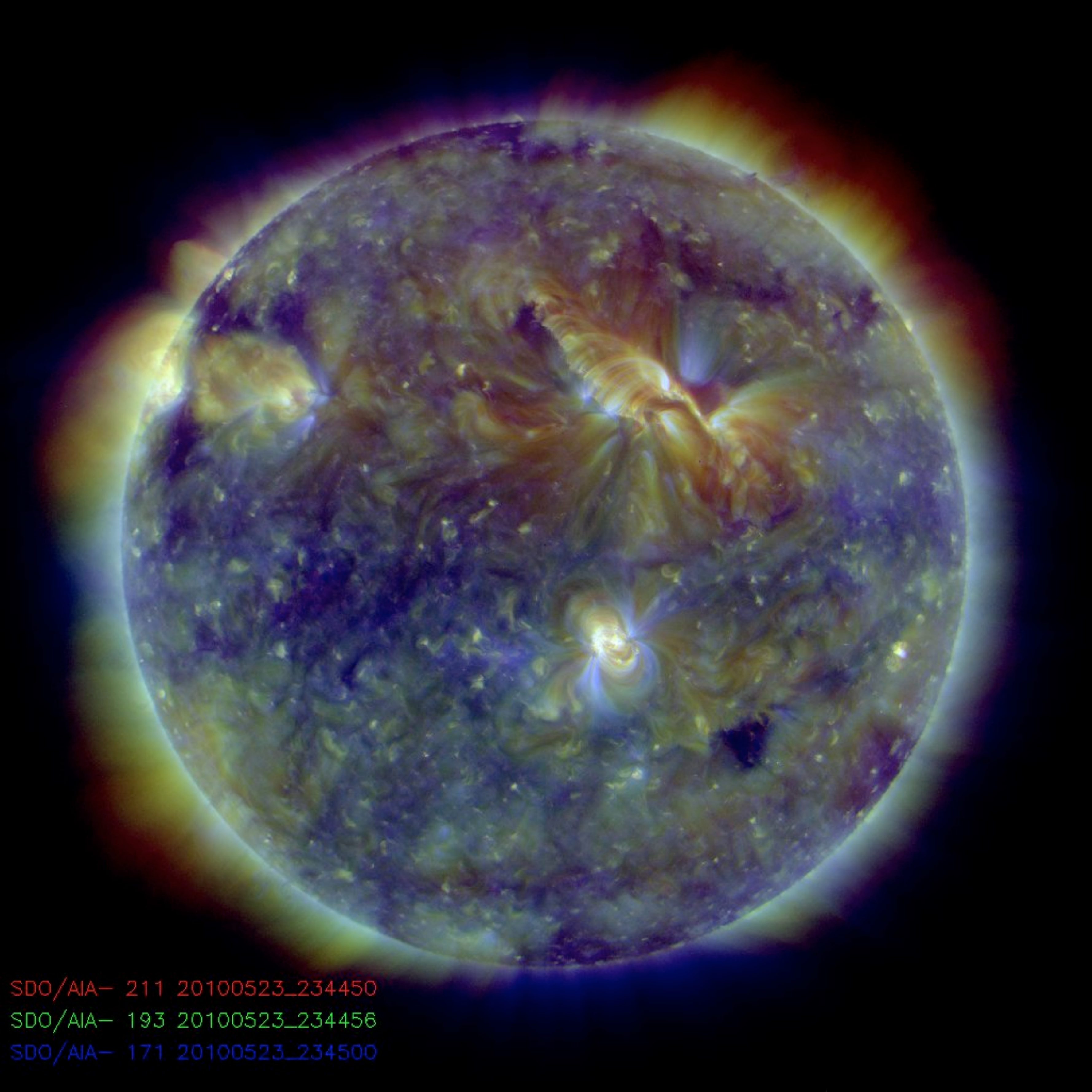}}
\caption{Composite EUV image of the low solar corona recorded by the Atmospheric Imaging Assembly (AIA) onboard the Solar Dynamics Observatory (SDO) on 23 May 2010. 
\newline
{\it Image credit: http://sdowww.lmsal.com/}}
\label{sdo_aia}
\end{figure}

The outermost part of the solar atmosphere is the corona, with electron densities ranging from $\sim$\,(1\,--\,2)\,$\times$\,10$^{14}$~m$^{-3}$ at its base height of $\sim$\,2,500~km above the photosphere, to $\lesssim$\,10$^{12}$~m$^{-3}$ for heights $\gtrsim$\,1~R$_{\odot}$ above the photosphere \citep{2005psci.book.....A}. The density varies across coronal holes which can have a base density of $\sim$\,(0.5\,--\,1)\,$\times$\,10$^{14}$~m$^{-3}$, or across streamer regions with higher densities of $\sim$\,(3\,--\,5)\,$\times$\,10$^{14}$~m$^{-3}$. Active regions that suspend and confine plasma in strong over-arching magnetic fields usually have the highest coronal densities of $\sim$\,2\,$\times$\,10$^{14}$\,--\,2\,$\times$\,10$^{15}$~m$^{-3}$. The temperature of the corona is generally $\gtrsim$\,1~MK, as indicated by emission from highly ionised iron lines, for example, which again appears to contradict the second law of thermodynamics given the much cooler layers of the chromosphere and photosphere below (the `coronal heating problem'). Its temperature structure is far from homogeneous, revealed in images such as that of Figure~\ref{sdo_aia} from the Solar Dynamics Observatory (SDO). Loop structures are observed at temperatures of 2\,--\,6~MK across regions of increased magnetic field density (such as above active regions/sunspots), and closed field regions are observed at temperatures of 1\,--\,2~MK across the quiet Sun, while open field regions of coronal holes have temperatures $\lesssim$\,1~MK. These high temperatures lead to EUV and X-ray emission due to ionisation and recombination processes from the interactions between photons, electrons, atoms and ions. The continuum and line emission of the solar corona result from the contributions of bound-bound transitions (excitations and de-excitations), bound-free absorption (photoionisation), free-free absorption (and its inverse bremsstrahlung emission), and electron scattering. It is the latter process, called Thomson scattering, that produces the white-light corona as visible during a solar eclipse or with the use of a coronagraph to occult the solar disk, which is six orders of magnitude brighter in optical wavelengths.
\newline
\indent The corona we observe comprises several parts: 
\begin{itemize}
\item The K-corona has a strongly polarised continuous emission spectrum due to Thomson scattering of photospheric light by the free electrons of the coronal gas, and it dominates within the first few R$_{\odot}$. It produces a polarised white-light continuum without the Fraunhofer lines which are broadened by Doppler shifts due to the fast electron motions at such high temperatures. The intensity of the K-corona gives the coronal electron density \citep{1991sia..book.1044K}.
\item The F-corona is due to scattering of sunlight by interplanetary dust particles, and contains the Fraunhofer lines. It is roughly equal in intensity to the K-corona at $\sim$\,4~R$_{\odot}$, and dominates at greater distances. 
\item The E-corona is due to emission from highly ionized coronal atoms such as iron and calcium.
\item The T-corona is caused by thermal (infrared) emission of the interplanetary dust. It is an unpolarised continuum, insignificant in the visible part of the spectrum.
\end{itemize}
In contrast to the chromosphere, solar interior, and indeed the heliosphere, the magnetic pressure in the corona dominates over the gas pressure and so governs the coronal plasma dynamics ($\beta<1$). The coronal structure we observe is thus shaped by the magnetic fields of the Sun, resulting in extended polar regions where there is mainly open magnetic field, and `helmet-streamers' spanning the equatorial latitudes where, except for coronal holes, the field is mostly closed. Since these features are magnetically governed, the shape of the corona varies greatly over the solar activity cycle: it appears rounder at solar maximum, when multiple streamers emerge at various latitudes distributed across the Sun; and it appears more elliptical at solar minimum, when only a few streamers are present, lying closer to the equator.
\newline
\indent Following \citet{1957SCoA....2....1C} the description of a static corona leads to an unreasonable pressure value at large distances from the Sun. This is outlined below, beginning with the assumption that the corona is in hydrostatic equilibrium:
\begin{equation}
\frac{dP}{dr}\; = \; - \rho \frac{GM_{\odot}}{r^2}
\end{equation}
The plasma density is $\rho=nm_p$, the pressure contribution from the protons and electrons is $P=2nk_BT$, and the coronal heat flux is $q=\kappa \nabla T$ with thermal conductivity $\kappa = \kappa_0 T^{5/2}$. In the absence of heat sources or sinks $\nabla \cdot q = 0$ so in a spherically symmetric system we can write:
\begin{equation}
\frac{1}{r^2} \frac{d}{dr} \left( r^2 \kappa_0 T^{5/2} \frac{dT}{dr} \right) \; = \; 0
\end{equation} 
Applying the boundary condition that the temperature tends to zero at large distances from the Sun, we obtain:
\begin{equation}
T \; = \; T_0 \left( \frac{r_0}{r} \right) ^{2/7}
\end{equation}
where $T_0=2$~MK is the temperature of the low corona at height $r_0=1.05$~R$_{\odot}$ from Sun centre. This would mean $T\approx4\times10^5$~K at Earth (1~AU\,$\approx$\,215~R$_{\odot}$), close to measured values. Rewriting in terms of pressure and integrating, results in:
\begin{equation}
P(r) \; = \; P_0 \exp \left( \frac{7}{5} \frac{GM_{\odot}m_p}{2kT_0r_0} \left[ \left( \frac{r_o}{r} \right)^{5/7} -1 \right] \right)
\end{equation}
which implies that as $r \rightarrow \infty$ the coronal pressure tends towards a finite constant value significantly larger than the pressure of the interstellar medium (ISM); $P \gg P_{ISM}$. This means the static coronal model is unphysical, and a dynamic model in which the material flows outward from the Sun must be considered, leading to a description of the solar wind.

\subsection{Solar Wind}

The solar wind is the constant out-stream of charged particles of plasma from the Sun's atmosphere due  to the persistent expansion of the solar corona. The wind consists mostly of electrons and protons at energies of $\sim$\,1~keV, observed in two regimes of propagation: the slow solar wind with speeds of $\sim$\,400~km~s$^{-1}$; and the fast solar wind with speeds of $\sim$\,800~km~s$^{-1}$, originating from regions of open magnetic field such as coronal holes. Thermal velocities of the particles are calculated at $\sim$\,260~km~s$^{-1}$ for coronal temperatures on the order of 3$\times$10$^6$~K, while the escape velocity in the Sun's gravitational field in the low corona can be $\sim$\,500~km~s$^{-1}$. The additional energy to accelerate the solar wind is imparted by the pressure gradient $P_{Sun} \gg P_{ISM}$ to attain the measured solar wind speeds \citep{1958ApJ...128..664P}. 
\begin{figure}[!t]
\centerline{\includegraphics{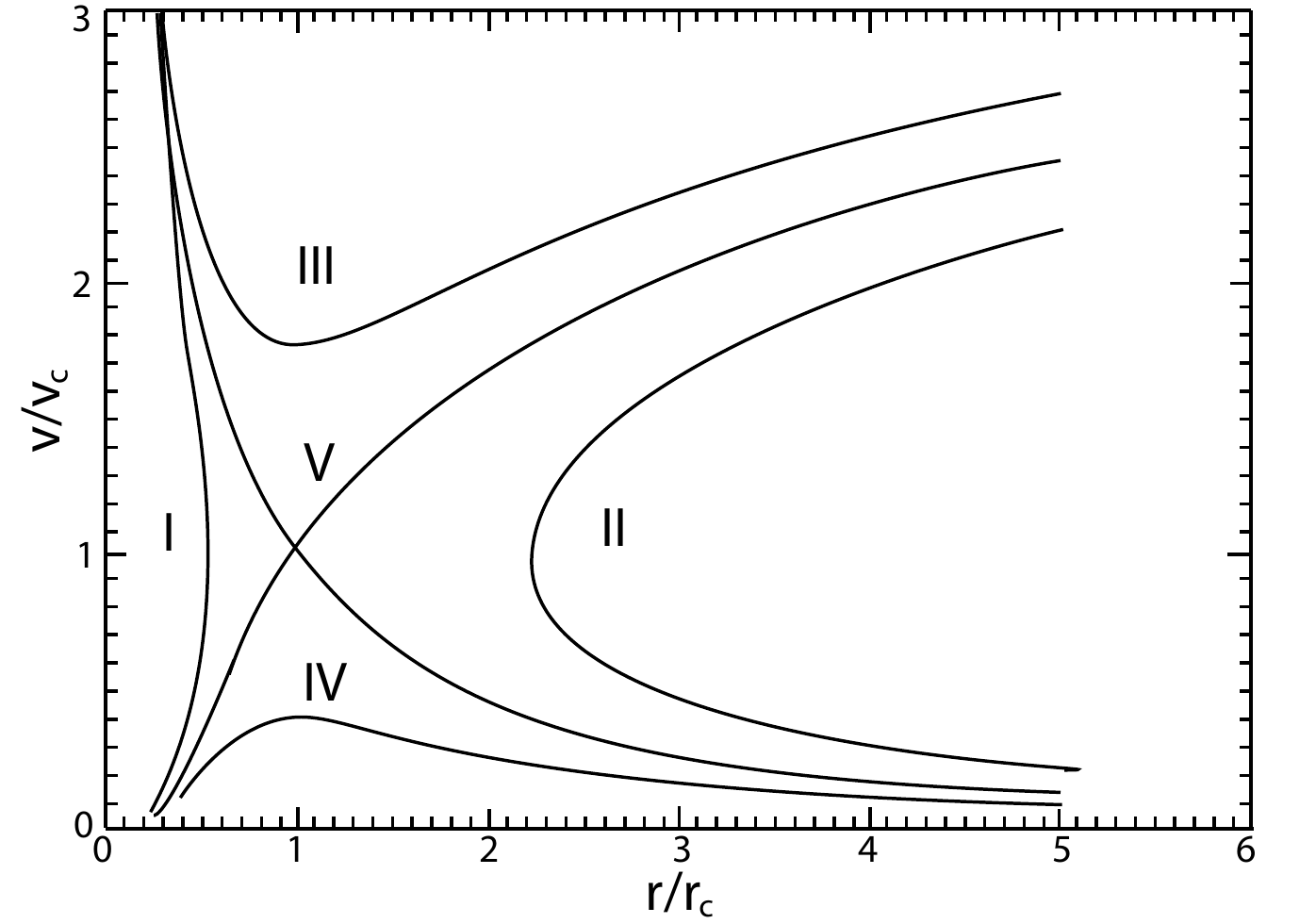}}
\caption{The five classes of Parker's solar wind solution for a steady, spherically symmetric, isothermal outflow.}
\label{parkers_solutions}
\end{figure}
The Parker model assumes the outflow is steady, spherically symmetric and isothermal. The momentum conservation equation of the corona takes the form:
\begin{equation}
\rho v \frac{dv}{dr} \; = \; - \frac{dP}{dr} - \rho \frac{GM_{\odot}}{r^2}
\label{eqn:momcons}
\end{equation}
Considering mass conservation $\dot{m} = 4 \pi r^2 \rho v = constant$, we obtain:
\begin{equation}
\frac{\partial}{\partial r} \left( r^2\rho v \right) \; = \; 0 \quad \Rightarrow \quad \frac{1}{\rho} \frac{\partial \rho}{\partial r} \; = \; - \frac{1}{v} \frac{\partial v}{\partial r} - \frac{2}{r}
\end{equation}
So for a perfect gas $P = R \rho T$ Equation~\ref{eqn:momcons} can be written:
\begin{equation}
%\rho v \frac{dv}{dr} + RT \left(- \frac{1}{v} \frac{\partial v}{\partial r} - \frac{2}{r} \right) +  \frac{GM_{\odot}}{r^2} \; = \; 0 \\
\quad \left( v- \frac{RT}{v} \right) \frac{\partial v}{\partial r} - \frac{2RT}{r} + \frac{GM_{\odot}}{r^2} \; = \; 0
\label{eqn:momprefin}
\end{equation}
A critical point occurs when $\partial_r v \rightarrow 0$ so we define:
\begin{equation}
r_c \; = \; \frac{GM_{\odot}}{2v_c^2} \quad \text{where} \quad v_c \; = \; \sqrt{RT}
\end{equation}
and rewrite Equation~\ref{eqn:momprefin} as:
\begin{equation}
\left( v^2 - v_c^2 \right) \frac{1}{v} \frac{\partial v}{\partial r} \; = \; 2 \frac{v_c^2}{r^2} \left( r-r_c \right)
\label{eqn:momfin}
\end{equation}
Integrating Equation~\ref{eqn:momfin} gives Parker's `solar wind solutions':
\begin{equation}
\left( \frac{v}{v_c} \right)^2 - \ln \left( \frac{v}{v_c} \right)^2 \; = \; 4 \ln \left( \frac{r}{r_c} \right) + 4 \frac{r_c}{r} + C
\end{equation}
where $C$ is a constant of integration, leading to five potential solutions as plotted in Figure~\ref{parkers_solutions}.
Solutions I and II are double-valued, with II being disconnected from the surface. Solution III is too large (supersonic) close to the Sun. Solution IV is called the `solar breeze' as it remains subsonic. Solution V is the standard solar wind solution, although the assumptions of radial expansion and isothermality are not completely true in reality, so it is only an approximate characterisation of the observed solar wind. Nonetheless it is sufficient to convey the dynamic expansion of the corona and ultimate supersonic regime of outflow, often described akin to a de Laval nozzle which is used to accelerate flows from subsonic to supersonic speeds \citep[as detailed in][for example]{2003ASSL..294.....G}. The presence of the solar wind was confirmed in 1959 by the Lunik I probe, and in 1962 by the Mariner II mission en route to Venus \citep{1962Sci...138.1095N}.
\begin{figure}[!t]
\centerline{\includegraphics[scale=0.3, clip=true, trim=0 10 0 0]{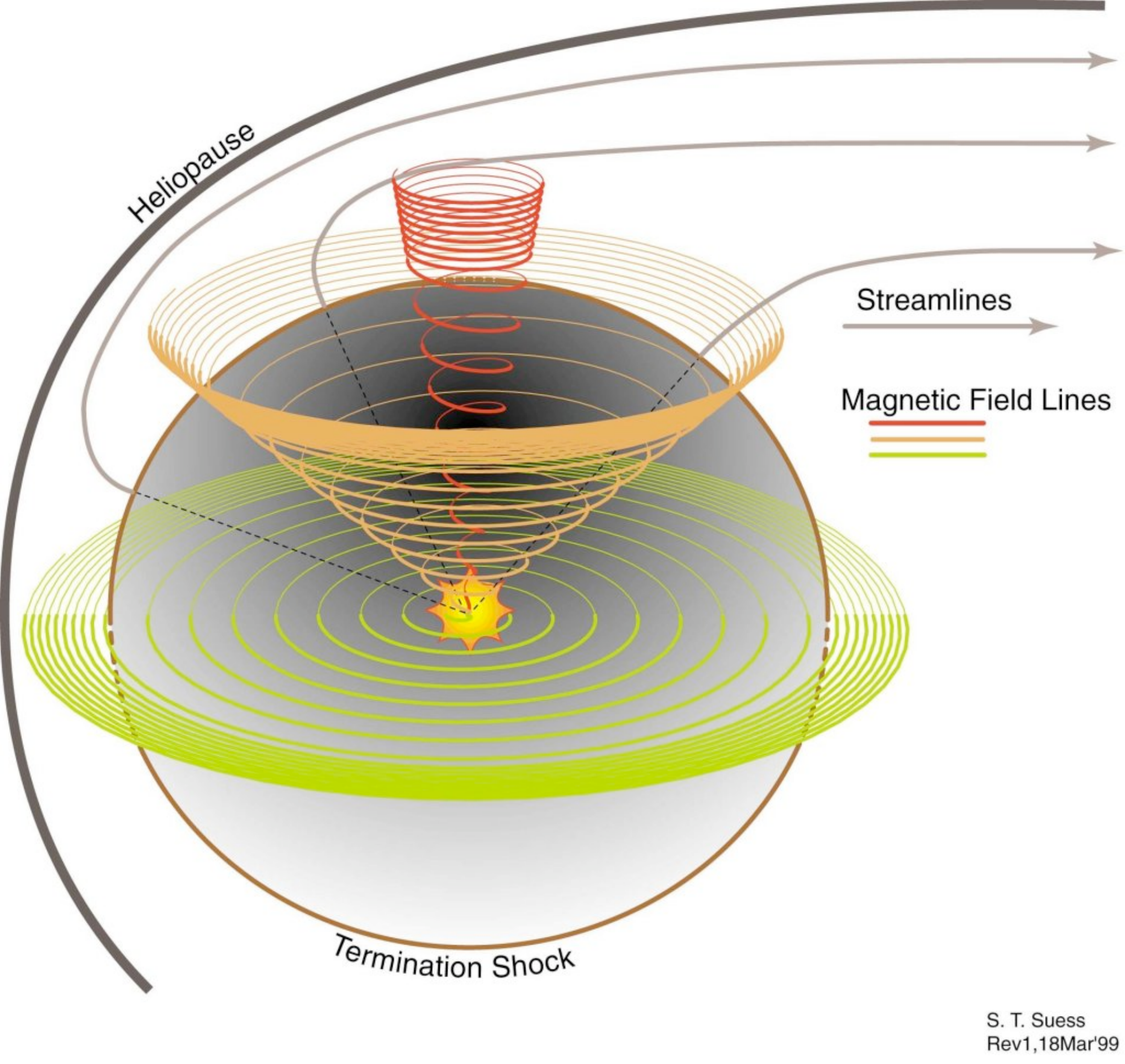}}
\caption{Schematic of the Parker Spiral in the heliosphere. The streamlines of the solar wind act to drag out the magnetic field lines of the Sun, which become wound up in an Archimedean spiral as a result of the Sun's rotation.
\newline
\emph{Image credit: Steve Suess, NASA/MSFC}.}
\label{parkerspiral}
\end{figure}
\newline
\indent Since the gas pressure dominates over the magnetic pressure in the solar atmosphere ($\beta>1$), the solar wind acts to drag out the magnetic field lines of the Sun which become wound up as a result of solar rotation to form the Parker Spiral (Figure~\ref{parkerspiral}). This is an Archimedean spiral drawn by the magnetic field lines as they are advected outward by the solar wind, described by the equation:
\begin{equation}
%\theta - \theta_0 \; = \; \Omega t \;=\; \Omega \left( \frac{r_0-r}{v} \right)
r-r_0 \; = \; \frac{v}{\Omega} \left( \theta - \theta_0 \right)
\end{equation}
where $\theta$ is the polar angle, $\Omega=2.7\times10^{-6}$~rad~s$^{-1}$ is the angular rotation rate of the Sun, $r$ is the distance, and $v$ is the solar wind speed \citep{1998assu.book.....Z, 2004pspi.book.....P}. The different speed streams can also lead to the formation of co-rotating interaction regions (CIRs) where the fast wind encounters the slow wind ahead of it in the Parker spiral, and can form shocks in the solar wind.
\newline
\indent The solar wind does not extend infinitely, but eventually terminates when it reaches the edge of the heliosphere. The point where the solar wind slows from supersonic to subsonic speeds is called the termination shock, the observation of which is reported in a series of papers discussing the data from Voyager II as it began to cross the shock in August 2007 \citep{2008Natur.454...63R, 2008Natur.454...67D, 2008Natur.454...75B}. Beyond the termination shock, the wind comes into pressure balance with the ISM to form the heliosheath, whose outer boundary is called the heliopause (Figure~\ref{parkerspiral}). In the heliosheath the continually slowing wind is compressed and becomes turbulent through its interaction with the ISM \citep{2009opher}. As the heliosphere moves through interstellar space, a bow shock is thought to form ahead of the heliopause as it encounters the ISM.

\subsection{Space Weather}

\begin{figure}[!t]
\centerline{\includegraphics[width=\linewidth]{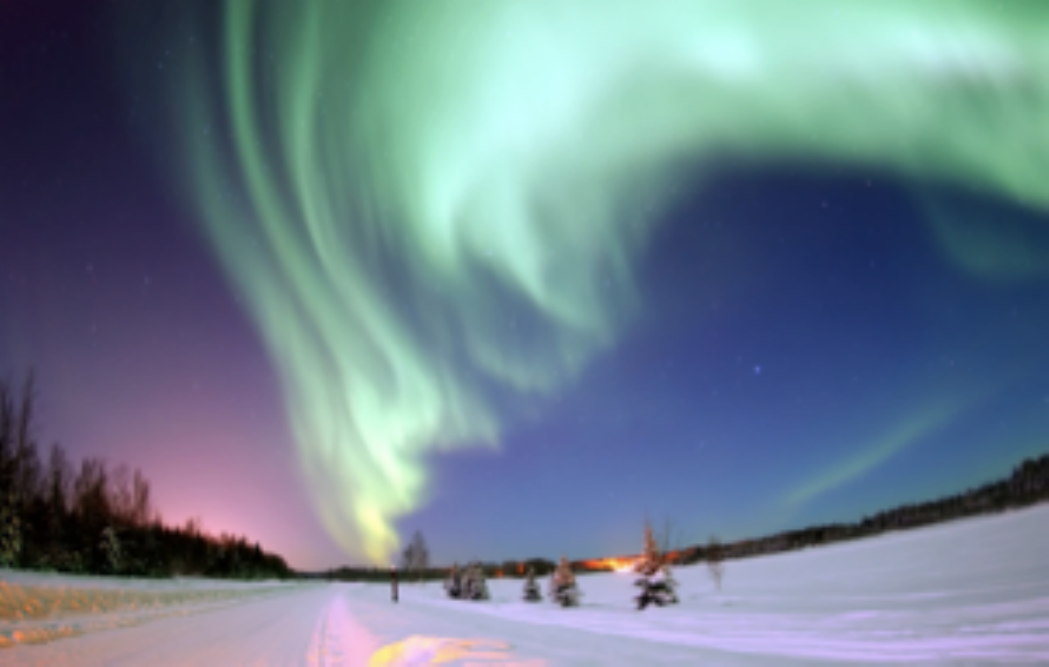}}
\caption{The Aurora Borealis, or Northern Lights, photographed above Bear Lake, Eielson Air Force Base, Alaska. Aurorae result from photon emissions of excited oxygen and nitrogen atoms in the Earth's upper atmosphere, as a result of geomagnetic storms and space weather. 
\newline
\emph{Image credit: Joshua Strang via Wikimedia Commons}.}
\label{aurora}
\end{figure}

Space weather is the name attributed to phenomena involving ambient plasma, magnetic field, radiation and other matter affecting the conditions in space as part of the vast field of heliophysics \citep{2010heliophysics}. It is predominantly due to the influences of the solar wind, flares and transients, and the interplanetary magnetic field (IMF). Earth's magnetosphere provides a natural shielding of the planet from the solar wind and space radiation, although large flares and coronal mass ejections (CMEs) that accelerate solar energetic particles (SEPs) can lead to geomagnetic storms at Earth which perturb the magnetosphere and cause increased electric currents in the ionosphere. Auroral sightings throughout history are an indicator of geomagnetic storm occurrences (e.g., Figure~\ref{aurora}), and by 1837 they were realised to be caused by electric currents in the upper atmosphere \citep{1837olmstead}. \citet{1852sabine} went on to show that there was a detailed correlation between the sunspot cycle and the frequency of auroral displays, implicating solar activity as the ultimate cause of the auroral phenomenon. A key event in the realisation of how strongly solar activity can influence us on Earth, was the Carrington-Hodgson flare that occurred on 1 September 1859 \citep{1859MNRAS..20...13C}, causing widespread sightings of aurorae down to latitudes as low as $\sim$\,18$^{\circ}$ and the loss of a significant portion of the telegraph service for many hours \citep{2006AdSpR..38..145G}. One model for the event comprises the ejection of two CMEs from the Sun on 27 August and 1 September 1859, whose interaction as the second CME ploughs through the first produced a shock responsible for the extreme nature of the 2\,--\,3 September auroral event \citep{2006AdSpR..38..130G}. More recently, a severe geomagnetic storm on 13 March 1989 caused the collapse of the Hydro-Qu{\'e}bec power network due to a transformer failure from the geomagnetically induced currents (GIC). Six million people were left without power for nine hours, with a substantial economic loss. The cause was a CME ejected from the Sun on 9 March 1989 impacting the Earth several days later. Other CMEs have similarly knocked out communication satellites such as the Canadian Aniks E1 and E2 and the international Intelsat K on 20 January 1994, and the AT\&T Telstar 401 on 7 January 1997. Moreover, the increased radiation of solar and space weather storms poses a risk to astronauts and high-altitude flight passengers, particularly when travelling over the poles. In the modern era of technological advancement and increased dependency on satellites communications, GPS networks and power distribution grids, the influence of space weather is an increasing cause for concern. A recent report by the US National Research Council \citep{2008sswe.rept.....C} indicates that the potential economic cost of a high-level geomagnetic storm could be up to \$2~trillion. Thus the monitoring and forecasting of potentially hazardous events is of great importance to society at large, with particular emphasis on CMEs and the dynamics governing their propagation through space.

\newpage
\section{Coronal Mass Ejections}

\emph{``We define a coronal mass ejection to be an observable change in coronal structure that (1) occurs on a time scale between a few minutes and several hours and (2) involves the appearance [and outward motion] of a new, discrete, bright, white-light feature in the coronagraph field of view."}\begin{flushright}-- \citep{1984JGR....89.2639H}\end{flushright}

%The differential rotation of the Sun gives rise to large shear forces which wrap and twist the magnetic field around on itself. Constant increasing of magnetic helicity can lead to magnetic buoyancy whereby field lines are driven to the surface and may erupt into the corona. Continued build-up of energy in the coronal magnetic field must be released through topological changes which restructure the field into a lower energy configuration. The explosive nature of these eruptions is attributed to magnetic reconnection which occurs in cases of large stressing and shearing of field lines due to footpoint motions, flux and/or helicity injection from below the surface causing a loss of equilibrium, or a combination of these and possible other instabilities such as field kinks or blast waves. Examples of these structures are coronal loops and arcades, and filaments (on-disk) and prominences (off-limb), while explosive eruptions are seen as flares and CMEs.

\begin{figure}[!p]
\centerline{\includegraphics[width=\linewidth]{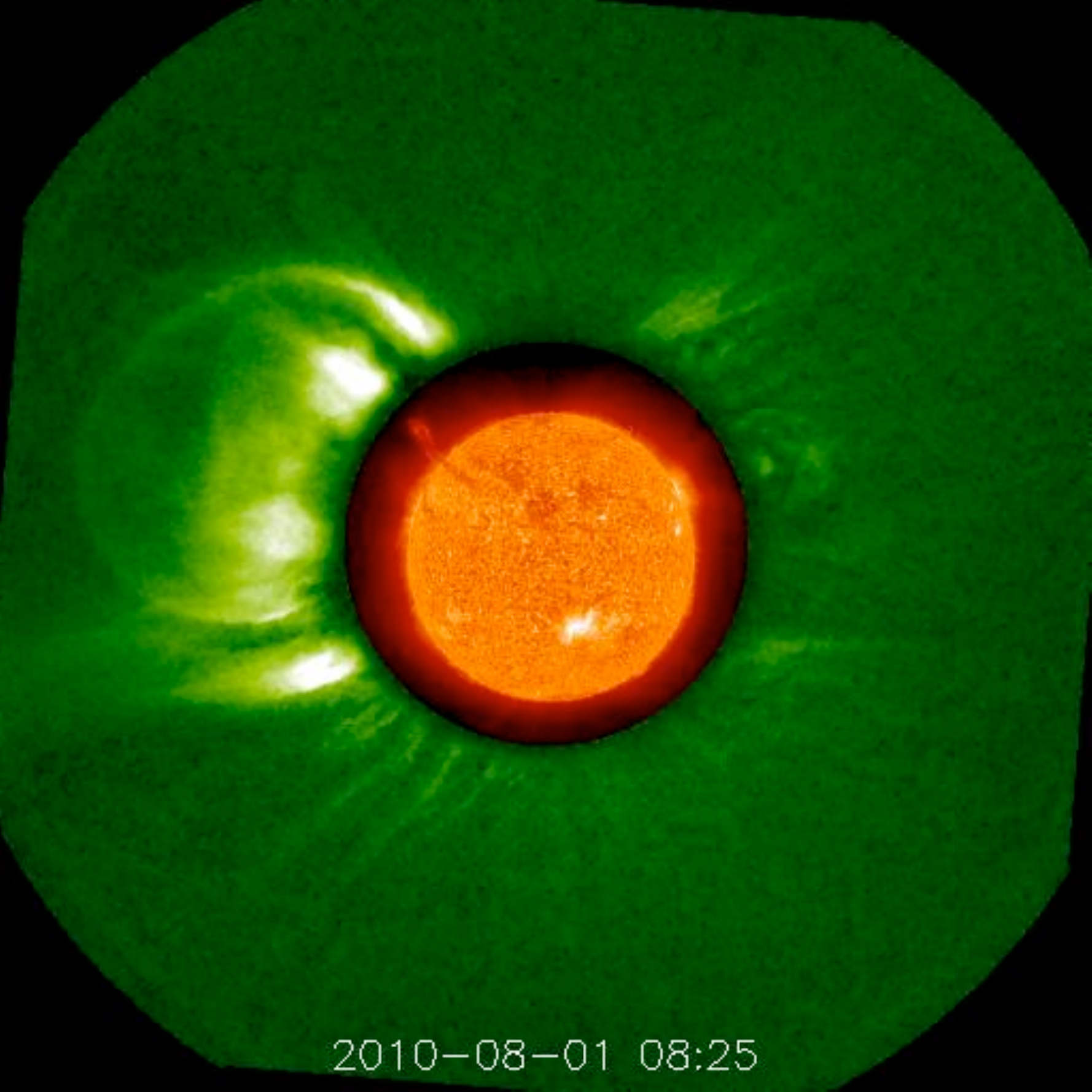}}
\caption{Observation of a CME and prominence lift-off from the EUVI and COR1 instruments of the SECCHI suite on board the STEREO-A spacecraft. The field-of-view extends to $\sim$\,4~R$_{\odot}$. The complexity of the magnetic field driving the eruption is clearly indicated by the twisted geometry of the bright ejecta. 
\newline
{\it Image credit: http://stereo.gsfc.nasa.gov/}}
\label{cme}
\end{figure}

Through the association of early observed flares, notably the Carrington event of 1859, and the detections of geomagnetic storms at Earth, the theory was put forward that plasma transients may be ejected from the Sun and possibly impact the Earth's magnetic field several days later \citep[e.g.,][]{1919Lindemann}. Observations of prominence disappearances provided evidence that the material may be rising through the corona with increasing velocity to eventually exceed the escape velocity of the Sun and erupt in to space \citep{1953sun..book..322K}. Theories were developed by scientists who postulated that the magnetic field should be affected by these ejections, as they might act to drag out field lines or sever completely from the Sun through magnetic reconnection, and potentially drive shock disturbances in the interplanetary gas \citep{1958PhRv..112..589P, 1962SSRv....1..100G}. It was only with the advent of space-borne coronagraphs that these transients were realised to be a common occurrence on the Sun, and their potential geomagnetic effects on Earth led them to become a topic of great interest. An example of such observations is shown in Figure~\ref{cme}.
\newline
\indent These coronal mass ejections (CMEs), as they became known, are the largest manifestation of the shedding of solar magnetic field during the Sun's 22 year cycle. Every 11 years the magnetic axis of the Sun flips, giving rise to periodic patterns in the activity called solar minimum and maximum. At solar minimum a CME may occur up to once a week but during solar maximum they can be as frequent as three a day. CMEs that travel directly towards or away from the observer are seen to encircle the full disk of the Sun and are thus named halo CMEs. In the case of a halo CME coming toward the Earth, the particle densities and energies involved can cause geomagnetic storms, especially if the orientation of the CME's impacting magnetic field is oppositely directed to that of Earth's magnetosphere since it can open up the field lines and penetrate deeper into Earth's atmosphere. This is referred to as space weather, and understanding this interaction is of considerable practical importance because technological systems, such as communications and navigation satellites, can suffer interruptions or damage. To this end, missions such as the Solar and Heliospheric Observatory \citep[SOHO;][]{1995SoPh..162....1D} and Solar Terrestrial Relations Observatory \citep{2008SSRv..136....5K} have been launched to study CMEs.
\newline
\indent CMEs are observed as a typical three-part plasma structure of a bright leading front, dark cavity, and bright core \citep{1985JGR....90..275I}. This configuration is indicative of a magnetic loop system erupting off the Sun as a flux rope (a helically twisted bundle of field lines) or magnetic bubble with plasma embedded within and coronal material being swept up ahead of it. Thus the forces acting on CMEs are described within the context of magnetohydrodynamics (MHD) as outlined below. %The ejected material is observed in white-light as a result of Thomson scattering of electrons, the process of which is outlined below. 

\subsection{Magnetohydrodynamic Theory}
\label{section:mhdtheory}

The interplay between the plasma and magnetic fields of the Sun, notably in phenomena such as flares and CMEs, may be described through the coupling of the equations of electromagnetism with the theory of fluid motions. MHD attempts to combine Maxwell's equations with the fluid equations through the relative dependence on the electron motion in the currents set up in the plasma and the effects of the magnetic fields. Thus we obtain the induction equation for magnetised plasma and describe how the field may undergo non-ideal (resistive) MHD processes such as magnetic reconnection - an important basis of many CME models.

\subsubsection{Maxwell's Equations}

Maxwell's equations describe the interaction of magnetic field {\bf B} and electric field {\bf E} according to:
\begin{align}
\label{AmperesLaw} \nabla \times {\bf B} \;&=\; \mu_0 {\bf j} + \frac{1}{c^2} \frac{\partial {\bf E}}{\partial t} \\
\label{solenoid} \nabla \cdot {\bf B} \;&=\; 0 \\
 \label{FaradaysLaw} \nabla \times {\bf E} \;&=\; -\frac{\partial {\bf B}}{\partial t} \\
\nabla \cdot {\bf E} \;&=\; \frac{1}{\epsilon_0} \rho 
\end{align}
where {\bf j} is the current density, $\rho$ is the charge density, $\mu_0$ is the magnetic permeability of a vacuum, $\epsilon_0$ is the permittivity of free space, and $c$ is the speed of light. The second term of Amp{\'e}re's law (Equation~\ref{AmperesLaw}) may be neglected if the typical plasma velocities are much less than the speed of light:
\begin{equation}
\nabla \times {\bf B} \;=\; \mu_0 {\bf j}
\label{ampereslaw2}
\end{equation}

\subsubsection{Fluid Equations}

The mass continuity equation states that matter is neither created nor destroyed:
\begin{equation}
\frac{\partial \rho}{\partial t} + \nabla \cdot \rho {\bf v} \;=\; 0
\end{equation}
where $\rho$ is the plasma density, and ${\bf v}$ the plasma velocity. This can be expanded to give:
\begin{equation}
\frac{\partial \rho}{\partial t} + ( {\bf v} \cdot \nabla) \rho + \rho \nabla \cdot {\bf v} \;=\; 0
\end{equation}
where for an incompressible fluid the convective time derivative is zero. (This is the derivative $D/Dt=\partial / \partial t + {\bf v} \cdot \nabla$ taken along a path moving with velocity ${\bf v}$.) So the mass continuity equation reduces to:
\begin{equation}
\nabla \cdot {\bf v} \;=\; 0
\end{equation}
The equation of motion $({\bf F} = m{\bf a})$ for a CME may be written:
\begin{equation}
\rho \frac{D{\bf v}}{Dt} \;=\; - \nabla p + {\bf j} \times {\bf B} + \rho {\bf g} - \frac{1}{2} \rho {\bf v}^2 A_{cme} C_D
\label{eqnmotion}
\end{equation}
where $p$ is the pressure, ${\bf j} \times {\bf B}$ is the Lorentz force, ${\bf g}$ is gravity, and the drag force depends on the cross-sectional area $A_{cme}$ and drag coefficient $C_D$. We neglect viscous forces.
\newline
\indent In addition, Ohm's law couples the plasma velocity to the electromagnetic fields by:
\begin{equation}
\label{OhmsLaw}
{\bf j} \;=\; \sigma ({\bf E} + {\bf v} \times {\bf B})
\end{equation}
where $\sigma$ is the electrical conductivity.

\subsubsection{The Induction Equation}

It is possible to eliminate the electric field {\bf E} by combining Amp{\'e}re's law (Equation~\ref{AmperesLaw}) and Ohm's law (Equation~\ref{OhmsLaw}):
\begin{equation}
{\bf E} \;=\; -{\bf v} \times {\bf B} + \frac{1}{\mu_0 \sigma} \nabla \times {\bf B}
\end{equation}
and substituting into Faraday's law (Equation~\ref{FaradaysLaw}) to obtain:
\begin{align}
\frac{\partial {\bf B}}{\partial t} \;&=\; \nabla \times \left({\bf v} \times {\bf B}\right) - \nabla \times \left(\eta \nabla \times {\bf B}\right) \\
&=\; \nabla \times \left({\bf v} \times {\bf B}\right) - \eta \nabla \times \left(\nabla \times {\bf B}\right) \\
&=\; \nabla \times \left({\bf v} \times {\bf B}\right) + \eta \left[\nabla^2 {\bf B} - \nabla \left(\nabla \cdot {\bf B}\right) \right]
\end{align}
where $\eta = 1/\mu_0 \sigma$ is the magnetic diffusivity.
Using the solenoidal constraint (Equation~\ref{solenoid}) provides the induction equation:
\begin{align}
\frac{\partial {\bf B}}{\partial t} \;&=\; \nabla \times \left({\bf v} \times {\bf B}\right) + \eta \nabla^2 {\bf B} \\
O \left( \frac{B}{t}\right) \; &\sim \; O \left( \frac{vB}{l} \right) + O \left( \frac{\eta B}{l^2} \right) \nonumber
\end{align}
This equation forms the basis of any model that considers magnetised plasma motion on a variety of length scales, e.g., from magnetic confinement devices on Earth, to the dynamo action of the Sun's magnetic field. How a magnetic field topology will respond to the forces of plasma motion, and vice versa, is governed by the ratio of the terms in the induction equation.

\subsubsection{The Magnetic Reynolds Number}

The magnetic Reynolds number is the ratio of the advection and diffusion terms in the induction equation:
\begin{equation}
R_m \;=\; \frac{\nabla \times ({\bf v} \times {\bf B})}{\eta \nabla^2 {\bf B}} \;\approx\; \frac{v_0 l_0}{\eta}
\end{equation}
for plasma speed $v_0$ and length scale $l_0$.
\newline
If $R_m \gg 1$ the induction equation is approximated by:
\begin{equation}
\frac{\partial {\bf B}}{\partial t} \;=\; \nabla \times ({\bf v} \times {\bf B}) 
\end{equation}
and the coupling of the magnetic field to the plasma motion is strong, so the topology of the field changes on the timescales of the plasma motion:
\begin{equation}
\frac{B}{\tau_{motion}} \;\approx\; \frac{v_0}{l_0}B \quad \Rightarrow \quad \tau_{motion}\;\approx\; \frac{l_0}{v_0}
\end{equation}
This is known as the `frozen-in' condition, whereby field lines are carried with the plasma motion. For example, in the corona the length scales are $l_0$\,$\sim$\,1,000~km, velocities are $v_0$\,$\sim$\,1,000~m~s$^{-1}$, and the magnetic diffusivity $\eta$\,$\sim$\,1~m$^2$~s$^{-1}$, resulting in a magnetic Reynold's number of $R_M$\,$\sim$\,10$^9$.
\newline
If $R_m \ll 1$ the induction equation is approximated by:
\begin{equation}
\frac{\partial {\bf B}}{\partial t} \;=\; \eta \nabla^2 {\bf B} 
\end{equation}
and the magnetic field can diffuse through the plasma and change its topology:
\begin{equation}
\frac{B}{\tau_{diffusion}} \;\approx\; \eta \frac{B_0}{{l_0}^2} \quad \Rightarrow \quad \tau_{diffusion}\;\approx\; \frac{{l_0}^2}{\eta}
\end{equation} 
Magnetic diffusion is an important condition for magnetic reconnection to occur such as in a current sheet where the length scales are small enough (on the order of metres) to account for the observed restructuring of magnetic field (which is on the order of seconds in flaring active regions for example).

\subsubsection{Magnetic Reconnection}

\begin{figure}[!t]
\centerline{\includegraphics[width=\linewidth, trim=0 0 0 300]{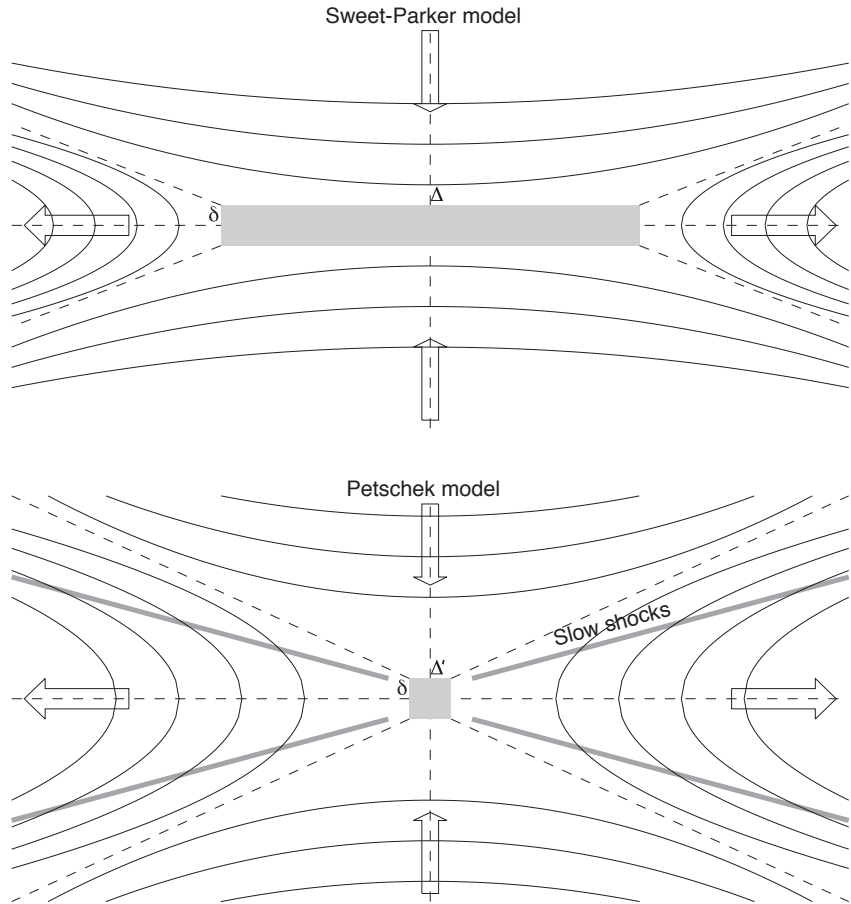}}
\caption{Geometry of the Sweet-Parker (top) and Petschek (bottom) reconnection models, reproduced from \citet{2005psci.book.....A}.}
\label{recxn_models}
\end{figure}

The large energies released in flares and CMEs and observed restructuring of the coronal magnetic field are often attributed to the phenomenon of magnetic reconnection occurring on the Sun. Magnetic reconnection is generally defined as a change in connectivity of field lines in time. In an ideal plasma the `frozen-in' condition is met and the magnetic field is coupled to the plasma motion. However, when regions of opposite polarity flux come together, a boundary layer will form to separate the two regimes of magnetic field in a form of pressure balance. The high resistivity of such a system counteracts the currents within and allows the occurrence of non-ideal MHD processes and the formation of structures having small spatial scales, e.g., a thin current sheet. This leads to a low magnetic Reynolds number and allows diffusion to occur (illustrated by the shaded regions in Figure~\ref{recxn_models}). The connectivity of field lines then changes to a more energetically favourable configuration, and in doing so will eject plasma along resulting outflows as the reconnected field lines relax to a new equilibrium. The outflows create a low pressure in the diffusion region which in turn allows continued inflow of plasma and magnetic field forming a runaway process of reconnection until the system comes to rest in a new topology. 
\newline
\indent A description of how magnetic reconnection occurs was put forward by \citet{1958IAUS....6..123S} and \citet{1957JGR....62..509P} as a two-dimensional incompressible MHD approximation. They estimated the rate of reconnection from a boundary layer analysis (top of Figure~\ref{recxn_models}). For a reconnection layer of length $\Delta$ and thickness $\delta$, the outflow must balance the inflow:
\begin{equation}
v_{in}\Delta \; = \; v_{out} \delta
\end{equation}
where $v_{in}$ is the inflow reconnection velocity, and $v_{out}$ is the outflow velocity, which by conservation of energy ($B_x^2/2\mu = \rho v_{out}^2 /2$) is equal to the Alfv{\'e}n velocity $v_A=B_0/\sqrt{\mu_0 \rho}$. From Ohm's law (Equation~\ref{OhmsLaw}) the configuration of the straight field lines ($\nabla \times B=0$) outside the layer is $E_z+v_RB_x=0$, and inside the layer (where there is a large current) is $E_z=\eta J_z$. Integrating Ampere's law (Equation~\ref{ampereslaw2}) around the layer gives $B_x=\mu J_z\delta$. Combining the results of these two laws gives:
\begin{equation}
v_{in} \; = \; \frac{E_z}{B_x} \; = \; \frac{\eta J_z}{\mu J_z \delta} \; = \; \frac{\eta}{\mu \delta}
\end{equation}
This can be written in terms of the outflow velocity (or interchangeably the Alfv{\'e}n velocity) by:
\begin{equation}
v_{in}^2 \; = \; \left( v_{out} \frac{\delta}{\Delta} \right) \left( \frac{\eta}{\mu \delta} \right) \; = \; v_{out}^2 \left( \frac{\eta}{v_A \mu \Delta} \right)
\end{equation}
The rate of reconnection is then written in terms of the Lunquist number $S$, which is the dimensionless ratio of an Alfv{\'e}n wave crossing timescale to a resistive diffusion timescale:%High Lundquist numbers indicate highly conducting plasmas, while low Lundquist numbers indicate more resistive plasmas.
\begin{equation}
\frac{v_{in}}{v_{out}} \; = \; \frac{1}{\sqrt{S}} \quad \text{where} \quad S \;=\; \frac{\mu \Delta v_A}{\eta}
\end{equation}
This is the Sweet-Parker result. The problem with the theory is that it predicts reconnection to take place on far too slow a timescale to reconcile with observations. Consider solar flares, for example, with $v_A$\,$\sim$\,1,000~km~s$^{-1}$, and $l$\,$\sim$\,10$^4$~km resulting in Sweet-Parker reconnection of tens of days when flare energy release is actually observed over minutes to hours.
\newline
\indent An extension to the Sweet-Parker model was put forward by \citet{1964NASSP..50..425P} in which the field lines do not have to reconnect along the entire length of the boundary, but could merge over a shorter length $\Delta'<\Delta$ (bottom of Figure~\ref{recxn_models}). The remaining length of the boundary is occupied by slow shocks, where the magnetic field tension accelerates the plasma to the Alfv{\'e}n velocity. The reconnection velocity in the Petschek model may be written:
\begin{equation}
v_{in}^2 \; = \; v_{out}^2 \left( \frac{\eta}{v_A \mu \Delta'} \right) \quad \Rightarrow \quad v_{in} \;=\; \frac{v_{out}}{\sqrt{S}} \sqrt{ \frac{\Delta}{\Delta'}}
\end{equation}
a factor of $\sqrt{\Delta/\Delta'}$ faster than the Sweet-Parker reconnection velocity. Issues with the description of the magnetic field and shock formation place limits on the plausibility of Petschek's formalism that $\Delta'$ is a free parameter which may be minimised to obtain the maximum reconnection velocity \citep{2001EP&S...53..417K}. Numerical simulations, such as that of \citet{1986mrt..conf...19B}, are in favour of the Sweet-Parker result, unless anomalous resistivity, for example, is considered. In either case, the process described by these theories forms the basis of many CME models in an effort to understand observations. % (resistivity enhanced by wave interactions).

\subsection{Theoretical CME Models}

\begin{figure}[!p]
\centering
\subfigure{\includegraphics[clip=true, scale=0.32, trim=0 230 0 230]{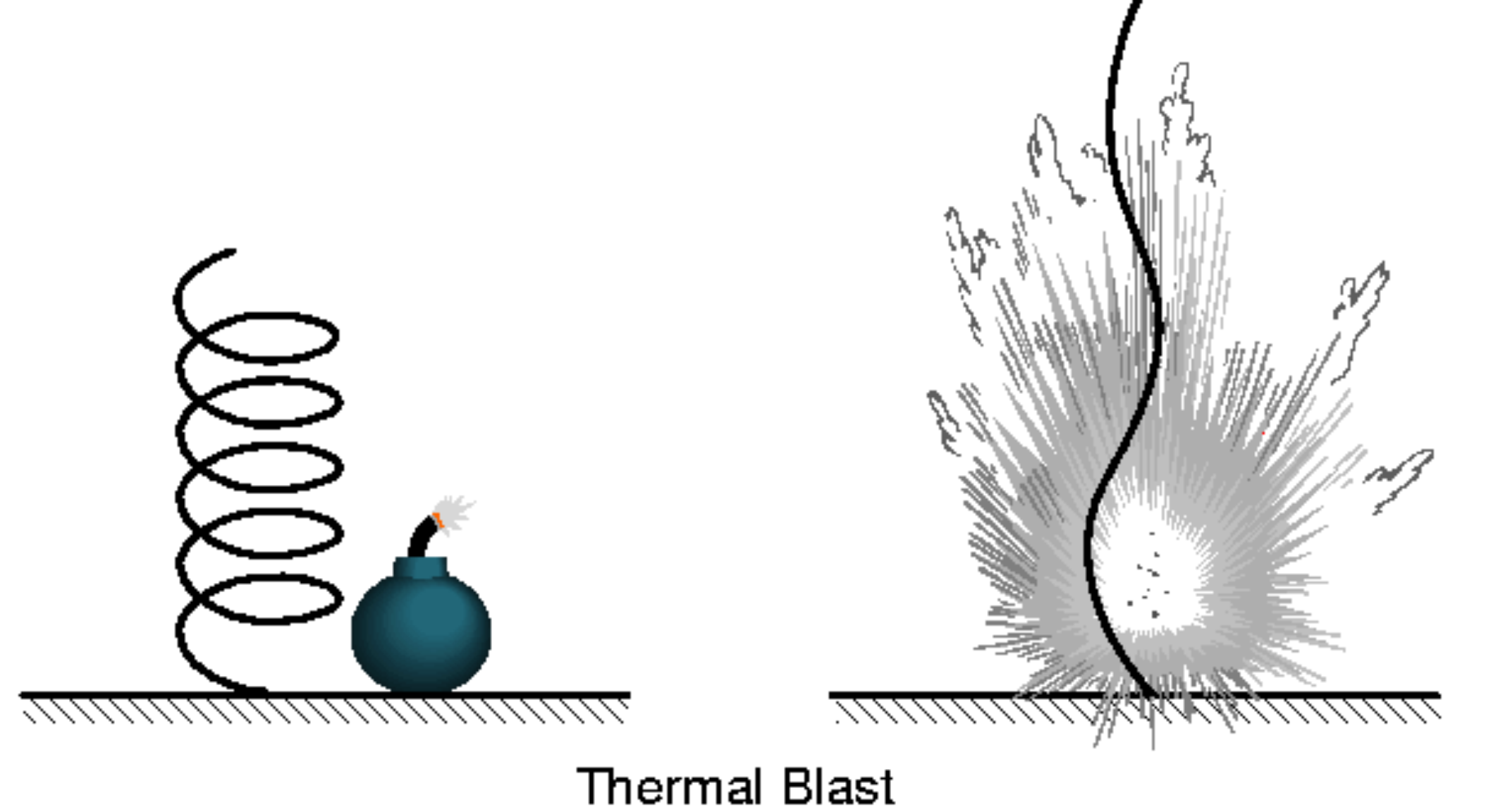}} \\
\subfigure{\includegraphics[clip=true, scale=0.3, trim=0 210 0 215]{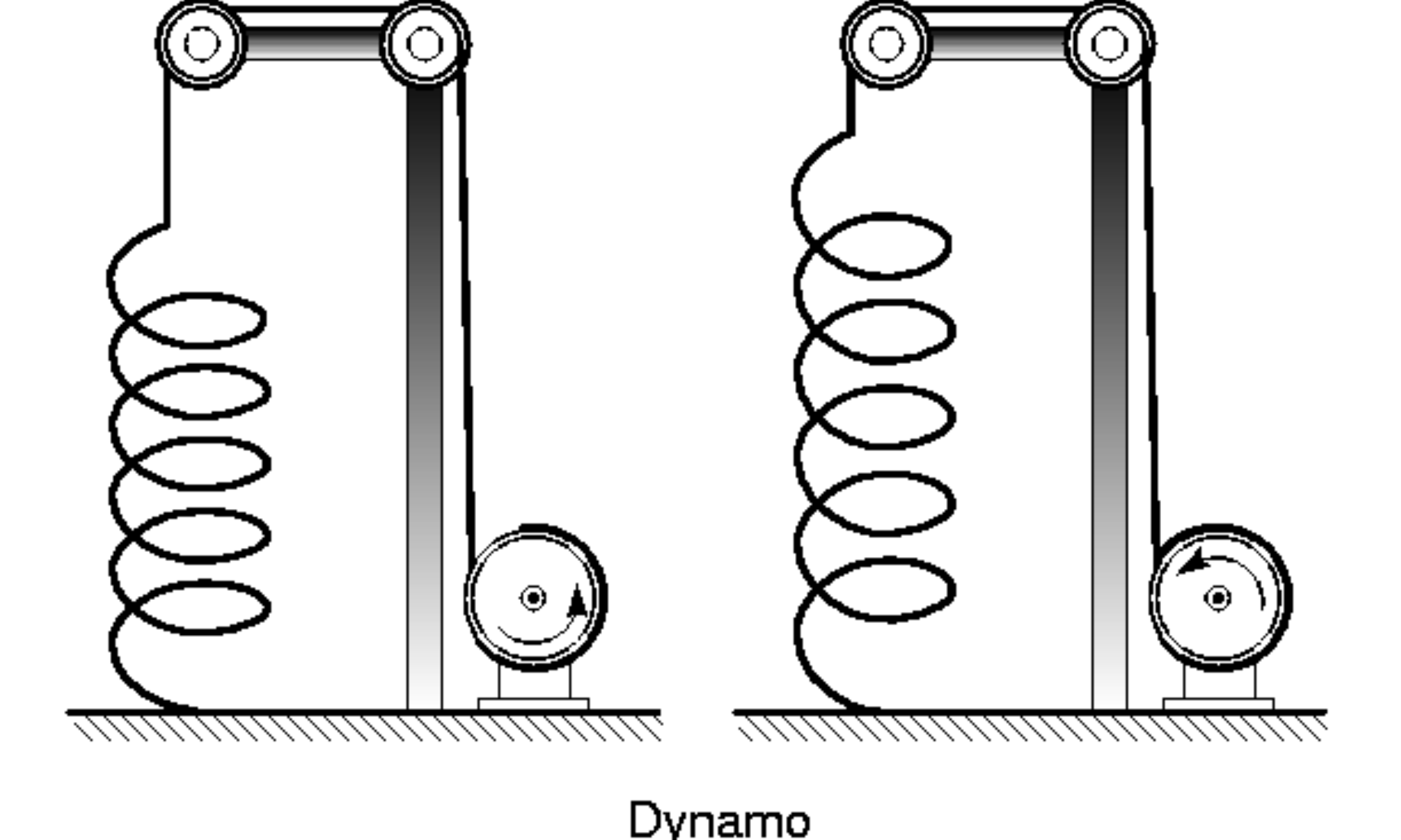}} \\
\subfigure{\includegraphics[clip=true, scale=0.33, trim=0 265 0 265]{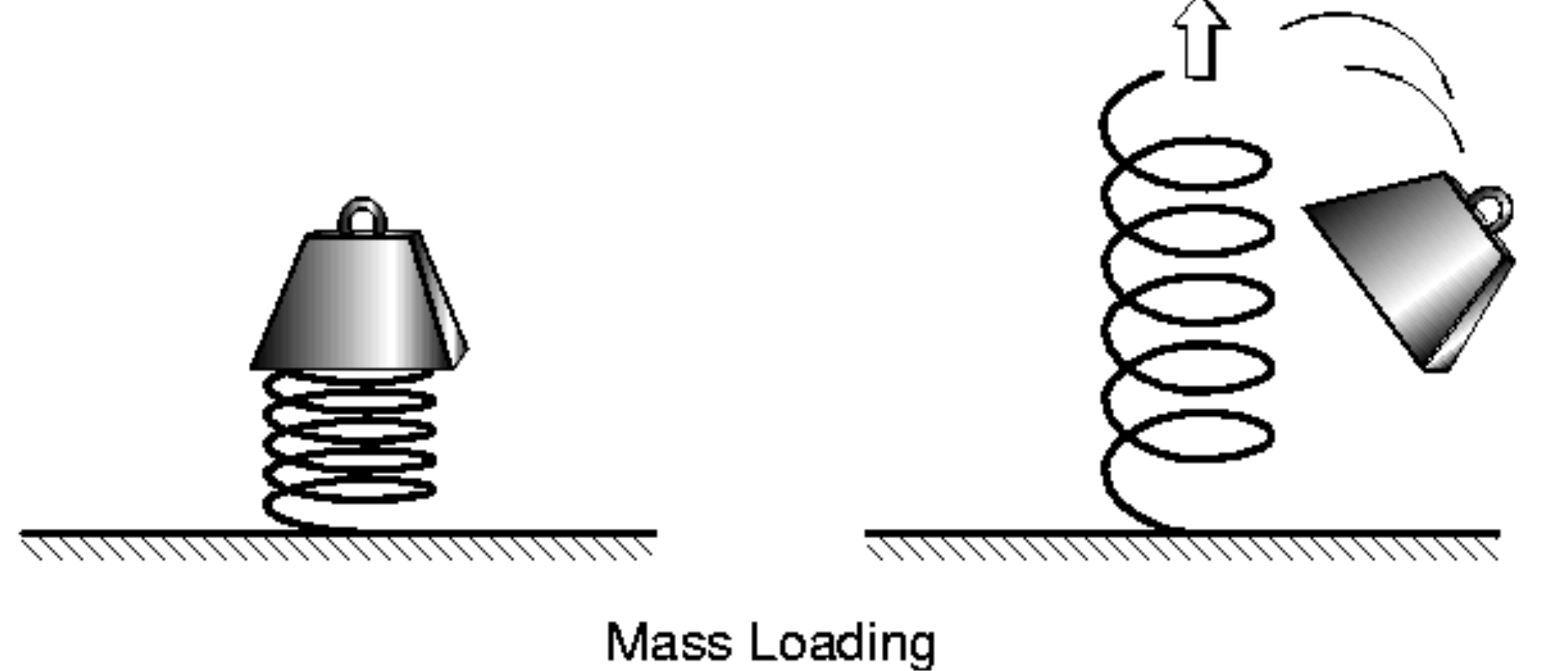}} \\
\subfigure{\includegraphics[clip=true, scale=0.65, trim=0 5 200 660]{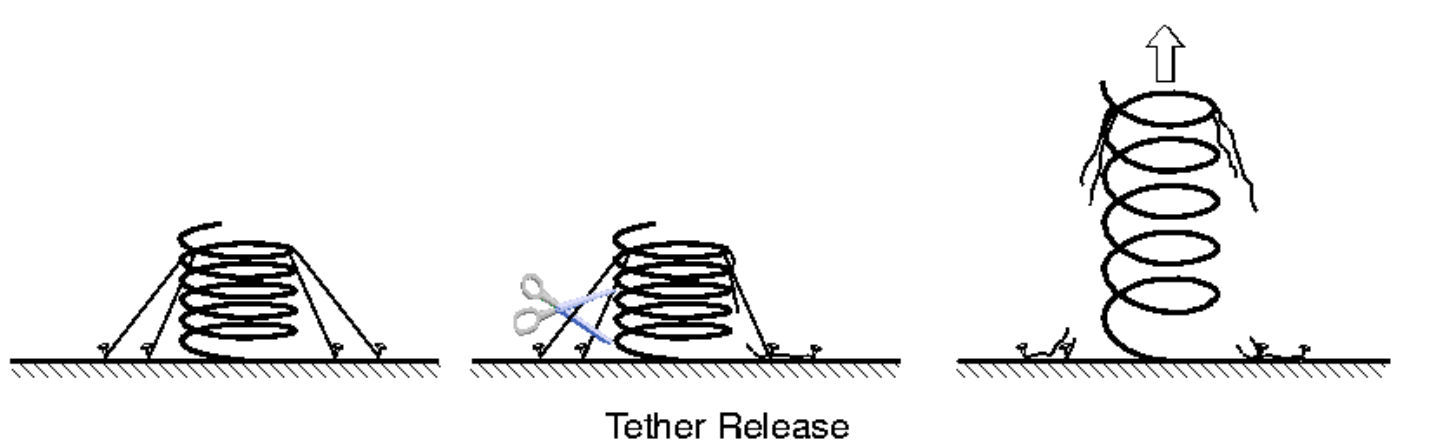}} \\
\subfigure{\includegraphics[clip=true, scale=0.44, trim=0 230 0 245]{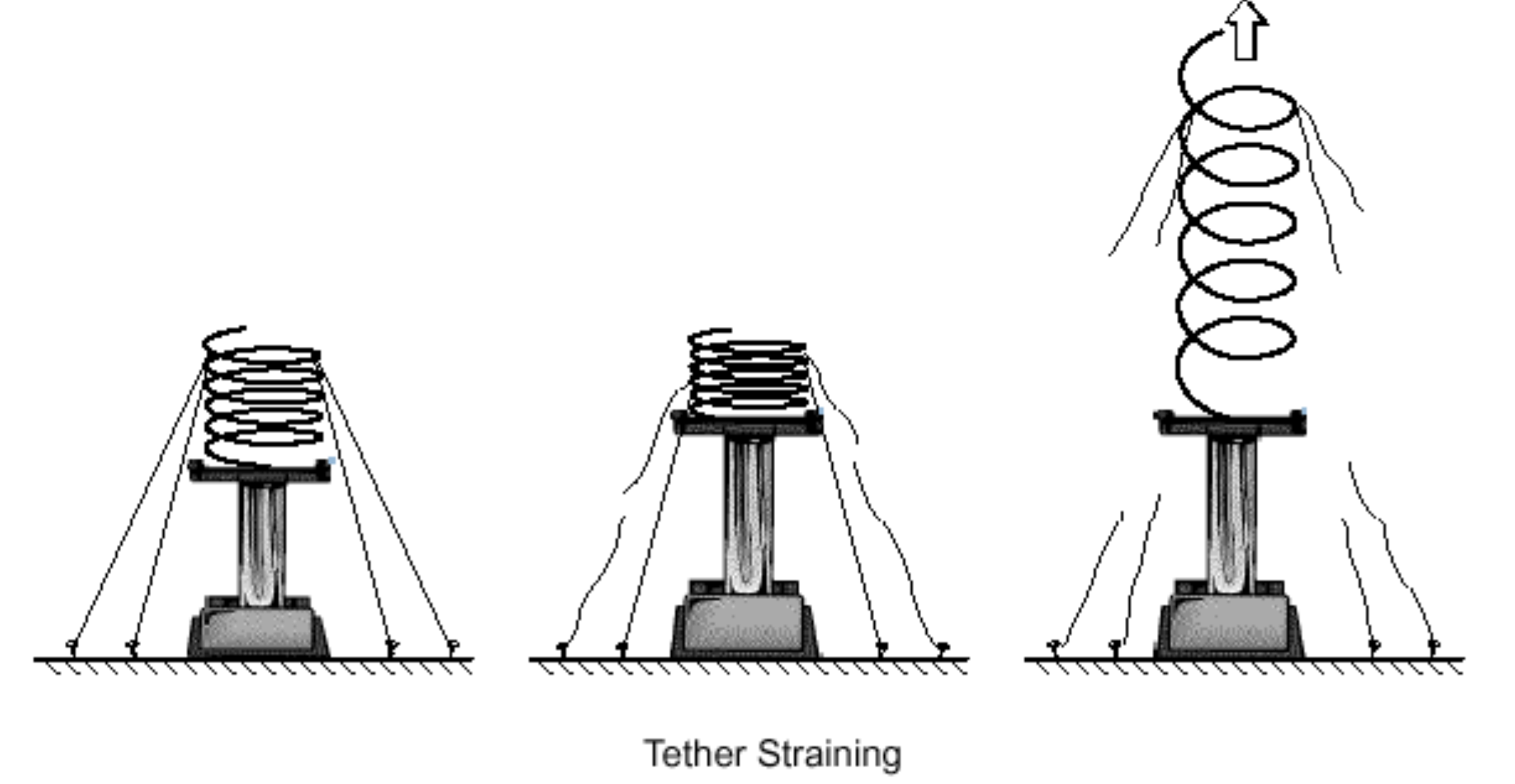}}
\caption{Illustrations of the different mechanical analogues of CME eruptions, reproduced from \citet{2001AGUGM.125..143K}.}
\label{mechanical_analogues}
\end{figure}

%How is energy imparted into the models? 
%Energy classes:: energy driven / thermal blast  / energy storage   -  Linker (2003)

It is well known that CMEs are associated with filament eruptions and solar flares \citep{2002ApJ...566L.117Z, 2002ApJ...581..694M} but the driver mechanism remains elusive. Several theoretical models have been developed in order to describe the forces responsible for CME initiation and propagation, all of which are based on the idea that some form of instability must trigger the eruption. These models may be explained in terms of the following mechanical analogues, illustrated in Figure~\ref{mechanical_analogues}.
\newline
\newline
{\it The Thermal Blast Model} proposes that the increased thermal pressure produced from a flare overcomes the magnetic field tension and blows it open to cause a CME. Observations, however, have shown that not all CMEs are preceded by a flare, nor even necessarily associated with a flare at all.
\newline
\newline
{\it The Dynamo Model} introduces the idea of magnetic flux injection or stressing of the field on a time-scale that is too fast for the system to dissipate the magnetic energy before it builds to a critical point and erupts.
\newline
\newline
{\it The Mass Loading Model} is concerned with the amount of material included in the eruption. Prominences, or regions of relatively higher electron density in the corona, overlaying a volume of lower density will erupt due to the Rayleigh-Taylor instability.
\newline
\newline
{\it The Tether Release Model} is based on the restraining of the outward magnetic pressure by the magnetic tension of the overlying field. As `tethers' are removed a loss-of-equilibrium occurs due to the magnetic pressure/tension imbalance and the system erupts.
\newline
\newline
{\it The Tether Straining Model} is a variant on the tether release model whereby an increase in magnetic pressure due to flux injection or field shearing eventually overcomes the tension forces and the `tethers' break and release the CME.
\newline

 %Firstly, we assume the plasma mass has zero net flux (that is no net generation or removal is at play) by the mass continuity equation:
%\begin{eqnarray}
%\frac{\partial \mathbf{\rho}}{\partial t} + \nabla \cdot \mathbf{\rho\mathbf{v}}& = & 0 
%\end{eqnarray}
%where $\rho$ is the density of the plasma, and $\mathbf{v}$ is its velocity.
%\begin{eqnarray}
%\rho \frac{D\mathbf{v}}{Dt} & = & \mathbf{j} \times \mathbf{B} - \nabla P + \rho \mathbf{g} - \frac{1}{2}\rho%\mathbf{v}^{2}A_{cme}C_{D}
%\end{eqnarray}
%with density $\rho$, pressure $P$, velocity $\mathbf{v}$, current density $\mathbf{j}$, magnetic field $\mathbf{B}$, and gravity $\mathbf{g}$. The CME drag force also considers the cross-sectional area $A_{cme}$ and drag coefficient $C_{D}$ \citep{2004SoPh..221..135C}.
The tether straining and release models are generally accepted as the most likely scenarios for CME initiation, being able to reproduce numerous observations of CMEs through the development of detailed 2D and 3D flux rope models, as discussed below. Within the context of MHD outlined in Section~\ref{section:mhdtheory}, we can describe the solar plasma as a fluid with the assumption that there is negligible viscosity, so the motion of the flux rope is governed by the forces of Equation~\ref{eqnmotion}. The Lorentz force is thought to be the dominant driver force in modelling CME eruptions, certainly during the early stages of propagation, before the drag force takes over and the CME propagates with the ambient solar wind through interplanetary space.

\subsubsection{Catastrophe Model}

\begin{figure}[!p]
\centerline{\includegraphics[width=\linewidth]{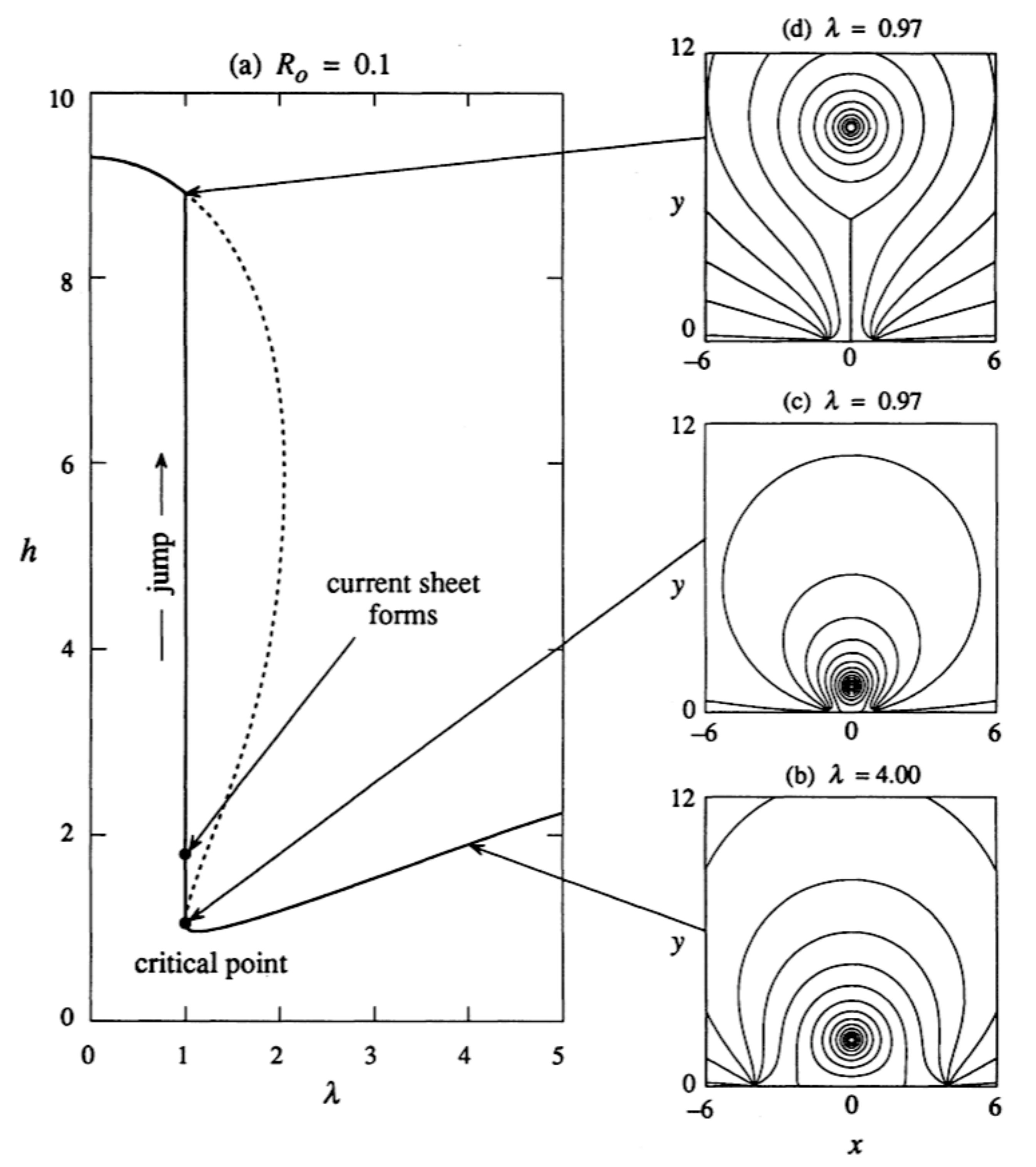}}
\caption{Theoretical evolution of a 2D flux rope, reproduced from \citet{1995ApJ...446..377F}. The flux rope footpoint separation $\lambda$ decreases, increasing the magnetic pressure until the flux rope becomes unstable and erupts away from the surface (\emph{b\,--\,d}). The resulting height evolution of the flux rope is illustrated in (\emph{a}).}
\label{2Dfluxrope}
\end{figure}

The 2D flux rope model is driven by a catastrophic loss of mechanical equilibrium as a result of footpoint motions in the photosphere \citep{1991ApJ...373..294F, 2002A&ARv..10..313P, 1990SoPh..126..319P, 1993ApJ...417..368I, 1995ApJ...446..377F, 2000mare.book.....P}. The model is illustrated in Figure~\ref{2Dfluxrope} as a coronal current filament channel and overlying magnetic field lines, in equilibrium due to the balance between the magnetic pressure and tension forces acting on the system. The description of the model's evolution in time may be split into a storage phase and an eruption phase. During the storage phase the footpoints of the system are slowly moved together such that the magnetic energy of the flux rope increases. The magnetic tension thus increases and causes the flux rope to move downwards, which builds up magnetic pressure until a critical footpoint distance is reached where equilibrium is lost. In the eruption phase the flux rope is accelerated upwards, stretching the magnetic field lines such that a current sheet forms behind it. If reconnection occurs in the current sheet then all the energy is released and the upward motion of the flux rope is unbounded. Otherwise it will come to equilibrium again at a greater height, or oscillate about this equilibrium height if it still has energy in excess of that required for the initial eruption. The configuration of this model outlined in \citet{2000mare.book.....P} predicts kinematics of the flux rope (provided it is `thin' so that its radius is less than the scale-length $\lambda_0$) prior to the formation of the current sheet (i.e. $h/\lambda_0 \leq 2$) according to:
\begin{equation}
\dot{h} \; \approx \; \sqrt{ \frac{8}{\pi} } v_{A0} \left[ \ln \left( \frac{h}{\lambda_0} \right) + \frac{\pi}{2} - 2 \tan^{-1}\left( \frac{h}{\lambda_0} \right) \right] ^{1/2} + \dot{h}_0
\end{equation}
%\begin{equation}
%h \; \simeq \; \lambda_0 + \dot{h}_0 t + \frac{4 v_{A0}}{5 \sqrt{3\pi}} \left( \frac{ \dot{h}_0}{\lambda_0} \right) ^{3/2} t^{5/2}
%\end{equation}
where $\dot{h}$ is the velocity of the flux rope, $\dot{h}_0$ is the initial perturbation velocity, $\lambda_0$ is the source separation at the critical point, and $v_{A0}$ is the Alfv\'en speed at $h=\lambda_0$. The kinematics may be further separated into an `early' phase when the time scale is less than the Alfv\'en time scale ($t \ll \lambda_0 / v_{A0}$), which gives:
\begin{equation}
\dot{h} \; \approx \; \dot{h}_0 + \frac{2v_{A0}}{\sqrt{3\pi}} \left( \frac{ \dot{h}_0}{\lambda_0} \right)^{3/2} t^{3/2}
\end{equation}
and a `late' phase when $h/ \lambda_0 \gg 1$, but $|\ln h|$ is still much smaller than $| \ln a|$, which gives:
\begin{equation}
\dot{h} \; \approx \; \sqrt{ \frac{8}{\pi}} v_{A0} \left[ \ln \left( \frac{h}{\lambda_0}\right)-\frac{\pi}{2}\right]^{1/2}
\end{equation}
After the formation of the current sheet the system becomes too complicated for the kinematics of the CME's continued propagation to be analytically derived.

\subsubsection{Toroidal Instability}

\begin{figure}[!t]
\centerline{\includegraphics[scale=0.3, clip=true, trim=0 10 0 10]{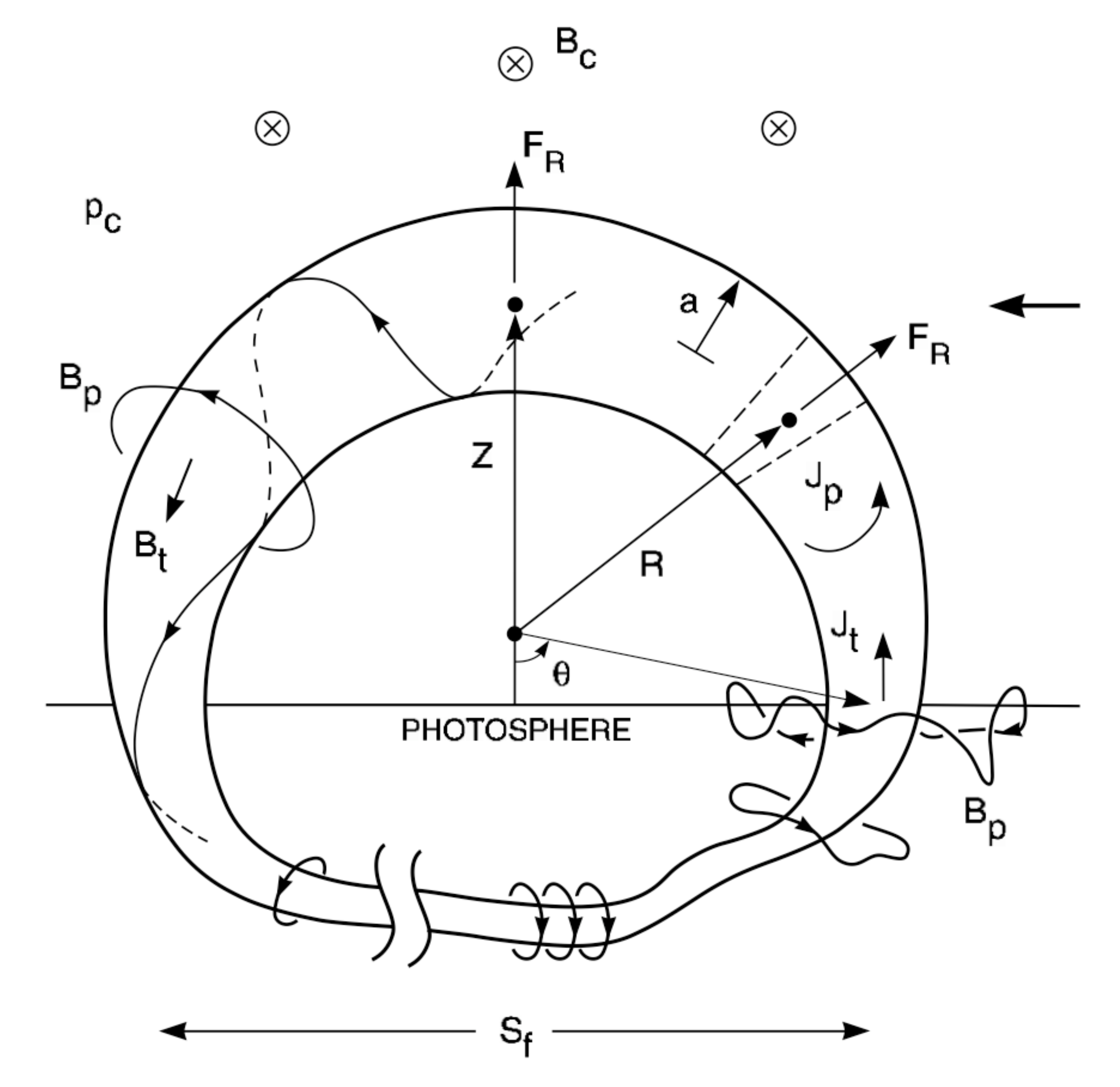}}
\caption{A schematic of the 3D flux rope model, reproduced from \citet{2003JGRA..108.1410C}, an extension of the 2D flux rope in Figure~\ref{2Dfluxrope} when viewed end-on as indicated by the arrow from the right. The flux rope is rooted below the photosphere, and surrounded by the ambient coronal magnetic field $\mathbf{B_c}$ and plasma density $\rho_c$. Components of the current density $\mathbf{J}$ and magnetic field $\mathbf{B}$ are shown, where subscripts `t' and `p' refer to the toroidal and poloidal directions respectively. The flux rope has a radius of curvature $R$, radius of cross-section $a$, apex height $Z$, footpoint separation $s_f$, and the radial force outward is $F_R$.}
\label{3Dfluxrope}
\end{figure}

An extension of the flux rope model to three-dimensions is illustrated in Figure~\ref{3Dfluxrope} \citep{1996JGR...10127499C, 2006PhRvL..96y5002K, 2003JGRA..108.1410C}. The eruption of the flux rope is triggered by an increase in the poloidal magnetic flux of the structure. The 3D flux rope consists of a current channel $\mathbf{J}$ and magnetic field $\mathbf{B}$, and has major radius $R$ and minor radius $a$ such that for $r<a$ the magnetic field lines are helical and can be described by their toroidal and poloidal components, but for $r>a$ the field is purely poloidal ($J_t=0$). The major radius $R$ is fixed, and the minor radius increases from $a_f$ at the footpoints to $a_a$ at the apex. The footpoints are assumed to be immobile because of the high density photosphere ($\sim$\,10$^{23}$~m$^{-3}$) relative to the corona ($\lesssim$\,10$^{16}$~m$^{-3}$). The poloidal field $B_p$ is also highly non-uniform in the photosphere since $\beta \gg 1$.
\newline
\indent The model may be directly compared to coronagraph observations as in \citet{2001ApJ...562.1045K}, where the leading edge of the CME front is located at $Z+2a$ with a width of $4a$ when viewed end-on, or a width of $2R+4a$ when viewed side-on. This definition arises from the fact that the poloidal field $B_p$ at $r=2a$ has decreased to about half the value of $B_{pa}$ at $r=a$ and is then comparable to the ambient coronal field $B_c$. This model sits well with observations, where the CME front corresponds to a plasma pileup ahead of the flux rope which appears as a darker cavity, and any erupting prominence material is suspended at the base of the flux rope and appears as the bright core of the CME. Background parameters such as coronal density and solar wind speeds are also specified in the model. The eruption is initiated by a poloidal flux injection that increases the toroidal current for a short period of time, increasing $B_{pa}$ while $R$ does not change significantly, such that the radial force $\mathbf{F_R}$ becomes more positive and exerts an upward net force on the structure. The eruption then proceeds through the corona as the external poloidal field decreases sufficiently rapidly in the direction of motion. The equation of motion (cf. Equation~\ref{eqnmotion}) may be written in terms of a radial force $F_R$, a gravitational force $F_g$ and drag force $F_d$, acting to cause the apex motion:
\begin{equation}
M \frac{d^2 Z}{dt^2} \;=\; F_R + F_g + F_d
\end{equation}
where the radial force $F_R$ results from the Lorentz magnetic force and pressure gradient of the system, and may be written:
\begin{equation}
F_R \; = \; \frac{ I_t^2}{c^2 R} f_R
\end{equation}
where $I_t$ is the toroidal current, $c$ is the speed of light, $R$ is the major radius as described above, and $f_R$ are the further collective pressure, magnetic and geometrical terms to be considered, detailed in \citet{2003JGRA..108.1410C}. This formalism shows how the toroidal current increase will add to the upward force on the structure. The change in current affects the inductance of the flux rope ($F_R \propto I_t^2 \propto L^{-2}$) and so, neglecting the gravity and drag terms, the acceleration may be expressed in terms of the geometrical size of the flux rope:
\begin{equation}
\frac{d^2 Z}{dt^2} \; \sim \; \frac{\Phi_p^2}{\left[R \ln \left( 8R/a_f \right) \right]^2} f_R
\end{equation}
\citet{2006PhRvL..96y5002K} show how the height of the flux rope during the very initial stages of the eruption may be approximated as a hyperbolic function:
\begin{equation}
h(\tau) \; = \; \frac{P_0}{P_1} \sinh (P_1\tau), \quad \quad h \; \equiv \; H/H_0 -1 \ll 1
\end{equation}
where $H$ is the height, and $H_0$ the initial height, of the flux rope; $\tau$ is the time normalised by the Alfv{\'e}n time; $P_0$ comprises initial parameters on the flux rope dynamics; and $P_1$ associates the external magnetic field profile. Their simulations show a fast rise and gradual decay phase of the CME accelerations due to the toroidal instability. However, \citet{2008ApJ...674..586S} demonstrate that tuning the initial parameters changes the acceleration profile from a fast initial rise to a more gradual rise phase.

\subsubsection{Breakout Model}

\begin{figure}[!p]
\centerline{\includegraphics[width=\linewidth]{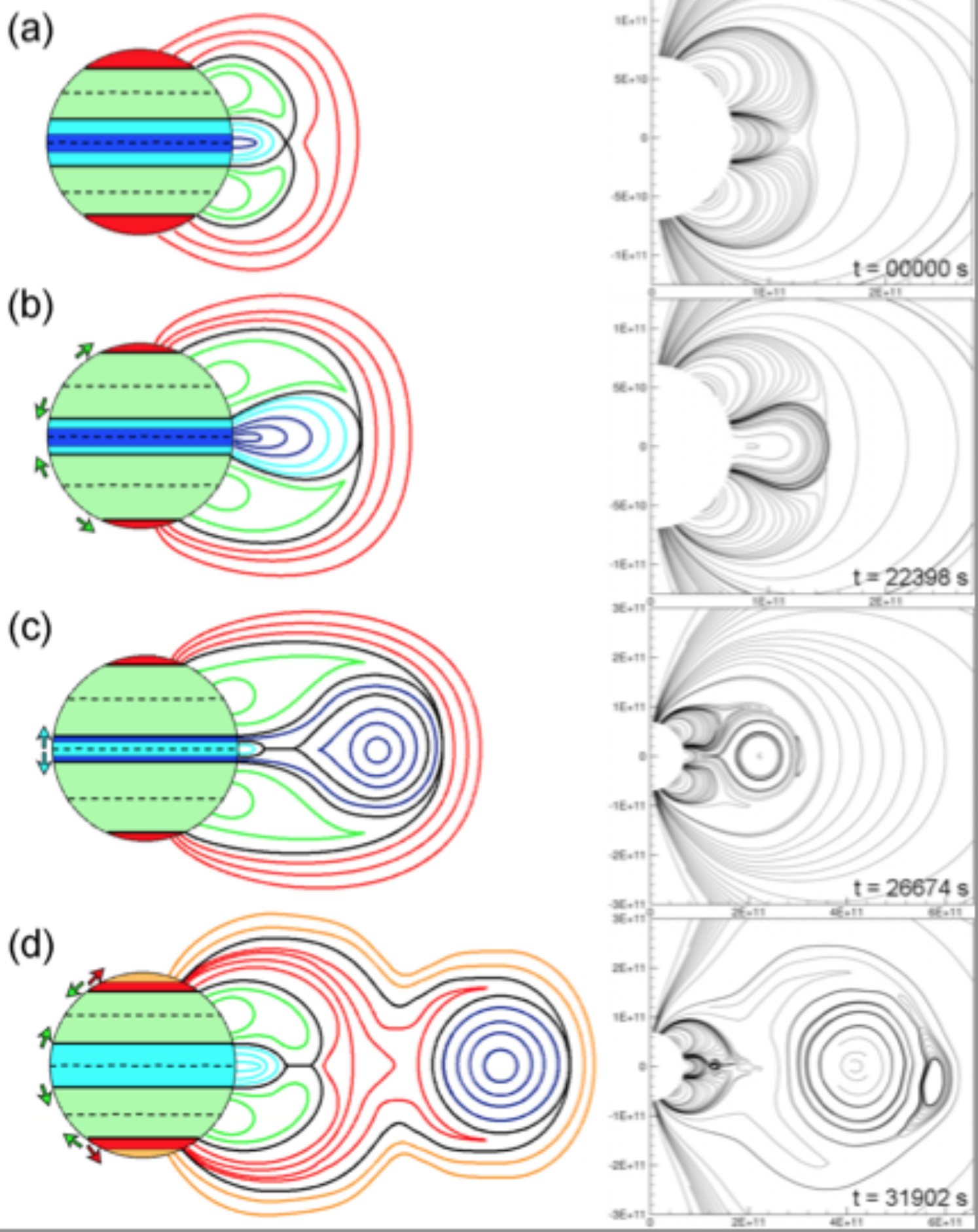}}
\caption{Schematic, and corresponding simulation snapshots, of the main stages of the axisymmetric 2.5D breakout model, reproduced from \citet{2008ApJ...683.1192L}. (a) shows the initial multipolar topology of the system, (b) shows the shearing phase which distorts the X-line and causes breakout reconnection to begin, (c) shows the onset of flare reconnection behind the eruption that disconnects the flux rope, and (d) shows the system restoring itself following the eruption.}
\label{breakout}
\end{figure}

In the magnetic breakout model the CME eruption is triggered by reconnection between the overlying field and a neighbouring flux system through the shearing of a multipolar topology, illustrated in Figure~\ref{breakout} \citep{1999ApJ...510..485A, 2004ApJ...617..589L, 2004ApJ...614.1028M}. It starts by shearing a potential field configuration consisting of a central arcade which will become the CME, two side arcades, and an overlying arcade, with a magnetic X-line separating the different topologies. This shearing adds magnetic pressure to the inner flux system and causes it to expand and distort the overlying field at the X-line, eventually forming a current sheet. As the current sheet grows, reconnection begins, transferring flux to the neighbouring arcades and creating a passage for the CME release as the central arcade erupts. A current sheet also forms beneath the erupting sheared field, creating a disconnected flux rope that escapes, and this is associated with flare reconnection. An increase in the rate of outward expansion drives a faster rate of breakout reconnection, yielding the positive feedback required for an explosive eruption. At given distances from the Sun, the simulation, which is intrinsically 2.5D but has recently been extended to 3D by \citet{2008ApJ...683.1192L}, produces a number of key observational properties to test against data \citep{2007ApJ...671L..77V}. Simulations run by \citet{2004ApJ...617..589L} produced kinematics of the CME front which showed constant acceleration, and they fit quadratics to the height-time data of the form:
\begin{equation}
h(t) \; = \; h_{0} + v_{0}(t-t_{0}) + \frac{1}{2}a(t-t_{0})^{2}
\end{equation}
However, the height-time profiles from the 3D simulations by \citet{2008ApJ...683.1192L} showed a more complex kinematic profile with separate rising and breakout phases of acceleration, proving a better match with observations. The velocity increases linearly from zero during the initial shearing and breakout phase, followed by an acceleration peak during the second stage of reconnection behind the CME. \citet{2008ApJ...680..740D} find a three-phase acceleration profile with a similar initial slow and fast acceleration profile during the breakout and flare reconnections, followed by a short interval of fast deceleration as the magnetic field configuration is reformed after the eruption.

\subsection{CME Observations}

In order to test the validity of the theoretical CME models, and subsequently understand the forces governing their eruption through the solar atmosphere, comparisons must be made with observations. Generally the kinematics of events are determined from white-light coronagraph images, obtained through Thomson scattered emission of the plasma, by tracking the structure as it moves through the field-of-view. 

\subsubsection{Thomson Scattering}

\begin{figure}[!t]
\centerline{\includegraphics[scale=0.4]{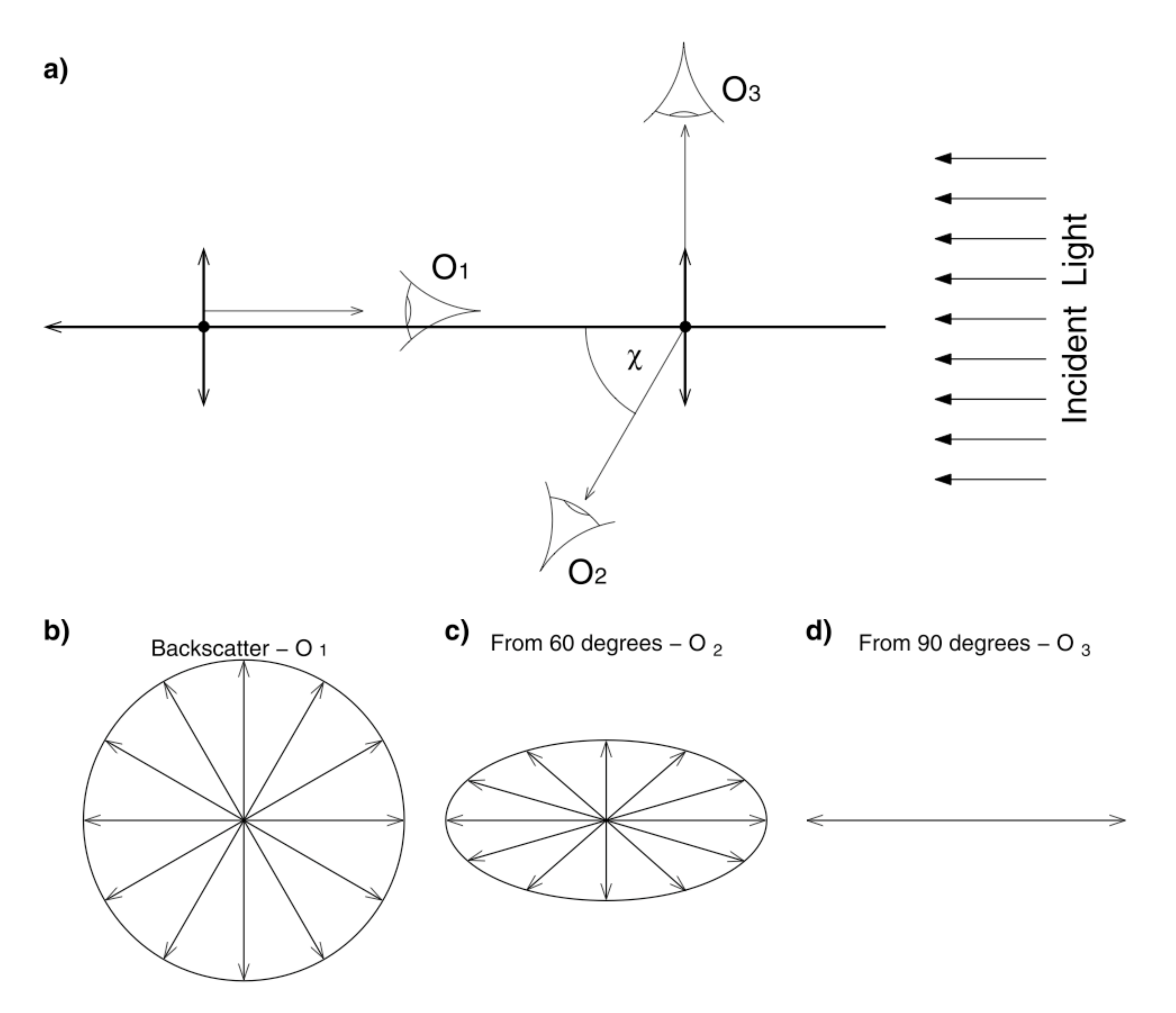}}
\caption{The Thomson scattering geometry for a single electron, reproduced from \citet{2009SSRv..147...31H}. {\bf a}) shows the electron and incident light with different observer positions indicated. {\bf b}), {\bf c}), {\bf d}) show the resultant unpolarised, partially polarised, and polarised scattering of light seen by observers O$_1$, O$_2$, O$_3$ respectively.}
\label{thomsonscattering}
\end{figure}

The low density, optically thin, plasma of the corona and solar wind is observable in white light through the process of Thomson scattering, whereby photons from the Sun are scattered by the free electrons of the coronal plasma \citep{1930ZA......1..209M, 1950BAN....11..135V, 1966gtsc.book.....B}. Essentially, when light is incident on an electron, the electric field of the light waves will cause the electron to accelerate and re-radiate light in the plane perpendicular to the incident wave (illustrated in Figure~\ref{thomsonscattering}). Depending on the angle $\chi$ to the observer, the scattered light may be unpolarised (Figure~\ref{thomsonscattering}b), partially polarised (Figure~\ref{thomsonscattering}c), or polarised (Figure~\ref{thomsonscattering}d). The tangential component of the scattered light intensity is isotropic, while the radial component varies as $\cos^2 \chi$, resulting in the following expression for the differential cross-section \citep{1975clel.book.....J}:
\begin{equation}
\frac{d \sigma}{d \omega} \; =\; \frac{1}{2} \left( \frac{e^2}{4\pi \epsilon_0 m_e c^2} \right)^2 \left(1+\cos^2\chi \right)
\end{equation}
where $d\omega$ is an element of solid angle at scattering angle $\chi$, and $\epsilon_0$ is the permittivity of free space.
Since the Sun is neither a point-source, nor is the intensity of light uniform across it, it is necessary to consider the three-dimensional geometry of the Thomson scattering process in detail to appreciate the effects it will have on CME observations, outlined in detail in \citet{1966gtsc.book.....B} and \citet{2009SSRv..147...31H}. In a full description of the corona the density profile must be considered, as well as the effect of limb darkening on the Sun whereby the light intensity from the photosphere decreases across the disk according to:
\begin{equation}
\label{eqn:limbdark}
I \;=\; I_0(1-u+u\cos\psi)
\end{equation}
where $I$ is the intensity observed at angle $\psi$ from the radial vector, $I_0$ is the radial intensity, and $u$ is the limb-darkening coefficient.
\begin{figure}[!p]
\centerline{\includegraphics[width=\linewidth]{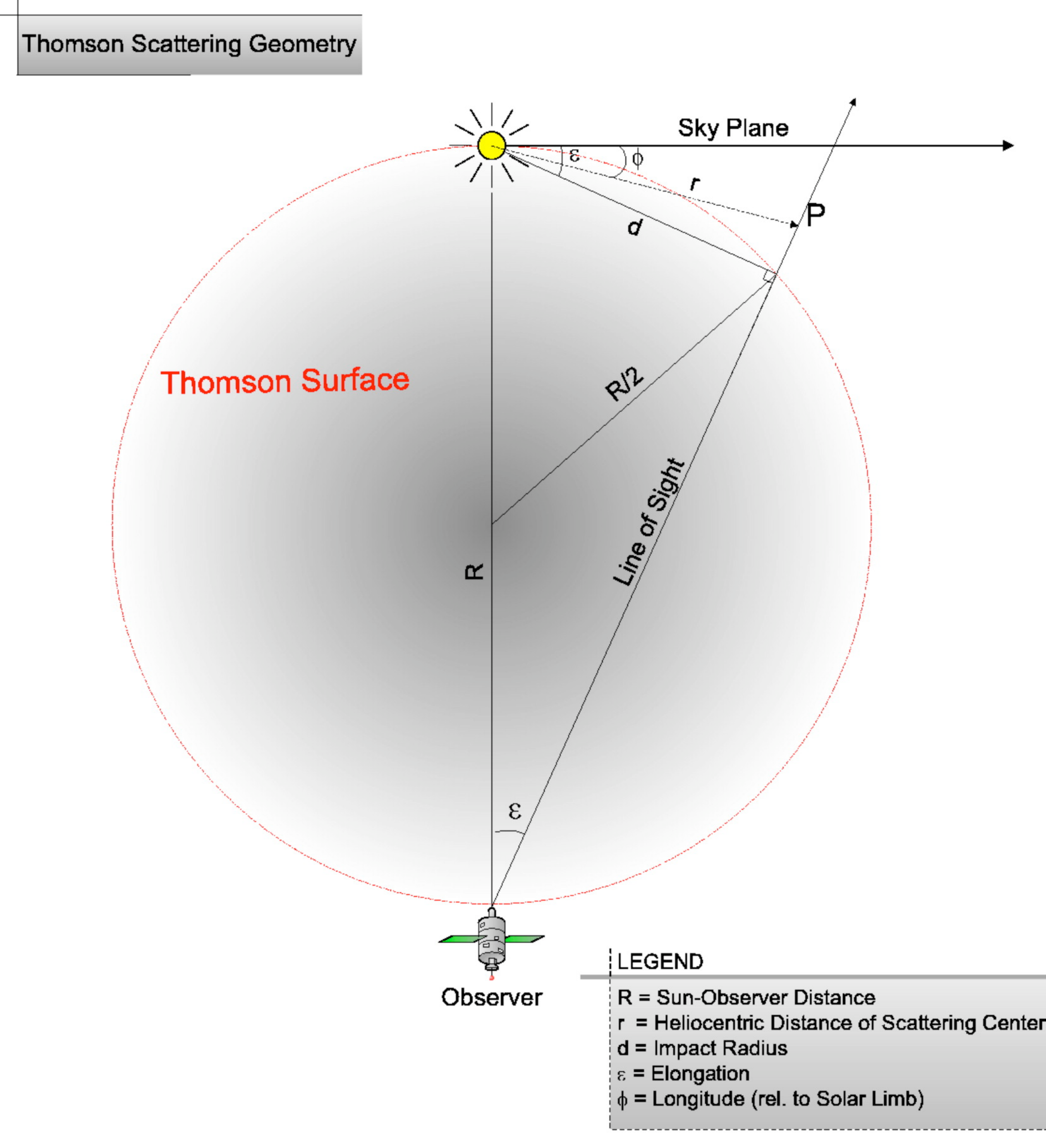}}
\caption{Schematic of the Thomson surface, being the sphere of all points which are located at an angle of 90$^{\circ}$ between the Sun and the observer, reproduced from \citet{2006ApJ...642.1216V}. An example line-of-sight is shown for an electron at point $P$, with radial distance $r$ from the Sun, at longitude $\phi$ relative to the solar limb.}
\label{thomson_sphere}
\end{figure}
\newline
\indent Describing the Thomson sphere as the locus of all points that make an angle of $\chi=90^{\circ}$ between the line-of-sight and the vector from the Sun to the scattering point $P$ (Figure~\ref{thomson_sphere}), the total intensity of scattered light is governed by three terms:
\begin{enumerate}
\item The scattering efficiency which is minimised on the Thomson sphere.
\item The incident intensity which is maximised on the Thomson sphere since that is where the line-of-sight is closest to the Sun.
\item The electron density in the scattering region which is maximised on the Thomson sphere since the solar wind density drops off with radial distance from the Sun.
\end{enumerate}
The combination of these effects makes the Thomson sphere an important consideration when interpreting CME observations, especially out through the wide-angle fields-of-view of the heliospheric imagers on the STEREO spacecraft. \citet{2006ApJ...642.1216V} note the following:
\begin{itemize}
\item CMEs that propagate along the solar limb and appear bright in near-Sun coronagraphs are unlikely to be detectable further out in the heliosphere.
\item Frontside events are always brighter than their backside counterparts, and ones at intermediate angles will exhibit approximately constant levels of brightness over a wide range of heliocentric distances.
\item The sky-plane assumption holds well for brightness observations out to at least $\sim$\,70~R$_{\odot}$.
\end{itemize}
If an expanding CME front moves along or crosses the Thomson sphere during its propagation, its observed brightness can change with regard to the changing location of its intersection with the sphere, as a result of the discussion above. This has implications for how the observed CME kinematics and morphology may be affected, certainly at large elongations from the Sun, and how CMEs may look quite distinct from different observers' points of view.
\newline
\indent An example of how the CME brightness can change due to the projection effects of an image and the consideration of Thomson scattering geometry is shown in Figure~\ref{fluxrope_intensity}. It is based upon a hollow, unkinked, flux rope model with an electron density profile that is peaked toward the outer surface of the flux rope, described in \citet{2000ApJ...533..481C}. To obtain the synthetic coronagraph image, the electron density is integrated in the direction perpendicular to the plane-of-sky weighted by the angular dependence of Thomson scattering. The line-of-sight integrated electron density is greatest where the helical lines bend around the toroidal surface or where they bunch up. Thus the most prominent feature of the synthetic image is the bright outer rim, and the projection of the interior of the flux rope conveys the relatively dark cavity (no prominence material is included hence there is no bright core). In reality, non-linear densities and kinked magnetic field topologies add to the complexities of these observations and increase the difficulties of tracking CMEs as they propagate and evolve through sequences of images.

\begin{figure}[!t]
\centerline{\includegraphics[scale=0.75, trim=0 0 80 540]{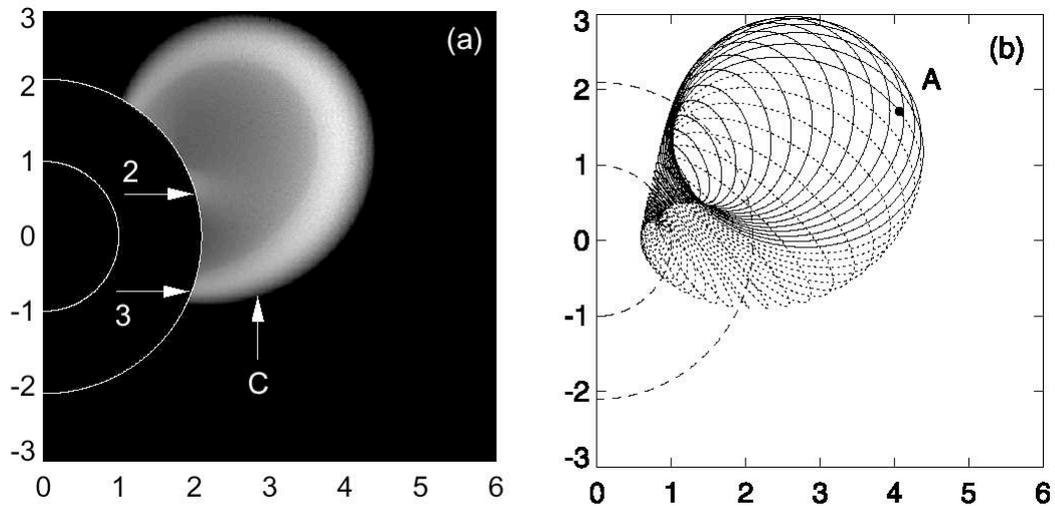}}
\caption{(a) Synthetic coronagraph image of the model flux rope, deduced from the integrated line-of-sight electron density, including a consideration of the Thomson scattering geometry. (b) The flux rope model represented by a mesh of helical field lines. Point A marks the true apex of the model. The axes are in units of R$_{\odot}$. Reproduced from \citet{2000ApJ...533..481C}.}
\label{fluxrope_intensity}
\end{figure}

\subsubsection{CME Kinematics \& Morphology}

\begin{figure}[!t]
\centerline{\includegraphics[width=\linewidth]{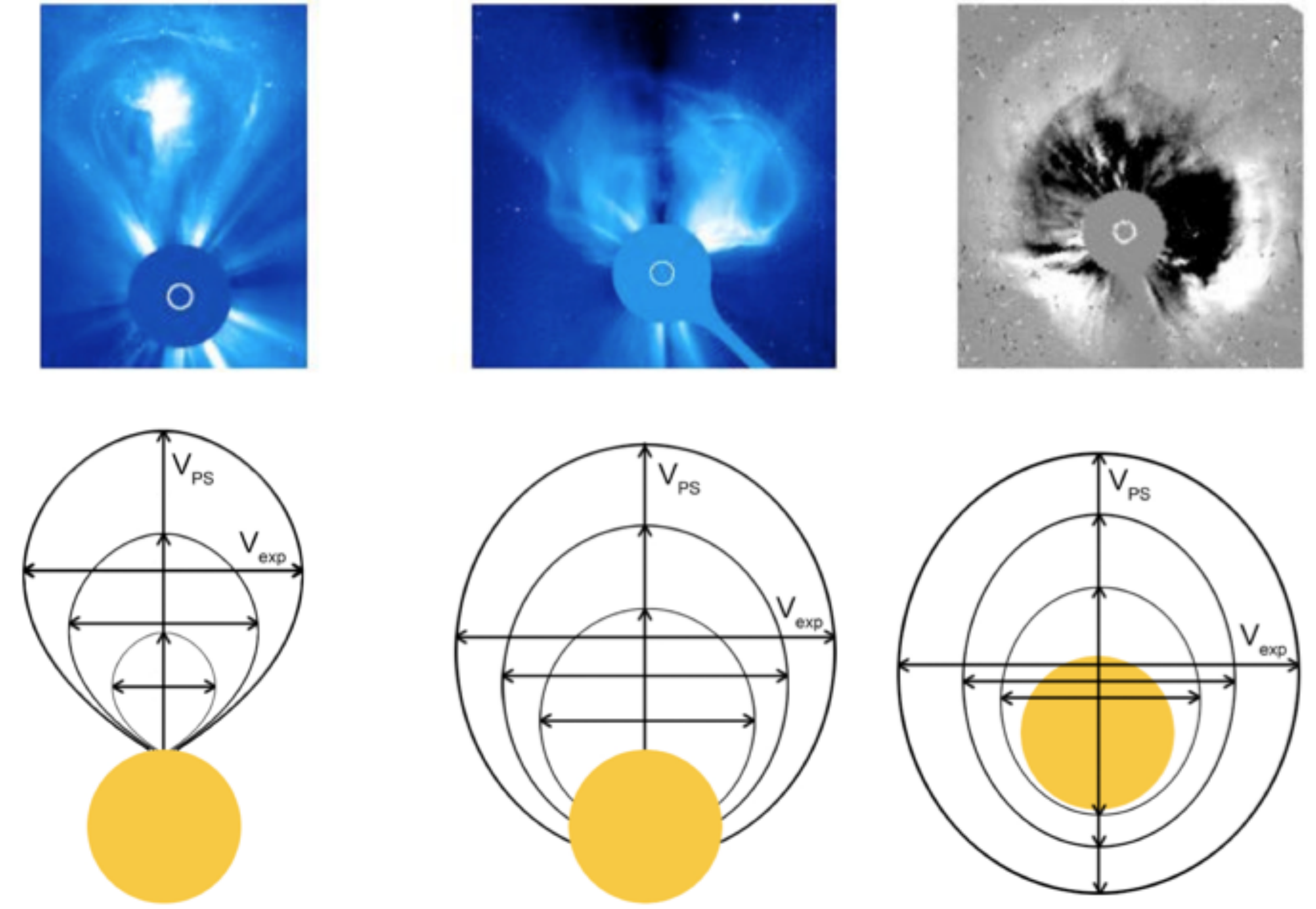}}
\caption{Three example CME observations (off the limb, partial halo, and full halo) and schematics showing how measurements of the plane-of-sky velocity $V_{PS}$ and expansion velocity $V_{exp}$ are skewed by projection effects, reproduced from \citet{2005AnGeo..23.1033S}.}
\label{schwenn_proj}
\end{figure}

A large number of CMEs have been studied since the advent of space-borne coronagraph observations, and they show speeds varying from tens up to a few thousand kilometres per second. Many of these exhibit a general multiphased kinematic evolution: CMEs tend to have an initial rise phase with possible high acceleration, and a subsequent constant-velocity cruise phase with another possible low acceleration or deceleration in their continued propagation through the heliosphere. This initially led \citet{1999JGR...10424739S} to distinguish CMEs as either `gradual' if their initial acceleration is low, or `impulsive' if it is high. However, statistics on a large sample of events do not show such a clear distinction but do indicate that slow CMEs tend to result from prominence lift-offs or streamer blowouts and speed up to the solar wind speed, while fast CMEs tend to result from flares and active regions and slow down to the solar wind speed \citep{2002ApJ...581..694M, 2000GeoRL..27..145G, 2003JGRA..108.1039G}. Statistical analyses can provide a general indication of CME properties, for example \citet{2006ApJ...649.1100Z} study 50 CMEs and find an average acceleration of 330.9~m~s$^{-2}$ with an average duration time of 180~minutes, and \citet{2000GeoRL..27..145G} study 28 CMEs and derive a formula for their acceleration $a$ related to their initial speed $u$ by $a=1.41-0.0035u$. However, plane-of-sky projection effects mean the measured kinematics are not representative of the true CME motion (Figure~\ref{schwenn_proj}), with \citet{2007ApJ...657.1117W} deducing that the error in CME leading-edge measurements grows roughly with the square of the distance from Sun centre within the first few solar radii and then varies approximately with the square root of the distance past $\sim$\,5~R$_{\odot}$. In an effort to overcome plane-of-sky effects, \citet{2003AdSpR..32.2637D} use a sample of 57 limb CMEs to derive an empirical relationship between their radial and expansion speeds as $V_{rad}=0.88V_{exp}$, and \citet{2005AnGeo..23.1033S} similarly use 75 events to derive a formula for their transit time to Earth $T_{tr}= 203-20.77 \ln \left( V_{exp} \right) $. However, \citet{2007A&A...469..339V} show the inherent difficulties in performing and trusting such corrections for CME projection effects. This means individual CMEs must be studied with rigour in order to satisfactorily derive the kinematics and morphology to be compared with theoretical models. Examples of such rigourous analyses are seen in the following kinematic studies. 
\begin{figure}[!t]
\centerline{\includegraphics[scale=0.4]{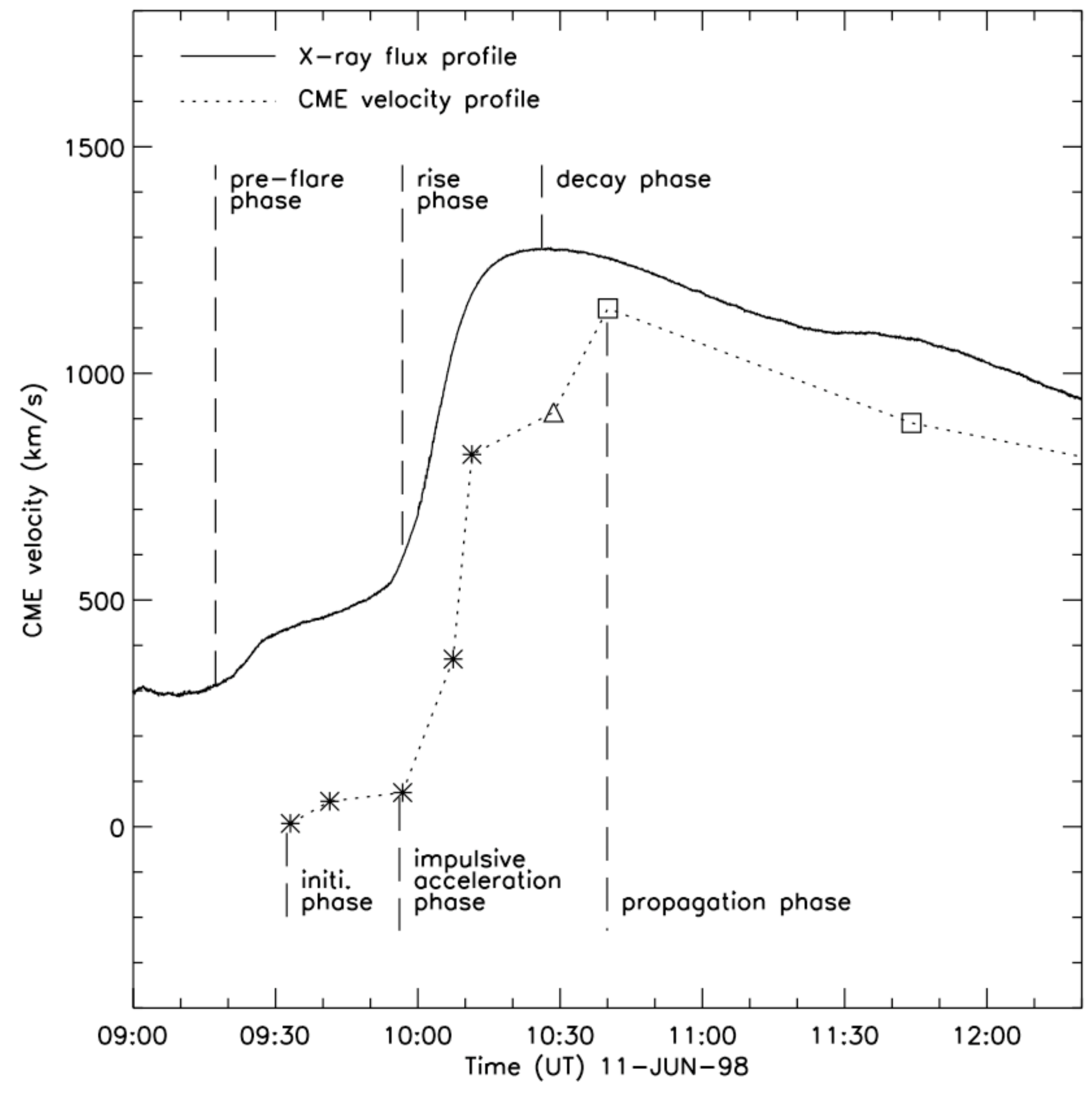}}
\caption{The CME velocity profile, and associated soft X-ray flare profile, for the event on 11 June 1998, reproduced from \citet{2001ApJ...559..452Z}. The profiles indicate a three-phase scenario of CME evolution: initiation, impulsive acceleration, and propagation. The datapoints are from LASCO/C1 (\emph{asterisks}), C2 (\emph{triangles}) and C3 (\emph{squares}).}
\label{zhang_kins}
\end{figure}
\begin{figure}[!t]
\centerline{\includegraphics[scale=0.25]{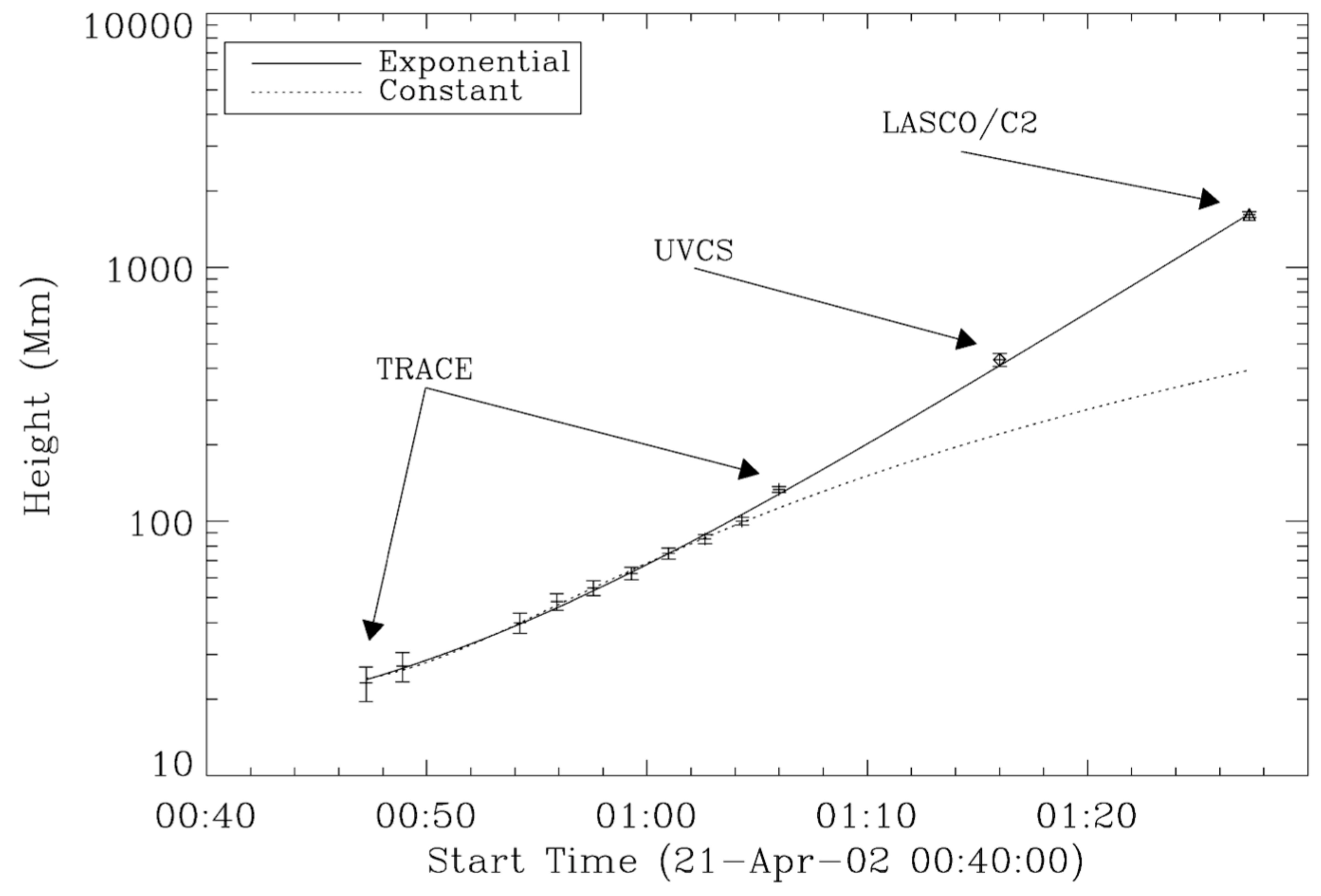}}
\caption{Height-time evolution of a CME observed with TRACE, UVCS and LASCO/C2 on 21 April 2002, reproduced from \citet{2003ApJ...588L..53G}. The datapoints have $\pm$\,5~pixel errorbars, and the best fits for an exponential varying and constant acceleration are plotted for comparison.}
\label{gallagher_inset}
\end{figure}
\begin{figure}[!p]
\centerline{\includegraphics[trim=0 70 0 80, scale=0.8]{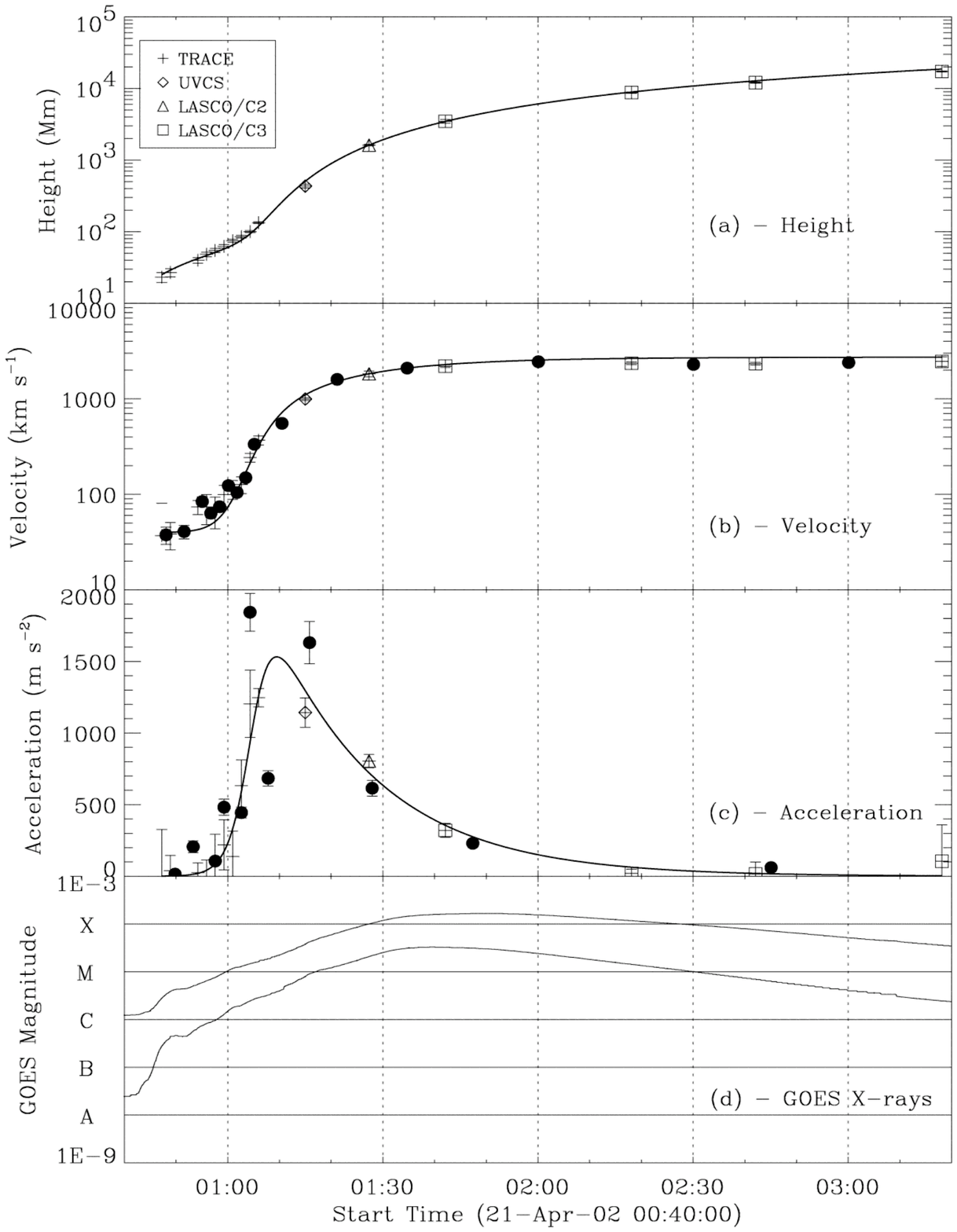}}
\caption{(a) Height-time, (b) velocity and (c) acceleration profiles for the CME on 21 April 2002, and (d) the GOES-10 soft X-ray flux for the associated X1.5 flare during the interval 00:47\,--\,03:20~UT, reproduced from \citet{2003ApJ...588L..53G}. A three-point difference scheme is used on the data points, and a first-difference scheme is plotted with filled circles. The solid line is the best fit of the exponentially increasing and decreasing acceleration of Equation~\ref{double_exp}.}
\label{gallagher_kins}
\end{figure}
\newline
\indent \citet{2001ApJ...559..452Z} study the temporal relationship between CMEs and flares, with an emphasis on the three-phased scenario of CME evolution: initiation, impulsive acceleration, and propagation (Figure~\ref{zhang_kins}). The height range of $\sim$\,1\,--\,3~R$_{\odot}$ was not possible to observe previous to the launch of the SOHO/LASCO suite. They obtained height-time measurements from running difference images, and compare the derived CME velocity profile with the soft X-ray flux of its associated flare, revealing a strong connection between the two phenomena.
\newline
\indent Following the loss of the innermost LASCO/C1 coronagraph, \citet{2003ApJ...588L..53G} include the use of low corona EUV observations for a CME on the 21 April 2002 in order to track a CME from the very low corona out to almost 30~R$_{\odot}$. They consider the separate sets of observations collectively in their fitting, which is not strictly appropriate since the observations are due to different mechanisms of emission from hot plasma in EUV images and Thomson scattering of light in coronagraph images. Following \citet{2002GeoRL..29j..41A} they first consider the simple case of constant acceleration in the early CME evolution (Figure~\ref{gallagher_inset}), of the form:
\begin{equation}
h(t) \; = \; h_0 + v_0t + \frac{1}{2}at^2
\end{equation}
and find that the best-fit to the data points does not adequately represent the observed profile. They then consider an exponentially varying acceleration, of the form:
\begin{equation}
h(t) \; = \; h_0 + v_0t + a_0 \tau^2 \exp \left(t/\tau\right)
\end{equation}
which does provide an acceptable fit to the datapoints. They track the CME front through running-difference images and measure its height-time profile from $\sim$\,20~Mm to over 10$^4$\,~Mm above the Sun's surface, almost 30~R$_{\odot}$ (Figure~\ref{gallagher_kins}), resulting in a rising and falling acceleration. They model this resulting acceleration profile with a combined function of exponentially increasing and decreasing terms:
\begin{equation}
a(t) \; =\; \left[ \frac{1}{a_r \exp \left(t/\tau_r\right)}+\frac{1}{a_d\exp\left(-t/\tau_d\right)}\right]^{-1}
\label{double_exp}
\end{equation}
where $a_r$ and $a_d$ are the initial accelerations, and $\tau_r$ and $\tau_d$ the $e$-folding times for the rise and decay phases, and show that the function provides a close approximation to the trend in the datapoints (for completeness they show both a three-point difference and first-difference scheme). They determine an early acceleration peak of $\sim$\,1,500~km~s$^{-1}$ and show it to correspond to the duration of the soft X-ray rise phase of the associated X1.5 flare, having implications for either the thermal blast model, or a magnetically-dominated process such as reconnection, causing the eruption. Either way they highlight the importance of low coronal observations, and precise measurements, for revealing different regimes of CME propagation.
\begin{figure}[!p]
\centerline{\includegraphics[width=\linewidth]{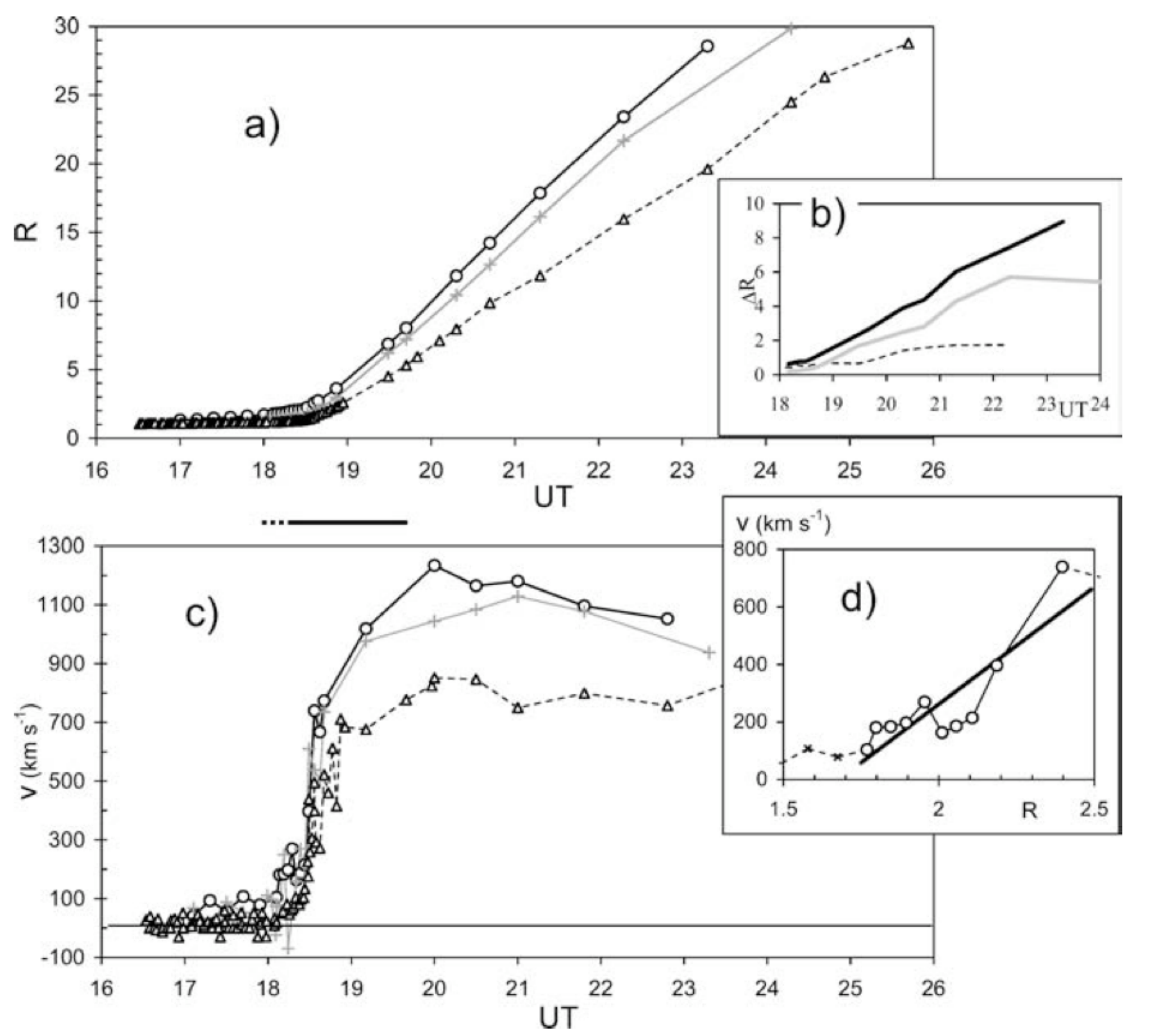}}
\caption{The kinematics of the CME leading edge ({\it circles}), cavity ({\it pluses}) and prominence ({\it triangles}) of the 15 May 2001 event, reproduced from \citet{2004SoPh..225..337M}. (a) shows the height-time profiles. (b) shows the distance between the leading edge and the prominence ({\it thick line}), the cavity and the prominence ({\it thin line}), and the leading edge and the cavity ({\it dashed line}). (c) shows the velocities determined by forward-difference technique upon the smoothed height-times. (d) shows the onset of acceleration against height, with a straight line fit. The horizontal bar between (a) and (c) indicates the period of the soft X-ray burst ({\it dotted} - precursor, {\it full} - main rise).}
\label{maricic_kins}
\end{figure}
\newline
\indent \citet{2004SoPh..225..337M} present a study of the initiation and development of a limb CME on 15 May 2001, tracking its leading edge, cavity and associated prominence from 0.32~R$_{\odot}$ out to $\sim$\,30~R$_{\odot}$ (Figure~\ref{maricic_kins}). They distinguish a pre-acceleration characterised by the slow rising motion of the prominence, suggestive of an evolution of the system through a series of quasi-equilibrium states. They offer numerous possible explanations for this: a slow shearing or merging motion of the arcade footpoints \citep[e.g.,][]{1990SoPh..130..399P, 2004ApJ...602..422L}, twisting of the embedded flux rope \citep{1990SoPh..129..295V}, emerging azimuthal flux \citep{2003JGRA..108.1410C}, mass loss from the prominence body \citep{1990SoPh..129..295V}, etc. This is followed by a rapid acceleration onset of the prominence, 380\,$\pm$\,50~m~s$^{-2}$, simultaneous with the CME leading edge acceleration, 600\,$\pm$\,150~m~s$^{-2}$. This simultaneity rules out the scenarios in which the prominence motion is merely a consequence of the disruption of the overlying magnetic structure, or that the prominence eruption itself drives the upper parts of the system (at least for this event, and the authors suggest a careful inspection of a larger sample of events). There is also a simultaneous soft X-ray burst which they deem an `acceleration precursor' as postulated in the catastrophic evolution of a flux rope/arcade system \citep{2000JGR...105.2375L}. A possible gradual deceleration of the CME in its later stages is also noted (-23\,$\pm$\,1~m~s$^{-2}$ for the leading edge) and attributed to the possible effects of aerodynamic drag in the solar wind discussed in \citet{1996JGR...101.4855C}, \citet{2001SoPh..202..173V} and \citet{2004A&A...423..717V}. \citet{2004SoPh..225..337M} also measure the relative height difference between the CME leading edge, cavity and prominence, and report a kind of self-similar expansion of the event.
\begin{figure}[!t]
\centerline{\includegraphics[width=\linewidth]{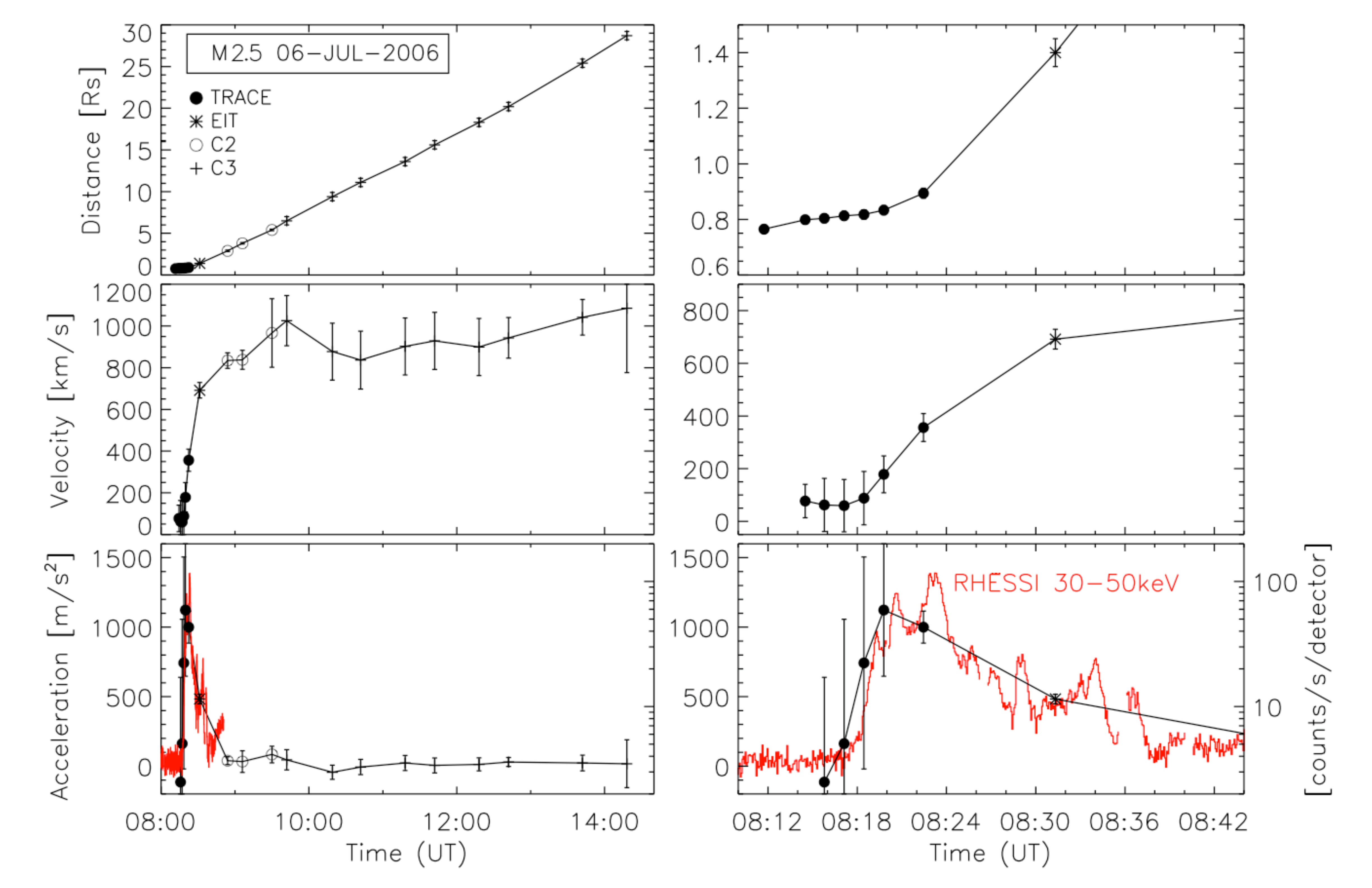}}
\caption{Kinematics of a CME observed with TRACE, EIT and LASCO/C2 and C3 on 6 July 2006, plotted over the full time range on the left, and zoomed in for the duration of the acceleration on the right, reproduced from \citet{2008ApJ...673L..95T}. The bottom panels include the hard X-ray flux of the associated flare (shown in red), peaking simultaneously with the acceleration.}
\label{temmer_kins}
\end{figure}
\newline
\indent \citet{2008ApJ...673L..95T} study the CME kinematics and associated flare flux profiles for two fast events of 17 January 2005 and 6 July 2006 (the latter is shown in Figure~\ref{temmer_kins}). Using running-difference techniques they track the rising loops of the flares and consider them as the early height-time profile of the subsequent CME eruption. This results in a clear early acceleration peak for each event, of $\sim$\,4,400~m~s$^{-2}$ and $\sim$\,1,100~m~s$^{-2}$ respectively. They find that the peaks of the hard X-rays correspond with the peak accelerations of the CMEs, indicating a strong feedback relationship indicative of magnetic reconnection occurring in a possible current sheet behind the CME to drive the eruption under the dominant Lorentz force.
\newline
\indent More recently \citet{2010A&A...516A..44L} performed the same type of analysis on two CMEs in order to specifically test the kinematics of each against the predictions of theoretical models (catastrophe, breakout and toroidal instability). Tracking the CME front through running-difference images, the events of 17 December 2006 and 31 December 2007 were found to have peak accelerations of $\sim$\,60~m~s$^{-2}$ and $\sim$\,1,500~m~s$^{-2}$ within the first $\sim$\,3~R$_{\odot}$ height range, simultaneous with the soft X-ray flux. The authors show the difficulty in distinguishing any single model as a basis for the eruption, since the predicted fits all fare relatively well in producing a profile within the scatter and error of the measured kinematics. The authors also acknowledge that the data is a two-dimensional projection of actual motion, and three-dimensional kinematics of CMEs are necessary for improved accuracy.
\newline
\indent The conclusions of such studies rely heavily on the kinematic uncertainties, since they are vital for providing a distinction between different models that attempt to fit the profiles. The above discussion highlights the importance of the image resolution, cadence and projection effects when performing CME analyses. It is also apparent that user-specific biases and different numerical methods for determining the kinematics and morphology can affect the resulting profiles and hence their interpretation. With these considerations in mind, the current aims of the community have been to develop methods for overcoming both the biases of individual users, and single viewpoint observations, in order to robustly determine the kinematics and morphology of CMEs with the greatest possible precision. These aims motivated the launch of the STEREO mission to provide twin-viewpoint observations of CMEs, with low-cadence observations of the inner corona, and off-limb imaging suites with fields-of-view extending out to the orbit of Earth at 1~AU.

%CME expansions?

%3D efforts?

\section{Outline of Thesis}

The work presented in this thesis improves the understanding of the kinematics and morphology of CMEs as they propagate through the solar corona. To date the quantification of these properties has been subject to various sources of error, most notably as a result of the innate difficulties in tracking CMEs with traditional image processing techniques. Moreover the projected two-dimensional nature of coronagraph observations has been a persistent source of error that scientists have striven to overcome when studying CMEs. This thesis outlines new methods of multiscale image processing and ellipse characterisation of the CME front in coronagraph observations so as to reduce the error on the derived kinematic and morphological parameters. This is extended to stereoscopic image data whereby an elliptical tie-pointing methodology for reconstructing the CME front in three-dimensions leads to a study of its true kinematics and morphology as it propagates into the heliosphere.
\newline
\indent Chapter 2 details the space-borne instrumentation used in this work for observing CMEs through the solar corona and heliosphere. Chapter 3 discusses the current CME catalogues in use and their limitations, leading to the implementation of new multiscale techniques and an ellipse characterisation for studying CME propagation. The efforts to automate this process for the development of a new cataloguing database are also discussed. These methods are the basis of the studies undertaken in Chapter 4, where a selection of CME events are presented and their kinematics and morphologies discussed in light of theory. Chapter 5 presents an important extension of these methods to perform a three-dimensional reconstruction of a CME front, which overcomes the issues of projection effects and so provides insight into the true kinematics and morphology of the eruption. A detailed discussion of the results and conclusions drawn from this work are also presented. Finally Chapter 6 presents the main conclusions of the thesis and details possible future work that could follow on from these new developments.	

\chapter{Instrumentation}
\label{chapter:instrumentation}

CMEs were initially observed in the early 1970s with the launch of the first space-borne coronagraph onboard the seventh Orbiting Solar Observatory (OSO~7) providing daily white-light coronal images with a field-of-view 2.8\,--\,10~R$_{\odot}$, observing about a dozen CMEs from 1971\,--\,1974 \citep{1975Koomen, 1972BAAS....4R.394T}. Skylab, a U.S. space station launched in 1973, housed a coronagraph with field-of-view 1.5\,--\,6~R$_{\odot}$ that imaged the corona every 6\,--\,8~hours and conclusively established these transient ejections as a common occurrence, observing $\sim$\,100 in 1973/74 \citep{1974ApJ...187L..85M}. Following Skylab, several more coronagraphs were flown: SOLWIND on satellite P78-1 observed over 1,500 CMEs from 1979\,--\,1985 \citep{1980ApJ...237L..99S}; the High Altitude Observatory Coronagraph/Polarimeter onboard Solar Maximum Mission (SMM) observed $\sim$\,1,350 CMEs from 1980\,--\,1989 \citep{1980SoPh...65...91M}; the Large Angle Spectrometric Coronagraph suite \citep[LASCO;][]{1995SoPh..162..357B} onboard the Solar and Heliospheric Observatory  \citep[SOHO;][]{1995SoPh..162....1D} has observed thousands of CMEs from 1995 to present; and the COR1/2 coronagraphs of the Sun-Earth Connection Coronal and Heliospheric Imaging suite \citep[SECCHI;][]{2008SSRv..136...67H} onboard the Solar Terrestrial Relations Observatory \citep[STEREO;][]{2008SSRv..136....5K} have been observing CMEs from 2006 to present. Data from the instruments of SOHO/LASCO and STEREO/SECCHI are used in the research presented throughout this thesis.

\section {SOHO/LASCO}

\begin{figure}[!t]
\centerline{\includegraphics[scale=1.25, clip=true, trim=0 0 0 0]{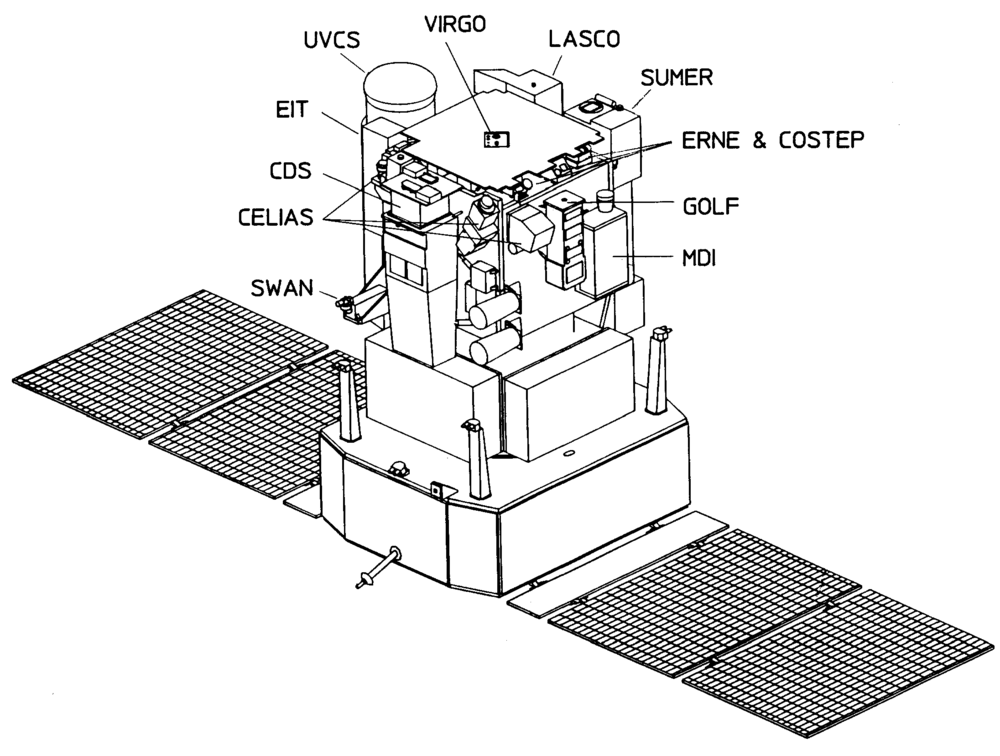}}
\caption{A schematic of the SOHO spacecraft and onboard instrument suites, reproduced from \citet{1995SoPh..162....1D}.}
\label{SOHO}
\end{figure}

SOHO is a joint European Space Agency (ESA) and National Aeronautics and Space Administration (NASA) mission, launched on 2 December 1995 to undertake scientific investigations of (1) helioseismology, the study of the interior solar structure, and (2) the physical processes that account for the heating and acceleration of the solar wind (or, more broadly, the nature of evolutionary change in the Sun's outer atmosphere). SOHO is situated in orbit about the Lagrangian L1 point approximately 1.5$\times$10$^{6}$~km sunward of the Earth for an uninterrupted view of the Sun. Onboard are twelve complementary science instruments: three helioseismology experiments to probe the Sun's inner structure through measurements of solar oscillations; three solar wind experiments to measure the in-situ properties of the ambient wind (densities, speeds, charge states, etc.); and six telescopes and spectrometers to study the solar disk and atmosphere. A schematic of SOHO and its instrument suites is shown in Figure~\ref{SOHO}.
\par
\begin{figure}[!t]
\centerline{\includegraphics[scale=0.3]{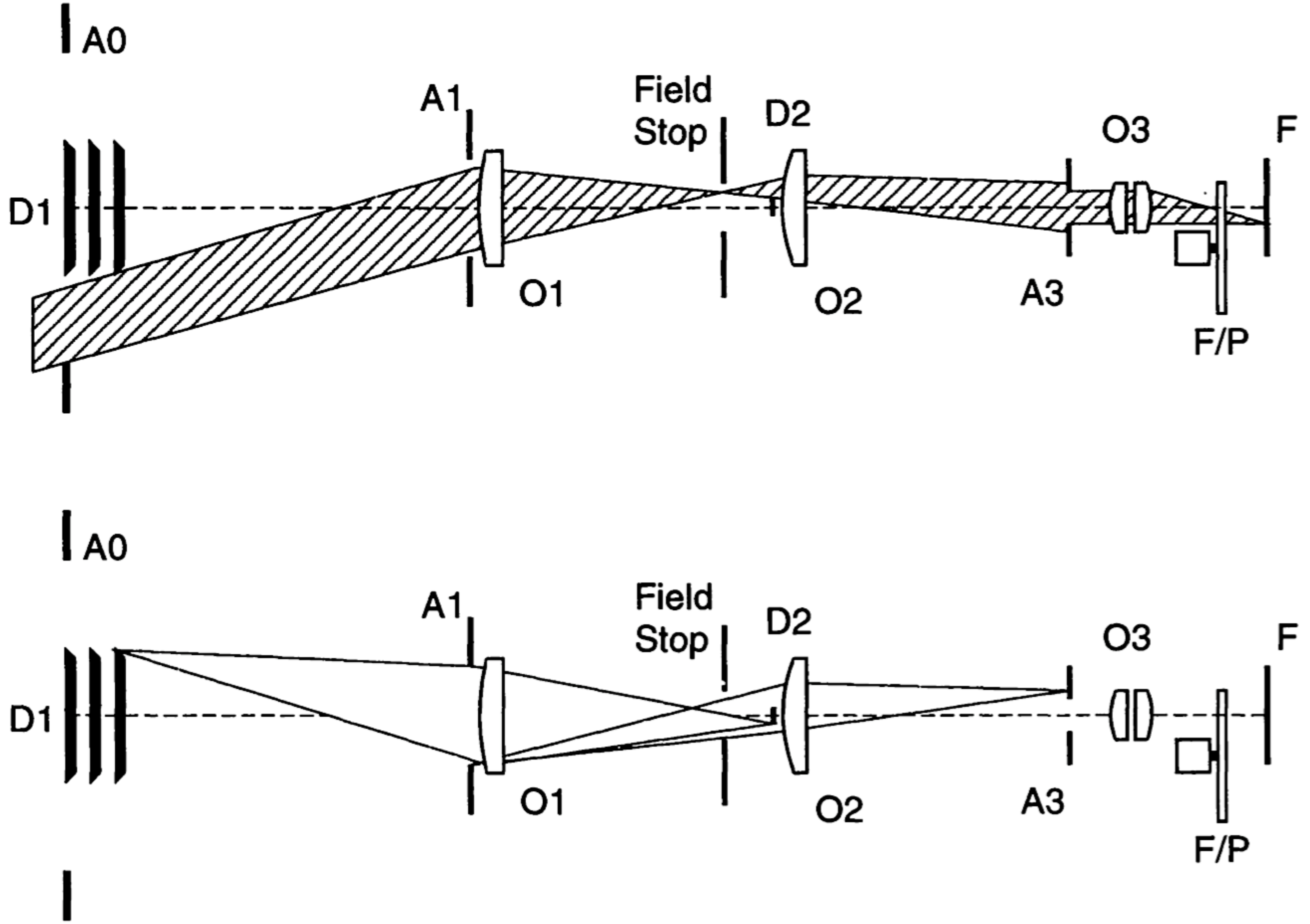}}
\caption{The externally occulted Lyot coronagraph, reproduced from \citet{1995SoPh..162..357B}, showing: front aperture A0 and external occulter D1; entrance aperture A1 and objective lens O1; the field stop; inner occulter D2 and field lens O2; Lyot stop A3 and relay lens O3 with Lyot spot; filter and polariser wheels F/P; and the focal plane F.}
\label{Lyot_external}
\end{figure}

The LASCO instrument suite is a set of three coronagraphs C1, C2 and C3 that image the solar corona from 1.1\,--\,3, 1.5\,--\,6 and 3.7\,--\,30~R$_{\odot}$ respectively (however the C1 coronagraph has not been in operation since 1998 when contact with the SOHO spacecraft was lost for several weeks). The coronagraph was invented by French astronomer Bernard Lyot in 1939 to artificially eclipse the Sun for observing the solar corona. It essentially blocks light rays from the centre of the telescope field-of-view by occulting the solar disk, in order to increase the relative intensity of the surrounding coronal light which is on the order of 10$^6$ times fainter. The externally occulted Lyot coronagraph design of LASCO/C2 and C3 is illustrated in Figure~\ref{Lyot_external}.
\begin{figure}[!p]
\centerline{\includegraphics[width=\linewidth]{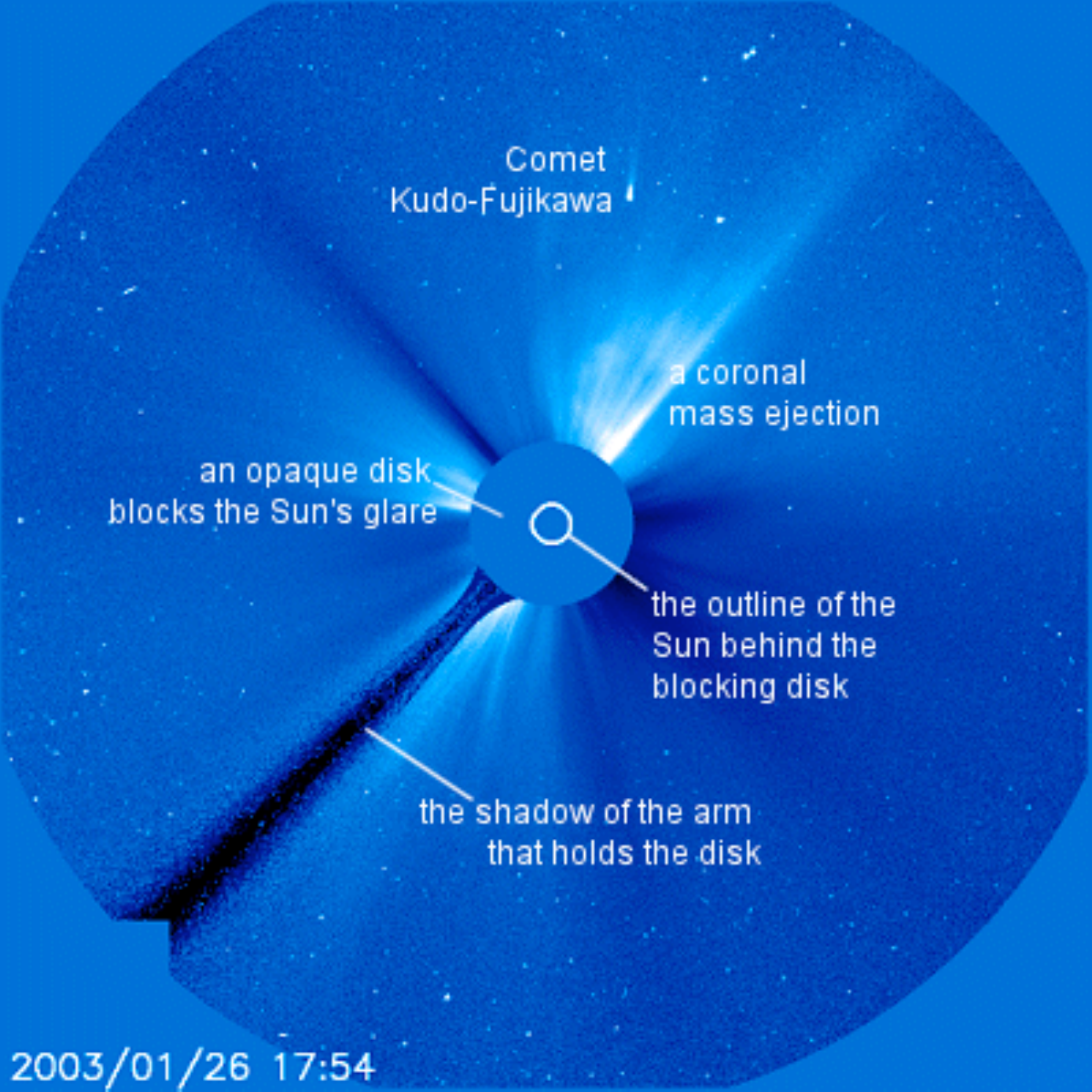}}
\caption{A LASCO/C3 image of the solar corona out to $\sim$\,30~R$_{\odot}$.}
\label{C3}
\end{figure}
The top diagram demonstrates how the optical assembly images the coronal light, while the bottom diagram demonstrates how stray light is suppressed. Light is incident through aperture A0 where the external occulter D1 eclipses the solar disk. The light then enters aperture A1 and is focused by the objective lens O1, through the field stop, onto the inner occulter D2 which anodises the bright fringe of the external occulter. Field lens O2 then collimates the light onto the Lyot stop A3 that intercepts the light rays diffracted off the entrance aperture A1. A relay lens O3 is placed behind A3 to focus the coronal image on to the plane F. O3 contains the Lyot spot for intercepting residual diffracted light from D1 and ghost images created by O1. In front of the focal plane F are the colour filters and linear polarising filters F/P. The colour filters distinguish specific bandpasses of the coronal light, in the ranges 400\,--\,850~nm for C2 and 400\,--\,1050~nm for C3. The polariser wheel is used to obtain total brightness $B$ or polarised brightness $pB$ images through combinations of polariser positions $I_{a}=-60^{\circ}$, $I_{b}=0^{\circ}$, and $I_{c}=60^{\circ}$, according to the equations \citep{1966gtsc.book.....B}:
\begin{align}
\label{eqn:totb}
B\;&=\; \frac{2}{3}(I_{a}+I_{b}+I_{c}) \\
\label{eqn:pb}
pB\;&=\;\frac{4}{3}[(I_{a}+I_{b}+I_{c})^{2}-3(I_{a}I_{b}+I_{a}I_{c}+I_{b}I_{c})]^{1/2}
\end{align}
A CCD is placed at the focal plane F and the final images are 1024\,$\times$\,1024 pixels, subtending an angle of 11.4~arcseconds per pixel in C2, and 56~arcseconds per pixel in C3 (see Section~\ref{sec:CCDs} for CCD details). A sample LASCO/C3 image is shown in Figure~\ref{C3}.

\section{STEREO/SECCHI}

STEREO is the third mission in NASA's Solar Terrestrial Probes program \citep{2008SSRv..136....5K}. It was launched on 25 October 2006, and employs two nearly identical space-based observatories; one ahead of Earth in its orbit, and the other behind, separating at $\pm$\,22$^{\circ}$ each year. This arrangement provides the first ever stereoscopic observations of the Sun and inner heliosphere. The main objectives of STEREO are to:
\begin{itemize}
\item Understand the causes and mechanisms of CME initiation.
\item Characterise the propagation of CMEs through the heliosphere.
\item Discover the mechanisms and sites of energetic particle acceleration in the low corona and the interplanetary medium.
\item Improve the determination of the structure of the ambient solar wind.
\end{itemize}

\begin{figure}[!t]
\centerline{\includegraphics[scale=0.8]{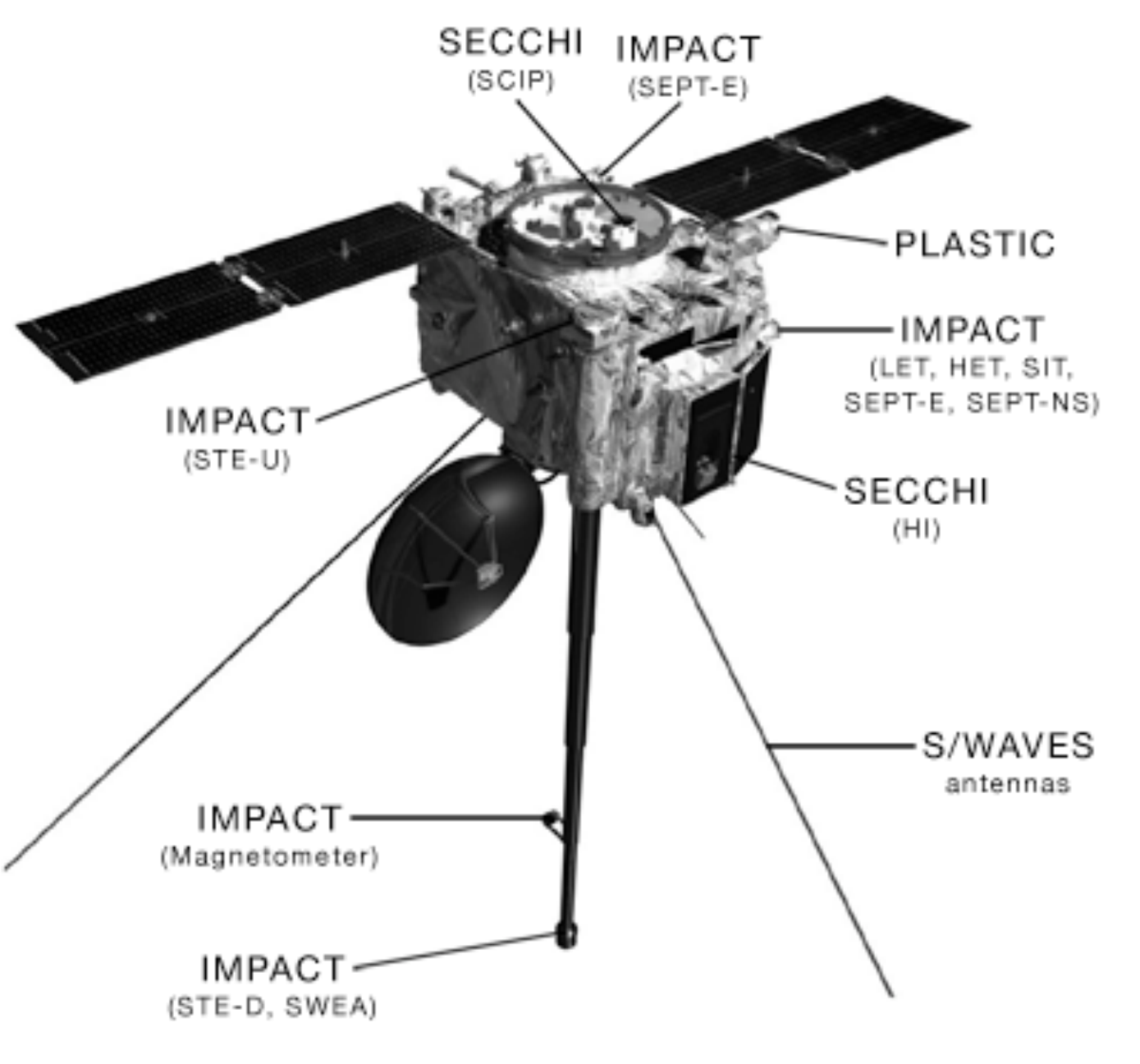}}
\caption{Payload diagram of one of the STEREO spacecraft, indicating the positions of the four instrument suites onboard: Sun-Earth Connection Coronal and Heliospheric Imagers (SECCHI); In-situ Measurements of Particles and CME Transients (IMPACT); Plasma and SupraThermal Ion Composition (PLASTIC); STEREO/WAVES radio burst tracker (SWAVES). 
\newline
\emph{Image credit: stereo.gsfc.nasa.gov}.}
\label{stereo}
\end{figure}
\par
STEREO hosts four instrument suites to achieve this, as illustrated in Figure~\ref{stereo}. The SECCHI suite comprises five scientific telescopes: firstly the Sun Centred Imaging Package (SCIP) consisting of an Extreme Ultraviolet Imager (EUVI) of the solar disk out to 1.7~R$_{\odot}$ and two coronagraphs (COR1/2) with fields-of-view 1.4\,--\,4 and 2\,--\,15~$R_{\odot}$; and secondly the Heliospheric Imagers (HI) consisting of two wide-angle visible light imagers positioned on the sides of the STEREO spacecraft for fields-of-view extending out to Earth at 1~A.U. (astronomical unit, based on the distance from the Earth to the Sun which is approximately 1.49$\times$10$^{8}$~km).

\subsection{EUVI}

\begin{figure}[!t]
\centerline{\includegraphics[clip=true, scale=0.25]{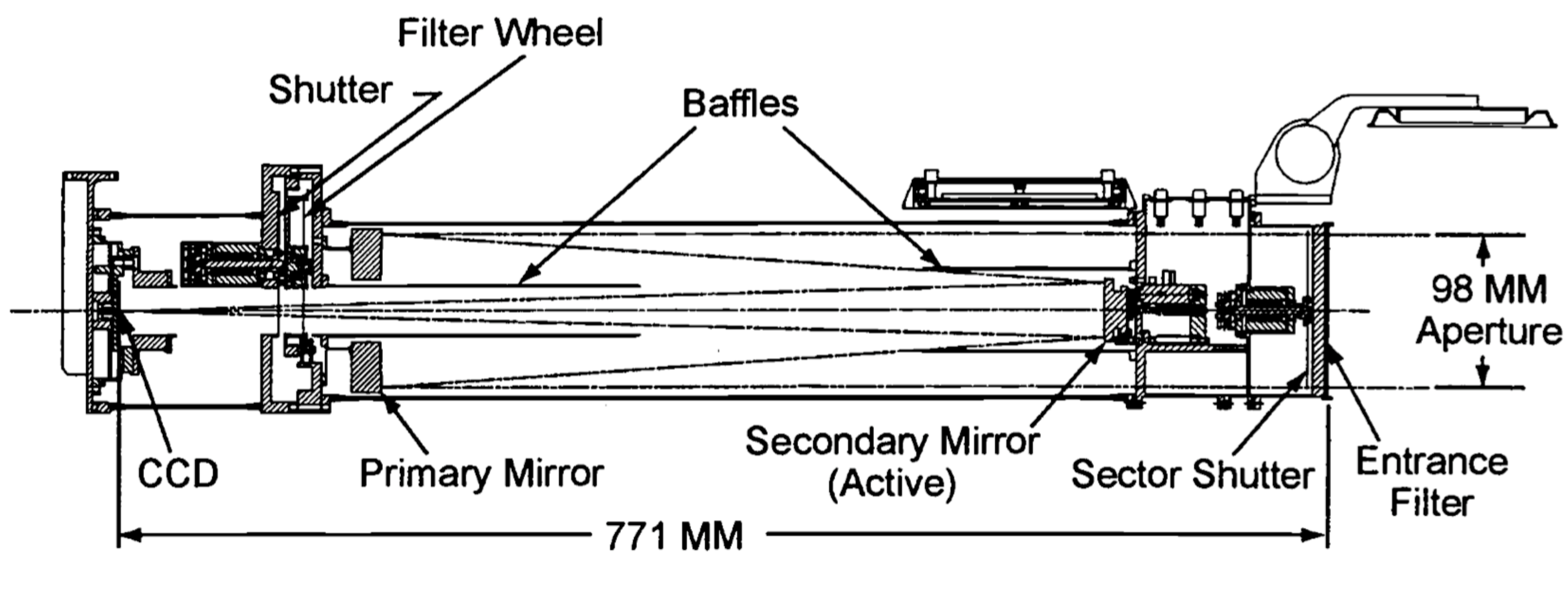}}
\caption{Diagram of the EUVI in the STEREO/SECCHI suite, reproduced from \citet{2000SPIE.4139..259H}.}
\label{euvi_schematic}
\end{figure}

The Extreme Ultraviolet Imager is a normal-incidence Ritchey-Chr\'{e}tien telescope that images the solar disk out to 1.7~R$_{\odot}$ at four wavelengths of emission that span a temperature range of 0.1 to 20~MK \citep{2004SPIE.5171..111W}. Radiation from the Sun enters through a thin aluminium filter of 150~nm thickness that suppresses most of the ultraviolet, visible and infrared wavelengths of light. The radiation  passes one of four quadrants that are each optimised for one of the EUV wavelength lines (listed in Table~\ref{table:euvi_wavelengths}). The primary and secondary mirrors direct the light through a filter wheel that has a redundant thin-film aluminium filter to remove the remainder of the visible and IR radiation. A shutter in the path controls the exposure time, and 2048\,$\times$\,2048 pixel images are produced by the CCD subtending an angle of 1.6~arcseconds per pixel.% (for details on the CCD see Section~\ref{sec:CCDs}). 

\begin{table}[!t] 
\caption{Summary of EUVI wavelengths.} % title of Table 
\centering      % used for centering table 
\begin{tabular}{c c}
\hline\hline                        %inserts double horizontal lines 
Principal emission lines & Wavelength \\ [0.5ex] % inserts table 
%heading
\hline                % inserts single horizontal line 
Fe~IX & 172~\AA  \\  % inserting body of the table 
Fe~XII & 194~\AA \\
Fe~XV & 284~\AA \\
He~II & 304~\AA \\        % [1ex] adds vertical space 
\hline     %inserts single line 
\end{tabular} 
\label{table:euvi_wavelengths}  % is used to refer this table in the text 
\end{table}

\subsection{COR1}

\begin{sidewaysfigure}[!p]
\centerline{\includegraphics[width=\linewidth]{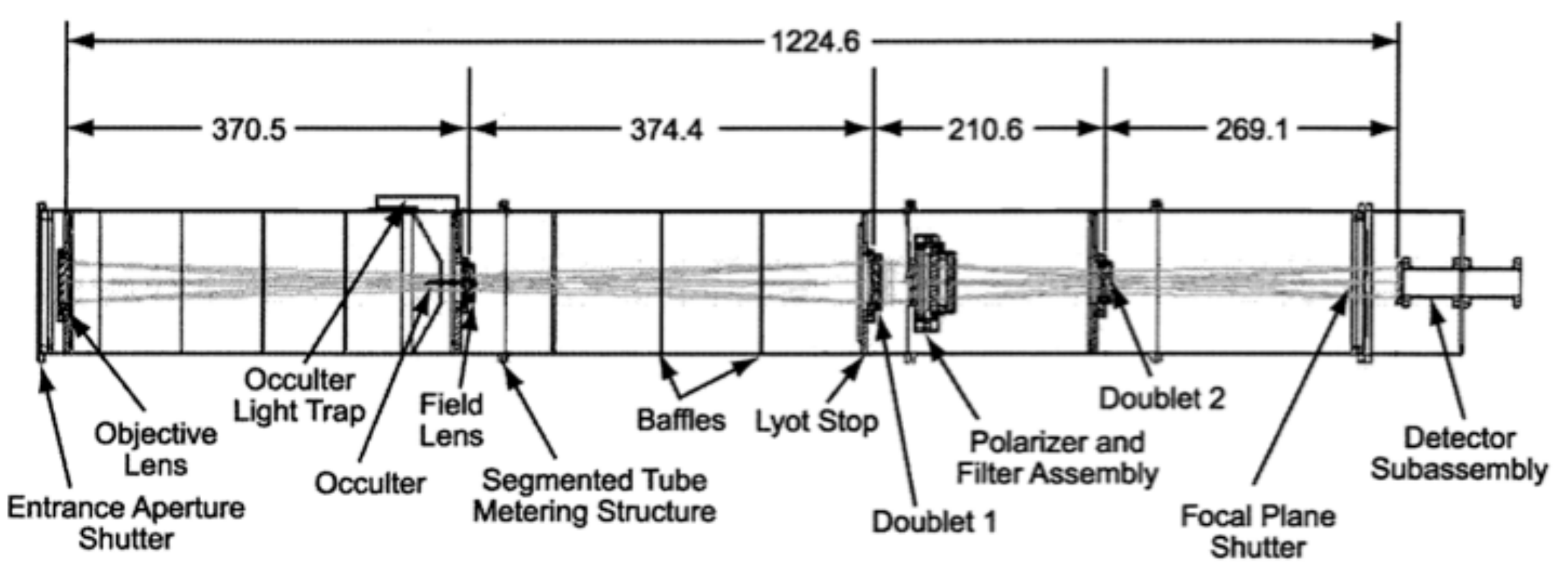}}
\caption{Diagram of the COR1 coronagraph in the STEREO/SECCHI suite, reproduced from \citet{2000SPIE.4139..259H}. The measurements are specified in millimetres.}
\label{cor1_schematic}
\end{sidewaysfigure}

COR1 is a classic Lyot internally occulting refractive coronagraph \citep{2003SPIE.4853....1T}. Light enters through the front aperture of the telescope and is focused by the objective lens onto the occulter, with a series of baffles in place to minimise scattering of light within the telescope (Figure~\ref{cor1_schematic}). The occulter is cone shaped to reject light from the centre of the field-of-view into a surrounding light trap. The field lens focuses the rest of the light down the telescope to the Lyot stop which removes light diffracted by the edge of the front aperture. A Lyot spot is also glued to the doublet lens immediately behind the Lyot stop in order to remove any ghosting of the objective lens. Two doublet lenses focus the light onto the CCD detector, with a bandpass filter 10~nm wide (centred on the H$\alpha$ line at 656~nm) and a linear polariser in between them. To extract both total brightness $B$ and polarised brightness $pB$ images, three sequential images are taken with polarisations of $I_{a}=-60^{\circ}$, $I_{b}=0^{\circ}$, and $I_{c}=60^{\circ}$, and combined using Equations \ref{eqn:totb} and \ref{eqn:pb}. The cut-on frequency (at 350~nm) of light through COR1 is set by the transmission of the BK7-G18 glass in the objective lens, and the cut-off frequency (at 1100~nm) is set by the band gap of the silicon CCD detector. The final images are 2\,$\times$\,2 binned onboard to 1024\,$\times$\,1024 pixels, subtending an angle of 7.5~arcseconds per pixel. The field is unvignetted except for a small area around the edge of the occulter and near the field stop in the corner of the images. The average radial brightness profile for both instruments is well below 10$^{-6}$~B/B$_{\odot}$ though some discrete ring-shaped areas of increased brightness in the COR1-Behind instrument are caused by features on the front surface of the field lens. %Software controlled heaters keep the instrument within the 0\,--\,40$^{\circ}$~C operational temperature range. Specialised composite coatings of oxides over silver with very low solar absorptivties are deposited onto exposed surfaces around the entrance aperture, door assembly, and objective lens holder, to help manage the intense solar fluxes experienced during the mission.

\subsection{COR2}

\begin{sidewaysfigure}[!p]
\centerline{\includegraphics[width=\linewidth]{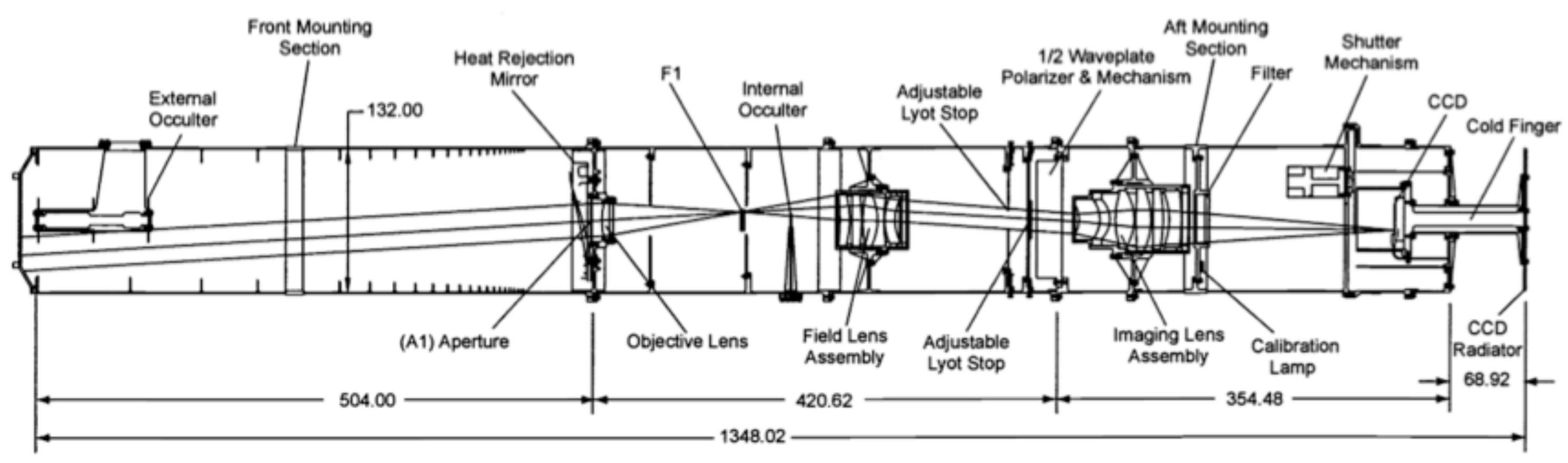}}
\caption{Diagram of the COR2 coronagraph in the STEREO/SECCHI suite, reproduced from \citet{2000SPIE.4139..259H}. The measurements are specified in millimetres.}
\label{cor2_schematic}
\end{sidewaysfigure}

%\begin{figure}[!t]
%\centerline{\includegraphics[clip=true, scale=0.5]{COR2A_image.pdf}}
%\caption{COR2 image from the STEREO-Ahead spacecraft of a coronal mass ejection, with the occulting disk and position of the Sun indicated (courtesy of \emph{STEREO NASA}).}
%\label{cor2a_image}
%\end{figure}

COR2 is an externally occulted Lyot coronagraph, similar to the LASCO/C2 and C3 telescopes (Figure~\ref{cor2_schematic}). An array of internal baffles sits behind the external occulter to reduce stray light entering the telescope, and an internal occulter and Lyot stop minimise diffraction effects. The final images are produced at three different polarisations as in COR1 for creating total brightness and polarised brightness images. The final images are 2048\,$\times$\,2048 pixels, subtending an angle of 14.7~arcseconds per pixel. The image is vignetted throughout the field-of-view, at a level of 40\,--\,50\% around the occulter pylon, and reaching a minimum of 20\% at about 10~R$_{\odot}$ before increasing again towards the image edge.%The instrument surfaces that are exposed to direct sunlight during the mission have a silver composite coating (CCAG) to prevent them overheating. All other surfaces are black anodised to minimise scattered light, and a heat rejection mirror is mounted in front of the objective lens to reflect incident solar radiation back through the A0 aperture and into space.%A COR2 total brightness image of a CME from STEREO-Ahead is illustrated in Fig.~\ref{cor2a_image}.

\subsection{Heliospheric Imagers}

\begin{figure}[!t]
\centerline{\includegraphics[clip=true, scale=0.5]{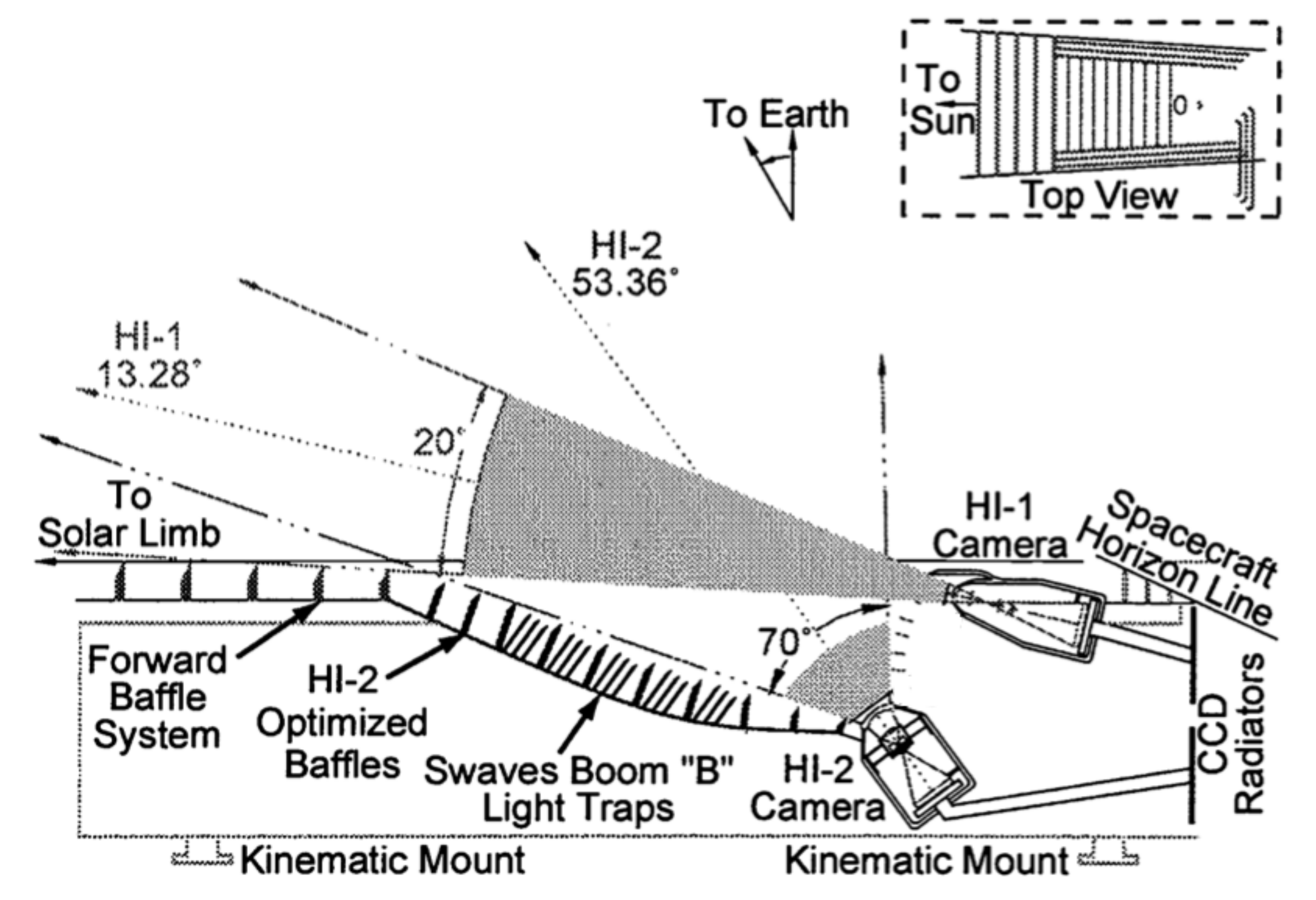}}
\caption{Diagram of the heliospheric imagers HI-1/2 in the STEREO/SECCHI suite, reproduced from \citet{2000SPIE.4139..259H}.}
\label{HI_schematic}
\end{figure}

\begin{figure}[!t]
\centerline{\includegraphics[clip=true, scale=0.35]{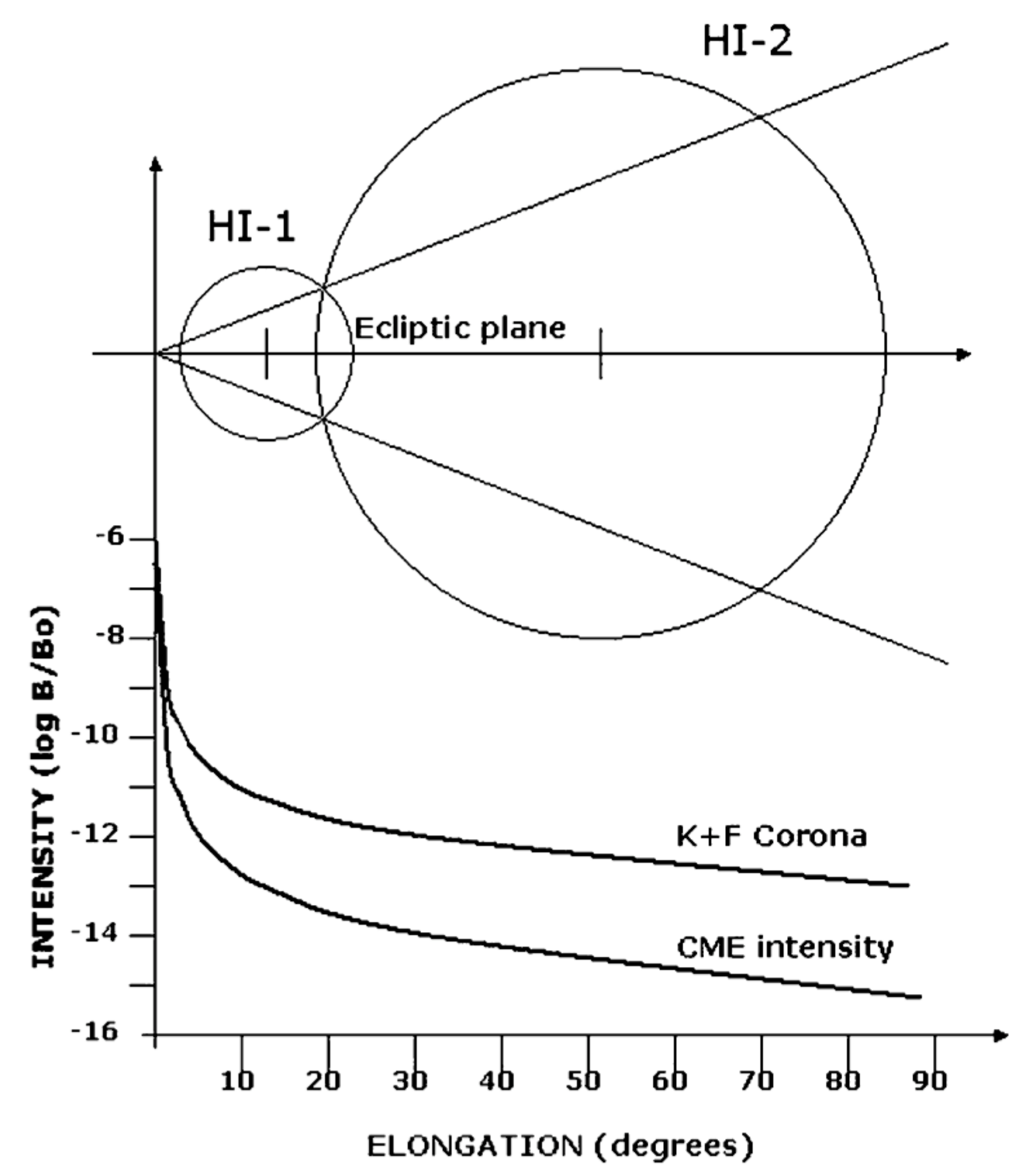}}
\caption{The intensity profile of a CME compared to the K \& F coronae observed at elongations up to 90$^{\circ}$, and the corresponding fields-of-view of the Heliospheric Imagers (HI1/2), reproduced from \citet{2000SPIE.4139..259H}.}
\label{HI_intensity}
\end{figure}

The Heliospheric Imagers \citep[HI1/2;][]{2009SoPh..254..387E} are two small, wide-angle, visible-light camera systems mounted to the side of each STEREO spacecraft to image along the Sun-Earth line from elongations of 4\,--\,88.7$^{\circ}$ (Figure~\ref{HI_schematic}). This has provided several new opportunities for CME research, notably the ability to track their evolution as they propagate through the inner heliosphere and potentially impact at Earth or one of either STEREO spacecraft which allows a comparison of in-situ data and white-light imagery of CMEs. The basic design of the HI comprises a number of occulting baffles that achieve the required level of light rejection for imaging the low intensities required to observe CMEs at large elongations (Figure~\ref{HI_intensity}). In order to image the low intensity CME signal sufficiently above the stellar background and zodiacal light (F-corona) whilst avoiding saturation by cosmic rays, a series of short exposures is taken and individually scrubbed of cosmic ray hits before being summed together. This increases the CME signal above the noise level of the relatively static F-corona, which may then be subtracted to reveal the CME intensity. The combination of summing (30 images for HI1, and 99 images for HI2) and 2\,$\times$\,2 binning increases the signal-to-noise ratio by about 14 times. The final images are 2\,$\times$\,2 binned onboard to 1024\,$\times$\,1024 pixels, subtending an angle of 70~arcseconds per pixel for HI1 and 4~arcminutes per pixel for HI2.

\section{CCD Detectors}
\label{sec:CCDs}

A charge-coupled device (CCD) is used in the LASCO and SECCHI instruments for detecting the incident photons and converting them to a digital output to generate images. Essentially a CCD converts light into electrons which are read and converted into numeric values used to display image intensities. The CCD is a small silicon chip divided into a grid of cells, or pixels. The electrons in the silicon atoms lie in discrete energy bands. In the ground state the outermost electrons lie in the valence band and can be excited to the conduction band by the absorption of a photon, via the photoelectric effect, leaving behind a `hole'. In a CCD an electric field is introduced to prevent recombination of the electron-hole pair. Thus an electric charge is accumulated proportional to the light intensity at that location. The charge is read out pixel-by-pixel to a charge amplifier which converts it to a voltage, then this voltage is digitised and stored in memory.

\begin{figure}[!t]
\centerline{\includegraphics[width = \textwidth, trim=20 20 20 20]{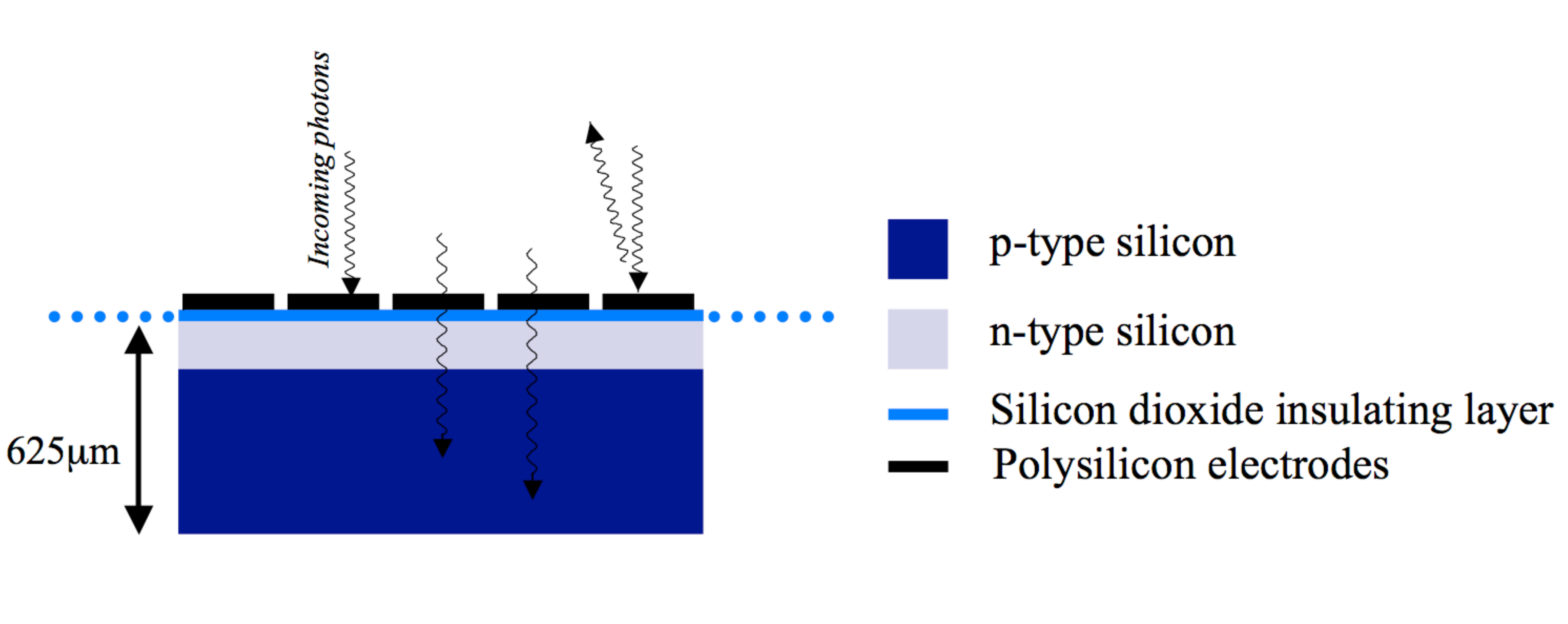}}
\caption{Illustration of a thick front-side illuminated CCD.
\newline
\emph{Image credit: www.ing.iac.es}.}
\label{frontsidedCCD}
\end{figure}
\par
A thick front-side illuminated CCD (Figure~\ref{frontsidedCCD}) is cheap to produce, but because photons are incident at the surface electrodes they can be reflected or absorbed, which gives low quantum efficiency (a measure of the percentage of photons detected: $QE = N_e / N_{\nu}$). The LASCO/C2 and C3 detectors are front-side illuminated Textronix CCDs that have a quantum efficiency of about 0.3\,--\,0.5 in the 500 to 700~nm spectral range. They are 1024\,$\times$\,1024 pixels in size, each pixel being a square measuring 21~$\mu$m on a side.
\par
\begin{figure}[!t]
\centerline{\includegraphics[width = \textwidth, trim=20 20 20 20]{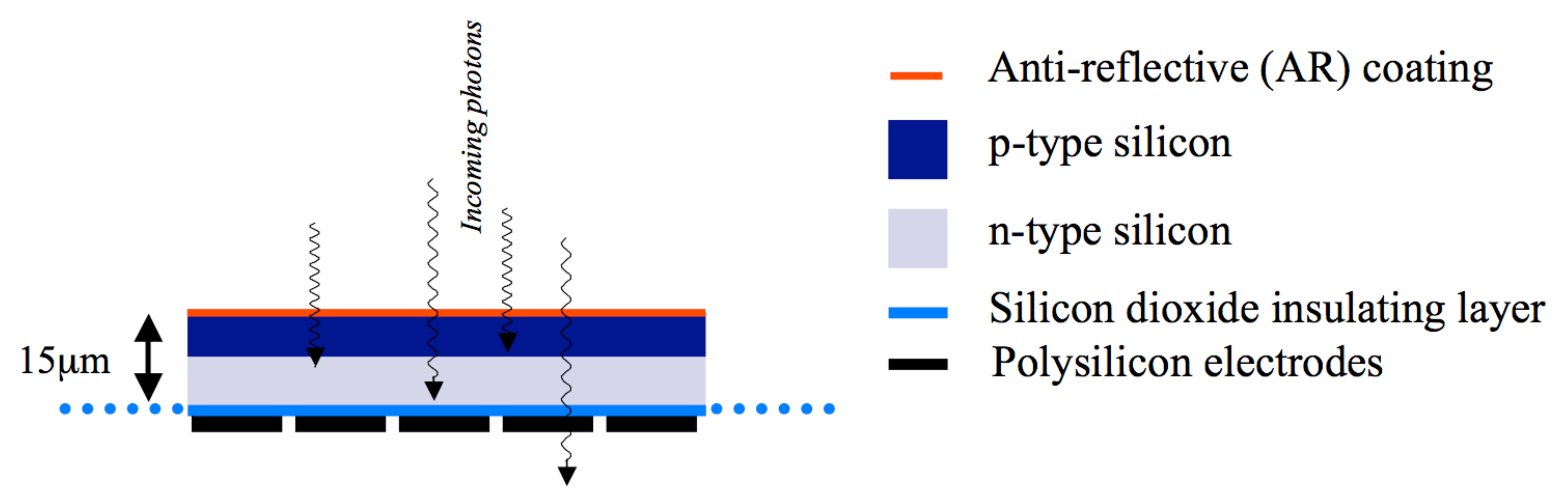}}
\caption{Illustration of a thinned back-side illuminated CCD. 
\newline
\emph{Image credit: www.ing.iac.es}.}
\label{CCD}
\end{figure}

To increase the quantum efficiency back-side illumination is used so the electrodes do not obstruct the photons. But the silicon in a back-side illuminated CCD must be chemically etched down (thinned) to a thickness of about 15~$\mu$m, which is an expensive process (Figure~\ref{CCD}). Silicon also has a high refractive index leading to strong photon reflection. It must therefore be coated with an anti-reflective material with a refractive index less than that of silicon (3.6) and preferably with an optical thickness of $1/4$ at a chosen wavelength of 550~nm (close to the middle of the optical spectrum). Hafnium dioxide is regularly used to significantly reduce the reflectivity of the CCD. Due to their high quantum efficiency, almost all current astronomical CCDs are thinned and back-side illuminated. Each of the SECCHI instruments uses a back-side illuminated E2V 42--40 CCD detector, that has a quantum efficiency of roughly 0.8 at 500~nm, 0.88 at 650~nm, 0.64 at 800~nm, and 0.34 at 900~nm. They are 2048\,$\times$\,2048 pixels in size, each measuring 13.5~$\mu$m on a side. This CCD has an operational temperature range of 153\,--\,323~K (-120\,--\,50$^{\circ}$~C).

Sources of noise in CCD imaging must be noted when performing image analysis. Thermal noise, or dark current, is due to thermal excitations of electrons in the CCD. A dark frame must be generated to correct for thermal noise by taking a closed shutter exposure of some known duration to study the effects on the resultant image, though this form of noise is minimal for space-borne instruments operating at temperatures of $\lesssim$\,200~K. 
\par
%While such temperatures are ideal for detecting faint coronal light they also mean the CCD is sensitive to energetic particles.
Hot pixels can result from energetic particles or cosmic rays causing ionisation in the silicon, because the resulting free electrons from these hits are indistinguishable from ones that are photo-generated. 
\par
CCD read-out noise can occur when charge is converted to voltage since electronic amplifiers are not perfect. A high charge transfer efficiency is also important during shift operations in the read-out process to minimise count errors. 
\par
Calibrations of CCD images must be performed to remove imperfections. CCDs are not always linear (measuring one count for one photon incident). A flat-field calibration removes variations in sensitivity across the surface of the CCD, due to silicon or manufacturing defects and vignetting effects. Flat-field images are normally generated in the lab by taking an exposure when the CCD is evenly illuminated by a light source, and dividing this into future images for linearity. 
\par
Similar to dark frames, bias frames may also be generated. A bias frame is a zero duration exposure taken with no light incident on the CCD (the shutter remains closed). Thus structures which appear in bias frames are as a result of defects in the CCD electronics and must be removed from future images.
\par
The charge capacity of a CCD pixel is limited and when full it can overflow, leading to blooming. While this is somewhat unavoidable when taking long exposures, especially if a bright star or comet comes into view for example, most CCD design ensures blooming only occurs in one direction.

\section{Coordinate Systems}

The pixel coordinates from the CCD images must be transformed to the relevant coordinate system for studying and interpreting observations, especially when comparing images from multiple viewpoints (such as STEREO and SOHO as discussed in Section~\ref{sohothirdeye}). First the pixel coordinates ($p_1,\,p_2,\,p_3,\,...$) must be transformed to intermediate world coordinates ($x_1,\,x_2,\,x_3,\,...$), meaning they are converted into the relevant units (e.g., metres or arcseconds) but are not necessarily corrected for the reference point of the observations nor geometric or projection effects:
\begin{equation}
x_i \; = \; s_i \sum_{j=1}^N m_{ij} \left( p_j - r_j \right)
\end{equation}
where $s_i$ is the scale function, $N$ is the number of axes, $m_{ij}$ is the transformation matrix, and $r_j$ is the reference pixel \citep{2002A&A...395.1077C}. These can then be transformed into one of the Sun-centred coordinate systems described below.

\begin{figure}[!t]
\subfigure[Stonyhurst Heliographic]{\includegraphics[scale=0.194, trim=20 48 0 0]{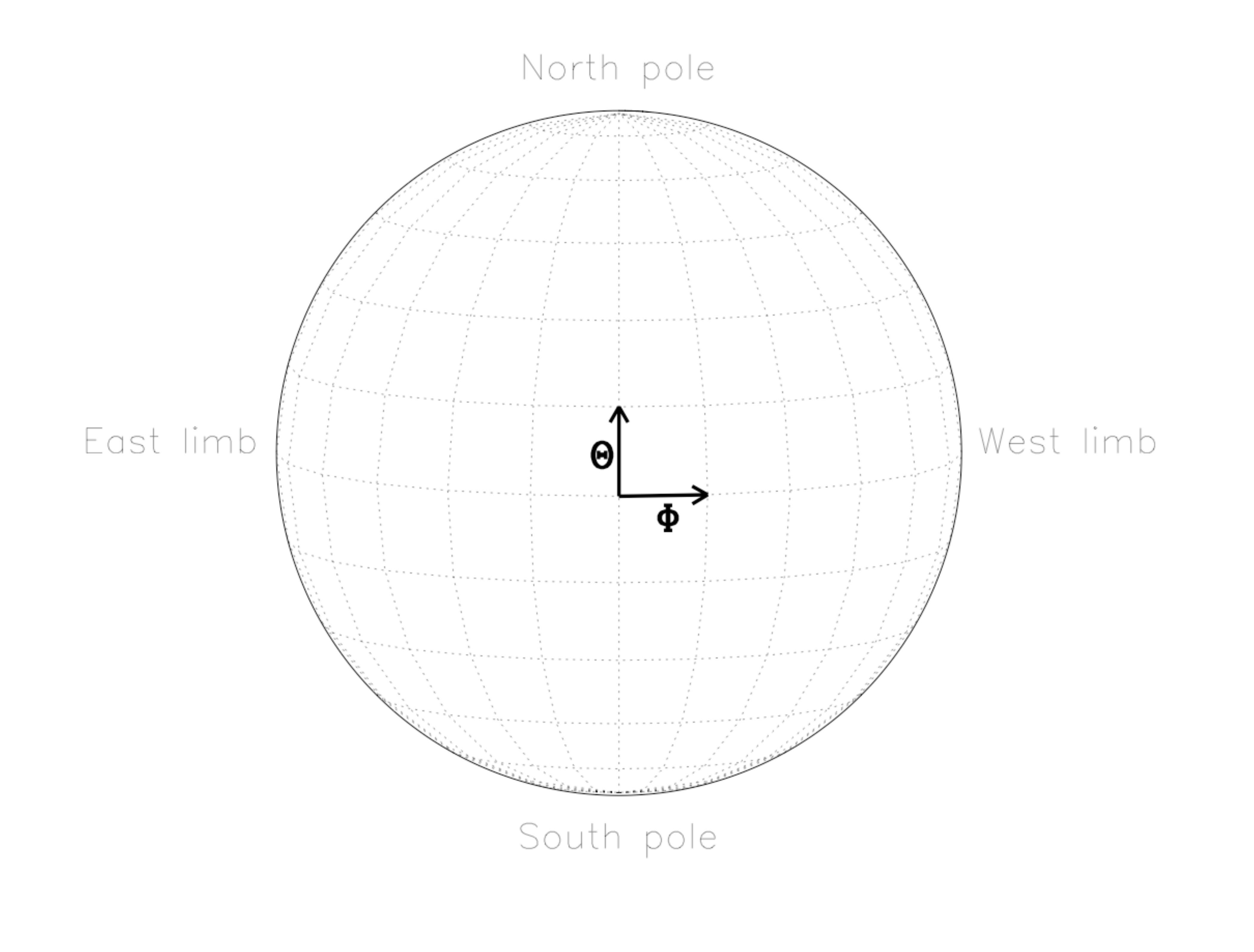}}
\subfigure[Heliocentric-Cartesian]{\includegraphics[scale=0.17, trim=0 0 0 0]{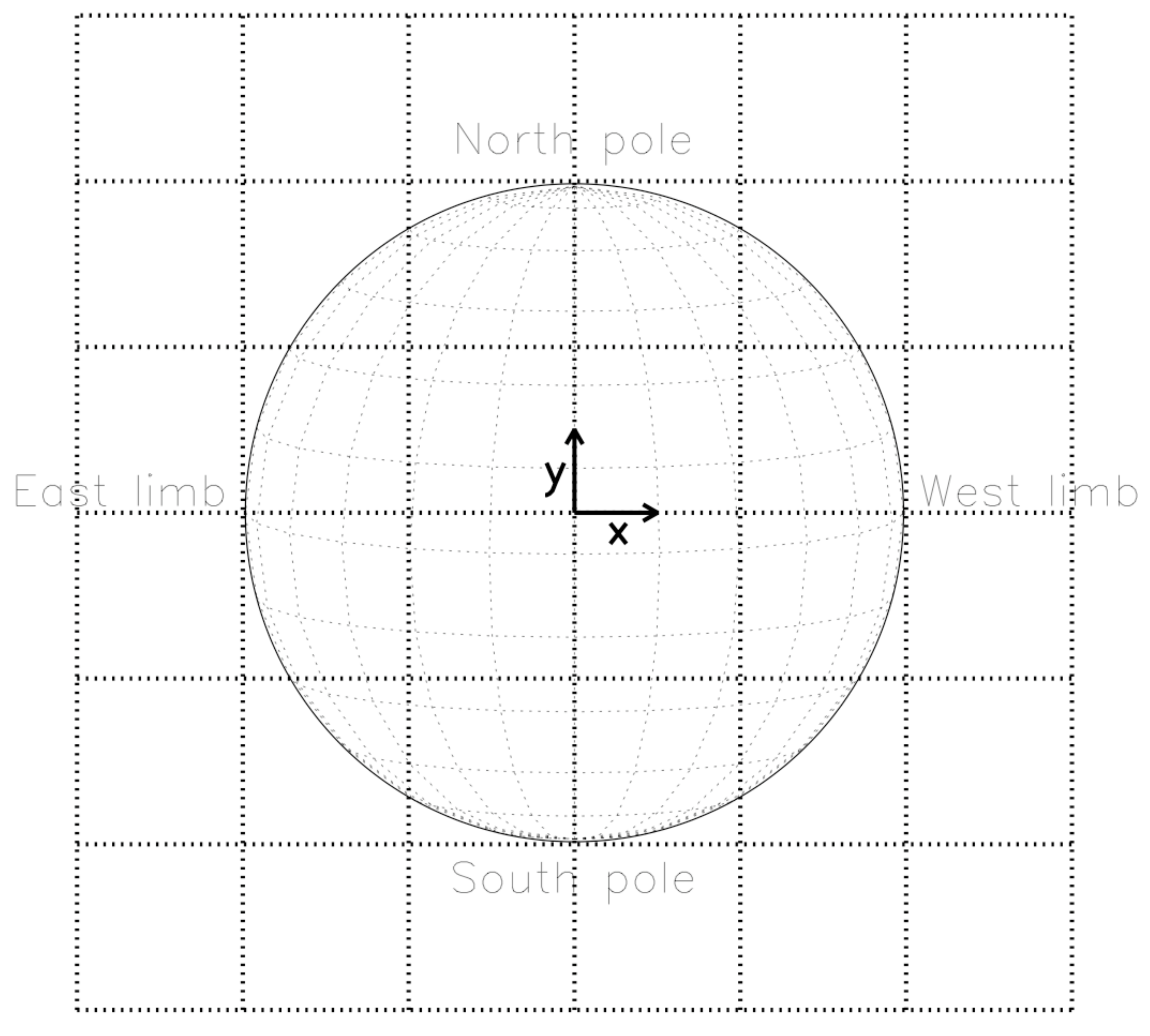}}
\caption{Schematics of two Sun-centred coordinate systems, reproduced from \citet{2006A&A...449..791T}. (a) Stonyhurst heliographic coordinates commonly used to locate features on-disk. (b) Heliocentric-cartesian coordinates commonly used for spatially localising features in the vicinity of the Sun.}
\label{coordinates}
\end{figure}

\subsection{Heliographic Coordinates} 

Features on the Sun are located by the coordinates of latitude, $\Theta$, and longitude, $\Phi$, with respect to the solar equator and rotational axis. In the Stonyhurst approach, the zero point of longitudinal measurements is set at the intersection of the solar equator and central meridian as seen from Earth (Figure~\ref{coordinates}a). In the Carrington approach, the central meridian is fixed to its observation on 9 November 1853, and the rotations since then are counted and labelled as the Carrington rotation number \citep{carrington1863}.

\subsection{Heliocentric Coordinates}

Heliocentric coordinates specify the location of a feature in space with respect to the centre of the Sun. The Heliocentric-Cartesian coordinate system ($x,\,y,\,z$) has the $z$-axis from Sun-centre along the Sun-observer line, the $y$-axis is perpendicular to this and lies in the plane containing solar north, and the $x$-axis is perpendicular to both $y$ and $z$ and increases towards solar west (Figure~\ref{coordinates}b). %These are also known as Heliographic Radial-Tangential-Normal (HGRTN) coordinates.
The Heliocentric-Radial coordinate system shares the same $z$-axis but measures features in cylindrical coordinates with radial distance $\rho$ from the $z$-axis, and position angle $\psi$ counter-clockwise from solar north.
\par
With observations from the STEREO/SECCHI instrument suite centred on the Sun and extending out along the full Sun-Earth line, the optimal coordinate systems are the heliocentric coordinates, which can be defined in three possible manners depending on the user's preference.

\subsubsection{Heliocentric Earth Equatorial (HEEQ)}

The HEEQ system is closely related to the Stonyhurst heliographic coordinates with the $z$-axis parallel to the solar rotational axis, and the $x$-axis towards the intersection of the solar equator and central meridian as seen from Earth, obtained by the following transformations:
\begin{align}
X_{HEEQ} \; &= \; r \cos \Theta \cos \Phi \nonumber \\
Y_{HEEQ} \; &= \; r \cos \Theta \sin \Phi \\
Z_{HEEQ} \; &= \; r \sin \Theta \nonumber
\end{align}
It is thus useful when considering features in space with respect to the Sun-Earth line and the Sun's rotational axis.

\subsubsection{Heliocentric Earth Ecliptic (HEE)}

The HEE system has the $x$-axis towards the Earth from Sun centre, and the $z$-axis is perpendicular to the plane of Earth's orbit around the Sun called the ecliptic, and the $y$-axis perpendicular to both $x$ and $z$. It is thus useful when considering features in space with respect to the Sun-Earth line and ecliptic.

\subsubsection{Heliocentric Aries Ecliptic (HAE)}

The HAE system has the $x$-axis towards the First Point of Aries (the direction to the point of intersection between Earth's equatorial plane and the plane of the ecliptic), the $z$-axis perpendicular to the ecliptic, and the $y$-axis perpendicular to both $x$ and $z$. It is thus useful when considering features in space with respect to a fixed point on the celestial sphere (outlined below).

\subsection{Helioprojective Coordinates}

\begin{figure}[!t]
\centerline{\includegraphics[scale=0.37, clip=true, trim=0 60 60 0]{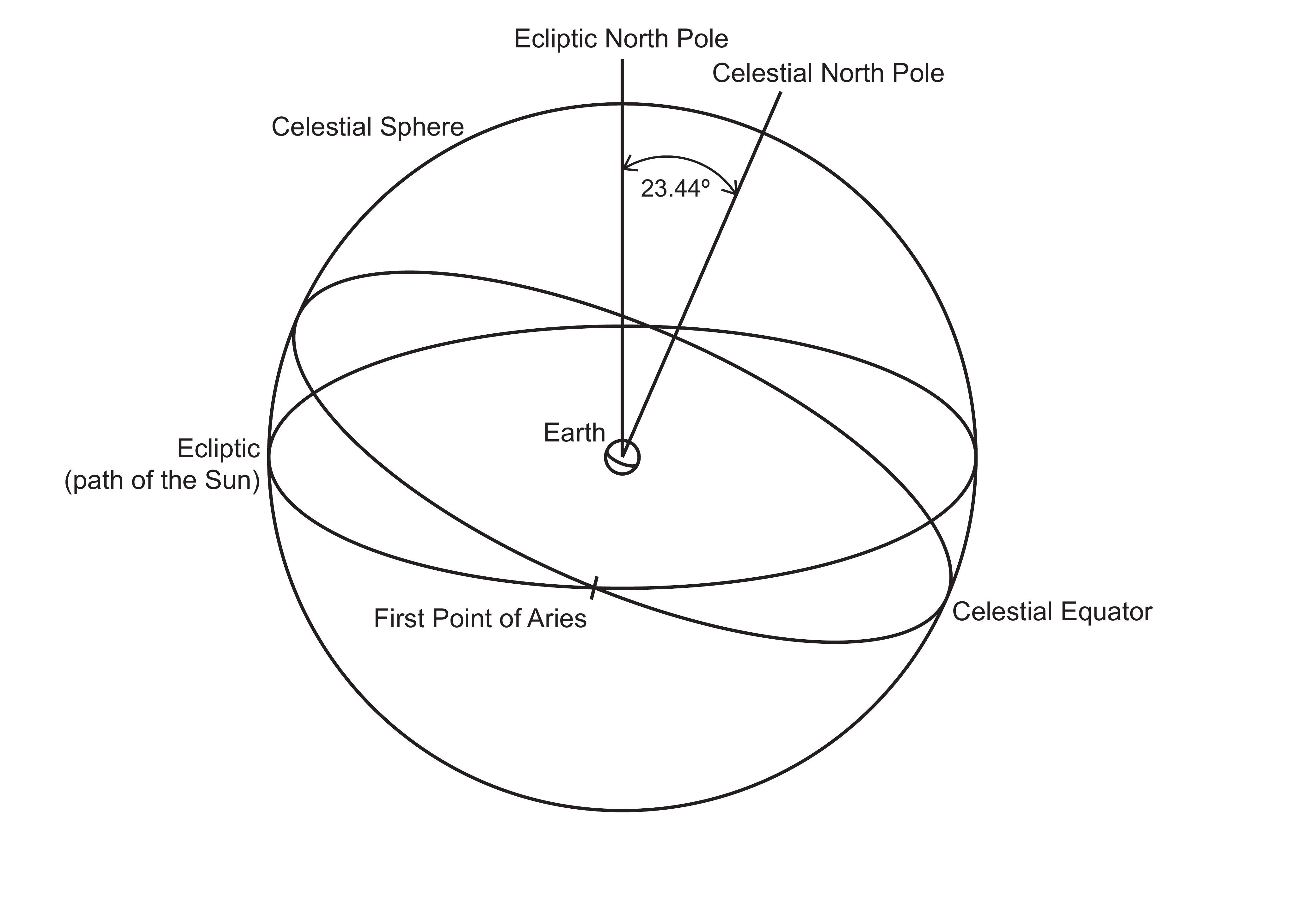}}
\caption{Schematic of the celestial sphere and ecliptic plane. The celestial equator is a projection of the Earth's equator, and the ecliptic is a projection of the Earth's orbit about the Sun.}
\label{ecliptic}
\end{figure}
When considered on a large scale it is more intuitive to take the projection of the Heliocentric-cartesian coordinates onto the celestial sphere; an imaginary sphere of arbitrarily large radius, centred on the Earth such that all observations may be considered as projections upon it (Figure~\ref{ecliptic}). This results in the Helioprojective-Cartesian coordinates ($\theta_x,\,\theta_y,\,\zeta$):
\begin{equation}
\theta_x \; \approx \; \left( \frac{180^{\circ}}{\pi} \right) \frac{x}{d} \; , \quad \theta_y \; \approx \; \left( \frac{180^{\circ}}{\pi} \right) \frac{y}{d} \; , \quad \zeta \; = \; D_{\odot}-d
\end{equation}
where $d$ is the distance between the observer and the feature, and $D_{\odot}$ is the distance between the observer and Sun centre. This similarly results in the Helioprojective-Radial coordinates ($\delta_{\rho},\, \psi,\, \zeta$):
\begin{equation}
\delta_{\rho} \; \equiv \; \theta_{\rho} - 90^{\circ} \quad \text{where} \quad \theta_{\rho} \; \approx \;  \left( \frac{180^{\circ}}{\pi} \right) \frac{\rho}{d}
\end{equation}
These sets of coordinates are thus useful when considering features in space with respect to the whole sky as seen from Earth.

\chapter{Detecting and Tracking CMEs}
\label{chapter:multiscale}

In coronagraph images CMEs are observed as outwardly moving regions of stronger brightness intensities relative to the background corona. Different approaches to thresholding the intensity of CMEs in these images have been employed in order to detect their appearance and track their motion through the field-of-view, leading to a cataloguing of their kinematics and morphology. However, these techniques suffer several drawbacks and, as such, different catalogues can vary significantly in their description of events. We introduce a method of multiscale analysis to overcome certain drawbacks of previous detection and tracking methods. In multiscale decompositions of images, noise and small-scale features are removed to leave only larger-scale features of interest such as CMEs. This allows them to be tracked through the image sequences in order to determine their changing kinematics and morphology (see Chapter~\ref{chapter:kinematics}). Unfortunately coronal streamers tend to appear on similar size scales to CMEs, making their automatic detection difficult. Streamers do, however, tend to remain static on timescales comparable to CME propagation through the field-of-view, and contain much less angular information than the typically curved structure of CMEs, so they may be removed through spatio-temporal filtering of multiscale CME images. This chapter discusses the previous CME detection catalogues, and outlines our use of new methods of multiscale filtering to detect the CME edges in single images. We discuss our efforts to extend this to an automated CME detection algorithm. We also outline an ellipse characterisation of the CME front for study.

\section{CME Detection Catalogues}
\label{sect:cmecatalogues}

\begin{figure}[!t]
\centerline{\includegraphics[scale=0.8, clip=true, trim=0 30 0 500]{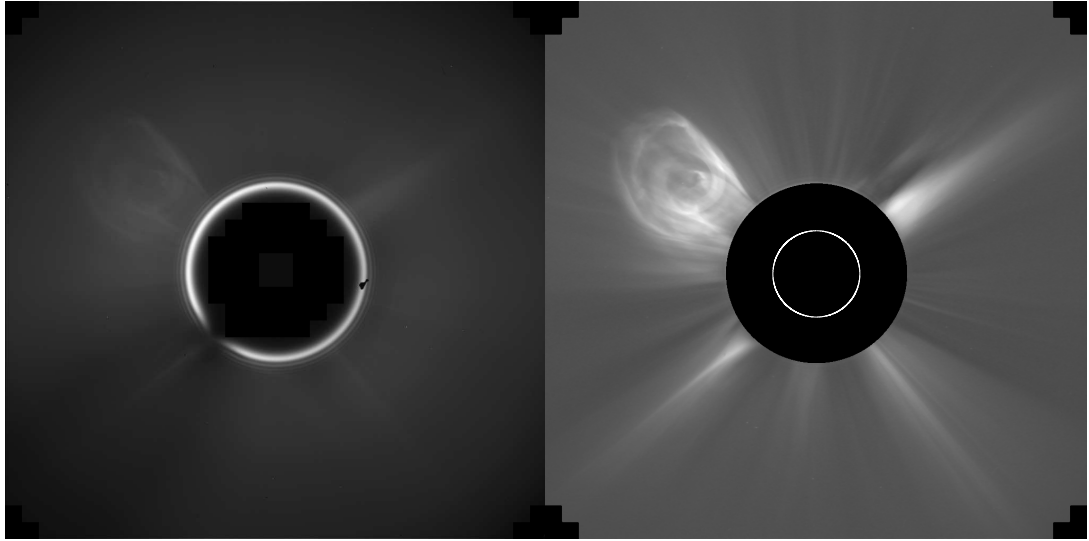}}
\caption{Raw (left) and pre-processed image (right) of a CME observed by LASCO/C2 on 1 April 2004. The pre-processing includes normalising the image statistics, subtracting the background, and masking the occulter disk. The white circle (right) indicates the relative size and position of the Sun behind the occulter.}
\label{init_rm}
\end{figure}

%The C2 and C3 coronagraphs of LASCO onboard SOHO give a field-of-view extending from approximately 2.2--30~R$_{\odot}$. The COR1/2 coronagraphs of the SECCHI suite onboard STEREO image the corona from 1.4--15~R$_{\odot}$. 
Current methods of CME detection have their limitations, mostly since these diffuse objects have been difficult to identify using traditional image processing techniques. These difficulties arise from the varying nature of the CME morphology, the scattering effects and non-linear intensity profile of the surrounding corona, the presence of coronal streamers, and the addition of noise due to cosmic rays and solar energetic particles (SEPs) that impact the coronagraph detectors. The images are also prone to numerous instrumental effects and possible data dropouts.  The following standard preprocessing methods are usually applied to optimise the images for CME studies. The coronagraph images are normalised with regard to exposure time in order to correct for temporal variations in the image statistics. A filter may be applied to remove pixel noise, for example to replace hot pixels with a median value of the surrounding pixel intensities, or to reduce the effects of background stars in the image. A correction for vignetting effects and/or lens distortion may be applied to the images. A background subtraction may also be applied, obtained from the minimum of the daily median pixels across a time span of a month. The occulting disc is normally masked, along with any data drop-outs in the images. These steps lead to a clear improvement in the image quality for CME study (Figure~\ref{init_rm}).

\subsection{CDAW}

The CME catalogue hosted at the Coordinated Data Analysis Workshop (CDAW\footnote{http://cdaw.gsfc.nasa.gov/CME\_list}) Data Center grew out of a necessity to record a simple but effective description and analysis of each event observed by SOHO/LASCO \citep{2009EM&P..104..295G}. The catalogue is wholly manual in its operation, with a user tracking the CME through C2 and C3 running-difference images and producing a height-time plot of each event. A linear fit to the height-time profiles provides a 1st-order estimate for the plane-of-sky velocity, and a quadratic fit then provides a 2nd-order velocity fit and an acceleration for the event. The central position angle and angular width of the CME are also deduced from the images, and the event flagged as a halo if it spans 360$^{\circ}$, partial halo if it spans $\ge$\,120$^{\circ}$, and wide if it spans $\ge$\,60$^{\circ}$. The catalogue itself lists each CME's first appearance in C2, central position angle, angular width, linear speed, 2nd-order speed at final height, 2nd-order speed at 20~R$_{\odot}$, acceleration, mass, kinetic energy, and measurement position angle (the angle along which the heights of the CME are determined). While the human eye is supremely effective at distinguishing CMEs in coronagraph images, errors may be introduced to the manual cataloguing procedure through the biases of different operators; for example, in deciding how the images are scaled, where along the CME the heights are measured, or whether a CME is even worth including in, or discarding from, the catalogue. In an effort to overcome such biases, different automated catalogues have been developed to perform robust CME detections over large data-sets. This is also of great benefit for future missions where the data rate is expected to be too high for manual cataloguing to remain feasible.

\subsection{CACTus}

\begin{figure}[!p]
\centering
\subfigure[The $(t,\, r)$ slice for a given angle $\theta$, with a mirrored illustration of the resulting CME intensity ridge detections.]{\includegraphics[clip=true, scale=0.8, trim=0 0 0 0]{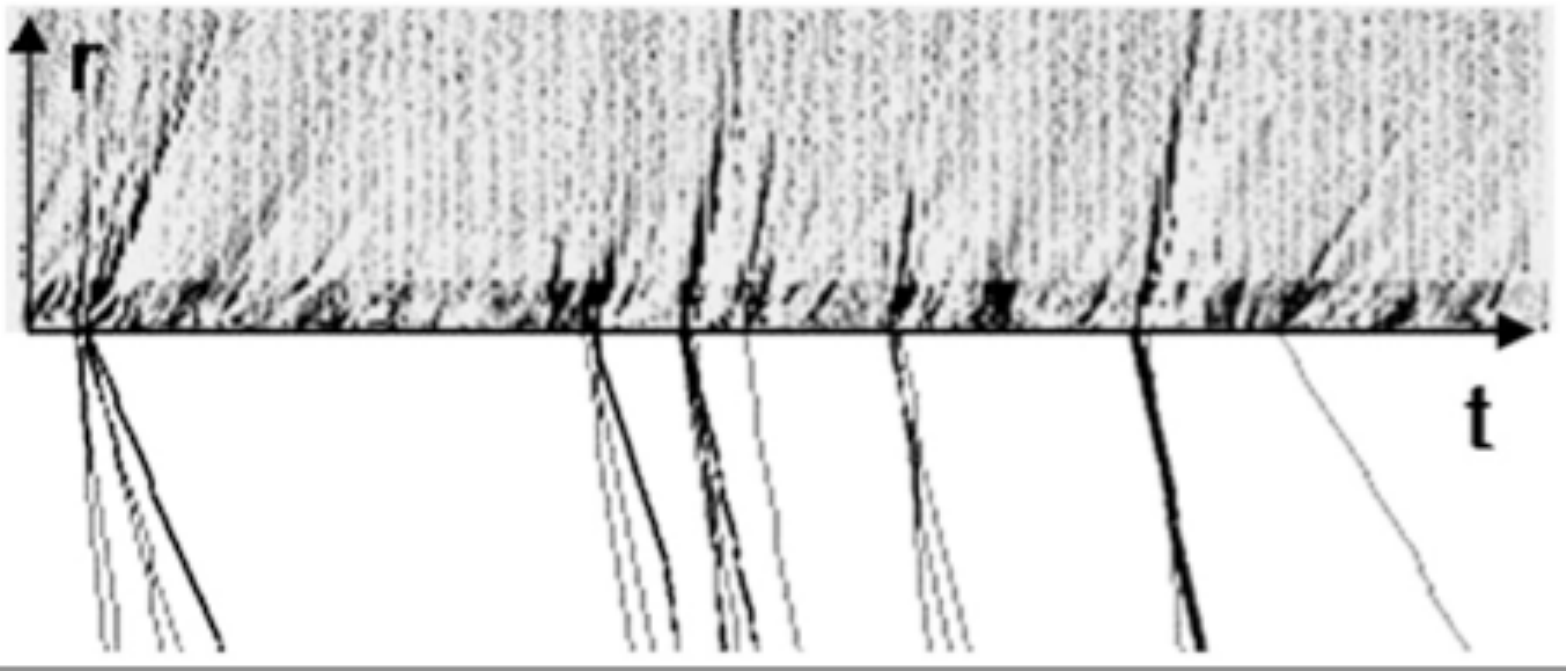}}
\label{cactusridges}
\subfigure[Left: the $(t,\, r)$ slice for a given angle $\theta$, with an example ridge drawn from onset time $t_0$ with duration $\Delta t$ across the field-of-view from $r_{min}$ to $r_{max}$. Right: the corresponding accumulator space $(t_0,\, \Delta t)$ where the ridge will appear as a point with a magnitude corresponding to the ridge intensity. This modified Hough transform is used to threshold the most significant ridges in the slice, automatically detecting the CME in the coronagraph image.]{\includegraphics[scale=0.25, clip=true, trim=0 0 0 0]{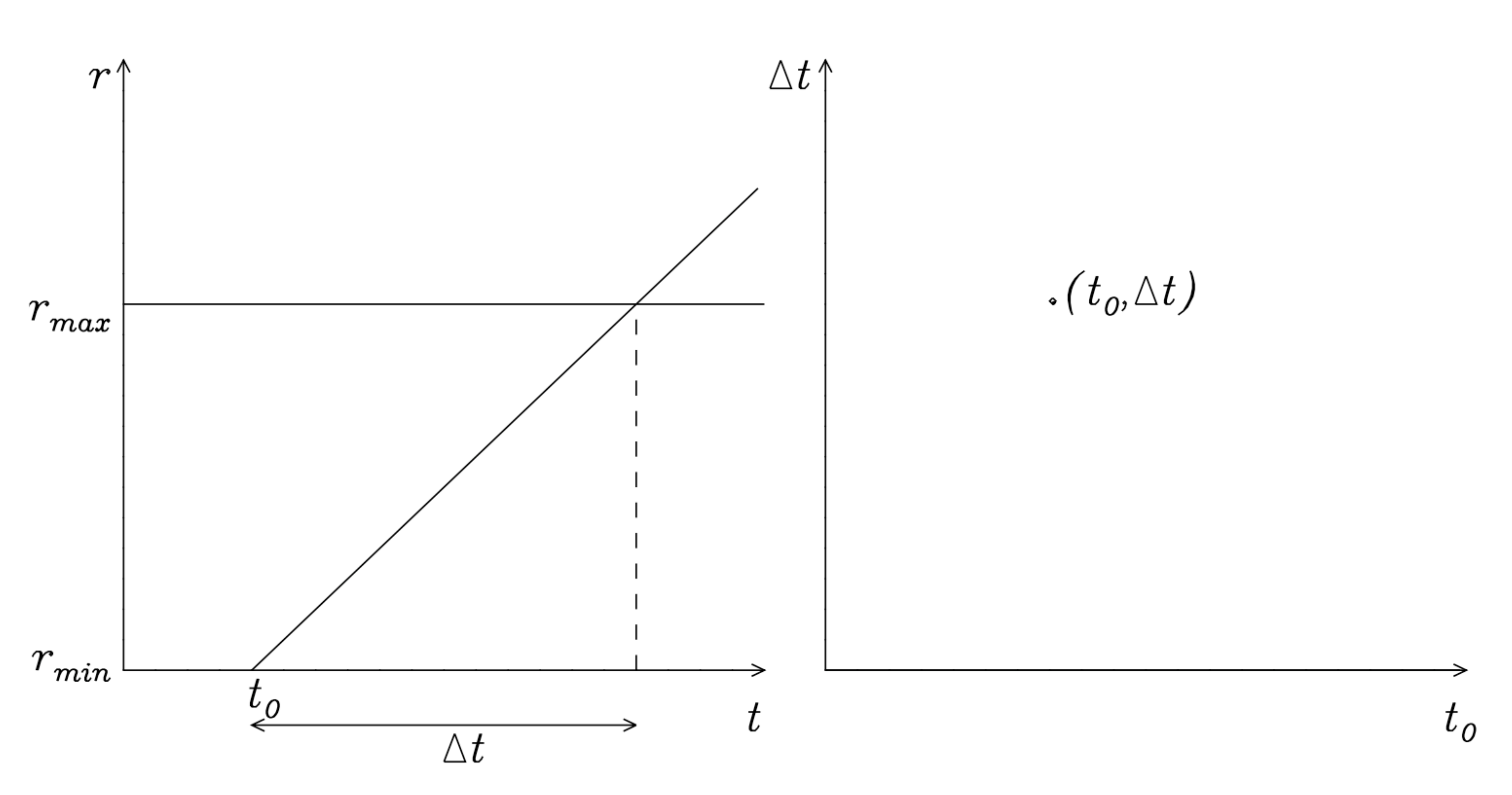}}
\caption{The top image (a) shows the detection of ridges in the $(t,\,r)$ stacks of the CACTus catalogue, through the use of the Hough transform detailed in the bottom image (b), reproduced from \citet{2004A&A...425.1097R}.}
\label{hough}
\end{figure}

The Computer Aided CME Tracking catalogue \citep[CACTus\footnote{http://sidc.oma.be/cactus/};][]{2004A&A...425.1097R} was the first automated CME detection algorithm, in operation since 2004. It is based upon the detection of CMEs as bright ridges in time-height slices $(t,\, r)$ at each angle $\theta$ around a coronagraph image. The images are preprocessed as standard, then a running-difference technique is applied and each image transformed into Sun-centred polar coordinates $(r,\; \theta)$, rebinned, and the C2 and C3 fields-of-view combined. These are then stacked in time, and for each angle the corresponding $(t,\,r)$ slice undergoes a modified Hough transform for detecting intensity ridges across it. This works by parameterising the $(t,\, r)$ slice by the variables $t_0$ and $\Delta t$, corresponding to the coordinate intersection point with the time axis, and the distance along the time axis respectively (together called the accumulator space; see Figure~\ref{hough}). So the equation of a line corresponding to an intensity ridge in the slice is given by:
\begin{equation}
r \; = \; \frac{r_{max}-r_{min}}{\Delta t} (t-t_0) + r_{min}
\end{equation}
Thresholding the most significant ridges in the resultant accumulator space filters out the progression of CMEs, with the variables for each ridge characterised by onset time $t_R$, the velocity $v_R \; \left( \sim \frac{1}{\Delta t} \right)$, for angle $\theta_R$, to give a characteristic variable $I_R=\left(v_R,\, \theta_R,\, t_R \right)$. A 3D scatter plot $(v,\, \theta,\, t)$ of all detected ridges $I_R$ is then integrated along the $v$-direction to identify clusters in the resulting $(\theta,\, t)$ map which illustrates the angular span and duration time of the detected CMEs in the coronagraph data. A median velocity across the angular span is quoted as the CME speed.
\newline
\indent The running-difference cadence, the ridge intensity threshold, and the imposed limit on how many frames a CME may exist (and indeed the definition of a CME) all affect how successful the detection can be. However, \citet{2004A&A...425.1097R} show the algorithm to be robust in reproducing well the detections of a human user by direct comparison with the CDAW catalogue. The main drawback of the CACTus catalogue for studying CMEs is the imposed zero acceleration of the detection algorithm, since the Hough transform thresholds the ridges as straight lines whose slopes provide a constant velocity. The velocity itself may also be underestimated since it is a median across the span of the CME. The angular spans are possibly over-estimated since side outflows in the images are enhanced by the running-difference and may also include streamer deflections. It is also difficult to distinguish when one CME has fully progressed from the field-of-view and another CME has entered it, so in some cases trailing portions of a CME are detected as separate events.

\subsection{SEEDS}

\begin{figure}[!p]
\centerline{\includegraphics[scale=0.45, clip=true, trim=0 0 0 0]{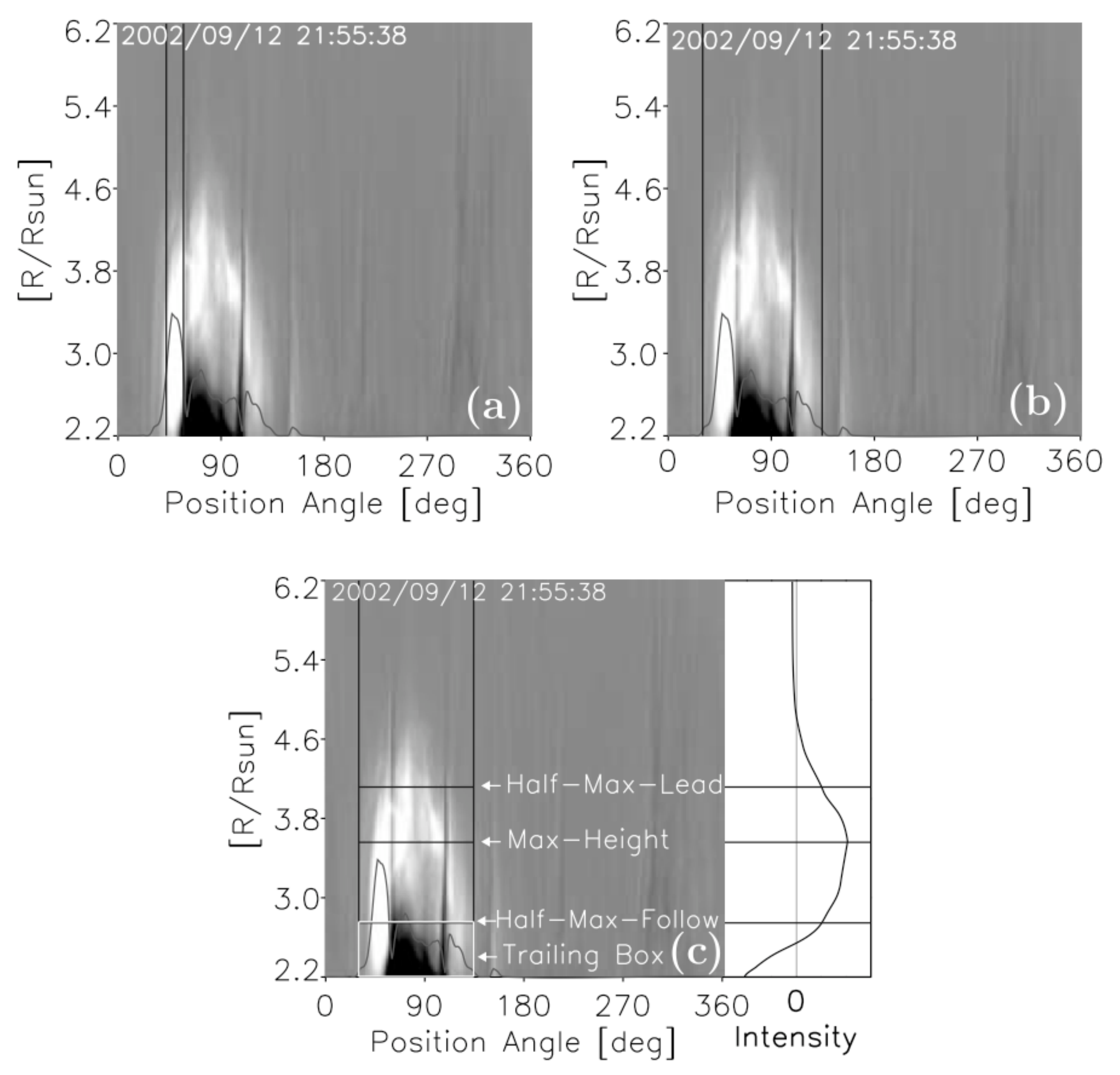}}
\caption{Example of the SEEDS CME detection and height determination, reproduced from \citet{2008SoPh..248..485O}. ({\bf a}) shows the running-difference image unwrapped into Sun-centre polar coordinates, showing a CME observed by LASCO/C2 on 12 September 2002. The black line distribution across the image represents the positive value intensity count along each angle, and the two vertical black lines mark the angular span at one standard deviation above the mean intensity. ({\bf b}) shows the new angular span following the region growing technique. ({\bf c}) shows the intensity within the angular span averaged across heights, and the `Half-Max-Lead' is taken as the CME height in the image.}
\label{seeds}
\end{figure}

The Solar Eruptive Event Detection System \citep[SEEDS\footnote{http://spaceweather.gmu.edu/seeds/};][]{2008SoPh..248..485O} is an automated CME detection algorithm for tracking an intensity thresholded CME front in running-difference images from LASCO/C2. The images are preprocessed as standard, unwrapped into Sun-centred polar coordinates $(r,\; \theta)$, and a normalised running-difference technique is applied using the following equation:
\begin{equation}
u_i \; = \; \left[ n_i - n_{i-1} \left( \frac{\bar{n}_i}{\bar{n}_{i-1}} \right) \right] \frac{\alpha}{\Delta t}
\end{equation}
where $u_i$ is the running-difference image, $\bar{n}$ is the mean of the pixels in the entire field-of-view of the image $n$, $\Delta t$ is the time difference between images (in minutes), $\alpha$ is a constant set to approximately the smallest time difference ($\Delta t$) between any image pair, where $i$ is the current image and $i-1$ the preceding image. This normalised difference ensures that the mean of the new image ($u_i$) will effectively be zero.
\newline
\indent The pixel intensities (positive values only) are then summed along angles and thresholded at a certain number of standard deviations above the mean intensity: $\mu+N\sigma$ (cf. Equation~\ref{eqn:threshold}) as in Figure~\ref{seeds}a. This determines the `core angles' of the CME, and a region growing technique based on a secondary threshold of intensities in the rest of the image is applied to open the angular span to include the full CME (Figure~\ref{seeds}b). Issues arise when streamer deflections occur that will offset the region growing technique and overestimate the CME angular width. An intensity average across the angles within the span of the CME is then determined, and where the forward portion of this intensity profile equals half its maximum value is taken as the CME height (Figure~\ref{seeds}c). The velocity and acceleration are determined from the heights through consecutive images and these results are output with the CME position angle and angular width in the SEEDS catalogue.
\newline
\indent Along with the issues of streamer deflections and the tracking being limited to C2 images, the choice of the `Half-Max-Lead' as the CME height is dependant on the overall CME brightness, and thus any brightness changes as the CME propagates will affect this measurement. This would add to the error on the height-time profile which, along with the error in time as a result of the running-difference technique, makes it difficult to accurately determine the velocity and acceleration.

\subsection{ARTEMIS}

\begin{figure}[!p]
\centerline{\includegraphics[scale=0.8, clip=true, trim=0 0 0 0]{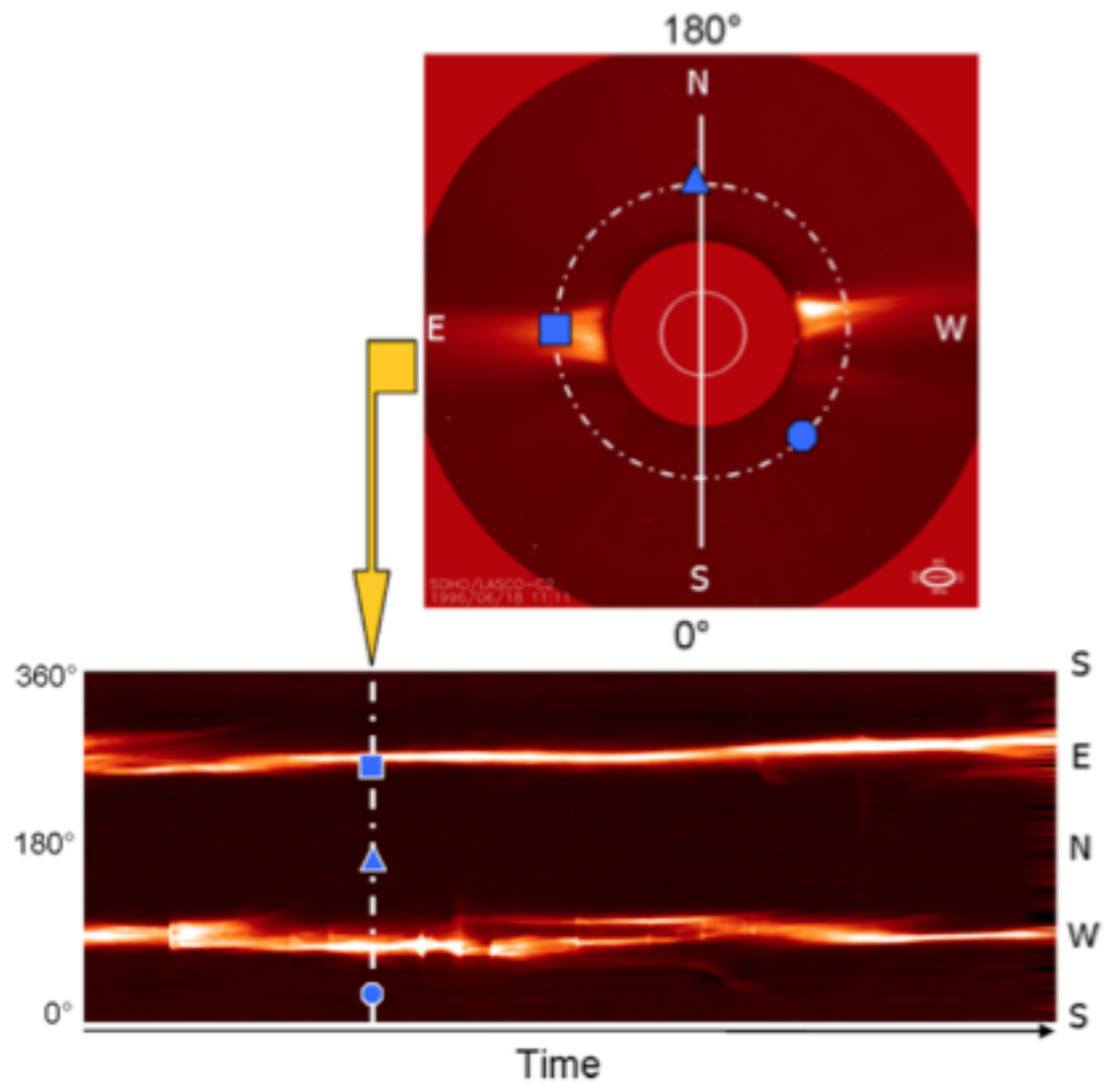}}
\caption{An example of how the synoptic maps are generated for the ARTEMIS catalogue, reproduced from \citet{2009SoPh..257..125B}. At a chosen height in the coronagraph image an annulus is unwrapped (indicated with the dashed line and blue square, circle and triangle) and these are then stacked together to illustrate how the intensity at that height changes through time. Vertical streaks represent transient events occurring on smaller time-scales than the more persistent streamers in the images.}
\label{artemis}
\end{figure}

The Automatic Recognition of Transient Events and Marseille Inventory from Synoptic maps \citep[ARTEMIS\footnote{http://www.oamp.fr/lasco/};][]{2009SoPh..257..125B} is an automated CME detection algorithm that works by identifying signatures of transients in synoptic maps. These maps are generated as $(t,\, \theta)$ slices for specific heights $r$ in the coronagraph images of LASCO/C2. The images are prepared through the standard preprocessing steps. Then at a specific height (e.g., $r=3$~R$_{\odot}$) the intensity is plotted across all angles $\theta$ for each image through time $t$ with transient events appearing as vertical streaks through the more persistent streamer intensities (Figure~\ref{artemis}). A method of image filtering and intensity thresholding is applied to distinguish the streaks in the synoptic map, and image segmentation then discards small features and closes off regions-of-interest (ROIs) to produce a binary map of the streaks. Specific parameters of these streaks are also computed, such as their total radiances, areas and centres of gravity. Merging with high-level knowledge helps to associate ROIs of the same CME if three criteria are met: the difference between the $x$-coordinates of the centres of two ROIs differs by less than two pixels; the difference between the $y$-coordinates of their centres differs by less than 60 pixels (corresponding to a 60$^{\circ}$ angular span); and the ratio of their radiances calculated at their centres (on the original synoptic map) ranges from 0.25 to 4. The result is a binary CME detection map in $(t,\, \theta)$ space for different heights in the corona: 3, 3.5, 4, 4.5, 5 and 5.5 R$_{\odot}$.
\newline
\indent With the CME detections in place, estimates of the velocity may be made. A first estimate is taken by testing a range of constant velocities 50\,--\,2,000~km~s$^{-1}$ to determine which best matches the shifting of the CME detection in synoptic maps at subsequent heights through the corona. The binary maps are shifted by an amount corresponding to velocity steps of 10~km~s$^{-1}$, such that the one which provides the maximum pixel value (with a minimum limit of 3) indicates the best velocity estimate of the event. A second estimate is taken by cross-correlating the detected CME ROIs on the original synoptic maps at 3~R$_{\odot}$ and 5.5~R$_{\odot}$ and inspecting the intensity shift in time (pixel shift in $x$-direction) to obtain the velocity estimate. A third estimate is taken by similar cross-correlation but specifically on each individual line of the ROIs to obtain a distribution of velocities across the angular span of the CME, the median of which is taken as the actual velocity. \citet{2009SoPh..257..125B} compare histograms of the three different velocity estimates for the ARTEMIS CME detections over a twelve year interval and find that, globally, the three estimates are highly consistent with each other.
\newline
\indent ARTEMIS is limited to the C2 field-of-view and it provides kinematics only in the 3\,--\,5.5~R$_{\odot}$ range. The velocity determinations themselves are not specific to either the CME front nor any other identifiable feature, and carry all the inaccuracies resulting from the image rebinning, intensity averaging, filtering and segmentation techniques in generating the final detection masks.
\newline
\par
Due to the drawbacks of each of the catalogues above, the motivation exists to study the kinematics and morphology of CMEs with as great an accuracy as possible in order to better compare with theory. To this end we outline below our application of multiscale analysis to remove small scale noise/features and enhance the larger scale CME in single coronagraph frames, allowing the CME front edges to be detected and a geometrical characterisation applied to study its propagation with increased accuracy for deriving the kinematics and morphology.

\section{Multiscale Filtering}

In this section a new multiscale method of analysing CMEs is described. The use of multiscale methods in astrophysics have proven effective at denoising spectra and images \citep{1995A&AS..112..179M, 1997A&AS..124..579F}, analysing solar active region evolution \citep{2008SoPh..248..311H}, and enhancing solar coronal images \citep{2003A&A...398.1185S, 2008ApJ...674.1201S}.  A particular application of multiscale decompositions uses high and low pass filters convolved with the image data to exploit the multiscale nature of the CME \citep{2008SoPh..248..457Y}. This highlights its intensity against the background corona as it propagates through the field-of-view, while neglecting small scale features (essentially denoising the data). It also leads to the use of non-maxima suppression to trace the edges in the CME images, and  \citet{2008SoPh..248..457Y} show the power of multiscale methods over previous edge detectors such as Roberts \citep{1975CGIP....4..248D} and Sobel \citep{1973pcsa.book.....D}. With these methods for defining the front of the CME we can characterise its kinematics (position, velocity, acceleration) and morphology (width, orientation) in coronagraph images. Multiscale analysis also has the benefit of working on independent images without any need for differencing, so the temporal errors involved are on the order of the exposure time of the instrument ($\sim$ a few seconds). %While these methods are not currently automated, they have great potential for such an implementation in the near future.
 %This will enable a large scale study of previous SOHO data, and current and future STEREO data. It is also of great importance to the space weather forecasting community, especially with the large amounts of data available from upcoming mission such as the Solar Dynamics Observatory (SDO).
%\par
%In Sect.~2 we discuss current image pre-processing methods from the coronagraphs onboard the SOHO \citep{Domingo97} and STEREO \citep{Kaiser08} spacecraft. We outline the implementation of multiscale methods and ellipse fitting to characterise the CME front. In Sect.~3 we present a sample of CME events analysed by these methods, and Sect.~4 is a discussion of these first results and conclusions.
%The continuous wavelet transform (CWT) of a signal $f(x)\in L^2(\mathbb{R})$ with respect to the mother wavelet $\psi(x) \in L^2(\mathbb{R})$ is defined by:
%\begin{equation}
%W(a,b)\,=\, \int_{-\infty}^\infty f(x) \psi_{a,b}(x)dx
%\end{equation}
%where $\psi_{a,b}(x)$ is a spatially localised function given by:
%\begin{equation}
%\psi_{a,b}(x)\,=\,\frac{1}{\sqrt{a}}\psi(\frac{x-b}{a})
%\end{equation}
%and $a,b \in L^2(\mathbb{R})$ correspond to the scaling (dilation) and shifting (translation) of $\psi(x)$. Note that the CWT spectrum obtained has no restriction as to how many scales are used, nor of the spacing between the scales. 
\newline
\indent The fundamental idea behind wavelet analysis is to highlight details apparent on different scales within the data. An example of this is the suppression of noise in images, which tends to occur only on the smallest scales. Wavelets have benefits over previous methods (e.g., Fourier transforms) because they are localised in space and are easily dilated and translated in order to operate on multiple scales, the basic equation being:
\begin{equation}
\psi_{a,b}(t)\,=\, \frac{1}{\sqrt{b}} \, \psi (\frac{t-a}{b})
\end{equation}
where $a$ and $b$ represent the shifting (translation) and scaling (dilation) of the mother wavelet $\psi$ which can take several forms depending on the required use. 
%For a computationally faster procedure we use a discrete wavelet transform (DWT) consisting of a scaling function $\phi (x)$ and corresponding wavelet $\psi (x)$ with finite support $[0,l]$ ($l$ being a positive number) given by:
%\begin{equation}
%\phi (x) \,=\, \sqrt{2} \sum_{s=0}^l h_s \phi (2x-s) 
%\end{equation}
%\begin{equation}
%\psi (x) \,=\, \sqrt{2} \sum_{s=0}^l g_s \phi (2x-s) 
%\end{equation}
%where $h$ and $g$ are constants called the low-pass and high-pass filter coefficients respectively. The DWT decomposes a signal into a set of wavelet components with dyadic scaling, i.e. the scale increases in powers of 2 (2$^1$, 2$^2$, etc).
\newline
\indent We explore a method of multiscale decomposition in 2D through the use of low and high pass filters; using a discrete approximation of a Gaussian, $\theta$, and its derivative, $\psi$, respectively \citep{2003A&A...398.1185S}. Since $\theta(x,y)$ is separable, i.e. $\theta(x,y)=\theta(x)\theta(y)$, we can write the wavelets as the first derivative of the smoothing function:
\begin{align}
\psi_{x}^{s}(x,y)\,&=\, s^{-2} \frac{\partial \theta(s^{-1}x)}{\partial x}\theta(s^{-1}y) \\
\psi_{y}^{s}(x,y)\,&=\, s^{-2} \theta(s^{-1}x)\frac{\partial \theta(s^{-1}y)}{\partial y}
\end{align}
where $s$ is the dyadic scale factor such that $s=2^j$ where $j=1,2,3,...,J$. Successive convolutions of an image with the filters produces the scales of decomposition, with the high-pass filtering providing the wavelet transform of image $I(x,y)$ in each direction:
\begin{align}
W_{x}^{s}I \,&\equiv\, W_{x}^s I(x,y)\,=\,\psi_{x}^s (x,y)*I(x,y) \\
W_{y}^{s}I \,&\equiv\, W_{y}^s I(x,y)\,=\,\psi_{y}^s (x,y)*I(x,y)
\end{align}
\begin{figure}[!p]
\centerline{\includegraphics[scale=0.8, clip=true, trim=0 130 0 150]{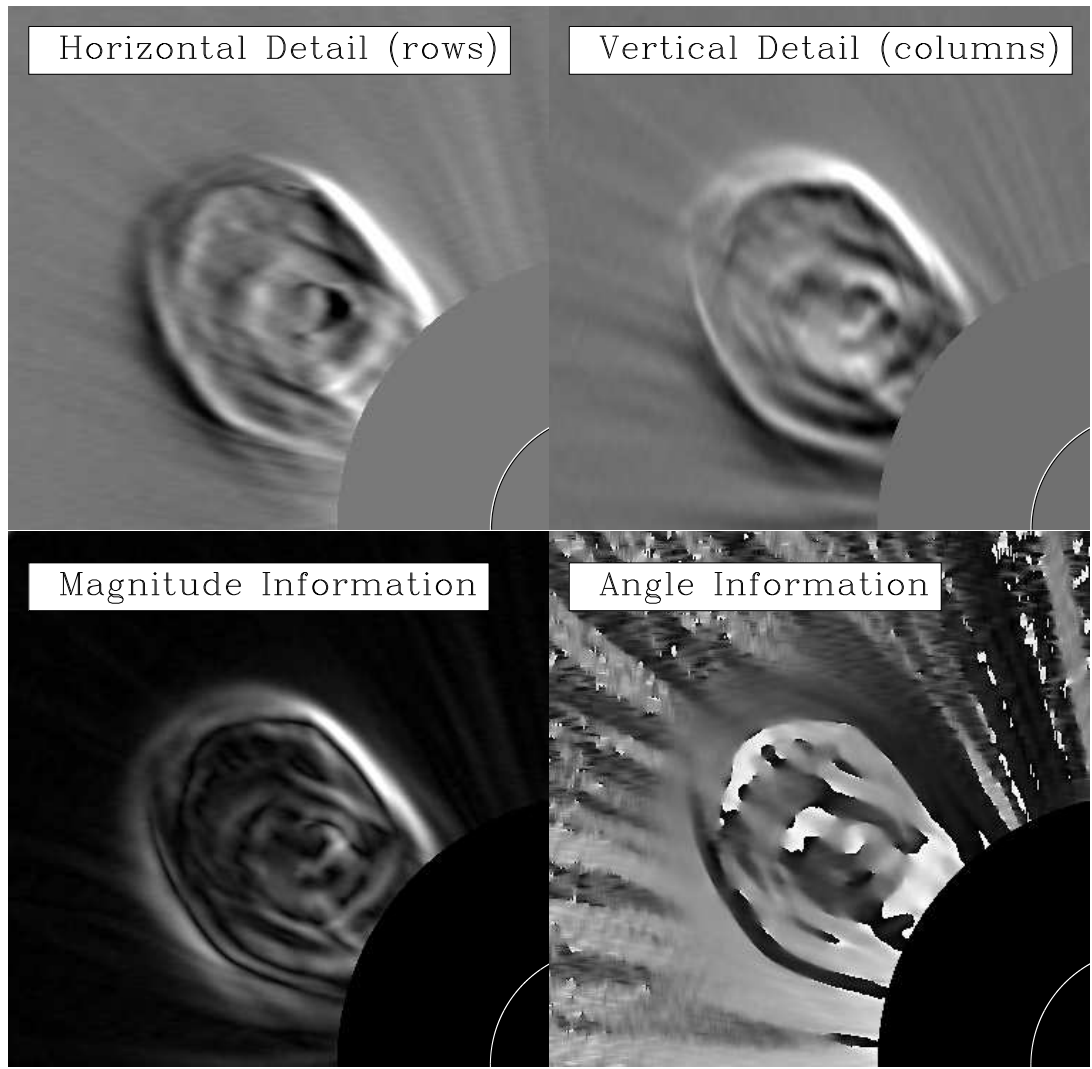}}
\caption{Top left, the horizontal detail, and top right, the vertical detail from the high-pass filtering at one scale of the multiscale decomposition (called the rows and columns respectively). Bottom left, the corresponding magnitude (edge strength) and bottom right, the angle information (0 -- 360$^{\circ}$) taken from the gradient space, for a CME observed in LASCO/C2 on 1 April 2004 \citep{2009A&A...495..325B}.}
\label{decomp}
\end{figure}
Akin to a Canny edge detector \citep{2008SoPh..248..457Y}, these horizontal and vertical wavelet coefficients are combined to form the gradient space, $\Gamma^s(x,y)$, for each scale: 
\begin{equation}
\Gamma^s (x,y)\, = \,\left[W_{x}^s I,~W_{y}^s I \right]
\end{equation}
The gradient information has an angular component $\alpha$ and a magnitude (edge strength) $M$:
\begin{align}
\alpha^s(x,y) \, &= \, tan^{-1}\left( W_{y}^s I~/~W_{x}^s I \right) \\
M^s(x,y) \, &= \, \sqrt{ ( W_{x}^s I ) ^2 + ( W_{y}^s I ) ^2 }
\end{align}
The resultant horizontal and vertical detail coefficients, and the magnitude and angular information are illustrated in Figure~\ref{decomp}.
\begin{figure}[!t]
\centering
\subfigure[00:40 UT]
{
	\label{arrow1}
	\includegraphics[scale=0.35, trim=105 130 0 130]{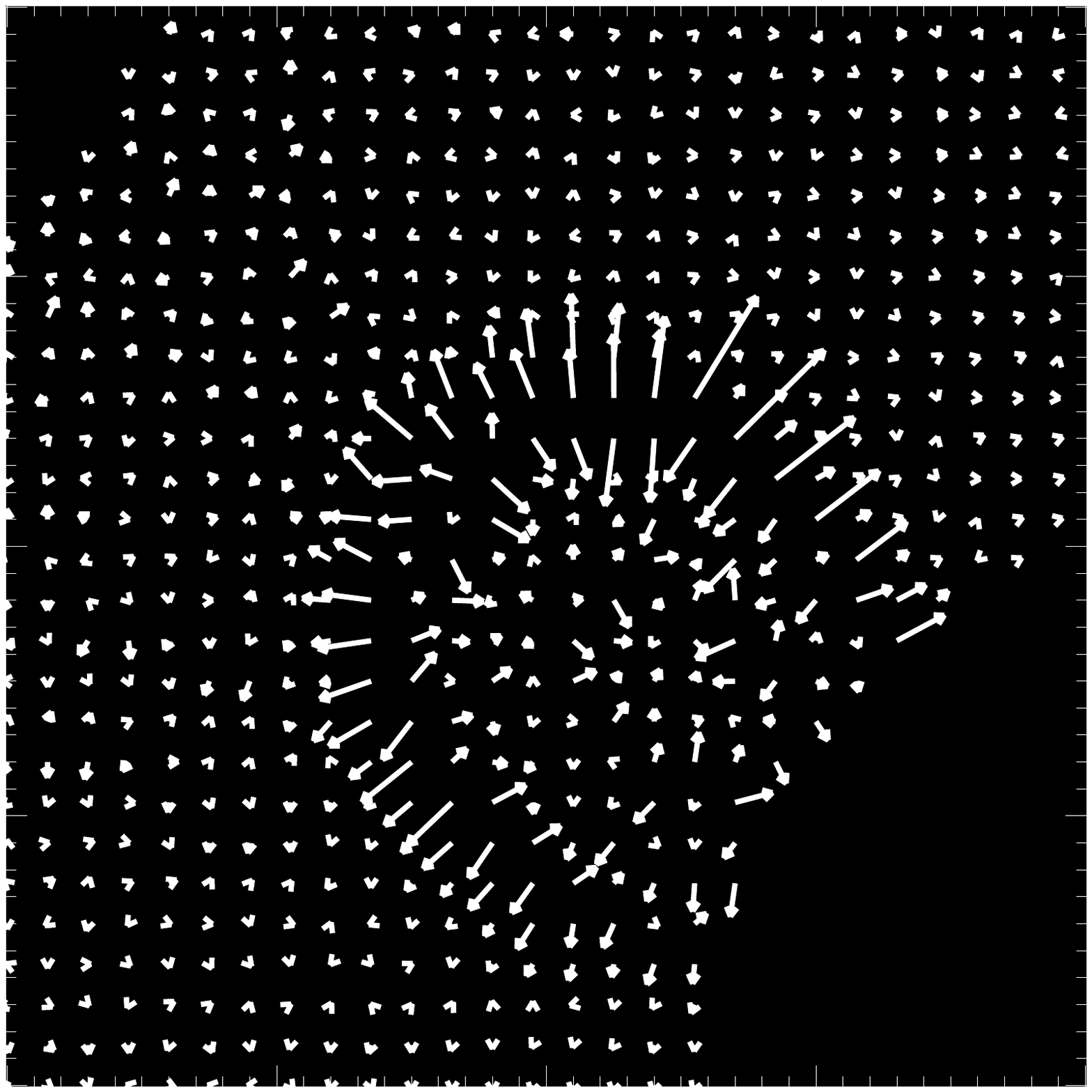}
		
}
\vspace{0cm}
\subfigure[01:00 UT]
{
	\label{arrow2}
	\includegraphics[scale=0.35, trim=55 130 80 130]{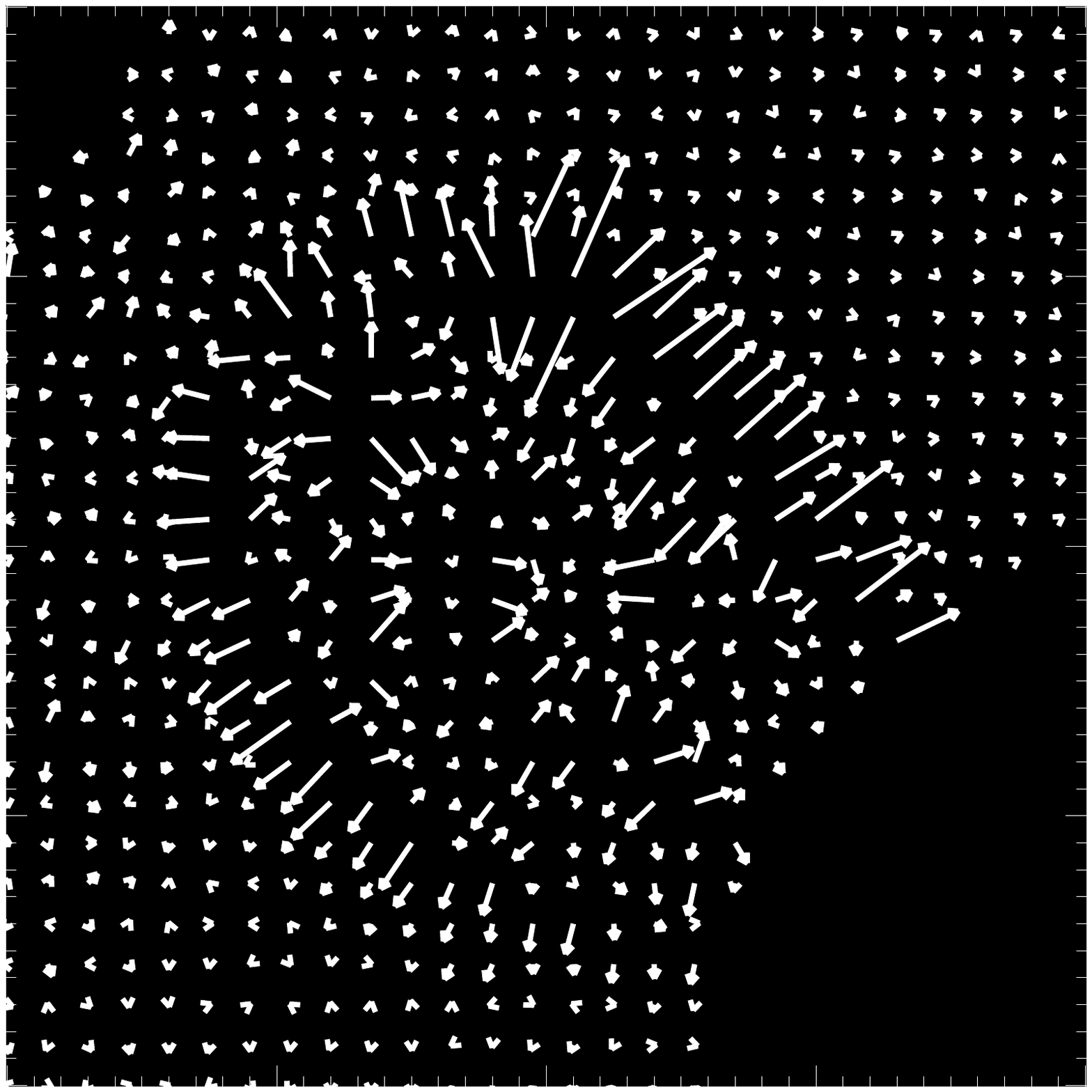}
}
\caption{The vectors plotted represent the magnitude and angle determined from the gradient space of the high-pass filtering at a particular scale. The CME of 2004 April 1 shown here is highlighted very effectively by this method \citep{2009A&A...495..325B}.}
\label{Vectors}
\end{figure} 
\newline
\indent At a particular scale the signal-to-noise ratio of the CME is highest and this is the optimum scale for determining the edges in the image. The angular component $\alpha$ of the gradient specifies a direction which points across the greatest intensity change in the data (an edge). A threshold is specified with regard to this gradient direction in order to chain pixels along maxima to highlight the edges. The changes in magnitude and angular information may then be implemented in a spatio-temporal filter for distinguishing those edges corresponding to the CME only. Overlaying a mesh of vector arrows on the data shows how the combined magnitude and angular information illustrate the progression of the CME. Each vector is rooted on a pixel in the gradient space, and has a length corresponding to the magnitude $M$ with an angle from the normal $\alpha$ (Figure~\ref{Vectors}). Using this information, it becomes possible to create a specific detection mask which is used to identify the edges along the CME front to study its propagation (as done for a sample of events in Chapter~\ref{chapter:kinematics}). However, for the cases of faint CMEs or strong streamer deflections, the filter is presently limited by exploiting the information from only one scale and ignoring all other scales, meaning it currently often requires the user to remove/include certain edges that the algorithm has mistakenly retained/discarded. Extending the algorithm to work on more than one scale may help alleviate this issue in order to develop a fully automated CME detection and characterisation routine, as outlined in the following section.

\section{Automated Multiscale CME Detection}
\label{automatedmultiscale}

For the most part, CMEs exist on size scales larger than noise and any small scale features in coronagraph images, such as stars or planets, that are redundant for studying CME propagation. This fact has led to the development and implementation of multiscale decompositions that highlight the CME in images from SOHO/LASCO and STEREO/SECCHI \citep{2008SoPh..248..457Y, 2009A&A...495..325B, 2003A&A...398.1185S}. However, coronal streamers (plasma outflows from open magnetic field regions on the Sun) can persist through coronagraph images with significant brightness intensities and tend to appear on similar scales as the CME in multiscale image analysis. If a CME propagates through an image with a strong streamer present, it becomes difficult to distinguish the two features by intensity thresholding alone, and this is one reason why differencing techniques have been widely used in CME analysis. In an effort to move away from differencing and the large errors involved in the subtraction of images from each other, since the goal is to obtain kinematics with the greatest precision, we endeavour to separate CME and streamer features from one another using multiscale methods alone. These efforts involve exploiting the angular distribution that exists across a curved CME front compared to the more linear streamers in single independent images. To do this, the coronagraph images must first be normalised for their radial gradient in intensity, since the drop-off across the field-of-view is too steep to effectively segment a single entire streamer from the inner to outer edge of an image. This can serve to enhance the noise at the edge of the images, but this is again suppressed by the multiscale analysis. Occasions when the CME propagates directly in front of, or behind, a streamer remain problematic, as do strong streamer deflections that can occur when a CME propagates into or expands alongside a streamer.

\subsection{Normalising Radial Graded Filter (NRGF)}

\begin{figure}[!t]
\resizebox{\hsize}{!}{\includegraphics[clip=true, trim=30 50 30 190]{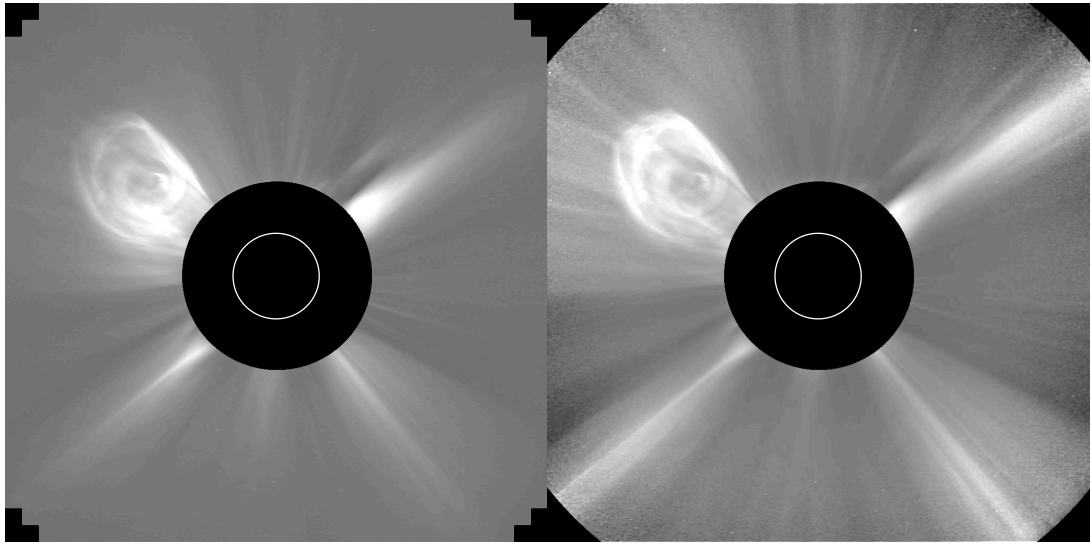}}
\caption{A normalised, background subtracted, LASCO/C2 image (left) of a CME on 1 April 2004, and the resulting NRGF image (right). The image radial intensity is scaled such that structure along streamers and the CME becomes visible across the field-of-view}
\label{im+filt}
\end{figure}

Since the brightness drop-off of the corona is large, falling from approximately $10^{-6}$\,--\,$10^{-9}$~B$_{\odot}$ across heights of 1\,--\,6~R$_{\odot}$ \citep{1998EP&S...50..493K}, a method for radially normalising coronagraph images to enhance features across this steep intensity gradient was developed by \citet{2006SoPh..236..263M}. It works by normalising the intensity in radial coordinates of the image according to the equation
\begin{eqnarray}
I'(r,\phi)\,=\,\frac{I(r,\phi)-I(r)_{<\phi>}}{\sigma(r)_{<\phi>}}
\end{eqnarray}
where $I'(r,\phi)$ is the processed and $I(r,\phi)$ is the original intensity at height $r$ and position angle $\phi$, and $I(r)_{<\phi>}$ and $\sigma(r)_{<\phi>}$ are the mean and standard deviation of intensities calculated over all position angles at height $r$. Figure~\ref{im+filt} shows the result when the NRGF is applied to a LASCO/C2 image of the 1 April 2004 CME.
\newline
\indent The multiscale decomposition introduced in \citet{2008SoPh..248..457Y} provides magnitude and angular information of the edges in the image. This information is combined to chain the strongest edges within the image on a scale that provides an optimal signal-to-noise ratio for studying the CME. \citet{2009A&A...495..325B} obtain the CME front edges in this manner, which are then used to fit an ellipse to characterise the CME propagation in the image sequence. In order to automate the algorithm, thresholds on the magnitude information (e.g., CME edges appear on larger scales than noisy features) and angular information (e.g., CME edges appear more curved than streamer edges) were investigated. The thresholding is strengthened by the inclusion of more than one scale in localising the CME and distinguishing it from the streamers, detailed below.

\subsection{Thresholding}

\begin{figure}[!t]
\resizebox{\hsize}{!}{\includegraphics[clip=true, trim=40 20 0 230]{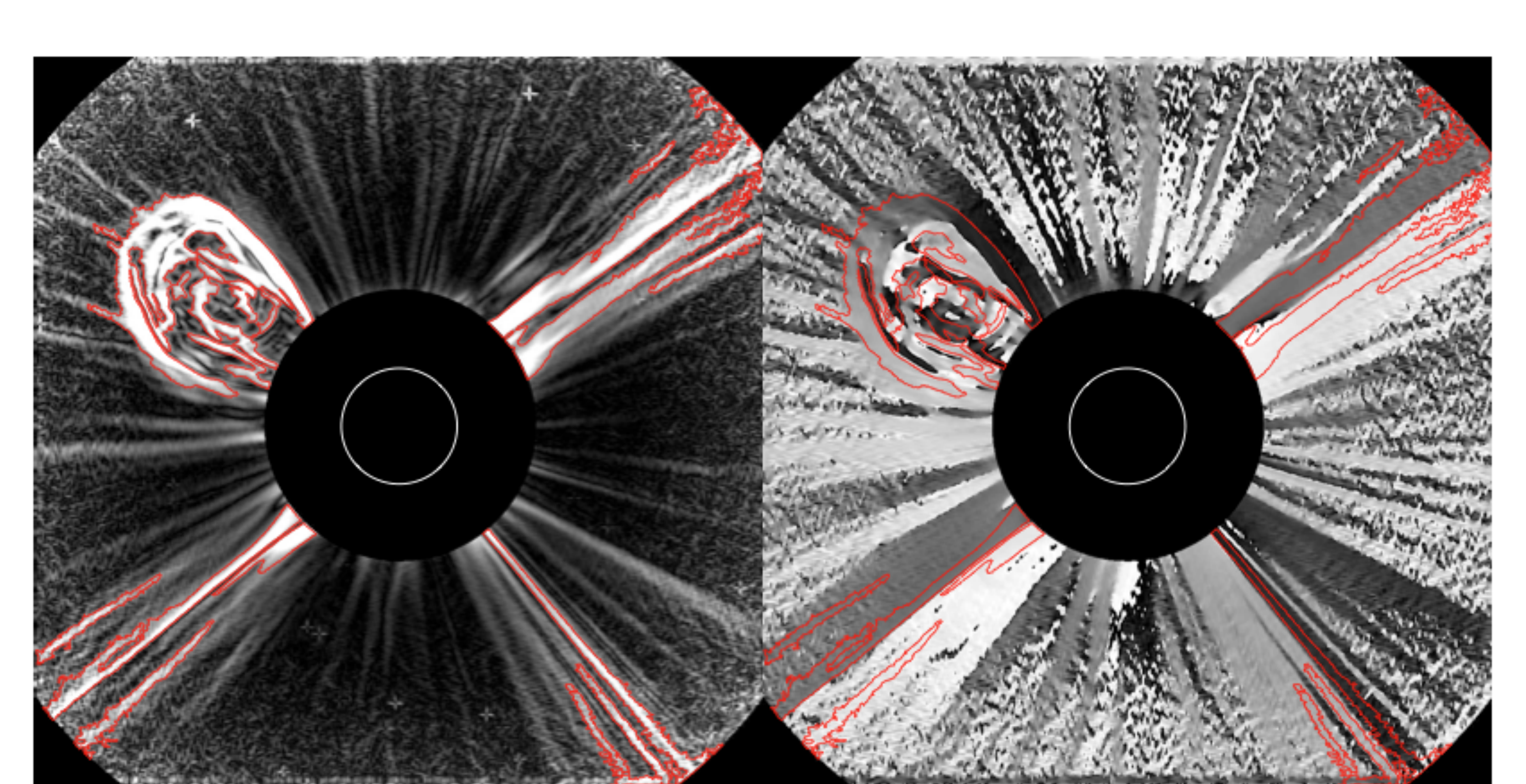}}
\caption{A chosen scale of the decomposed NRGF image provides a magnitude image of the edge strengths displayed on the left, which is thresholded at one standard deviation from the mean intensity to obtain contoured regions of interest that could contain a CME (sample contours indicated in red). As shown, the streamers have edges which appear on the same scale as the CME edges in this image. The angular information from the decomposition is displayed on the right, and the contoured regions of interest overlaid for comparison. The grey scale indicates angles from 0\,--\,360$^{\circ}$ and it is clear that streamers tend to have a linear grey scale while the CME has a gradient of greys across the scale.}
\label{modalpmapcontour}
\end{figure}

The magnitude information corresponds to the strength of the edges in the image, and so can be thresholded to discard the small scale noise. For the NRGF image (Figure~\ref{im+filt}), a hard threshold $T$ is set at one standard deviation $\sigma$ of the mean $\mu$ of the image intensity to contour regions of interest that may be a CME according to the equations:
\begin{equation}
T \,=\, \mu + N \sigma \,=\, \underbrace{ \frac{1}{n}\sum_{i=1}^n x_i}_{\mbox{mean}} \,+ N \underbrace{ \sqrt{ \frac{1}{n} \sum_{i=1}^n \left( x_i - \mu \right) ^2 } }_{\mbox{standard deviation}}
\label{eqn:threshold}
\end{equation}
where $x_i$ are the pixel intensity values of the image, and we choose the number of standard deviations $N=1$. The left image of Figure~\ref{modalpmapcontour} illustrates this thresholding with a sample of contoured regions (outlined in red) on the multiscale decomposition of a CME observed in LASCO/C2 on 1 April 2004.
\begin{figure}[!t]
\centerline{\includegraphics[scale=0.7]{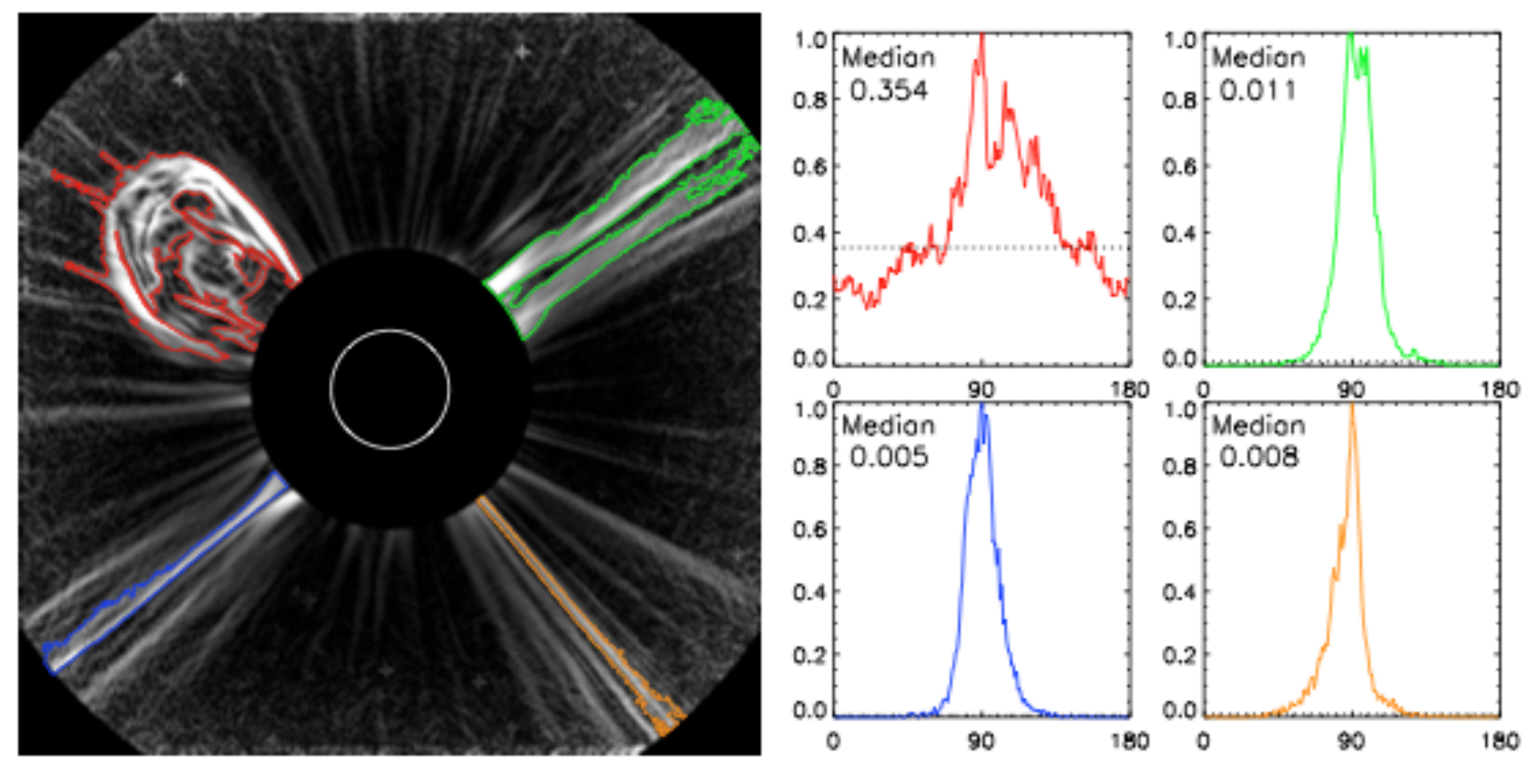}}
\caption{Left: four contoured regions (at one standard deviation of the mean image intensity) highlighted on the magnitude information from the multiscale decomposition of the 1 April 2004 CME. Right: the corresponding angular distribution of each region, normalised and folded into the 0\,--\,180$^{\circ}$ range (centred on 90$^{\circ}$). The angular distribution may be thresholded with respect to its median value to distinguish regions corresponding to CMEs from those along streamers.}
\label{ccdistributions}
\end{figure}
\newline
\indent It is apparent from the right of Figure~\ref{modalpmapcontour} that the CME will contain edges whose normals are widely distributed across 0\,--\,360$^{\circ}$ compared to the more linear, radially directed, streamer edges. The angular distribution of each region is determined and then normalised and folded into 0\,--\,180$^{\circ}$ range, centred on 90$^{\circ}$, due to the symmetry of the edge normals. This is illustrated for four selected contour regions in Figure~\ref{ccdistributions}. The resulting angular distributions are then thresholded with regard to their median value, since the distribution of angles across the CME will be wider and have a higher median value than for a distribution of angles along a streamer.
\newline
\indent This thresholding is repeated across four scales of the multiscale decomposition, neglecting smaller scales dominated by noise and larger scales that are overly smoothed. Assigning a score to the regions that may contain a CME at each scale, it becomes possible to build a detection mask as in Figure~\ref{cme_mask}. 
\begin{figure}[!t]
\centerline{\includegraphics[scale=0.55, clip=true, trim=40 20 0 230]{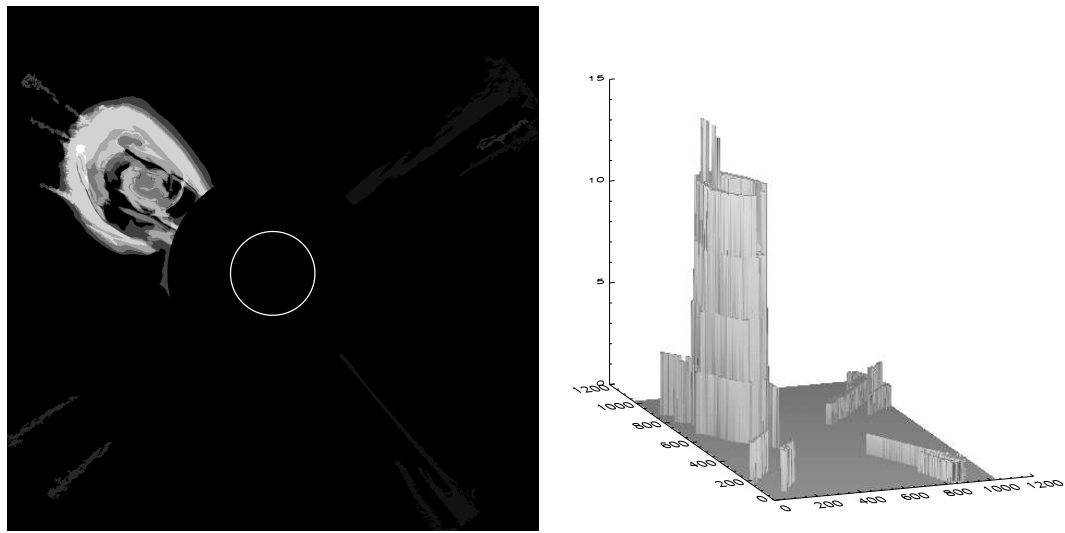}}
\caption{The resulting CME detection mask from combining the thresholded regions of strongest magnitude and angular distribution at four scales of the decomposition. The 3D representation on the right illustrates the pixel values of the mask.}
\label{cme_mask}
\end{figure}
\newline
\indent The scoring system is chosen arbitrarily to work best with the chosen thresholds, and these may be changed and refined as an analysis of more CMEs is done. For example, the current thresholds from working on a sample of $\sim$\,10 CMEs are as follows:
\begin{enumerate}
\item The magnitude information is thresholded at one standard deviation (1$\sigma$) of the mean intensity across the image.
\item The 15 largest contoured regions across the image are investigated (there are rarely more than $\sim$\,5 streamers of similar intensity to a CME, and we allow for disjointed contours along structures). 
\item If the median angular value is $>$\,20\% of the angular distribution peak then the region is deemed a CME and assigned a score of 3 (the pixels in that region of the mask are given the value 3). 
If it is $>$\,10\% the score is 2 (potential CME structure), or $>$\,5\% the score is 1 (weak CME structure or portion thereof).
\item The final CME mask through the combination of scores at each scale results in a dominant region that localises the CME front in the image and can be used to characterise the front, or input into a spatio-temporal filter if subsequent CME images are available in order to refine the masked region whenever streamers are still present. 
\end{enumerate}
The resultant set of detection masks for three of the frames of the CME observed on 1 April 2004 are shown in Figure~\ref{20040401_cme_masks}.

\begin{figure}[!p]
\centerline{\includegraphics[scale=1]{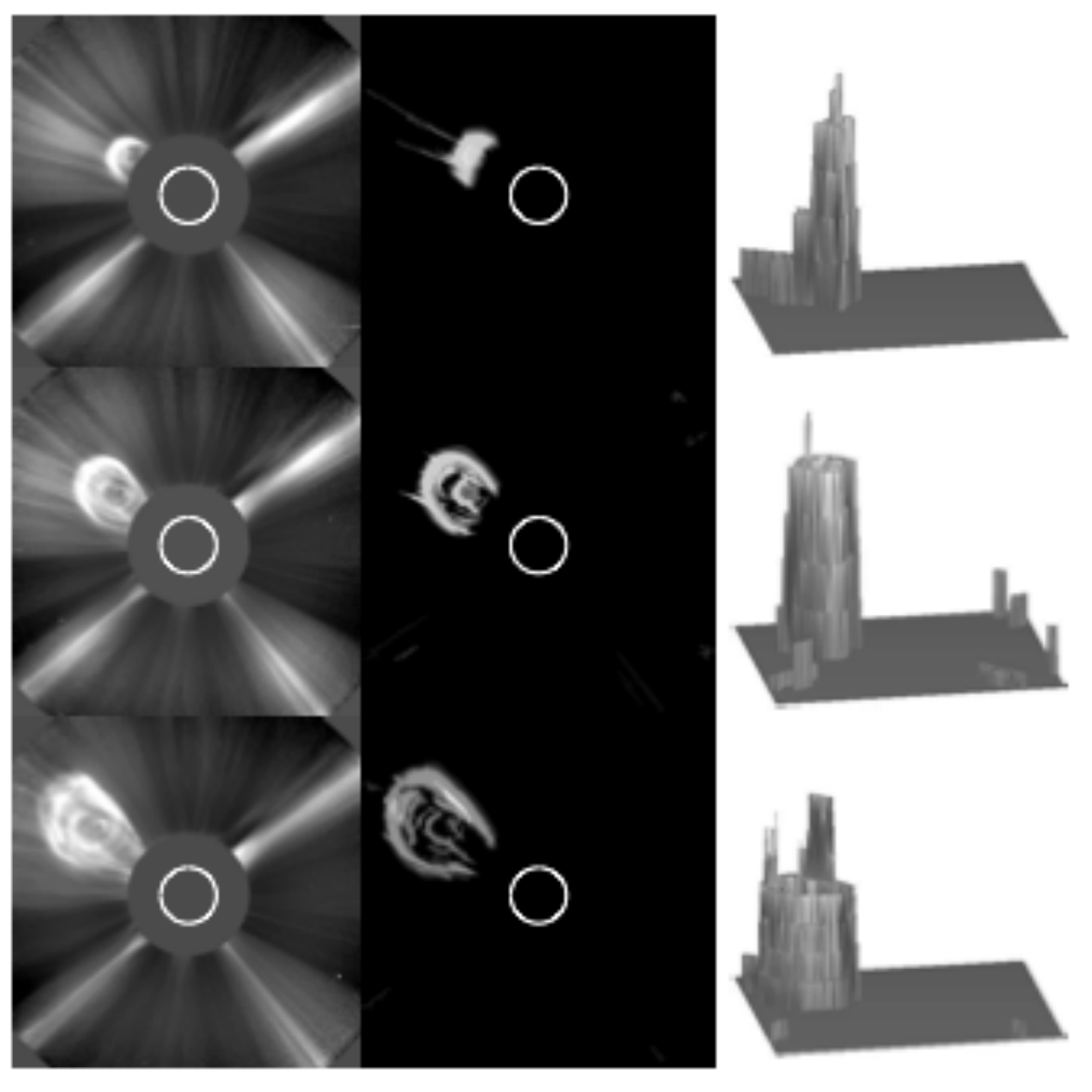}}
\caption{The NRGF (left) and resulting detection masks (middle and right) for different frames of a CME observed by LASCO/C2 at times of 00:00~UT (top), 00:40~UT (middle) and 01:20~UT (bottom) on 2 April 2004. The location of the CME front is highlighted very efficiently by this method, although the detection masks may contain artefacts of the chosen thresholds which must be discarded when characterising the CME front.}
\label{20040401_cme_masks}
\end{figure}

\subsection{Faint CMEs and Streamer Interactions/Deflections}

\begin{figure}[!p]
\centerline{\includegraphics[scale=0.8]{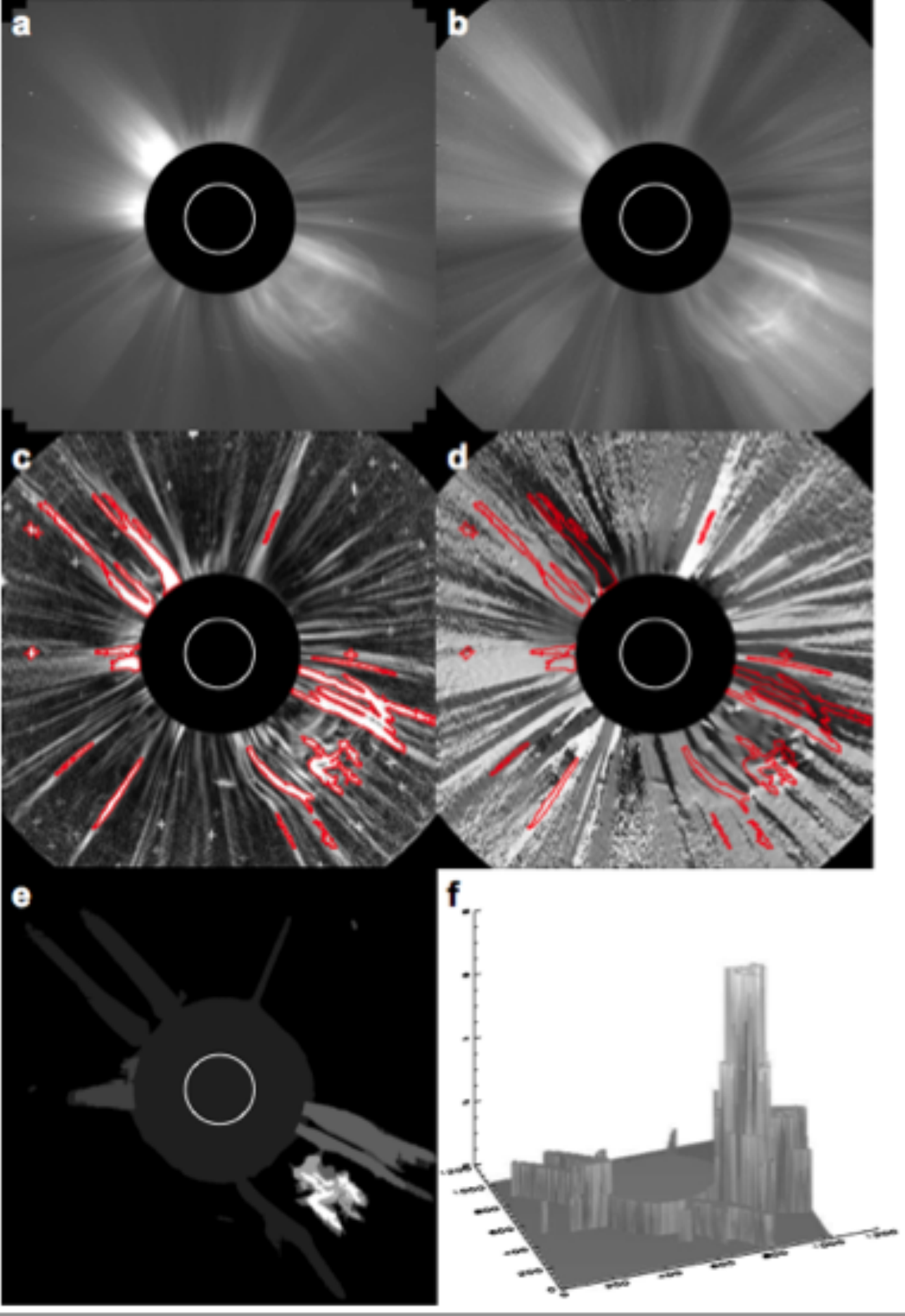}}
\caption{Example of the difficulty in detecting the faint CME observed by LASCO on 23 April 2001. If the CME is far-sided and/or of low intensity it becomes difficult to threshold its edges in the image compared to the coronal streamers. ({\bf a}) is the pre-processed CME image. ({\bf b}) is the NRGF image. ({\bf c}) and ({\bf d}) show the magnitude and angular information from the multiscale decomposition with intensity contours overlaid in red. ({\bf e}) and ({\bf f}) show the resulting CME detection mask in 2D and 3D respectively.}
\label{faintCME}
\end{figure}

%\begin{figure}[!p]
%\centerline{\includegraphics[scale=0.88, clip=true, trim=0 90 90 55]{20000211_streamers.pdf}}
%\caption{Example of the difficulty in distinguishing the CME from streamer deflections observed by LASCO on 11 February 2000. If the angular distribution along a streamers is increased due to its deflection as the CME propagates then it becomes difficult to distinguish it from the CME.}
%\label{streamerdeflection}
%\end{figure}

Due to the nature of the hard thresholds in place on the magnitude and angular information, there are problems which arise when the algorithm mistakenly disregards a CME or includes a streamer, or portions thereof. Firstly, if a CME is faint enough that the intensity falls below the 1$\sigma$ magnitude threshold, it will not be detected as a region of interest in the image. Secondly, if the CME interacts with a streamer, the two features may be contoured together and this will skew the angular distribution and affect the detection mask. And thirdly, if the CME causes a significant streamer deflection, it will lead to a wider distribution of angles along the streamer and the algorithm may thus detect it as part of the CME. This is why the above scoring system was introduced in an effort to minimise these effects, which are highlighted in Figure~\ref{faintCME} for a CME observed on 23 April 2001. The event is too faint compared with the streamers across it for it to be easily distinguished in the image, and parts of the streamers are then mistakenly included in the final detection mask. This is where the current algorithm requires a user to specify which parts of the edges correspond to the CME for characterisation. Such limitations in current wavelet analysis of CMEs may be overcome by extending these algorithms to work with ridgelets or curvelets that better suit the curved form of a typical CME front as discussed in \citet{2010gallagher}. Furthermore, it should be noted that a relatively small number of CMEs exhibit a narrow, spiky or strongly kinked structure which may not be readily distinguished from streamers in an automated fashion, nor have an adequately high angular distribution for automatic detection in the images. These CMEs may also not be satisfactorily characterised with an ellipse, discussed in the following section, however they are not typical of the commonly observed curved nature of CMEs and thus represent a small class of events outside the limits of these methods as they currently stand.

\section{Characterising the CME Front}
\label{sect:characterisation}

Using a model such as an ellipse to characterise the CME front across a sequence of images, has the benefit of providing the kinematics and morphology of a moving and/or expanding structure. The ellipse's multiple parameters, namely its changeable axis lengths and tilt angle, is adequate for approximating the varying curved structures of CMEs. \citet{1997ApJ...490L.191C} suggest an ellipse to be the two-dimensional projection of a flux rope, and \citet{2006ApJ...652.1740K} use ellipses to parameterise CMEs and explore their geometrical properties. It also serves as the observed projection of the base of the cone model applied to CME images \citep{2005JGRA..11008103X, 2004JGRA..10903109X, 2002JGRA..107.1223Z}. We fit ellipses to the points determined to be along the CME front by means of a Levenberg-Marquardt least squares algorithm. A kinematic analysis then provides height, velocity and acceleration profiles; while the ellipse's changing morphology provides the tilt angle and angular width (see the example in Figure~\ref{ellipse3}). Measuring these properties in the observed data is vitally important for accurate comparison with theoretical models. 
\begin{figure}[!t]
\centerline{\includegraphics[scale=0.3, clip=true, trim=100 60 100 40]{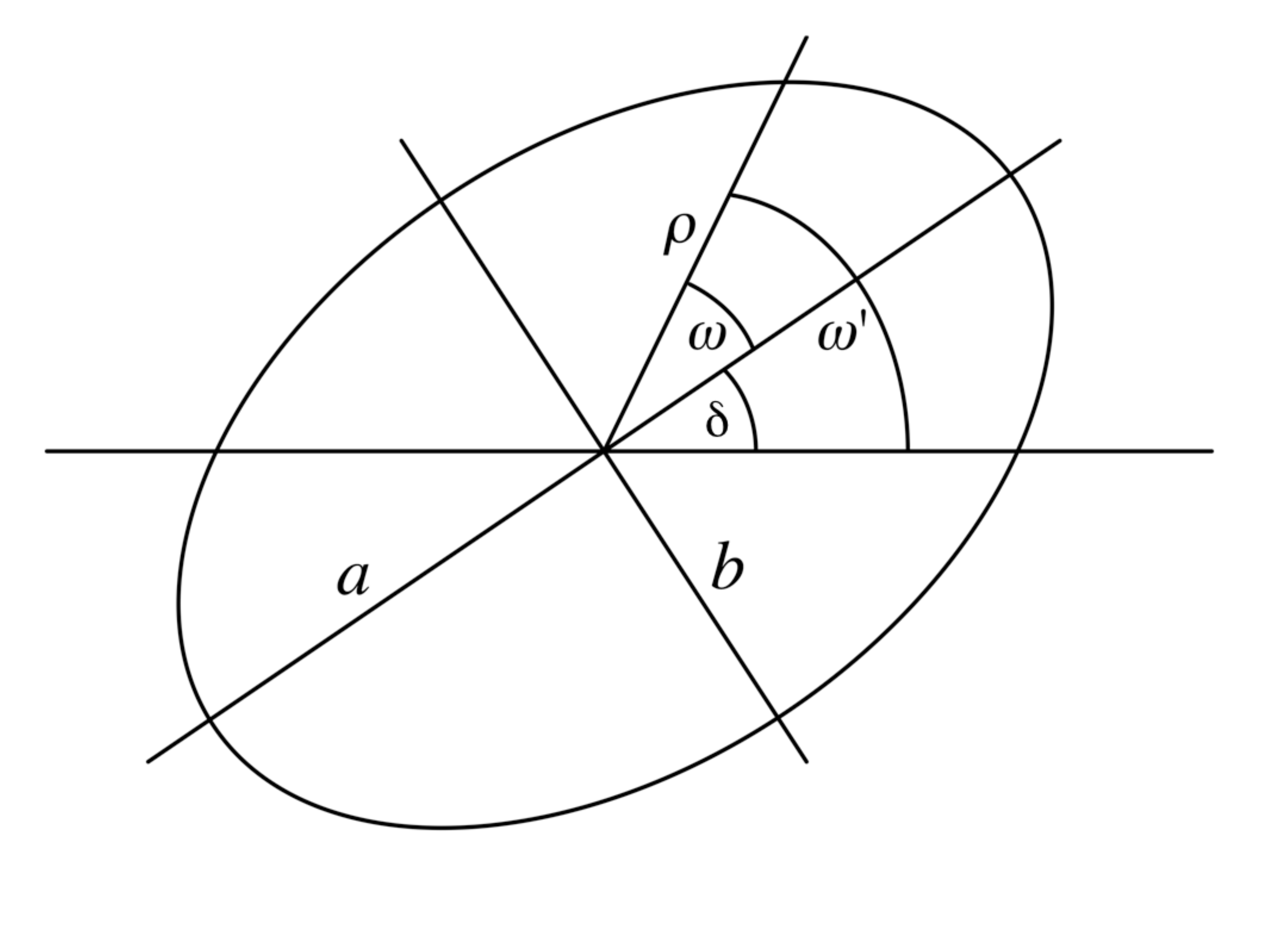}}
\caption{Ellipse tilted at angle $\delta$ in the plane, with semimajor axis $a$, semiminor axis $b$, and radial line $\rho$ inclined at angle $\omega$ to the semimajor axis.}
\label{ell_tilt}
\end{figure} 
\newline
\indent Following \citet{1145021}, we may determine the polar equation of a tilted ellipse by starting with the standard equation for an ellipse with centre point $(x_0, y_0)$, semimajor axis $a$, and semiminor axis $b$:
\begin{eqnarray}
\frac{(x-x_0)^2}{a^2}+\frac{(y-y_0)^2}{b^2}\,=\,1
\end{eqnarray}
This is written in polar coordinates by $x=\rho\cos\omega$, $y=\rho\sin\omega$ and centred on the origin $(x_0=0, y_0=0)$ to give:
\begin{equation}
\frac{\rho^2\cos^2\omega}{a^2}+\frac{\rho^2\sin^2\omega}{b^2}\,=\,1
\end{equation}
where $\rho$ is a radial line from the centre to any point on the ellipse, at an angle $\omega$ to the semimajor axis $a$ (Figure~\ref{ell_tilt}). Allowing for a tilt angle $\delta$ on the ellipse, we may define $\omega'=\omega+\delta$ to obtain:
\begin{equation}
\label{eqn:inclinedellipse}
\rho^2\,=\,\frac{a^2b^2}{(\frac{a^2+b^2}{2})-(\frac{a^2-b^2}{2})\cos(2\omega'-2\delta)}
\end{equation}
This gives a first approximation which can then be used to iteratively solve the ellipse parameters until a best fit to the points along the CME front is obtained (Figure~\ref{ellipse3}).

\begin{sidewaysfigure}[!p]
\centerline{\includegraphics[width=\linewidth]{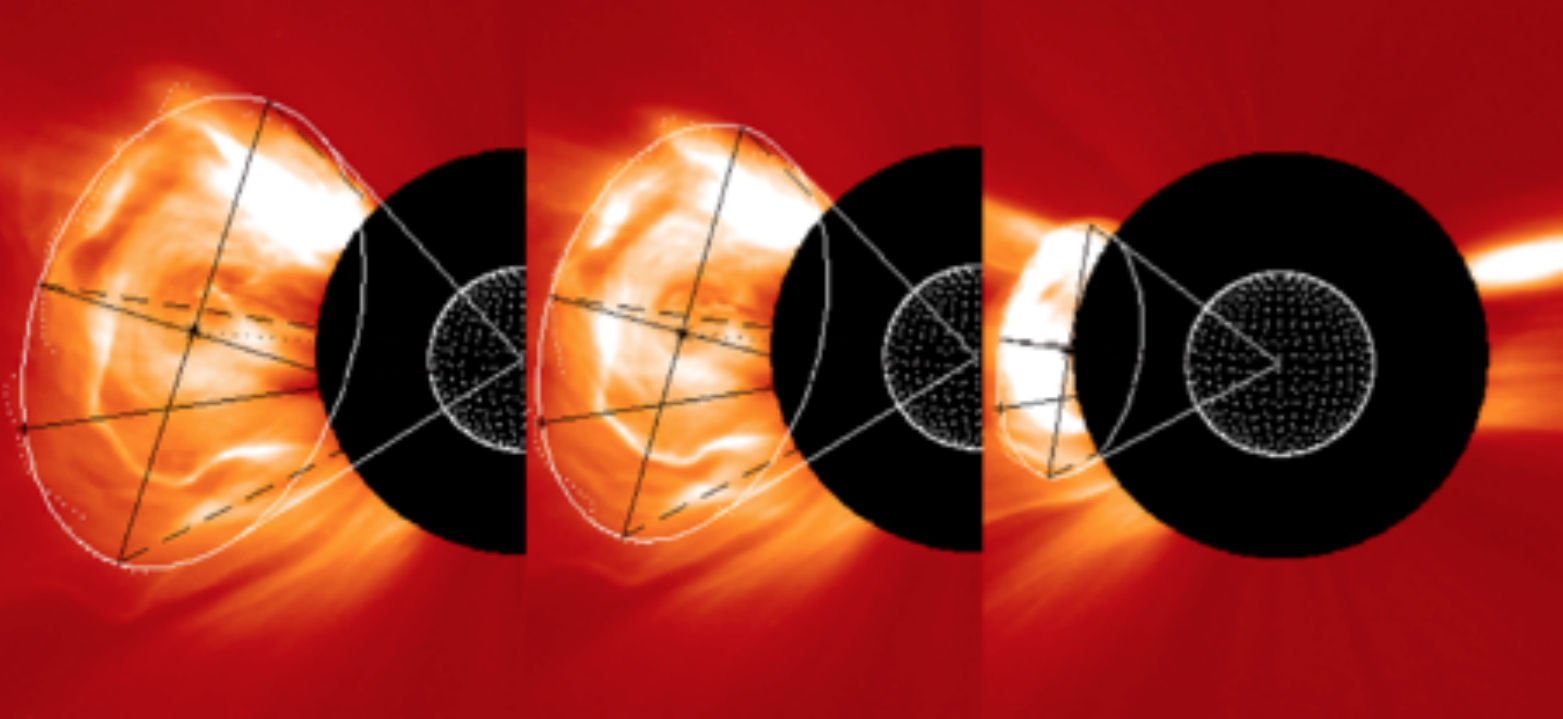}}
\caption{A depiction of three frames sampled from the ellipse characterisation of a CME observed by LASCO/C2 on 24 January 2007. The white dots along the CME front indicate the multiscale edge detection to which the ellipse is fit. The black cross on the ellipse indicates the furthest point measured from Sun centre which provides the height-time profile for determining the kinematics. The opening angle from Sun centre measures the CME width.}
\label{ellipse3}
\end{sidewaysfigure}

%\subsection{Levenberg-Marquardt Algorithm?!?}

\chapter{The Kinematics and Morphology of CMEs using Multiscale Methods}
\label{chapter:kinematics}
% Byrne et al. A&A 2009

%aims
\hrule height 1mm
\vspace{0.5mm}
\hrule height 0.4mm 
\noindent 
\\
The diffuse morphology and transient nature of CMEs make them difficult to identify and track using traditional image processing techniques. We apply multiscale methods to enhance the visibility of the faint CME front. This enables an ellipse characterisation to objectively study the changing morphology and kinematics of a sample of events imaged by SOHO/LASCO and STEREO/SECCHI. The accuracy of these methods allows us to test the CMEs for non-constant acceleration and expansion. This chapter is founded on work published in Byrne {\it et al., Astronomy \& Astrophysics} (2009).
\vspace{4mm}
\hrule height 0.4mm
\vspace{0.5mm}
\hrule height 1mm 
 
%methods
%We exploit the multiscale nature of CMEs to extract structure with a multiscale decomposition, akin to a Canny edge detector. Spatio-temporal filtering highlights the CME front as it propagates in time. We apply an ellipse parameterisation of the front to extract the kinematics (height, velocity, acceleration) and changing morphology (width, orientation).
%results
%The kinematic evolution of the CMEs discussed in this paper have been shown to differ from existing catalogues. These catalogues are based upon running-difference techniques which can lead to overestimating CME heights. Our resulting kinematic curves are not well fitted with the constant acceleration model. It is shown that some events have high acceleration below $\sim$5~R$_{\odot}$. Furthermore, we find that the CME angular widths measured by these catalogues are overestimated, and indeed for some events our analysis shows non-constant CME expansion across the plane-of-sky.

\newpage
\section{Introduction}
\label{multiscalekinsintro}

\begin{sidewaysfigure}[!p]
\centerline{\includegraphics[width=\linewidth]{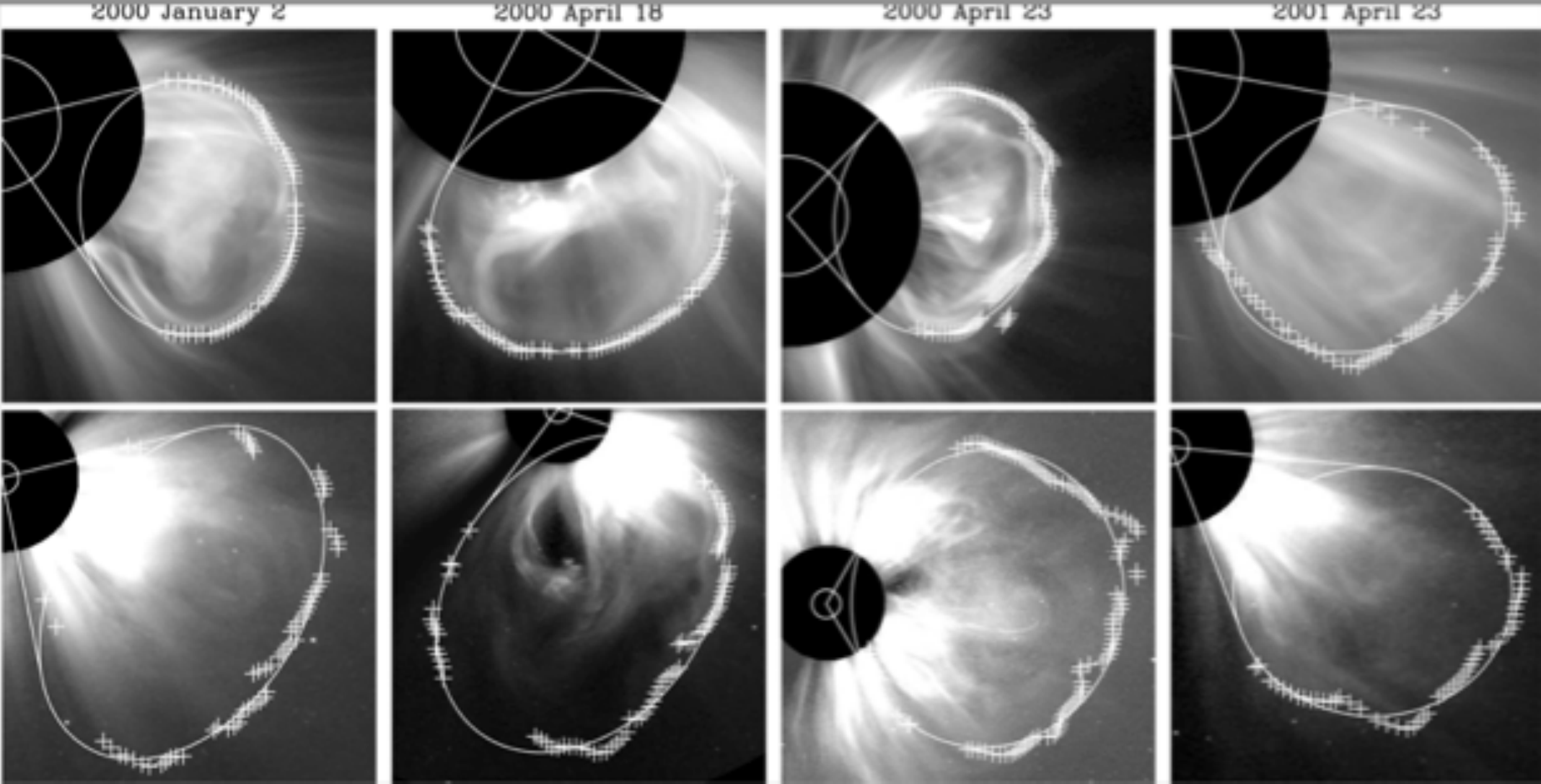}}
\caption{A sample of ellipse fits to the multiscale edge detection of certain events, reproduced from \citet{2009A&A...495..325B}. For each event the upper and lower image show LASCO/C2 and C3.}
\label{cme_images1}
\end{sidewaysfigure}

\begin{sidewaysfigure}[!p]
\centerline{\includegraphics[width=\linewidth]{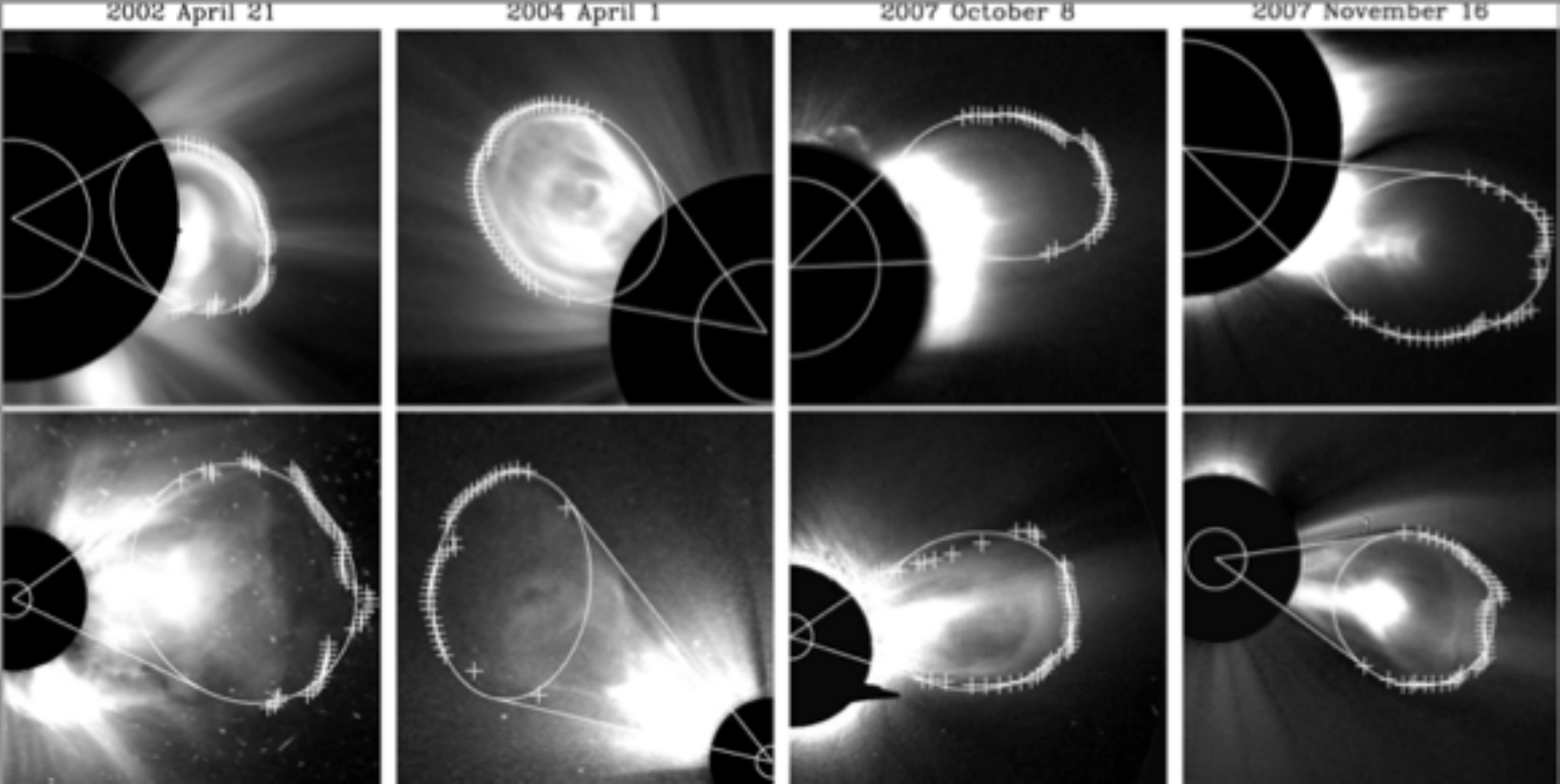}}
\caption{A sample of ellipse fits to the multiscale edge detection of certain events, reproduced from \citet{2009A&A...495..325B}. The two left event images show LASCO/C2 and C3 in the upper and lower panels respectively, while the two right events show SECCHI/COR1 and COR2.}
\label{cme_images2}
\end{sidewaysfigure}

To date, most CME kinematics are derived from difference images; a technique based either on the subtraction of a single pre-event image (fixed-difference) or the subtraction of each image from the next in an event sequence (running-difference). These techniques are applied in order to highlight regions of changing intensity, increasing the relative brightness of the CME against the background coronal features. However, drawbacks do exist. Numerical differencing can enhance noise to a level comparable to the signal. The noise can be suppressed to a certain degree by using a standard box-car or median filter, but this will also smooth out CME features such as structure along the CME front and its environs. An additional issue resulting from differencing is the introduction of spatio-temporal cross-talk in difference frames. Since it is used to highlight non-stationary features in both space and time, then the differencing of subsequent images of a moving feature will show a signature at the position where the feature was initially observed and a signature at the position that the feature has moved to when next observed. Since the signature of motion in the difference images is heavily dependent on the time between frames and how many pixels the feature has moved, it may be considered to blend spatial and temporal information in a non-trivial manner: an effect referred to as spatio-temporal cross-talk. This can serve to blur out CME features and introduce ambiguity in estimating positions and times, critical to accurately deriving the kinematics of the event. Furthermore, user bias is introduced by the choice of intensity scaling and thresholding when determining the location of CME features by point-and-click methods or automated detection algorithms, as discussed in Section~\ref{sect:cmecatalogues}.
\newline
\indent In this work we apply multiscale methods for analysing CMEs as described in Chapter~\ref{chapter:multiscale}, which has the benefit of working on independent images without any need for differencing. Once the edges of the CME front are resolved, an ellipse characterisation is applied to determine the CME kinematics (position, velocity, acceleration) and morphology (width, orientation) in coronagraph images. CME height measurements are taken as the height of the furthest point on the ellipse from Sun centre. The angular width is taken as the opening angle of the ellipse from Sun centre, and the tilt of the ellipse is given by the calculated angle $\delta$. (Note that in cases where the code produces an extremely large and oblate ellipse with one apex approximating the CME front, the width and tilt information is deemed redundant. Hence the resulting analysis of some events can have less data points included in the width and tilt plots than in the height-time plots.) Following previous concerns on the errors in CME heights \citep[e.g.,][]{2007ApJ...657.1117W}, multiscale methods allow us to determine the kinematics to a high degree of accuracy in order to improve confidence in their interpretation and comparison to theory. These methods also show potential for future automation. In this chapter a sample of CMEs observed by SOHO/LASCO and STEREO/SECCHI are studied, namely the gradual events of 2 January 2000, 18 April 2000, 23 April 2001, 1 April 2004, 8 October 2007 and 16 November 2007, and the impulsive events of 23 April 2000  and 21 April 2002 (see Figures~\ref{cme_images1} and \ref{cme_images2}). We compare our results with the catalogues of CDAW, CACTus and SEEDS (note ARTEMIS was not included due to difficulties interpreting the entries in its database that correspond to the chosen CMEs).

\section{Error Analysis}

The front of the CME is determined through the multiscale decomposition and consequent rendering of a gradient magnitude space. At scale 3 of the decomposition the smoothing filter is $2^3$ pixels wide, which we use as our 3$\sigma$ error estimate in edge position. This error is input to the ellipse fitting algorithm for weighting the ellipse parameters, and a final error output is produced for each ellipse fit. In the case of a fading leading edge the reduced amount of points along the front will increase the error on our analysis. The final errors are displayed in the height-time plots of the CMEs, and are used in the velocity and acceleration calculations. The derivative is a 3-point Lagrangian interpolation, so there is an enhancement of error at the edges of the data sets as explained below. It should also be noted that in the 3-point Lagrangian counter-intuitively results in larger errorbars on the velocity and acceleration profiles in cases where the number of height-time measurements are increased (i.e., for smaller cadences such as the inner coronagraphs). This is due to the algorithm's inverse dependence on the spacing between points.

\subsection{3-Point Lagrangian Interpolation}
\label{3pointlagrangian}

3-point Lagrangian interpolation is used on the discrete set of given data points in order to determine the first and second derivatives corresponding to the velocity and acceleration of the CME height-time measurements in a more robust manner than simple forward, reverse or centre difference techniques. Considering three data points, $(x_0, y_0)$, $(x_1, y_1)$, $(x_2, y_2)$, the Lagrangian interpolation polynomial is given by:
\begin{eqnarray}
L(x) \,=\, \sum_{j=0}^k y_j l_j(x) \quad \mbox{where} \quad l_j(x) \,=\, \prod_{i=0, i\neq j}^k \frac{x-x_i}{x_j-x_i}
\end{eqnarray}
\begin{align}
\Rightarrow \, L(x) \,&=\, y_0l_0(x)+y_1l_1(x)+y_2l_2(x) \nonumber \\
&=\, y_0 \left( \frac{x-x_1}{x_0-x_1}\frac{x-x_2}{x_0-x_2} \right) + y_1 \left( \frac{x-x_0}{x_1-x_0}\frac{x-x_2}{x_1-x_2} \right) + y_2 \left( \frac{x-x_0}{x_2-x_0}\frac{x-x_1}{x_2-x_1} \right) \nonumber 
\end{align}
So the derivative is determined to be:
\begin{align}
L' \,&\equiv \, \frac{\partial L(x)}{\partial x} \\
 &=\, y_0 \frac{2x-x_1-x_2}{\left(x_0-x_1\right)\left(x_0-x_2\right)} + y_1 \frac{2x-x_0-x_2}{\left(x_1-x_0\right)\left(x_1-x_2\right)} + y_2\frac{2x-x_0-x_1}{\left(x_2-x_0\right)\left(x_2-x_1\right)} \nonumber
\end{align}
And the edge point $x=x_0$ (and similarly for $x=x_n$) is weighted as follows:
\begin{equation}
d_{0} \,=\, y_0 \frac{2x_{0}-x_1-x_2}{\left(x_0-x_1\right)\left(x_0-x_2\right)} + y_1 \frac{x_{0}-x_2}{\left(x_1-x_0\right)\left(x_1-x_2\right)} + y_2 \frac{x_{0}-x_1}{\left(x_0-x_2\right)\left(x_1-x_2\right)} 
\end{equation}
In the case where the points are equally spaced this is simply:
\begin{equation}
d_{0} \,=\, \frac{1}{2} \left[ -3y_{0}+4y_{1}-y_{2} \right]
%d_{n} \,&=\, \frac{1}{2} \left[ 3y_{n}-4y_{n-1}+y_{n-2} \right]
\end{equation}
The error propagation equation is used to determine the errors on the resulting derivative points in $L' \equiv f(L(x),x)$:
\begin{align}
\sigma_{L'}^2 \,&=\, \sigma_L^2 \left(\frac{\partial L'}{\partial L}\right)^2 + \sigma_x^2\left(\frac{\partial L'}{\partial x}\right)^2+... \\
&=\, \frac{\sigma_L^2}{\partial x^2}+\frac{\sigma_x^2}{\partial x^2}\left(\frac{\partial L}{\partial x}\right)^2
\end{align}
Or more appropriately written in this context as:
\begin{equation}
\sigma_d^2 \,=\, \frac{\sigma_{y_{n+1}}^2+\sigma_{y_{n-1}}^2}{dx^2} + \frac{\sigma_{x_{n+1}}^2+\sigma_{x_{n-1}}^2}{dx^2}\left(\frac{dy}{dx}\right)^2
\end{equation}
So the errors on the end points become:
\begin{align}
\sigma_{d_0}^2 \,&=\, \frac{9\sigma_{y_0}^2+16\sigma_{y_1}^2+\sigma_{y_2}^2}{\left(x_2-x_0\right)^2} + \frac{\sigma_{x_2}^2+\sigma_{x_0}^2}{\left(x_2-x_0\right)^2} \left( \frac{3y_0-4y_1+y_2}{x_2-x_0}\right)^2 \\
\sigma_{d_n}^2 \,&=\, \frac{9\sigma_{y_n}^2+16\sigma_{y_{n-1}}^2+\sigma_{y_{n-2}}^2}{\left(x_n-x_{n-2}\right)^2} + \frac{\sigma_{x_{n-2}}^2+\sigma_{x_n}^2}{\left(x_{n-2}-x_n\right)^2} \left( \frac{3y_n-4y_{n-1}+y_{n-2}}{x_{n-2}-x_n}\right)^2
\end{align}
This effect is reflected in the larger errorbars on the end points of the derived kinematics of Section~\ref{sec:multiscaleresults}.
\newline
\indent The errors on the heights are used to constrain the best fit to a constant acceleration model of the form:
\begin{equation}
h(t)\,=\,at^2+v_0t+h_0
\end{equation}
where $t$ is time and $a$, $v_0$ and $h_0$ are the acceleration, initial velocity and initial height respectively. This provides a linear fit to the derived velocity points and a constant fit to the acceleration. An important point to note is the small time error (taken to be the image exposure time of the coronagraph data) since the analysis is performed upon the observed data frames individually. Previous methods of temporal-differencing would increase this time error. With these more accurate measurements we are better able to determine the velocity and acceleration errors, leading to improved constraints upon the data and providing greater confidence in comparing to theoretical models.

\section{Results}
\label{sec:multiscaleresults}

This section outlines events which have been analysed using our multiscale methods. We use data from the LASCO/C2 and C3, and SECCHI/COR1 and COR2 instruments, and preprocess the images as discussed in Section~\ref{sect:cmecatalogues}. Events were selected on the basis that they were strong candidates to test for non-constant acceleration and expansion due to their high signal-to-noise ratio out through the full coronagraph fields-of-view. The ellipse fitting algorithm applied to each event gives consistent heights of the CME front measured from Sun centre to the maximum height on the ellipse, and these lead to velocity and acceleration profiles of our events. The ellipse fitting also provides the angular widths and orientations, as shown below. The velocity (where the ranges are from initial to final speeds), acceleration and angular width results of each method are highlighted in Tables~\ref{table:vel}, \ref{table:accel} and \ref{table:aw}. In each instance we include the values from CACTus, CDAW and SEEDS. Note that CACTus lists a median speed of the CME; CDAW provide the speed at the final height and from the velocity profile we infer the speed at the initial height; and the SEEDS detection applies only to the LASCO/C2 field-of-view but doesn't currently provide a velocity range or profile. Note also that the CMEs of 8 October 2007 and 16 November 2007 are analysed in SECCHI images by CACTus and our multiscale methods (marked by asterisks in the Tables), while CDAW and SEEDS currently only provide LASCO analysis, and so may be observing a different part of the CME structure so it is difficult to compare directly. Overall, it is clear that many of the CACTus, CDAW and SEEDS results lie outside the results of our analysis. Notably, since CDAW involves labour intensive analysis by qualified scientists, this is a prime indication that robust methods of determining the kinematics and morphology of CMEs are needed, which our methods of multiscale analysis and characterisation provide, especially given our rigourous error analysis.

\begin{table}[!t] 
\caption{Summary of CME velocities as measured by CACTus, CDAW, SEEDS and multiscale methods. Asterisks indicate analysis of SECCHI data rather than LASCO.} % title of Table 
\centering      % used for centering table 
\begin{tabular}{c | c c c c}
\hline\hline                        %inserts double horizontal lines 
Date & CACTus & CDAW & SEEDS & Multiscale \\ [0.5ex] % inserts table 
%heading
& km~s$^{-1}$ & km~s$^{-1}$ & km~s$^{-1}$ & km~s$^{-1}$\\
\hline                    % inserts single horizontal line 
02 Jan 2000 & 512 & 370 -- 794 & 396 & 396 -- 725  \\   % inserting body of the table 
18 Apr 2000 & 463 & 410 -- 923 & 339 & 324 -- 1049  \\ 
23 Apr 2000 & 1041 & 1490 -- 898 & 595 & 1131 -- 1083  \\ 
23 Apr 2001 & 459 & 540 -- 519 & 501 & 581 -- 466 \\ 
21 Apr 2002 & 1103 & 2400 -- 2388 & 702 & 2195 -- 2412 \\
01 Apr 2004 & 487 & 300 -- 613 & 319 & 415 -- 570 \\
08 Oct 2007 & 235* & 85 -- 331 & 103 & 71 -- 330* \\
16 Nov 2007 & 337* & 210 -- 437 & 154 & 131 -- 483* \\ [1ex]       % [1ex] adds vertical space 
\hline\hline     %inserts single line 
\end{tabular} 
\label{table:vel}  % is used to refer this table in the text 
\end{table}

\begin{table}[!t] 
\caption{Summary of CME accelerations as measured by CACTus, CDAW, SEEDS and multiscale methods. Asterisks indicate analysis of SECCHI data rather than LASCO.} % title of Table 
\centering      % used for centering table 
\begin{tabular}{c | c c c c}
\hline\hline
Date & CACTus & CDAW & SEEDS & Multiscale \\ [0.5ex] % inserts table 
%heading 
 & m~s$^{-2}$ & m~s$^{-2}$ & m~s$^{-2}$ & m~s$^{-2}$ \\
\hline                    % inserts single horizontal line 
02 Jan 2000 & 0 & 21.3 & $-$5.8 & 14.7~$\pm$~3.6  \\   % inserting body of the table 
18 Apr 2000 & 0 & 23.1 & 17.5 & 32.3~$\pm$~3.5  \\ 
23 Apr 2000 & 0 & $-$48.5 & $-$8.9 & $-$4.8~$\pm$~20.6  \\ 
23 Apr 2001 & 0 & $-$0.7 & $-$1.4 & $-$4.8~$\pm$~4.1 \\ 
21 Apr 2002 & 0 & $-$1.4 & 33.5 & 32.5~$\pm$~26.6 \\
01 Apr 2004 & 0 & 7.1 & 12.9 & 4.4~$\pm$~2.0 \\
08 Oct 2007 & 0* & 3.4 & 2.4 & 5.7~$\pm$~0.9* \\
16 Nov 2007 & 0* & 4.9 & 11.0 & 13.7~$\pm$~1.7* \\ [1ex]       % [1ex] adds vertical space 
\hline\hline     %inserts single line 
\end{tabular}
\label{table:accel}  % is used to refer this table in the text 
\end{table}

\begin{table}[!t] 
\caption{Summary of CME angular widths as measured by CACTus, CDAW, SEEDS and multiscale methods. Asterisks indicate analysis of SECCHI data rather than LASCO.}  % title of Table 
\centering      % used for centering table 
\begin{tabular}{c | c c c c}
\hline\hline
Date & CACTus & CDAW & SEEDS & Multiscale \\ [0.5ex] % inserts table 
%heading
 & degrees & degrees & degrees & degrees \\ 
\hline                    % inserts single horizontal line 
02 Jan 2000 & 160 & 107 & 96 & 50 -- 95  \\   % inserting body of the table 
18 Apr 2000 & 106 & 105 & 108 & 68 -- 110  \\ 
23 Apr 2000 & 352 & 360 & 130 & 96 -- 130  \\ 
23 Apr 2001 & 124 & 91 & 74 & 55 -- 60 \\ 
21 Apr 2002 & 352 & 360 & 186 & 53 -- 65 \\
01 Apr 2004 & 66 & 79 & 58 & 44 -- 38 \\
08 Oct 2007 & 52* & 82 & 59 & 23 -- 60* \\
16 Nov 2007 & 68* & 78 & 54 & 40 -- 55* \\ [1ex]       % [1ex] adds vertical space 
\hline\hline     %inserts single line 
\end{tabular}
\label{table:aw}  % is used to refer this table in the text 
\end{table}

\subsection{Arcade Eruption: 2 January 2000}
This CME was first observed in the south-west at 06:06 UT on 2 January 2000 and appears to be a far-side event associated with an arcade eruption consisting of one or more bright loops. 
\par
\begin{figure}[!p]
\centerline{\includegraphics[scale=0.8, clip=true, trim=0 120 0 10]{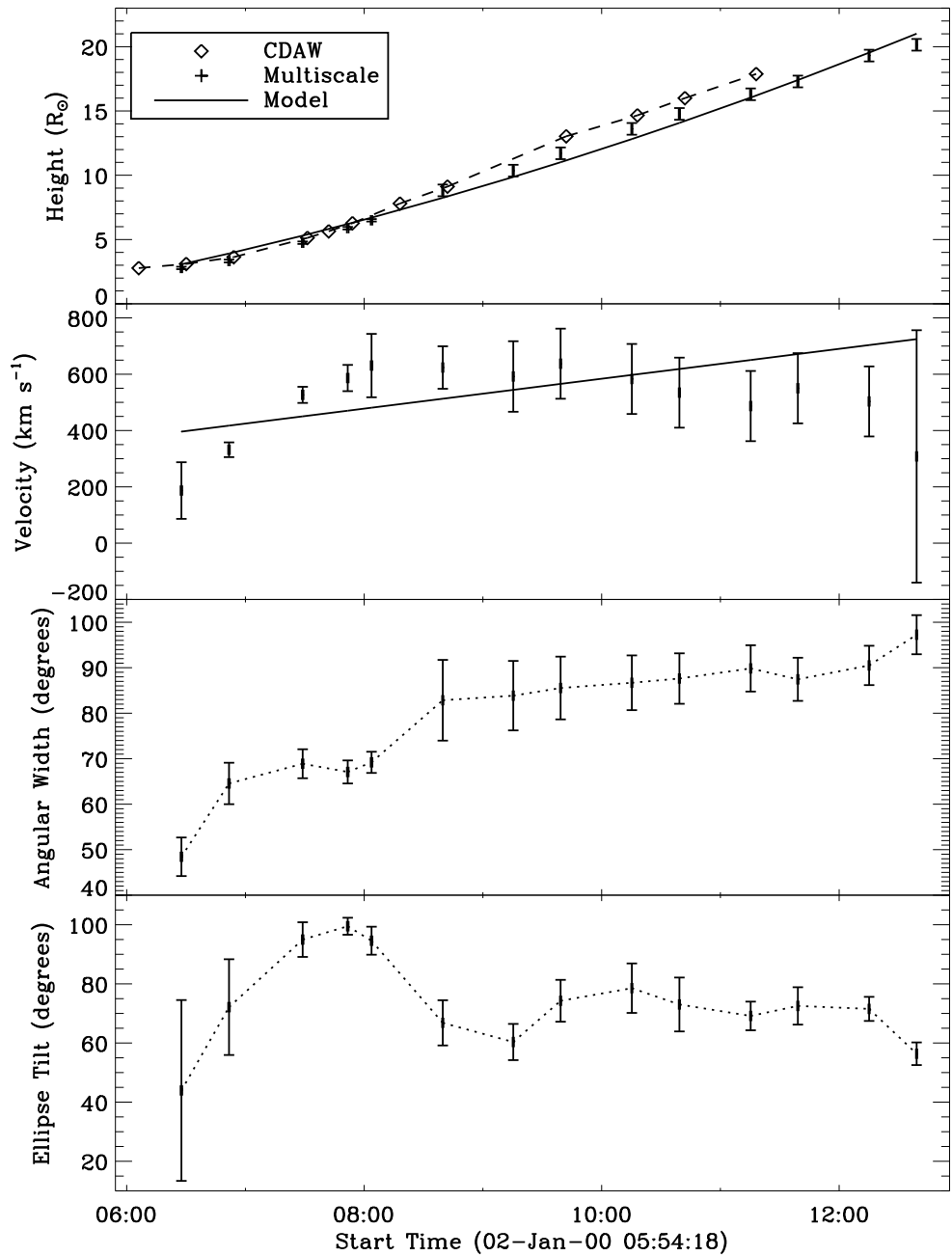}}
\caption{Kinematic and morphological profiles for the ellipse fit to the multiscale edge detection of the 2 January 2000 CME observed by LASCO/C2 and C3. The plots from top to bottom are height, velocity, angular width, and ellipse tilt. The CDAW heights are over-plotted with a dashed line. The height and velocity fits are based upon the constant acceleration model.}
\label{20000102_kins_angs}
\end{figure}

The height-time plot has a trend not unlike that of CDAW (overplotted in top Figure~\ref{20000102_kins_angs} with a dashed line). However, the offset of the CDAW heights -- which puts them outside our error bounds -- may be due to how the difference images are scaled for display. This is a problem multiscale methods avoid. From Figure~\ref{20000102_kins_angs}, the velocity fit was found to be increasing from 396 to 725~km~s$^{-1}$, giving an acceleration of 14.7\,$\pm$\,3.6~m~s$^{-2}$. The ellipse fit spans approximately 50\,--\,70$^{\circ}$ of the field-of-view in the inner portion of C2, and expands to over 95$^{\circ}$ in C3. This expansion may simply be attributed to the inclusion of one or more loops in the ellipse fit as the arcade traverses the LASCO/C2 and C3 fields-of-view.  The orientation of the ellipse as a function of time is shown in the bottom of Figure~\ref{20000102_kins_angs}. It can be seen that the orientation angle of the CME increases to approximately 100$^{\circ}$ before decreasing toward 60$^{\circ}$. 
\newline
\indent The constant acceleration model is not a sufficient fit to the data in this event. The kinematics produced from the multiscale edge detection would be better fit with a non-linear velocity and a non-constant acceleration. This would show the CME to have a period of decreasing acceleration in the C2 field-of-view, levelling off to zero in C3 (if not decelerating further).

\subsection{Gradual/Expanding CME: 18 April 2000}
\begin{figure}[!p]
\centerline{\includegraphics[scale=0.8, clip=true, trim=0 120 0 10]{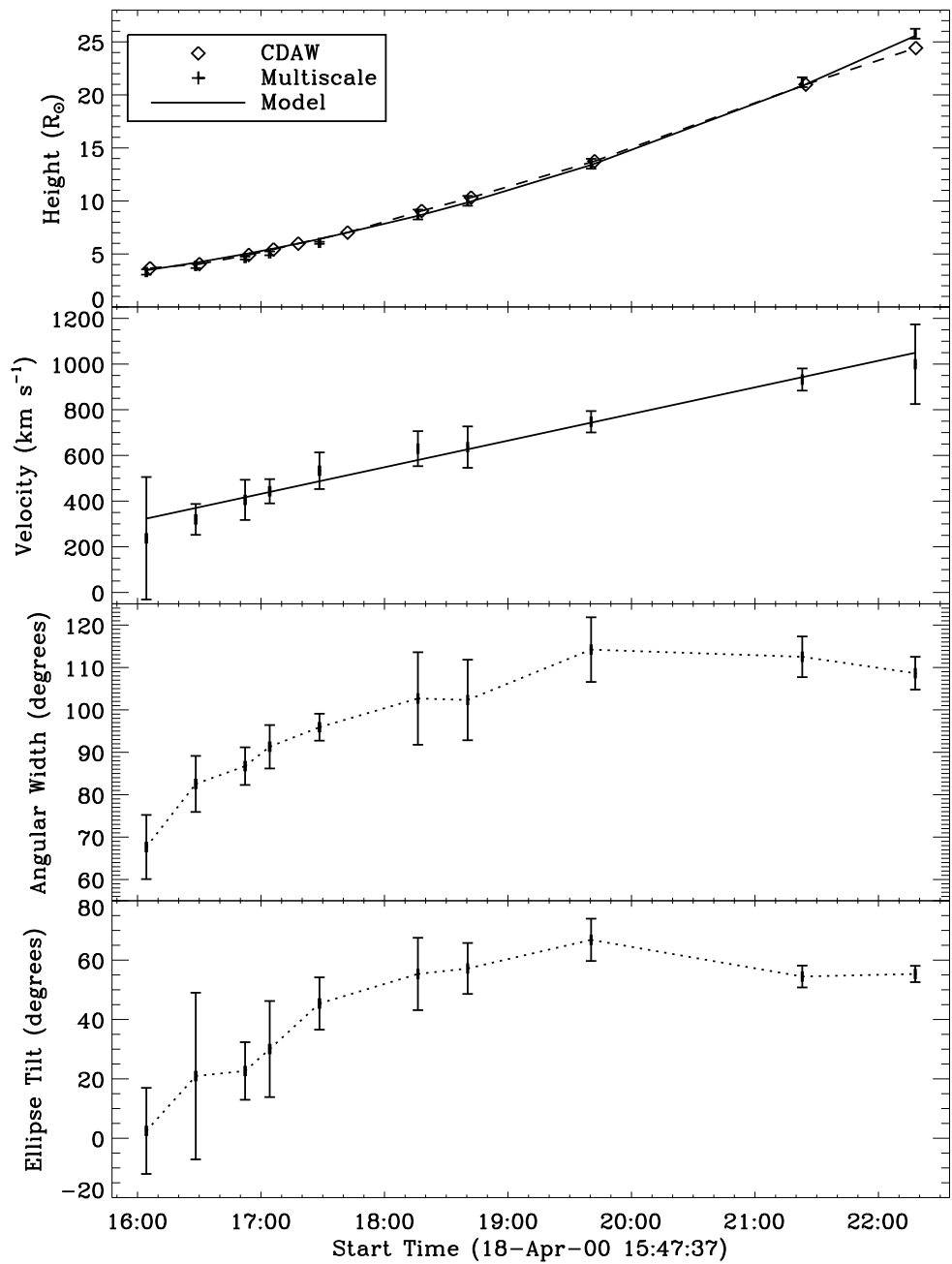}}
\caption{Kinematic and morphological profiles for the ellipse fit to the multiscale edge detection of the 18 April 2000 CME observed by LASCO/C2 and C3. The plots from top to bottom are height, velocity, angular width, and ellipse tilt. The CDAW heights are over-plotted with a dashed line. The height and velocity fits are based upon the constant acceleration model.}
\label{20000418_kins_angs}
\end{figure}

This CME was first observed off the south limb at 16:06 UT on 18 April 2000 and exhibits a flux rope type structure.
\newline
\indent The height-time plot for this event has a trend similar to that of CDAW (overplotted in top Figure~\ref{20000418_kins_angs} with a dashed line). The velocity fit was found to be linearly increasing from 324 to over 1,000~km~s$^{-1}$, giving an acceleration of 32.3\,$\pm$\,3.5~m~s$^{-2}$. The ellipse fit spans from 68$^{\circ}$ of the field-of-view in the inner portion of C2, to approximately 110$^{\circ}$ in C3.  The orientation of the ellipse as a function of time is shown to increase from just above 0$^{\circ}$ to over 60$^{\circ}$ in Figure~\ref{20000418_kins_angs}. 
\newline
\indent This CME is fitted well with the constant acceleration model but shows an increasing angular width, implying a type of over-expansion across the field-of-view.

\subsection{Impulsive CME: 23 April 2000}
\begin{figure}[!p]
\centerline{\includegraphics[scale=0.8, clip=true, trim=0 120 0 10]{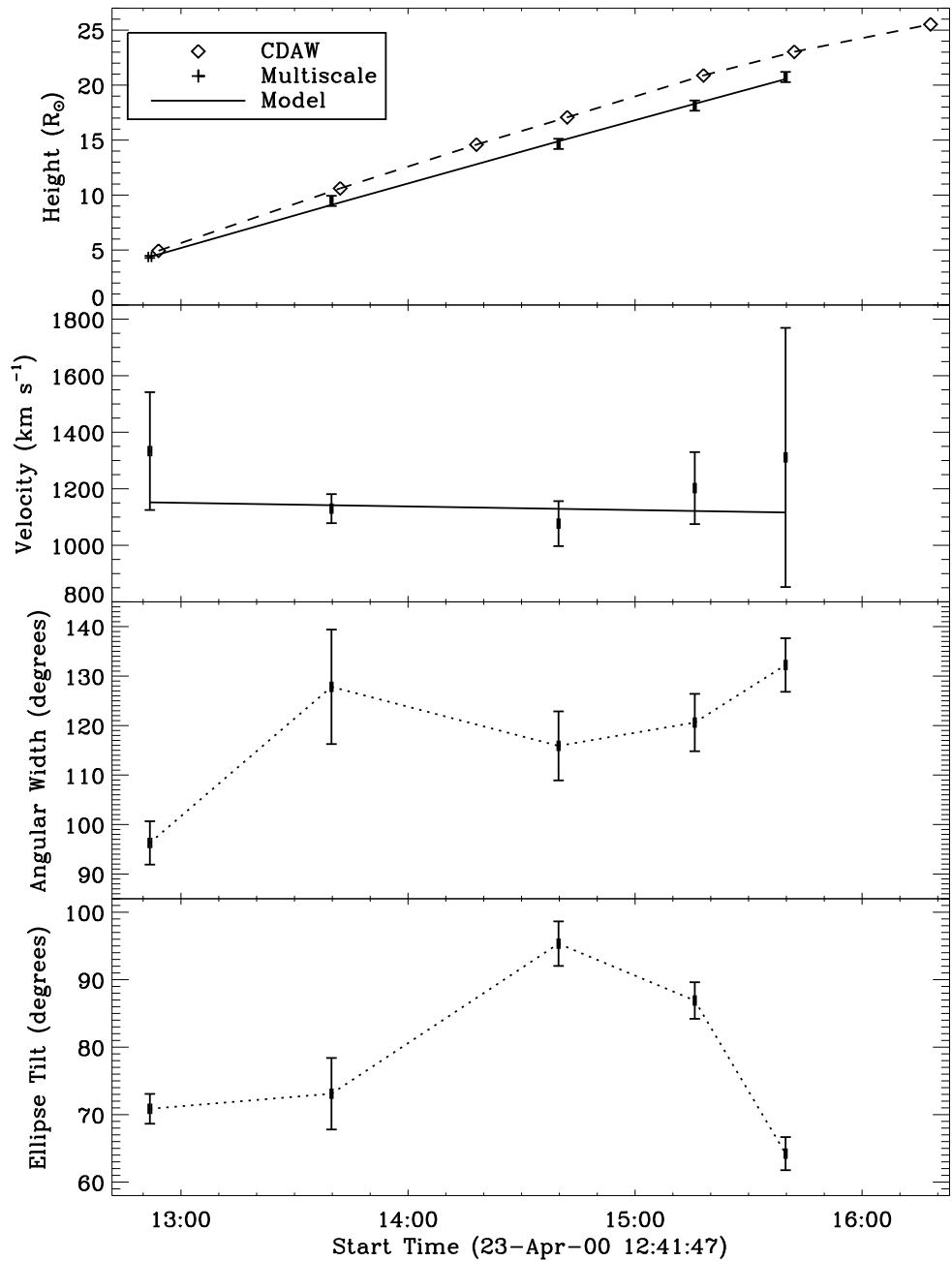}}
\caption{Kinematic and morphological profiles for the ellipse fit to the multiscale edge detection of the 23 April 2000 CME observed by LASCO/C2 and C3. The plots from top to bottom are height, velocity, angular width, and ellipse tilt. The CDAW heights are over-plotted with a dashed line. The height and velocity fits are based upon the constant acceleration model.}
\label{20000423_kins_angs}
\end{figure}

This impulsive CME was first observed in the west at 12:54 UT on 23 April 2000 and exhibits strong streamer deflection.  
\newline
\indent The height-time plot derived using our methods has a trend which diverges from that of CDAW (overplotted in top Figure~\ref{20000423_kins_angs} with a dashed line). The velocity fit was found to be linearly decreasing from 1,131 to 1,083~km~s$^{-1}$, giving a constant deceleration of $-4.8$\,$\pm$\,20.6~m~s$^{-2}$. The CME is present for one frame in C2 with an ellipse fit spanning 96$^{\circ}$, increasing to approximately 120\,--\,130$^{\circ}$ in the C3 field-of-view, and the orientation of the ellipse as a function of time is shown to rise from 71$^{\circ}$ to 95$^{\circ}$ then fall to 64$^{\circ}$ (see bottom Figure~\ref{20000423_kins_angs}). 
\newline
\indent This event is modelled satisfactorily with a constant deceleration. However, due to the impulsive nature of the CME there are only a few frames available for analysis, making it difficult to constrain the kinematics. 

\subsection{Faint CME: 23 April 2001}
\begin{figure}[!p]
\centerline{\includegraphics[scale=0.8, clip=true, trim=0 120 0 10]{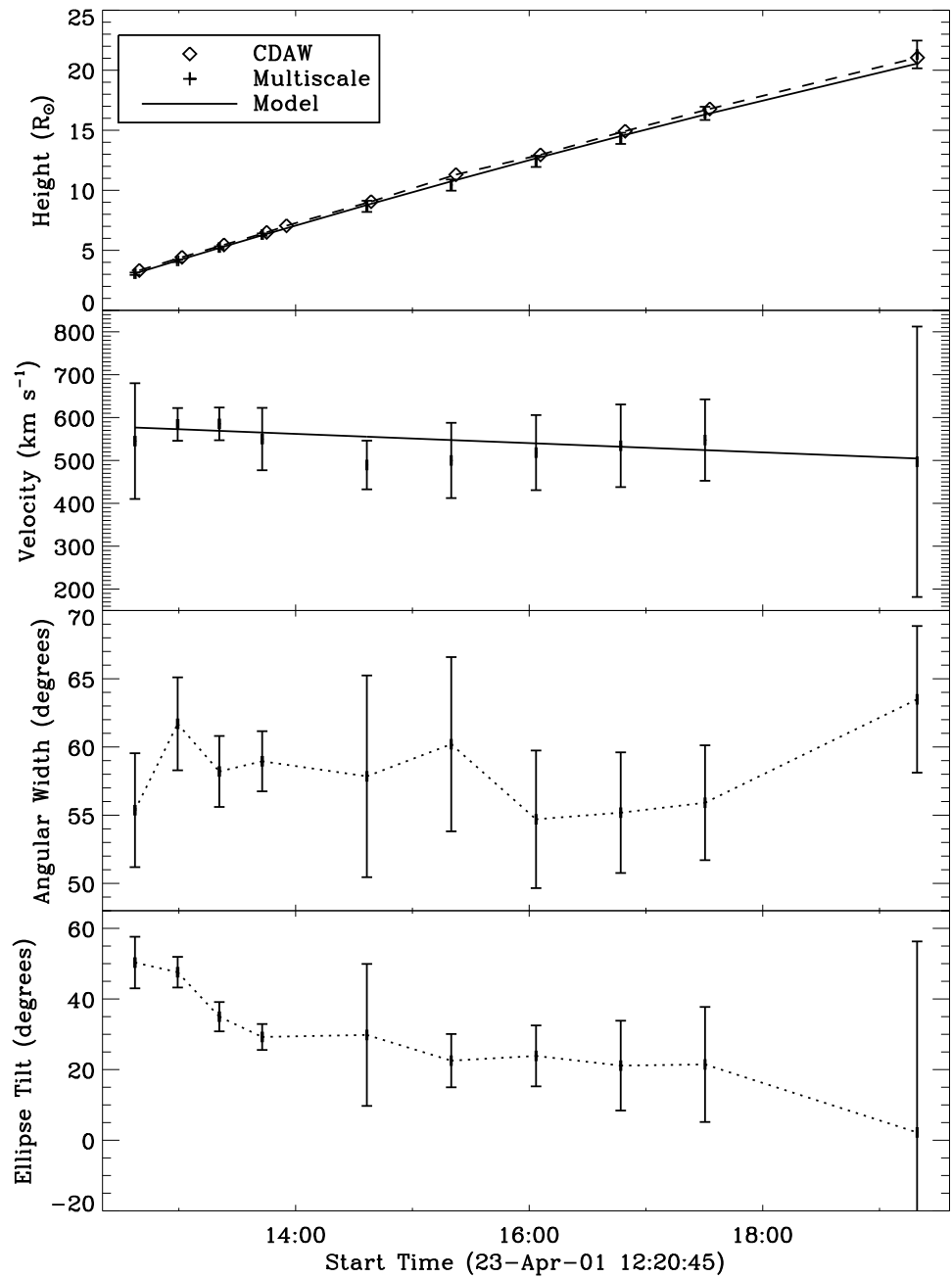}}
\caption{Kinematic and morphological profiles for the ellipse fit to the multiscale edge detection of the 23 April 2001 CME observed by LASCO/C2 and C3. The plots from top to bottom are height, velocity, angular width, and ellipse tilt. The CDAW heights are over-plotted with a dashed line. The height and velocity fits are based upon the constant acceleration model.}
\label{20010423_kins_angs}
\end{figure}

This CME was first observed in the south-west at 12:39 UT on 23 April 2001 and exhibits some degree of streamer deflection. 
\newline
\indent The height-time plot has a similar trend to CDAW (overplotted in top Figure~\ref{20010423_kins_angs} with a dashed line). The velocity fit was found to be linearly decreasing from 581 to 466~km~s$^{-1}$, giving a deceleration of $-$4.8\,$\pm$\,4.1~m~s$^{-2}$. The ellipse fit spans approximately 55\,--\,60$^{\circ}$ of the field-of-view throughout the event, and the orientation of the ellipse as a function of time is shown to decrease from approximately 50$^{\circ}$ to almost 0$^{\circ}$ (see Figure~\ref{20010423_kins_angs}).
\newline
\indent This CME is fitted well with the constant acceleration model.

\subsection{Fast CME: 21 April 2002}
\begin{figure}[!p]
\centerline{\includegraphics[scale=0.8, clip=true, trim=0 120 0 10]{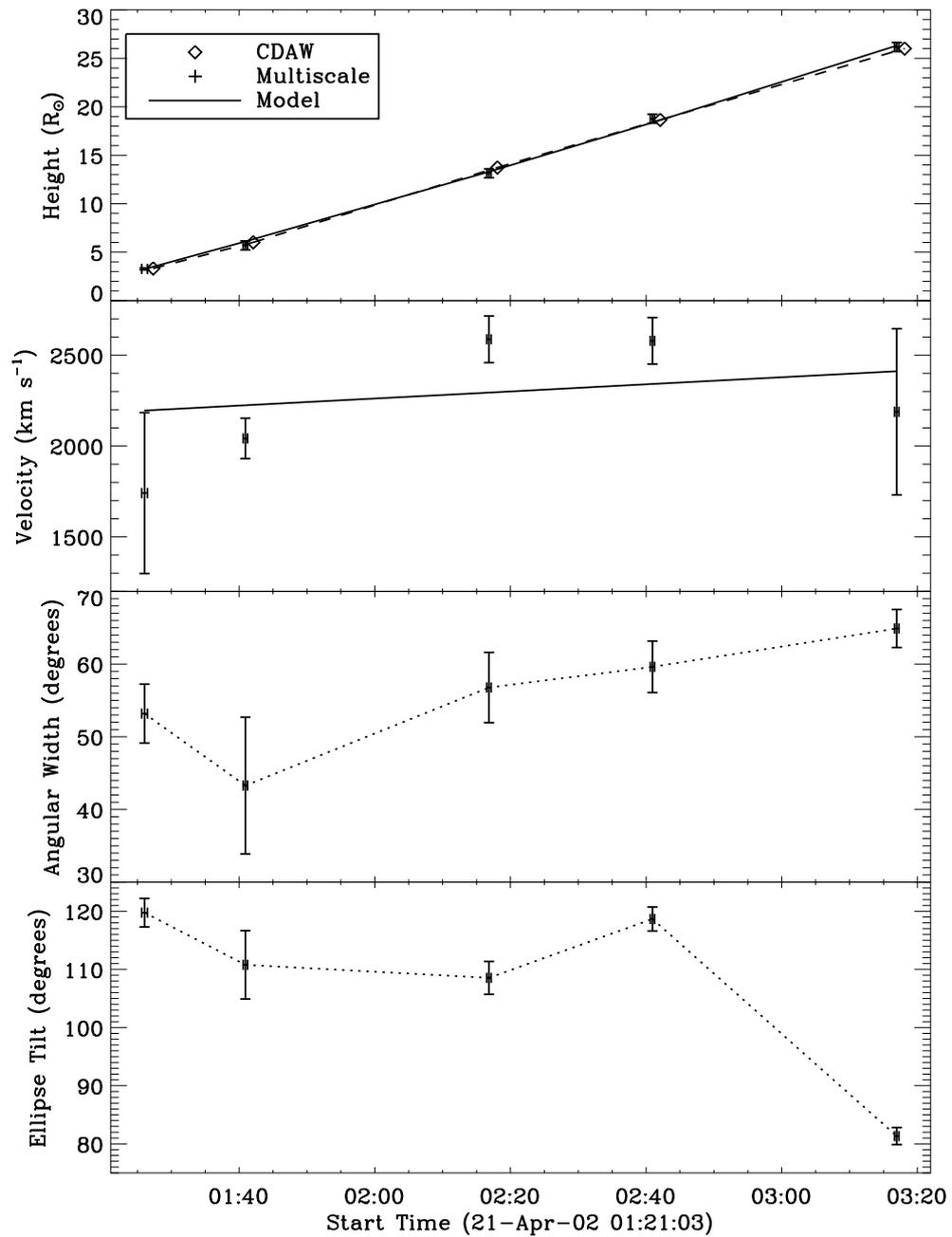}}
\caption{Kinematic and morphological profiles for the ellipse fit to the multiscale edge detection of the 21 April 2002 CME observed by LASCO/C2 and C3. The plots from top to bottom are height, velocity, angular width, and ellipse tilt. The CDAW heights are over-plotted with a dashed line. The height and velocity fits are based upon the constant acceleration model.}
\label{20020421_kins_angs}
\end{figure}

This CME was first observed in the west from 01:27 UT on 21 April 2002. 
\newline
\indent The height-time plot follows a similar trend to that of CDAW (overplotted in top Figure~\ref{20020421_kins_angs} with a dashed line). The velocity fit was found to be linearly increasing from 2,195 to 2,412~km~s$^{-1}$, giving a constant acceleration fit of 32.5\,$\pm$\,26.6~m~s$^{-2}$. The ellipse fit spans 53$^{\circ}$ in C2, and shows an increasing trend to 65$^{\circ}$ in C3.  The orientation of the ellipse as a function of time is shown to scatter about 115$^{\circ}$ though it drops to approximately 81$^{\circ}$ in the final C3 image.
\newline
\indent The kinematics of this event are not modelled satisfactorily by the constant acceleration model, since the fits do not lie within all error bars. The argument for a non-linear velocity profile, with a possible early decreasing acceleration, is justified for this event, although the instrument cadence limits the data set available for interpretation. The previous analysis of \citet{2003ApJ...588L..53G} resulted in a velocity of $\sim$\,2,500~km~s$^{-1}$ past $\sim$\,3.4~R$_{\odot}$ which is consistent with our results past $\sim$\,6~R$_{\odot}$ in Figure~\ref{20020421_kins_angs}.

\subsection{Flux-Rope/Slow CME: 1 April 2004}
\begin{figure}[!p]
\centerline{\includegraphics[scale=0.8, clip=true, trim=0 120 0 10]{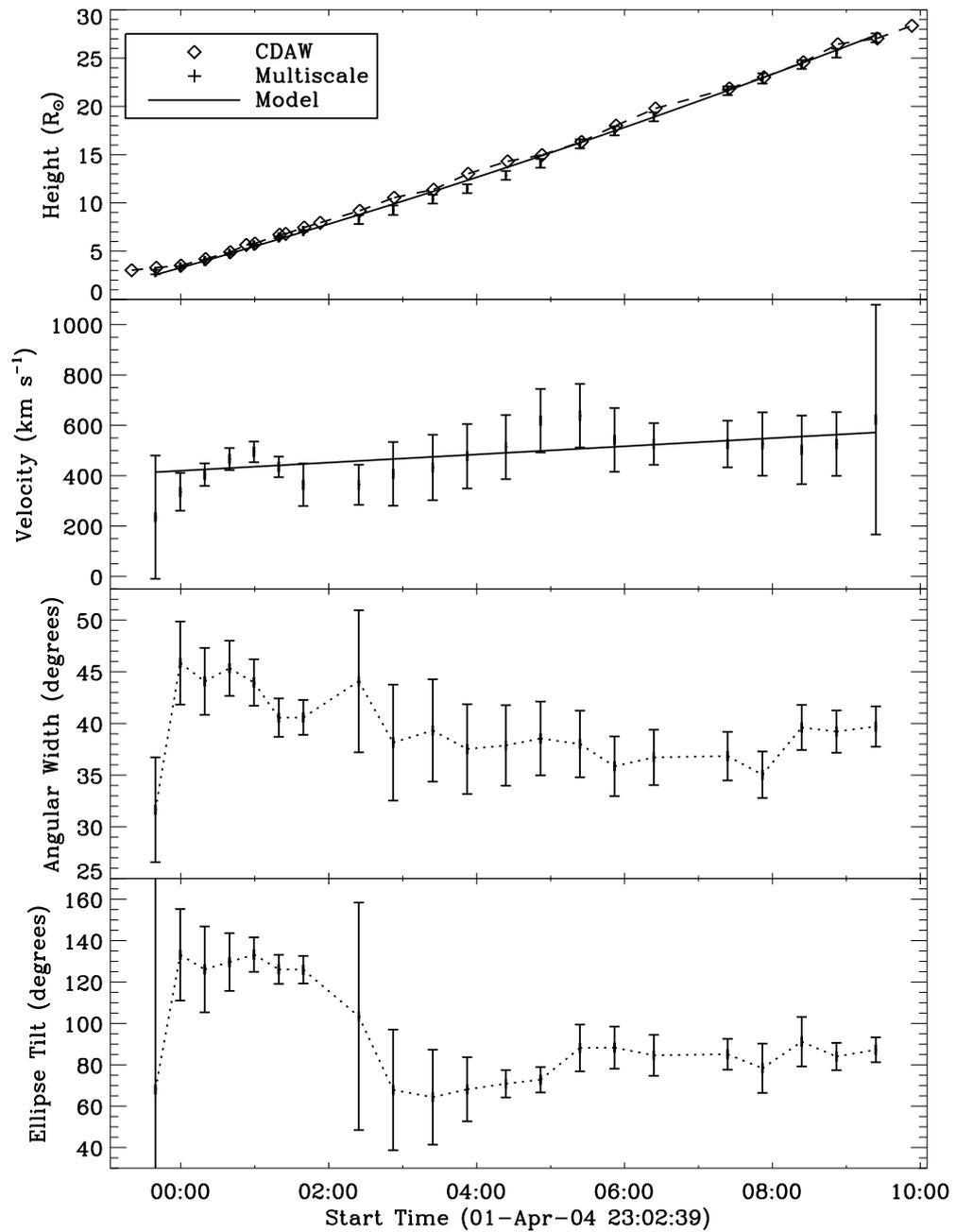}}
\caption{Kinematic and morphological profiles for the ellipse fit to the multiscale edge detection of the 1 April 2004 CME observed by LASCO/C2 and C3. The plots from top to bottom are height, velocity, angular width, and ellipse tilt. The CDAW heights are over-plotted with a dashed line. The height and velocity fits are based upon the constant acceleration model.}
\label{20040401_kins_angs}
\end{figure}

This CME was first observed in the north-east from approximately 23:00 UT on 1 April 2004, is in the field-of-view for over 9 hours, and exhibits a bright loop front, cavity and twisted core. 
\newline
\indent The height-time plot follows a similar trend to that of CDAW (overplotted in top Figure~\ref{20040401_kins_angs} with a dashed line). The velocity fit was found to be linearly increasing from 415 to 570~km~s$^{-1}$, giving an acceleration of 4.4\,$\pm$\,2.0~m~s$^{-2}$. Note also that the kinematics of this event exhibit non-linear structure clearly seen in the velocity and acceleration profiles. The ellipse fit spans approximately 44$^{\circ}$ in C2, stepping down to approximately 38$^{\circ}$ in C3.  The orientation of the ellipse as a function of time is shown to jump down from approximately 130$^{\circ}$ in C2 to approximately 70\,--\,80$^{\circ}$ in C3.
\newline
\indent This event shows unexpected structure in the velocity and acceleration profiles which indicates a complex eruption not satisfactorily modelled with constant acceleration.

\subsection{STEREO-B Event: 8 October 2007}
\begin{figure}[!p]
\centerline{\includegraphics[scale=0.8, clip=true, trim=0 120 0 10]{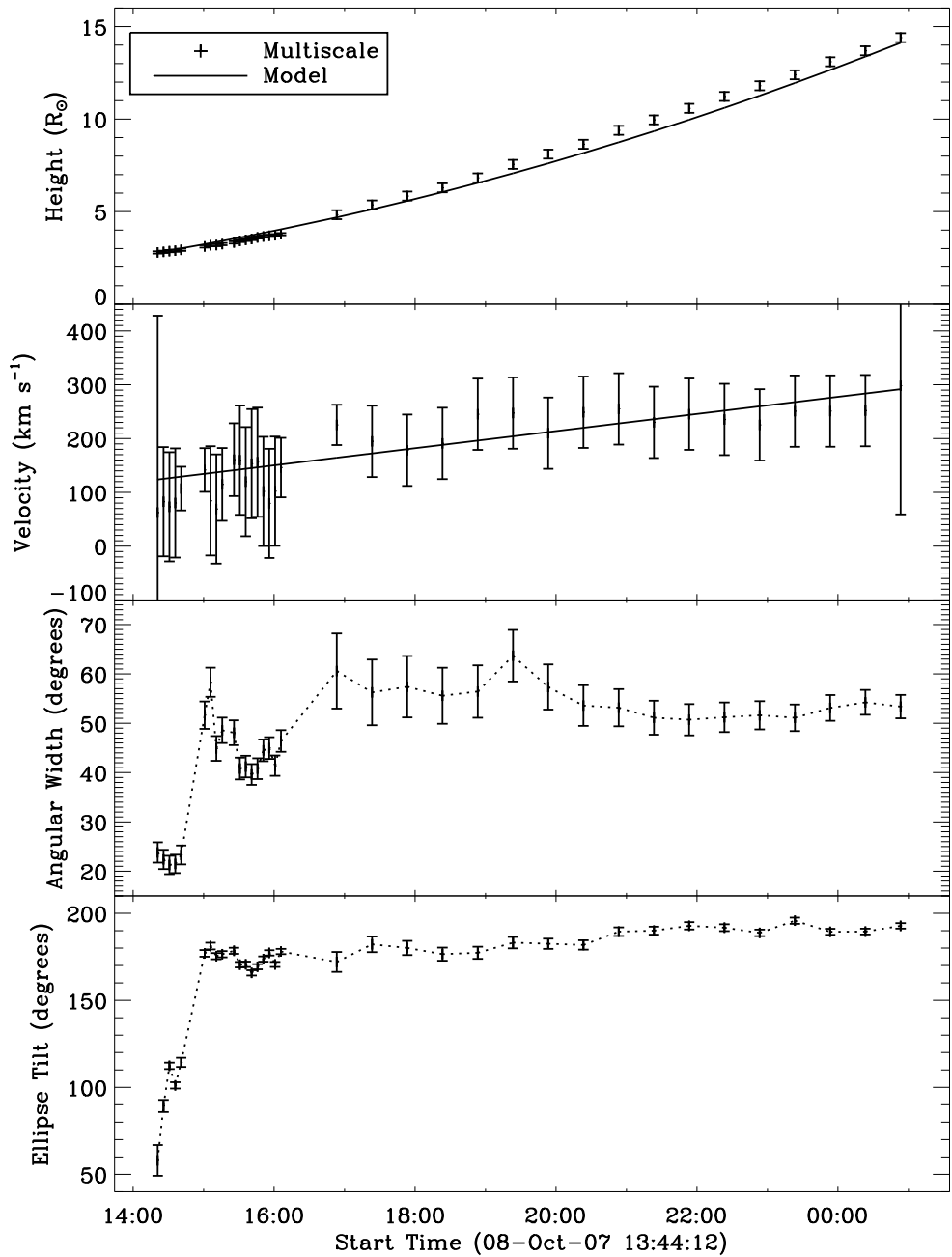}}
\caption{Kinematic and morphological profiles for the ellipse fit to the multiscale edge detection of the 8 October 2007 CME observed by SECCHI/COR1 and COR2 on STEREO-B. The plots from top to bottom are height, velocity, angular width, and ellipse tilt. The height and velocity fits are based upon the constant acceleration model.}
\label{20071008_kins_angs}
\end{figure}

This CME was first observed in the west from approximately 12:00 UT on 8 October 2007, and is best viewed from the STEREO-B spacecraft. It is noted that the kinematics as measured by SOHO and STEREO will be different due to projection effects \citep{2007A&A...469..339V, 2008JGRA..11301104H}. On this date STEREO-B was at an angular separation of 16.5$^{\circ}$ from Earth.
\newline
\indent The height-time plot for this event is shown in Figure~\ref{20071008_kins_angs}. The velocity fit was found to be linearly increasing from 71 to 330~km~s$^{-1}$, giving an acceleration of 5.7\,$\pm$\,0.9~m~s$^{-2}$. The ellipse fit in COR1 spans approximately 23$^{\circ}$ stepping up to a scatter about 40\,--\,50$^{\circ}$ which rises slightly to 50\,--\,60$^{\circ}$ in COR2.  The orientation of the ellipse as a function of time is shown to increase from 55\,--\,110$^{\circ}$ then jumps to an approximately steady scatter about 180\,--\,190$^{\circ}$. %This is representative of the behaviour of the CME as it visibly appears to 'roll' out of the low COR1 data before propagating steadily outward in COR2. 
\newline
\indent This CME is fitted well with the constant acceleration model.

\subsection{STEREO-A Event: 16 November 2007}
\begin{figure}[!p]
\centerline{\includegraphics[scale=0.8, clip=true, trim=0 120 0 10]{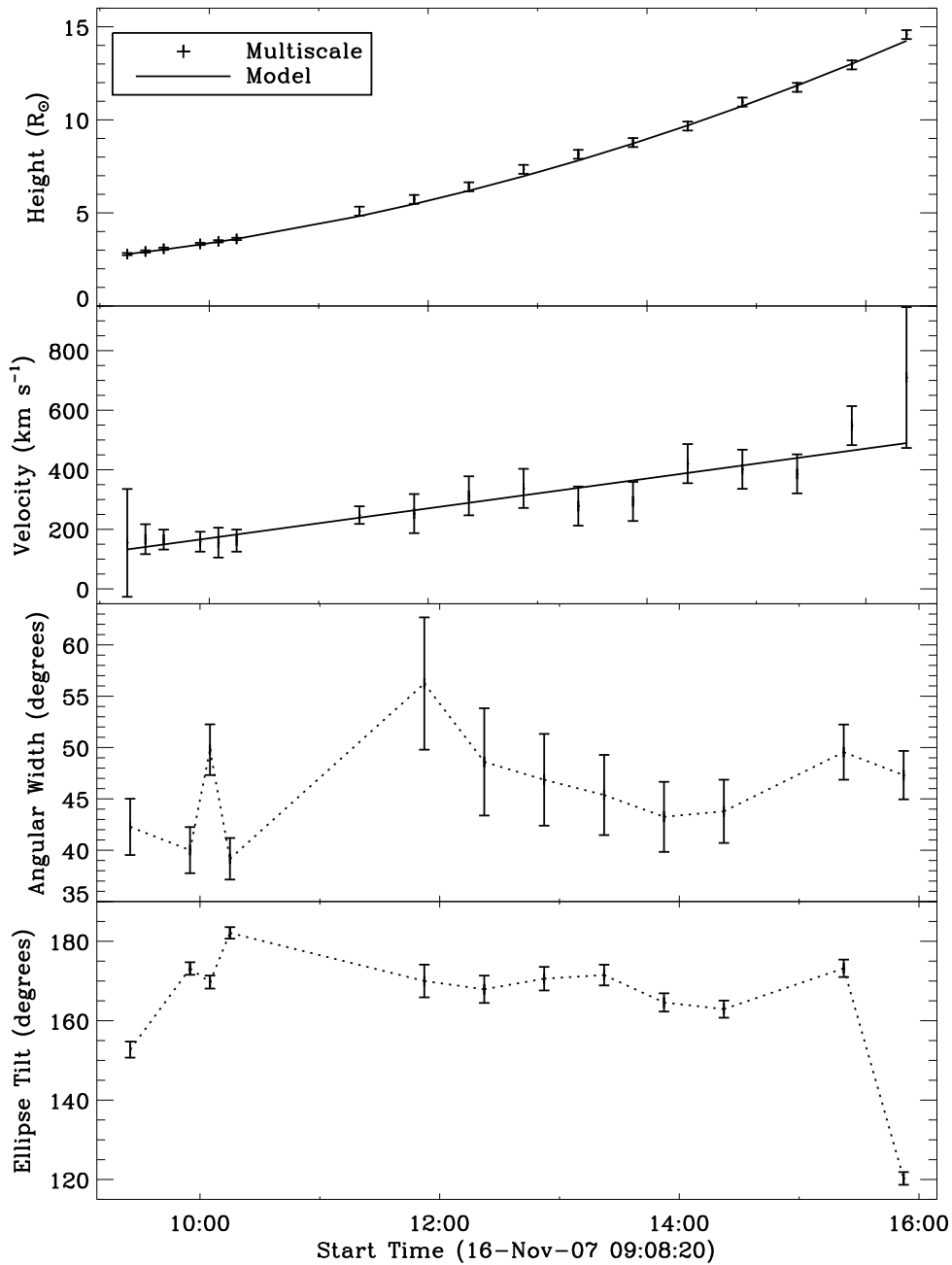}}
\caption{Kinematic and morphological profiles for the ellipse fit to the multiscale edge detection of the 16 November 2007 CME observed by SECCHI/COR1 and COR2 on STEREO-A. The plots from top to bottom are height, velocity, angular width, and ellipse tilt. The height and velocity fits are based upon the constant acceleration model.}
\label{20071116_kins_angs}
\end{figure}

This CME was first observed in the west from approximately 08:26 UT on 16 November 2007, and is best viewed from the STEREO-A spacecraft. On this date STEREO-A was at an angular separation of 20.3$^{\circ}$ from Earth.
\newline
\indent The height-time plot for this event is shown in Figure~\ref{20071116_kins_angs}. The velocity fit was found to be linearly increasing from 131 to 483~km~s$^{-1}$, giving an acceleration of 13.7\,$\pm$\,1.7~m~s$^{-2}$. The ellipse fit in COR1 spans approximately 40\,--\,50$^{\circ}$ stepping up slightly to a scatter about 45\,--\,55$^{\circ}$ in COR2.  The orientation of the ellipse as a function of time is shown to start at 153$^{\circ}$ and end at 120$^{\circ}$ with the mid points scattered about 170$^{\circ}$. %This is representative of the behaviour of the CME as it visibly appears to 'roll' out of the low COR1 data before propagating steadily outward in COR2. 
\newline
\indent This CME is fitted well with the constant acceleration model.

\section{Discussion \& Conclusions}

A variety of theoretical models have been proposed to describe CMEs, especially their early propagation phase. Observational studies, such as those outlined above, are necessary to determine CME characteristics. We argue that the results of previous methods are limited in this regard due mainly to large kinematic errors which fail to constrain a model, an artefact of CME detection based upon either running- (or fixed-) difference techniques or other operations. Current methods fit either a linear model to the height-time curve, implying constant velocity and zero acceleration (e.g., CACTus) or a second order polynomial, producing a linear velocity and constant acceleration (e.g., CDAW, SEEDS). The implementation of a multiscale decomposition provides a time error on the scale of seconds (the exposure time of the instrument) and a resulting height error on the order of a few pixels. The height-time error is used to determine the errors of the velocity and acceleration profiles of the CMEs. It was shown that for certain events the results of CACTus, CDAW and SEEDS can differ significantly from our methods, as illustrated in the Tables of Section~\ref{sec:multiscaleresults}.
\newline
\indent Our results clearly confirm that the constant acceleration model may not always be appropriate. The 2 January 2000 and 21 April 2002 CMEs are good examples of the possible non-linear velocity profile and consequent non-constant acceleration profile (see Figure~\ref{20000102_kins_angs} and Figure~\ref{20020421_kins_angs}). Indeed these events are shown to have a decreasing acceleration, possibly to zero or below, as the CMEs traverse the field-of-view. Simulations of the breakout model outlined in \citet{2004ApJ...617..589L} resulted in constant acceleration fits which do not agree with these observations. It may be further noted that the events of 23 April 2001 and 1 April 2004 show a possible decreasing acceleration phase early on, though within errors this cannot be certain (see Figure~\ref{20010423_kins_angs} and Figure~\ref{20040401_kins_angs}). Furthermore, the structure seen in some events would indicate that the CME does not progress smoothly. The velocities of the 1 April 2004 CME in Figure~\ref{20040401_kins_angs} and the 16 November 2007 CME in Figure~\ref{20071116_kins_angs} show non-smooth profiles and may imply a form of bursty reconnection or other staggered energy release driving the CME. Other profiles such as Figure~\ref{20000102_kins_angs} and to a lesser extent Figures~\ref{20010423_kins_angs} and \ref{20020421_kins_angs} may show a stepwise pattern, indicative of separate regimes of CME progression. None of the current CME models indicate a form of non-smooth progression, although the flux-rope model does describe an early acceleration regime giving a non-linear velocity to the eruption \citep[see Figure~11.5 in][]{2000mare.book.....P}.
\newline
\indent It may be concluded that the angular widths of the events are indicative of whether the CME expands radially or otherwise in the plane-of-sky. For the CMEs studied above, the observations of 18 April 2000, 23 April 2000, and 21 April 2002 show an angular width expansion (see Figure~\ref{20000418_kins_angs}, Figure~\ref{20000423_kins_angs}, and Figure~\ref{20020421_kins_angs}). These events also show high velocities, obtaining top speeds of up to 1,000~km~s$^{-1}$, over 1,100~km~s$^{-1}$ and 2,500~km~s$^{-1}$ respectively, and may therefore indicate a link between the CME expansion and speed. Furthermore, it is suggested by \citet{2006ApJ...652.1740K} that the flux-rope model can account for different observed expansion rates due to the axial versus broadside view of the erupting flux system. %especially if an untwisting of the flux-rope occurs during its evolution.
\newline
\indent The observed morphology of the ellipse fits may be further interpreted through the tilt angles plotted in Section~\ref{sec:multiscaleresults}. In knowing the ellipse tilt and the direction of propagation of the CME it is possible to describe the curvature of the front. For the events above, the changing tilt and hence curvature is possibly significant for the 18 April 2000, 1 April 2004, and 8 October 2007 events (see bottom Figure~\ref{20000418_kins_angs}, Figure~\ref{20040401_kins_angs}, and Figure~\ref{20071008_kins_angs}). The elliptical flux rope model of \citet{2006ApJ...642..541K} was shown to have a changing orientation of the magnetic axis which results in a dynamic radius of curvature of the CME, possibly accounting for these observed ellipse tilts.
\newline
\indent The work outlined here is an initial indication that the zero and constant acceleration models in CME analysis are not an accurate representation of all events, and the overestimated angular widths are not indicative of the true CME expansion. The ellipse characterisation has provided additional information on the system through its changing width and orientation. This work will be further explored and developed with STEREO data whereby the combined view-points can give additional kinematic constraints and lead to a correction for projection effects through 3D reconstructions (discussed in Chapter~\ref{chapter:3D}).

%%%%%%%%%%%%%%%%%%%%%%%%%%%%%%%%%%%%%%%%%%%%end A&A

\chapter{Propagation of an Earth-Directed CME in Three-Dimensions}
\label{chapter:3D}

\hrule height 1mm
\vspace{0.5mm}
\hrule height 0.4mm 
\noindent 
\\
CMEs are long known to be significant drivers of adverse space weather at Earth, but the physics governing their propagation is not fully understood. The launch of the STEREO mission in 2006 has provided new insight into their motion in the heliosphere, although the mechanisms governing their evolution remain unclear due to difficulties in reconstructing their true 3D structure. Here we use a new elliptical tie-pointing technique to reconstruct a full CME front in 3D, enabling us to quantify an early acceleration profile, non-radial motion, increasing angular width and `pancaking' of the CME front as it propagates from 2\,--\,46~R$_{\odot}$. Beyond 7~R$_{\odot}$, we show that its motion is determined by aerodynamic drag in the solar wind and, using our reconstruction as input for a 3D MHD simulation, we determine an accurate arrival time at the L1 point near Earth. This chapter is founded on research published in Byrne {\it et al., Nature Communications} (2010).
\vspace{4mm}
\hrule height 0.4mm
\vspace{0.5mm}
\hrule height 1mm 

\newpage
\section{Introduction}

It is predominantly believed that magnetic reconnection is responsible for the destabilisation of magnetic flux ropes on the Sun, which then erupt through the corona into the solar wind to form CMEs \citep{2006GMS...165...43M}. There is much debate as to the specific processes which trigger the eruption of CMEs, and different models exist to explain these \citep{1995ApJ...446..377F, 1996JGR...10127499C, 1999ApJ...510..485A, 2006PhRvL..96y5002K, 2007ApJ...671L..77V}. In the low solar atmosphere, it is postulated that high latitude CMEs undergo deflection since they are often observed at different position angles with respect to their associated source region locations \citep{2009SoPh..259..143X}. It has been suggested that field lines from polar coronal holes may guide high-latitude CMEs towards the equator \citep{angeo-27-4491-2009}, or that the initial magnetic polarity of a flux rope relative to the background magnetic field influences its trajectory \citep{2005A&A...432..331C, 2001SoPh..203..119F}. During this early phase, CMEs are observed to expand outwards from their launch site, though plane-of-sky measurements of their increasing sizes and angular widths are ambiguous in this regard \citep{2009CEAB...33..115G}. This expansion has been modelled as a pressure gradient between the flux rope and the background solar wind \citep{2004ApJ...600.1035R, 1999JGR...104..493O}. At larger distances in their propagation, CMEs are predicted to interact with the solar wind and the interplanetary magnetic field. Studies that compare in-situ CME velocity measurements with initial eruption speeds through the corona show that slow CMEs must be accelerated toward the speed of the solar wind, and fast CMEs decelerated \citep{2009SoPh..256..149M, 2003JGRA..108.1039G}. It has been suggested that this is due to the effects of drag acting on the CME in the solar wind \citep{2006SoPh..233..233T, 2004SoPh..221..135C}. However, the quantification of drag, along with that of both CME expansion and non-radial motion, is currently lacking, due primarily to the limits of observations from single fixed viewpoints with restricted fields-of-view.
\newline
\indent Efforts to reconcile 2D plane-of-sky images with the true 3D morphology of CMEs have been underway since they were first observed in the 1970s. The inherent difficulties in this are predominantly due to the single, fixed-position imagers with restricted fields-of-view, as well as the difficulties in observing the optically thin coronal plasma of these dynamic events. Before the launch of STEREO, there was limited ability to infer the 3D CME morphology from the available observations such as SOHO/LASCO. Coronagraphs mainly measure the Thomson scattered light of the free electrons in the coronal plasma, providing white-light images of CMEs against the plane-of-sky that are not trivial to deconvolve,  and the projected 2D nature of these images introduces uncertainties in kinematical and morphological analyses \citep{2007A&A...469..339V}. Some efforts were based upon a pre-assumed geometry of the CME, such as the cylindrical model \citep{2004A&A...422..307C} or the cone model \citep{2005JGRA..11008103X, 2002JGRA..107.1223Z}, whose shapes were simply oriented to best match the 2D observations. Others used either a comparison of multiple events to infer a statistical relationship between plane-of-sky measurements and true CME motion \citep{2005AnGeo..23.1033S, 2005A&A...440..373H}, or a comparison of observations with in-situ data and/or signatures on-disk \citep{2008SoPh..250..347D, 2008JGRA..11301104H}. One prominent method was the use of 3D polarisation analysis of LASCO images \citep{2004Sci...305...66M}, whereby the line-of-sight averaged distance from the plane-of-sky is determined from the brightness ratio of polarised to unpolarised electron scattered emissivity (K-corona). However, this lacks in details such as whether the feature is truly unique along the line-of-sight, and, if so, is it towards or away from the observer with respect to the plane-of-sky. Polarisation analysis itself is only acceptable up to heights of $\sim$\,5~R$_{\odot}$, since beyond this distance the dust-scattered F-corona may no longer be considered unpolarised \citep{1966gtsc.book.....B}. These issues motivated the launch of the STEREO mission to further our understanding of CMEs.

\section{The STEREO Era}

\begin{figure}[!t]
\centerline{\includegraphics[scale=0.35]{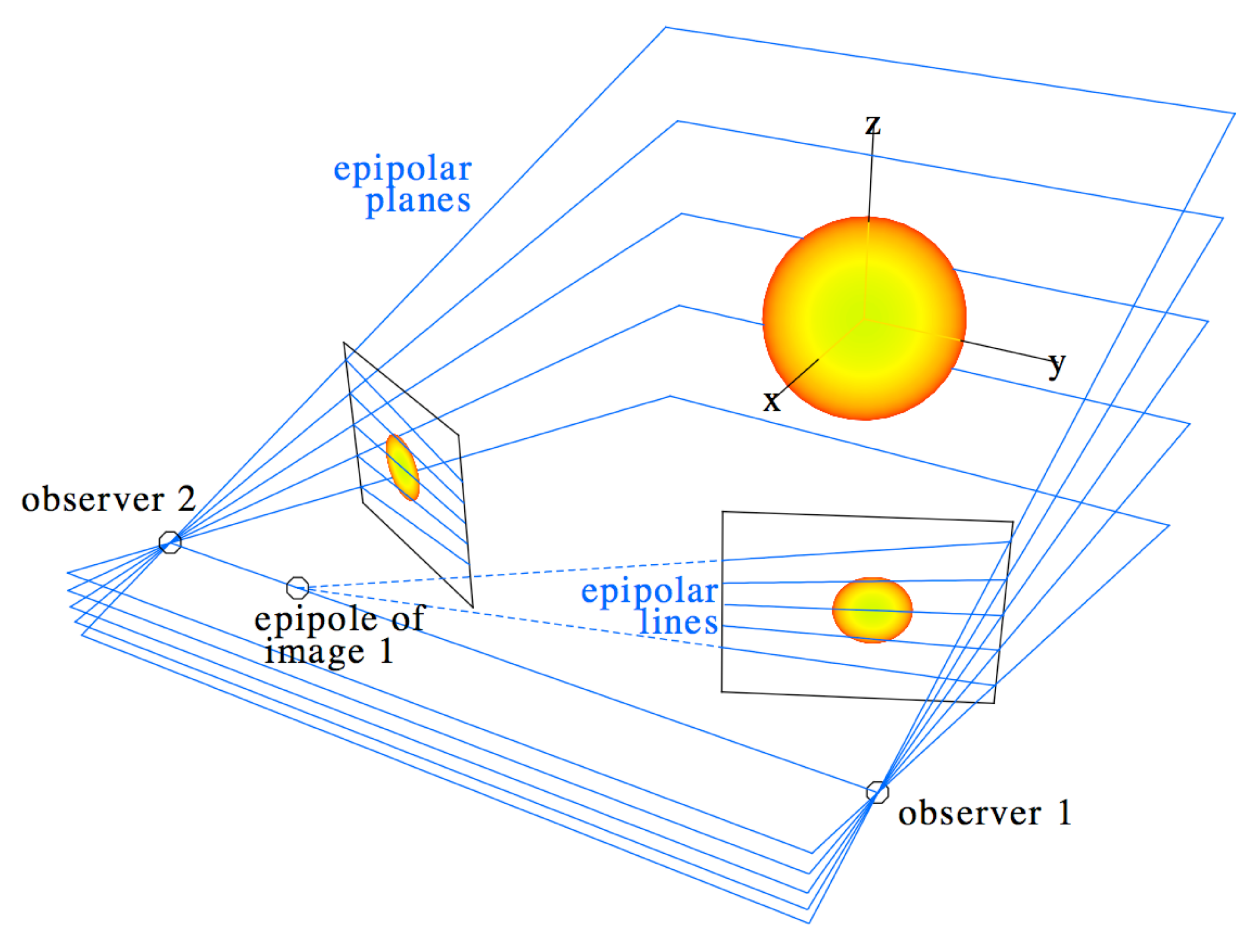}}
\caption{Schematic of the epipolar geometry used to relate the observations from the two STEREO spacecraft, reproduced from \citet{2006astro.ph.12649I}. This geometry enables us to localise features in 3D space by the triangulating lines-of-sight across epipolar planes.}
\label{epipolar}
\end{figure}

The two near-identical spacecraft of the STEREO mission provide simultaneous observations of CMEs from independent viewpoints to better observe their true morphology. Unfortunately there are limitations on how much 3D information can be extracted from the combined plane-of-sky observations, especially when the object is optically thin and its boundaries ill-defined. In order to determine the morphology of an object in 3D from only two viewpoints, techniques must be applied within the context of an epipolar geometry, as illustrated in Figure~\ref{epipolar} \citep{2006astro.ph.12649I}. This epipolar coordinate system for considering the 3D space observed from two independent viewpoints is built up as follows. A line is drawn to connect the two observers, called the stereo base line. The two observer locations and any third object point or location in the observing space then define a plane. Numerous object points will define numerous planes that share an intersection with the stereo base line. These are the epipolar planes of Figure~\ref{epipolar}. The plane-of-sky as seen by each observer then intersects the epipolar planes such that they appear as epipolar lines across the image, and will converge on a point along the stereo base line referred to as the epipole of that image. So if a line-of-sight from observer 1 is drawn across an epipolar plane, it will appear as a single point on image 1, but as a complete line across the corresponding epipolar plane in image 2 as seen by observer 2, who is then able to triangulate upon an object in 3D space by the intersecting lines-of-sight. This technique is known as tie-pointing.
\begin{figure}[!t]
\centerline{\includegraphics[scale=0.2]{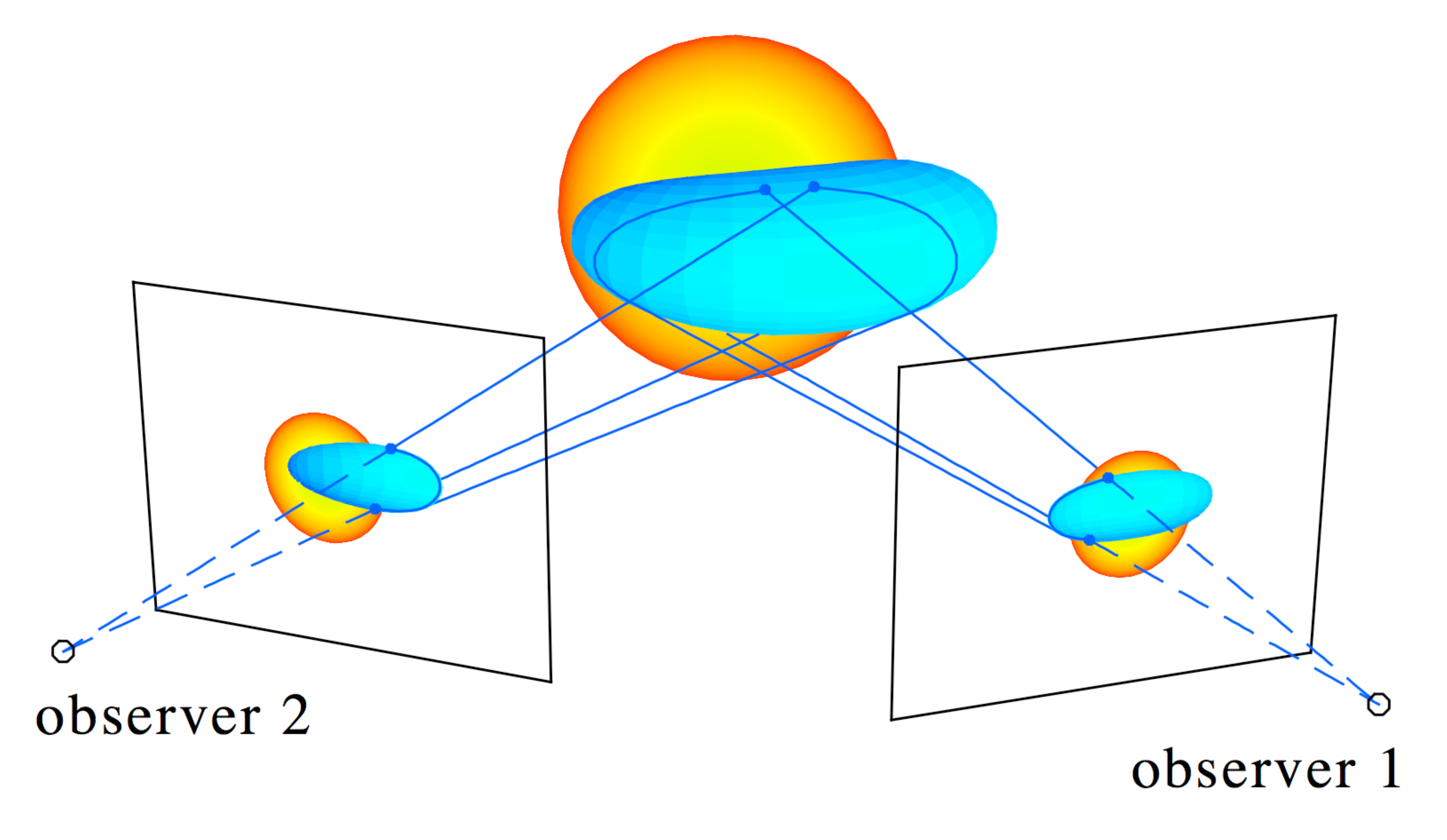}}
\caption{Schematic of how tie-pointing a curved surface within an epipolar geometry is limited in its ability to resolve the true feature, since lines-of-sight will be tangent to different edges of the surface and not necessarily intersect upon it. Reproduced from \citet{2006astro.ph.12649I}.}
\label{epipolarcurved}
\end{figure}
\newline
\indent The technique of tie-pointing lines-of-sight across epipolar planes is best for resolving a single feature, such as a coronal loop on-disk \citep{2008ApJ...679..827A}. Under the assumption that the same feature may be tracked in coronagraph images many CME studies have also employed tie-pointing techniques \citep{2009SoPh..256...57L, 2009SoPh..259..213S, 2009SoPh..256..183T, 2008SoPh..252..385M, 2009SoPh..259..163W}. However, when measuring the kinematics of the CME front this technique alone doesn't hold true, since it is inevitable that the same part of the curved front cannot be confidently resolved from both viewpoints once the CME has traversed a certain distance in space, nor similarly once the spacecraft have moved beyond a certain angular separation during the mission (Figure~\ref{epipolarcurved}). Furthermore, triangulating CME observations using only the COR images confines the kinematic and morphological analyses to within the 20~R$_{\odot}$ field-of-view. The additional use of the heliospheric imagers allows a study of CMEs out to distances of $\sim$\,1~AU, however (instrumental effects aside) a 3D analysis can only be carried out if the CME propagates along a trajectory between the two spacecraft so that it is observed by both HI instruments. Otherwise, assumptions of its trajectory have to be inferred from either its association with a source region on-disk \citep{2008SoPh..252..373H} or its trajectory through the COR data \citep{2009SoPh..256..149M}, or derived by assuming a constant velocity through the HI fields-of-view \citep{2009GeoRL..3608102D}. Triangulation of CME features using time-stacked intensity slices at fixed latitude, named `J-maps' due to the characteristic propagation signature of a CME, has also been developed \citep{2010SoPh..tmp...49D, 2010ApJ...710L..82L}. This technique is hindered by the same limitation of standard tie-pointing techniques; namely that the curvature of the feature is not considered, and the intersection of sight-lines may not occur upon the surface of the observed feature.
\begin{sidewaysfigure}[!p]
\centerline{\includegraphics[width=\linewidth]{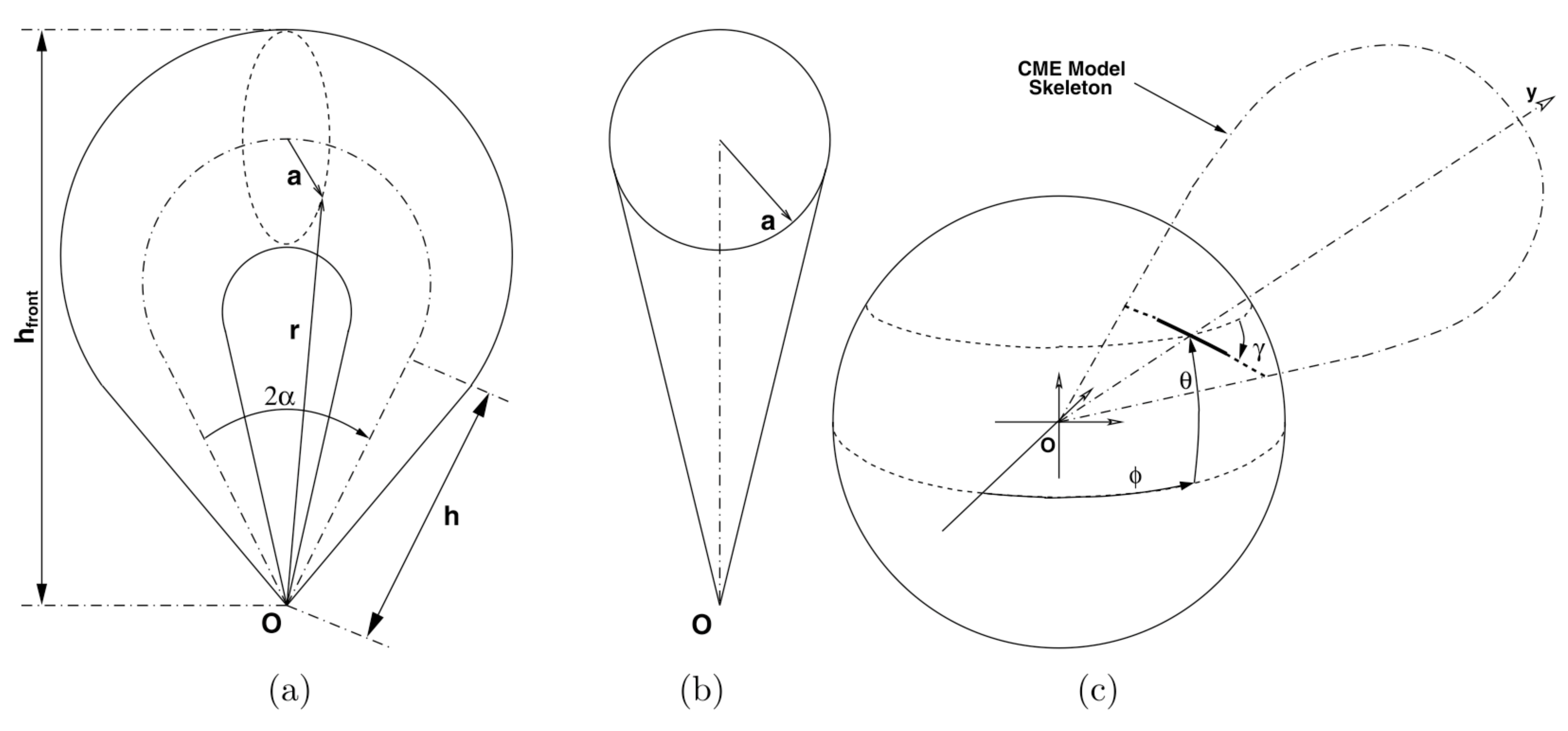}}
\caption{Schematic of the Graduated Cylindrical Shell (GCS) model of a flux rope CME, reproduced from \citet{2009SoPh..256..111T}. Indicated are the model parameters of front height {\bf h$_{front}$}, leg height {\bf h}, angle between the legs $2\alpha$, cross-sectional radius {\bf a}, and distance from Sun centre {\bf O} to a point on the edge of the shell {\bf r}. Two views of the GCS model are shown; ({\bf a}) `face-on', and ({\bf b}) `end-on'. The positional parameters of longitude $\phi$, latitude $\theta$, and orientation angle $\gamma$ are illustrated in ({\bf c}).}
\label{thernisien}
\end{sidewaysfigure}
\newline
\indent An alternative to tie-pointing is a method called forward modelling which presumes a given shape of the CME and seeks to match it with observational data. \citet{2006ApJ...652..763T} employ a graduated cylindrical shell which is warped to form a flux rope model overlaid on CME images (Figure~\ref{thernisien}). The parameters governing the model's shape and orientation may be changed by the user to fit the model to STEREO-Ahead and Behind data simultaneously and obtain a 3D flux rope characterisation of the CME as it propagates, though this may not always be appropriate \citep{2009ApJ...695L.171J}. \citet{2009SoPh..256..131B} outline a similar forward model which assumes one of three pre-assigned shapes: a hemispherical cap, a flux rope, or a cloud-like model. However, in each of these methods the predetermined shape of the CME model has a spherical cross-section and must adhere to some quasi-similarity (self-invariance) over the sequence of images. So while forward modelling better accounts for the curved nature of the CME being observed, the inherent restrictions of the imposed model still limit the analysis of the true 3D structure and dynamics of the CME as it propagates.

\section{Elliptical Tie-Pointing}
\label{sect:ellipticaltiepointing}

In the epipolar geometry outlined above, 3D information may be gleaned from two independent viewpoints of a feature using tie-pointing techniques to triangulate lines-of-sight in space. However, when the object is known to be a curved surface, sight-lines will be tangent to it and not necessarily intersect upon it (Figure~\ref{epipolarcurved}). Consequently CMEs cannot be reconstructed by tie-pointing alone, but rather their localisation may be constrained by intersecting sight-lines tangent to the leading edges of a CME \citep{2004GeoRL..3121802P, 2009SoPh..256..167D}. Following the multiscale edge detection and ellipse characterisation outlined in Chapter~\ref{chapter:multiscale}, it is possible to extract the intersection of a given epipolar plane through the ellipse fits of both the STEREO-Ahead and Behind images. This defines a quadrilateral in 3D space which localises the ellipse characterisation of the CME front in that plane.
\begin{figure}[!p]
\centerline{\includegraphics[scale=0.8]{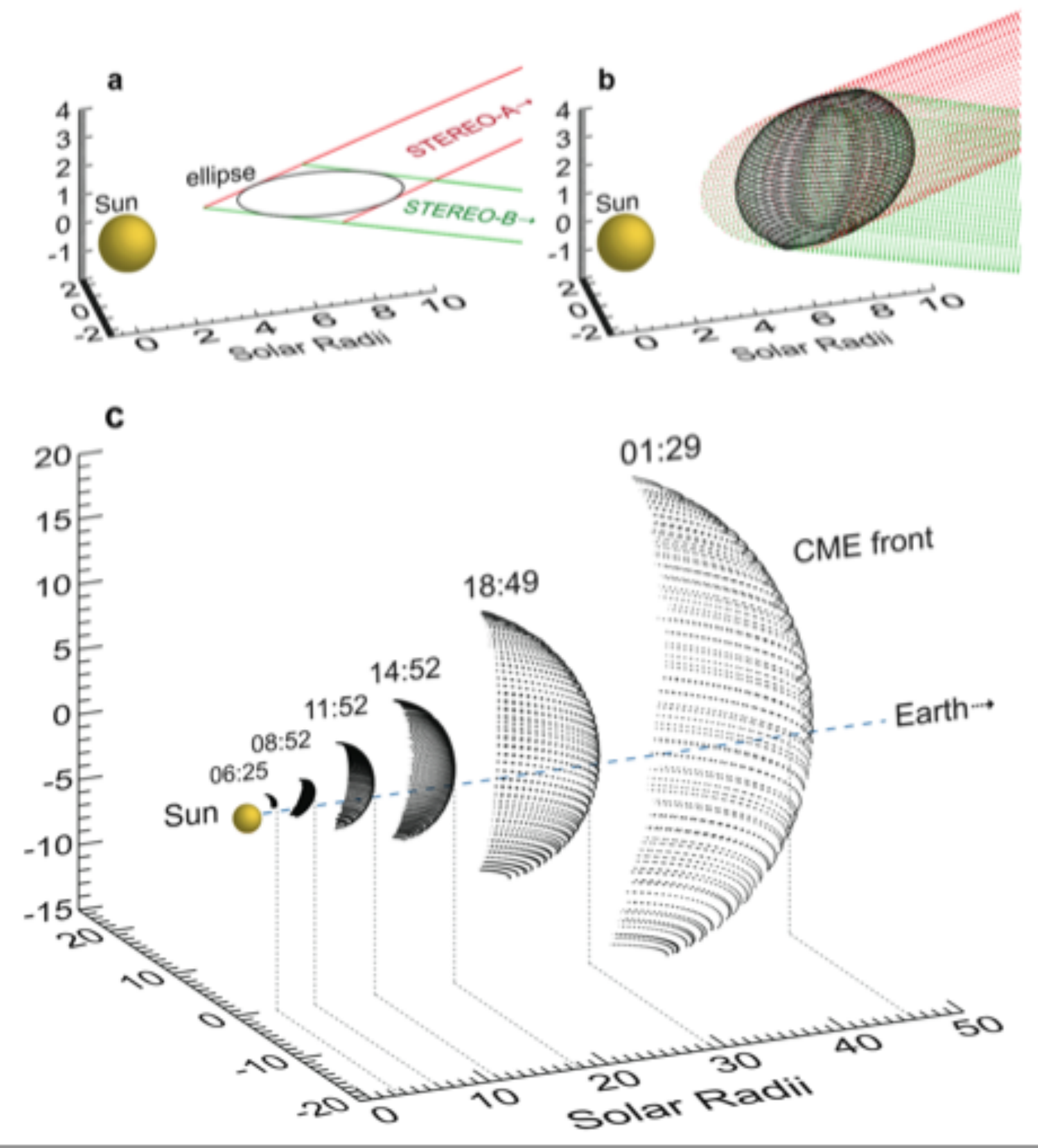}}
\caption{The elliptical tie-pointing technique developed to reconstruct the 3D CME front, shown here for the 12 December 2008 event. One of any number of epipolar planes will intersect the ellipse characterisation of the CME at two points in each image from STEREO-A and B. ({\bf a}) illustrates how the resulting four sight-lines intersect in 3D space to define a quadrilateral that constrains the CME front in that plane. Inscribing an ellipse within the quadrilateral such that it is tangent to each sight-line provides a slice through the CME that matches the observations from each spacecraft. ({\bf b}) illustrates how a full reconstruction is achieved by stacking multiple ellipses from the epipolar slices to create a model CME front that is an optimum reconstruction of the true CME front. ({\bf c}) illustrates how this is repeated for every frame of the eruption to build the reconstruction as a function of time and view the changes to the CME front as it propagates in 3D. While the ellipse characterisation applies to both the leading edges and, when observable, the flanks of the CME, only the outermost part of the reconstructed front is shown here for clarity.}
\label{ellipticaltiepointing}
\end{figure}
\newline
\indent Inscribing an ellipse within the quadrilateral such that it is tangent to all four sides (detailed below) provides a slice through the CME that matches the observations from each spacecraft (Figure~\ref{ellipticaltiepointing}a). A full reconstruction is achieved by stacking ellipses from numerous epipolar slices (Figure~\ref{ellipticaltiepointing}b). Since the positions and curvatures of these inscribed ellipses are constrained by the characterised curvature of the CME front in the stereoscopic image pair, the modelled CME front is considered an optimum reconstruction of the true CME front. This is repeated for every frame of the eruption to build the reconstruction as a function of time and view the changes to the CME front as it propagates in 3D (Figure~\ref{ellipticaltiepointing}c).
\newline
\indent Following \citet{2002SWJPAM..1.6H, 2005AJMAA..2.1H}, we inscribe an ellipse within a quadrilateral using the following steps (see Figure~\ref{ellipseinscribed}):
\begin{figure}[!t]
\centerline{\includegraphics[scale=0.75]{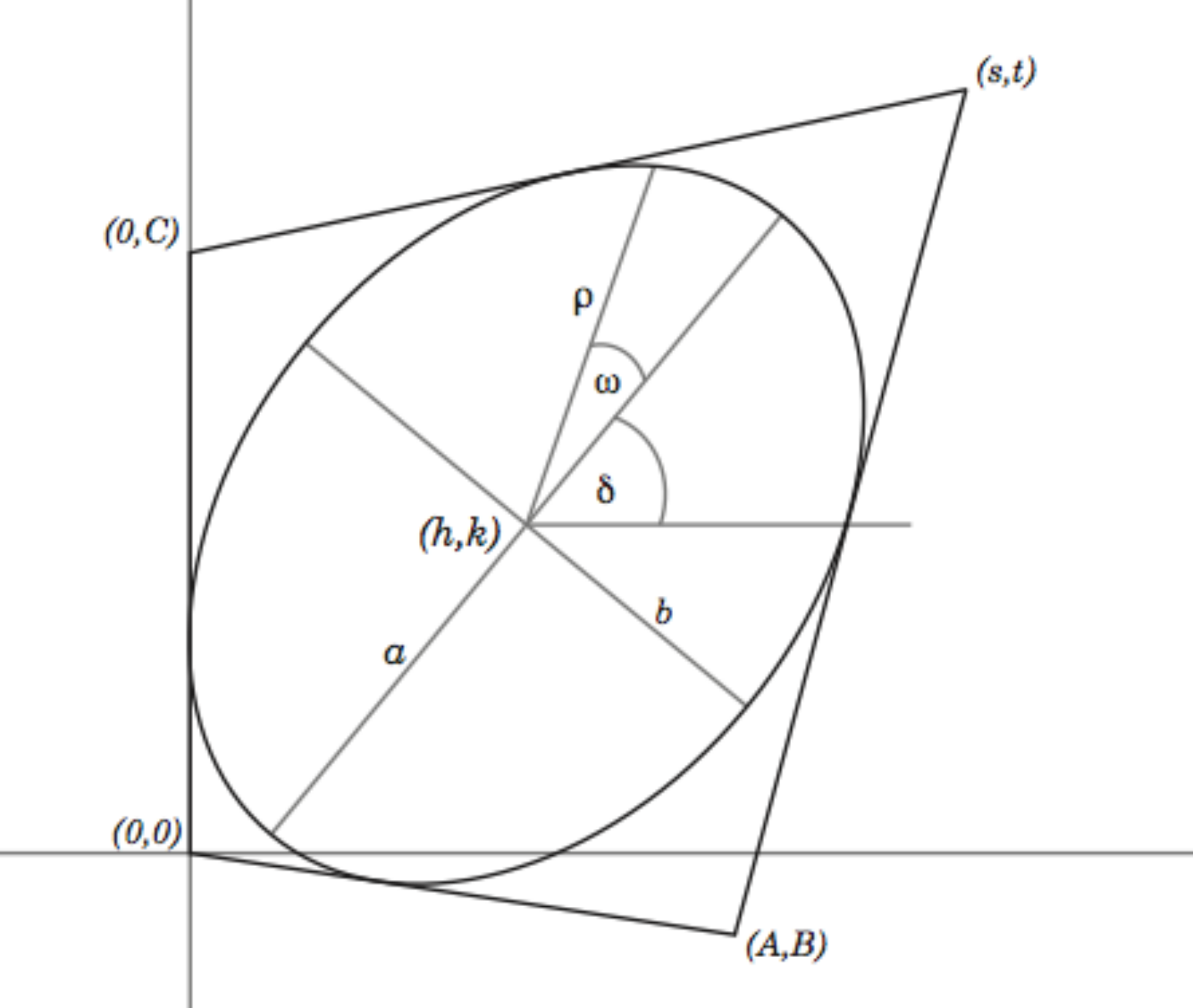}}
\caption{An ellipse inscribed within a convex quadrilateral. An isometry of the plane is applied such that the quadrilateral has vertices $(0,0)$, $(A,B)$, $(0,C)$, $(s,t)$. The ellipse has centre $(h,k)$, semimajor axis $a$, semiminor axis $b$, tilt angle $\delta$, and is tangent to each side of the quadrilateral.}
\label{ellipseinscribed}
\end{figure}
\begin{enumerate}
\item Apply an isometry to the plane such that the quadrilateral has vertices $(0,0)$, $(A,B)$, $(0,C)$, $(s,t)$, where $A>0$, $C>0$, $s>0$ and $t>B$. (Note, in the case of an affine transformation we set $A=1$, $B=0$ and $C=1$, with $s$ and $t$ variable.)
\item Set the ellipse centre point $(h, k)$ by fixing $h$ somewhere along the open line segment connecting the midpoints of the diagonals of the quadrilateral and hence determine $k$ from the equation of a line, for example:
\begin{eqnarray}
h =  \frac{1}{2}\left(\frac{s}{2}+\frac{A}{2}\right), \quad
k = \left(h-\frac{s}{2}\right)\left(\frac{t-B-C}{s-A}\right) + \frac{t}{2}
\end{eqnarray}
\item To solve for the ellipse tangent to the four sides of the quadrilateral, we can solve for the ellipse tangent to the three sides of a triangle whose vertices are the complex points
\begin{eqnarray}
z_{1} = 0, \quad
z_{2} = A+Bi, \quad
z_{3} = -\frac{At-Bs}{s-A}i
\end{eqnarray}
and the two ellipse foci  are then the zeroes of the equation
\begin{eqnarray}
p_{h}(z) \;=\; \left(s-A\right)z^{2}-2\left(s-A\right)\left(h-ik\right)z-\left(B-iA\right)\left(s-2h\right)C
\end{eqnarray}
whose discriminant can be denoted by $r(h)=r_{1}(h)+ir_{2}(h)$ where
\begin{align} \nonumber
r_1 \;=\; &4 \left(\left(s-A\right)^{2}-\left(t-B-C\right)^{2}\right)\left(\frac{h-A}{2}\right)^{2} \\ \nonumber
 &+4 \left(s-A\right)\left(A\left(s-A\right)+B\left(B-t\right)+C\left(C-t\right)\right)\left(\frac{h-A}{2}\right) \\
 &+ \left(s-A\right)^{2}\left(A^{2}-\left(C-B\right)^{2}\right) \\ \nonumber
r_2 \;=\; &8\left(t-B-C\right)\left(s-A\right)\left(\frac{h-A}{2}\right)^{2} \\ \nonumber
&+ 4\left(s-A\right)\left(At+Cs+Bs-2AB\right)\left(\frac{h-A}{2}\right) \\
&+ 2A\left(s-A\right)^{2}\left(B-C\right)
\end{align}
Thus we need to determine the quartic polynomial $u(h)=|r(h)|^{2}={r_1(h)}^{2}+{r_2(h)}^{2}$ and we can then solve for the ellipse semimajor axis, $a$, and semiminor axis, $b$, from the equations
\begin{eqnarray}
a^{2}-b^{2} \;=\; \sqrt{ \frac{1}{\left(16\left(s-A\right)^{4}\right)}u(h)} 
\end{eqnarray}
\begin{eqnarray}
a^{2}b^{2} \;=\; \frac{1}{4}\left(\frac{C}{\left(s-A\right)^{2}}\right)\left(2\left(Bs-A\left(t-C\right)\right)h - ACs\right)\left(2h-A\right)\left(2h-s\right) 
\end{eqnarray}
by parameterising $R=a^{2}-b^{2}$ and $W=a^{2}b^{2}$ to obtain
\begin{eqnarray}
a \;=\; \sqrt{ \frac{1}{2}\left(\sqrt{R^{2}+4W}+R\right)}, \quad
b \;=\; \sqrt{ \frac{1}{2}\left(\sqrt{R^{2}+4W}-R\right)}
\end{eqnarray}
\item Knowing the axes we can generate the ellipse and float its tilt angle $\delta$ until it sits tangent to each side of the quadrilateral, using the inclined ellipse equation~(\ref{eqn:inclinedellipse}) introduced in Section~\ref{sect:characterisation}.
%\begin{eqnarray}
%\rho^{2} \;=\; \frac{a^{2}b^{2}}{\left(\frac{a^{2}+b^{2}}{2}\right)-\left(\frac{a^{2}-b^{2}}{2}\right)\cos\left(2\omega'-2\delta\right)}
%\end{eqnarray}
%where $\omega'=\omega+\delta$ and $\omega$ is the angle from the semimajor axis to a radial line $\rho$ on the ellipse.
\end{enumerate}
\noindent

\subsection{SOHO as a Third Perspective}
\label{sohothirdeye}

The elliptical tie-pointing technique was used to reconstruct the front of a CME observed by STEREO on 26 April 2008 in order to test its efficacy by comparing it with observations from SOHO - a third perspective on the event. The CME appears as a halo from STEREO-B, so a running-difference technique is used to highlight the faint CME front in the images. The front is defined in the images by a point-and-click methodology and characterised with an ellipse fit (outlined in Section~\ref{sect:characterisation}). From STEREO-A the event appears off the east limb and shows a strong streamer deflection to the south-east (in fact the CME would probably be considered only as the northern portion of the erupting material in STEREO-A if it were not shown by STEREO-B to expand further south).
\begin{figure}[!t]
\centerline{\includegraphics[width=\linewidth]{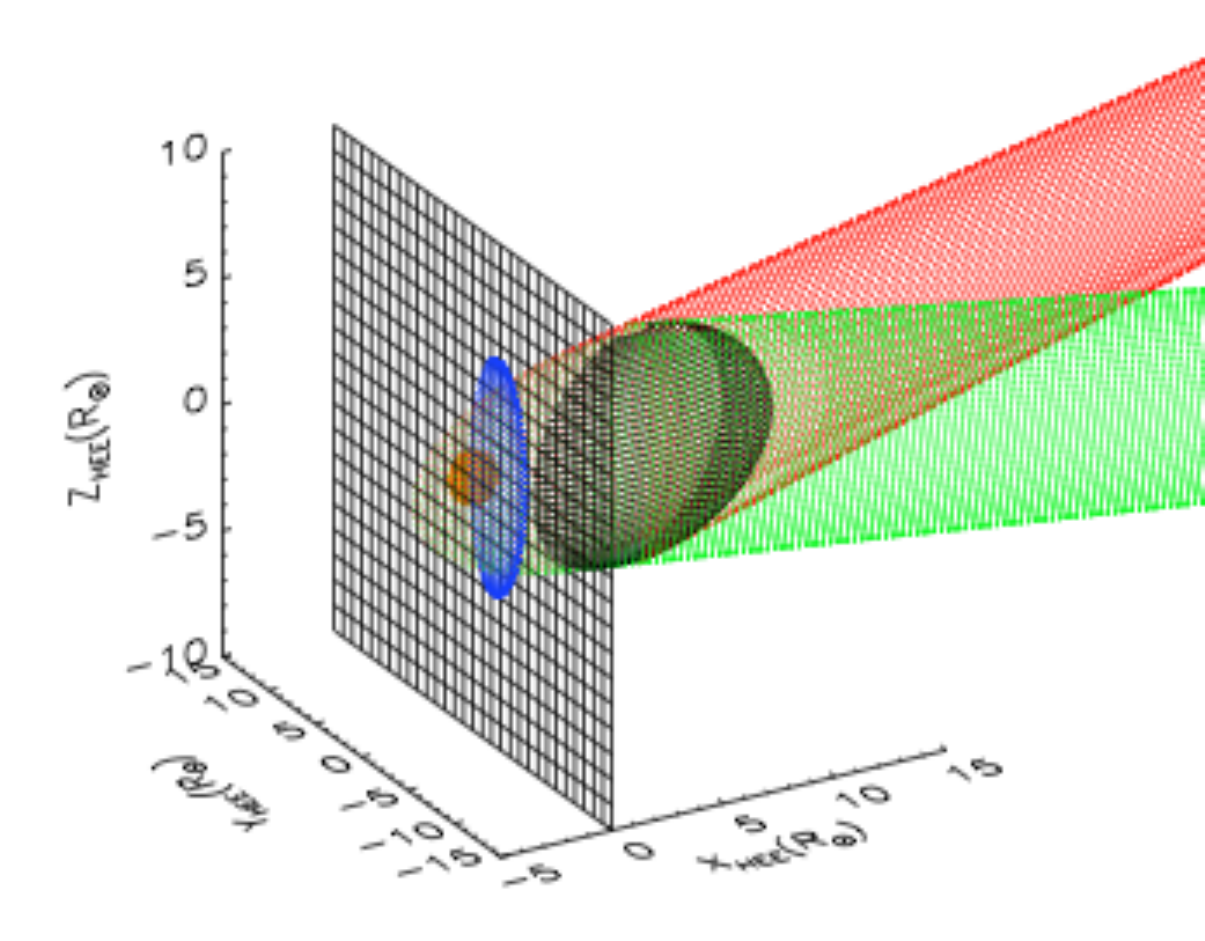}}
\caption{The back-projection of a STEREO 3D CME front reconstruction onto the SOHO/LASCO plane-of-sky, from the observations by STEREO-A (red) and STEREO-B (green) at 16:22 UT on 26 April 2008.}
\label{backprojsoho}
\end{figure}
\newline
\indent With the ellipse characterisations determined for the CME front in the simultaneous images from COR1 and COR2 onboard STEREO-A and B, the elliptical tie-pointing technique is performed and the CME front reconstructed in 3D. This reconstruction is then back-projected onto the LASCO plane-of-sky in order to compare it with observations of the CME from SOHO's vantage point at L1 (Figure~\ref{backprojsoho}). This back-projection is performed by standard geometry of lines-of-sight from the observer position $O(x_0,\,y_0,\,z_0)$ through the 3D reconstruction points $P(x_i,\,y_i,\,z_i)$ and determining where they intersect the plane-of-sky $Q(x_j=0,\,y_j,\,z_j)$ as follows:
\begin{align}
\tan \alpha \; &= \; \frac{y_i - y_0}{x_i - x_0} = \frac{y_j - y_0}{x_j - x_0} = \frac{y_j - y_i}{x_j - x_i} \quad\Rightarrow \quad y_j = x_i \left(\frac{y_i-y_0}{x_i-x_0}\right) + y_i \\
\tan \beta \; &= \; \frac{z_i - z_0}{x_i - x_0} = \frac{z_j - z_0}{x_j - x_0} = \frac{z_j - z_i}{x_j - x_i} \quad \Rightarrow \quad z_j = x_i \left(\frac{z_i-z_0}{x_i-x_0}\right) + z_i 
\end{align}
\newline
\indent Due to the different instrument cadences of the SECCHI and LASCO coronagraphs, frames which lie closest in time were chosen for comparison. Figure~\ref{backproj3} shows the COR2 frames from STEREO-A and B at 16:22~UT on 26 April 2008 with the ellipse characterisations of the CME front (left and right panels), and the back-projected 3D front reconstruction as compared with the LASCO/C2 frame from SOHO at 16:30~UT (middle panel). The reconstruction from the STEREO observations adequately fits with the SOHO observations given the time offset, and so gives credence to the  elliptical tie-pointing technique.

\begin{sidewaysfigure}[!p]
\centerline{\includegraphics[width=\linewidth]{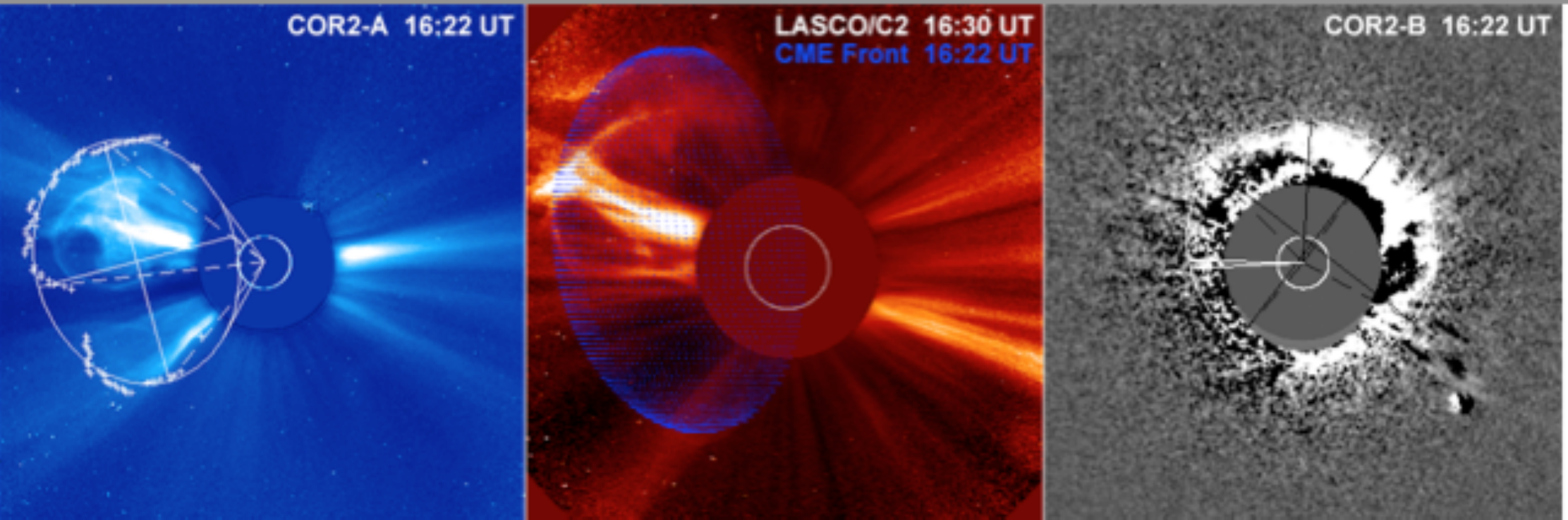}}
\caption{The back-projection of a STEREO 3D CME front reconstruction onto the SOHO/LASCO plane-of-sky, at 16:22 UT on 26 April 2008. The left panel shows the multiscale edge detection and ellipse characterisation of the CME front observed off the east limb from STEREO-A. The right panel shows the running-difference point-and-click ellipse characterisation of the CME front observed as a halo event from STEREO-B. . The middle panel shows the LASCO/C2 observation of the CME at 16:30~UT (8 minutes later than the STEREO observations) with the back-projected 3D CME front reconstruction overplotted for comparison.}
\label{backproj3}
\end{sidewaysfigure}

%\begin{figure}[!p]
%\centerline{\includegraphics[width=\linewidth]{backproj.pdf}}
%\caption{The back-projection of a STEREO 3D CME front reconstruction onto the SOHO/LASCO plane-of-sky, at 16:22 UT on 26 April 2008. The top left panel shows the multiscale edge detection and ellipse characterisation of the CME front observed off the east limb from STEREO-A. The top right panel shows the running-difference point-and-click ellipse characterisation of the CME front observed as a halo event from STEREO-B. The bottom left panel shows the elliptical tie-pointing of the STEREO-A and B lines-of-sight (red and green respectively) and the corresponding back-projection of the reconstructed CME front onto the LASCO plane-of-sky (blue). The bottom right panel shows the LASCO/C2 observation of the CME at 16:30~UT (8 minutes later than the STEREO observations) with the back-projected 3D CME front reconstruction overplotted for comparison.}
%\label{backproj}
%\end{figure}

\section{Earth-Directed CME}

\begin{figure}[!p]
\includegraphics[width=\linewidth]{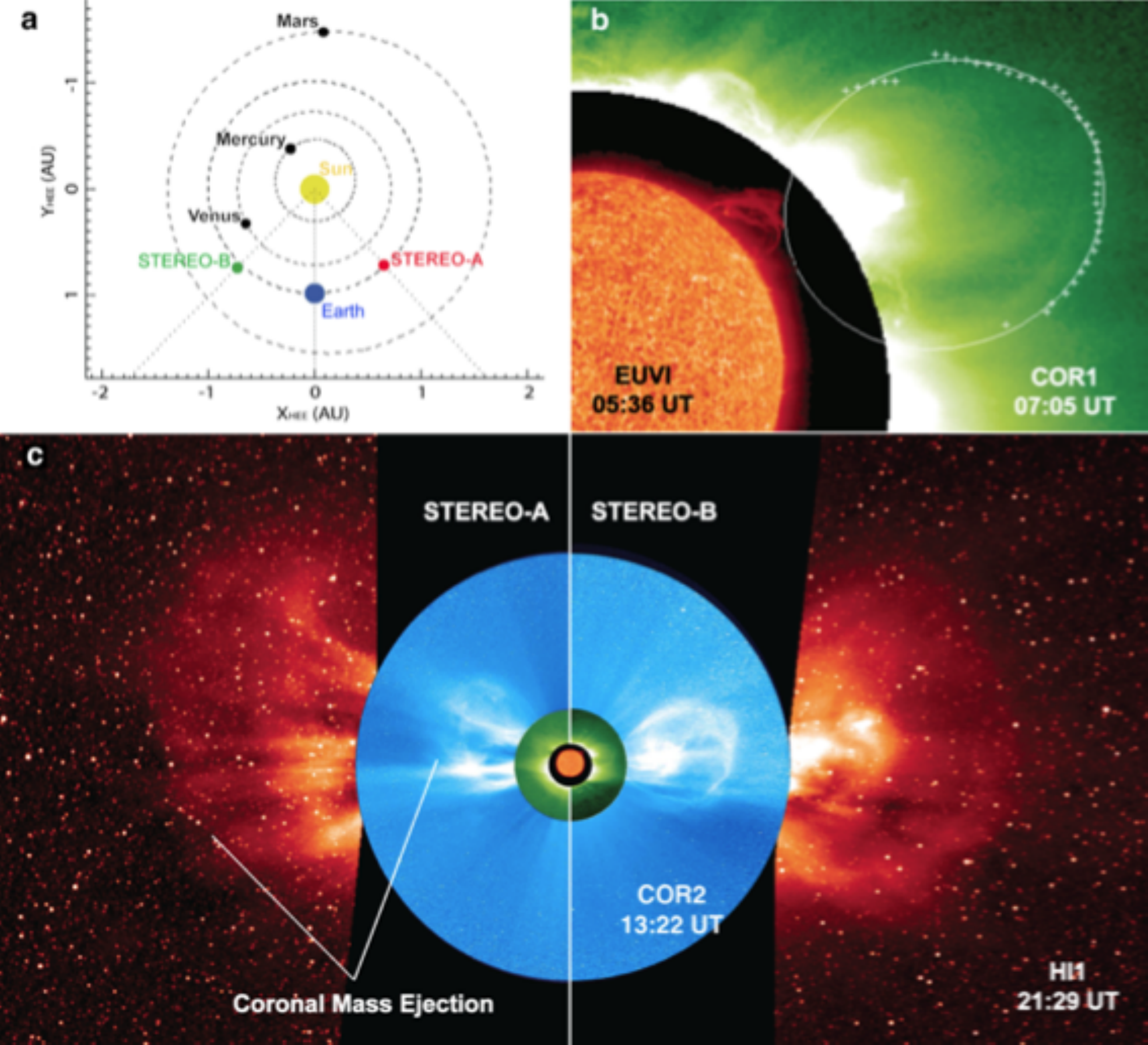}
\caption{Composite of STEREO-Ahead and Behind images from EUVI, COR1, COR2, and HI1 \citep{2010NatCo...1E..74B}. ({\bf a}) indicates the STEREO spacecraft locations, separated by an angle of 86.7$^{\circ}$ at the time of the event. ({\bf b}) shows the prominence eruption observed in EUVI-B off the NW limb from approximately 03:00~UT which is considered to be the inner material of the CME. The multiscale edge detection and corresponding ellipse characterisation are overplotted in COR1. ({\bf c}) shows that the CME is Earth-directed, being observed off the east limb in STEREO-A and the west limb in STEREO-B.}
\label{12dec2008CME}
\end{figure}

\begin{figure}[!p]
\includegraphics[width=\linewidth]{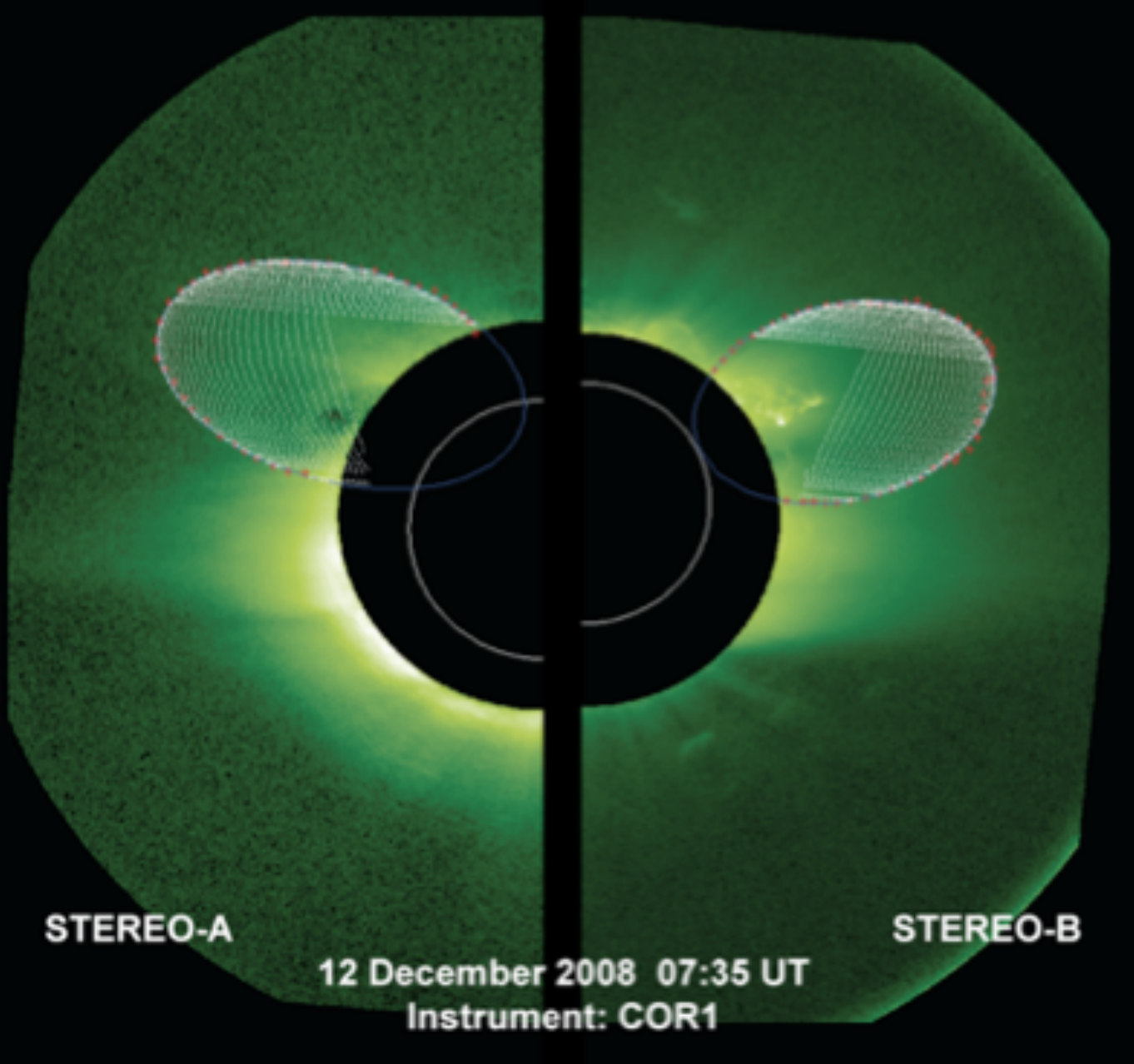}
\caption{STEREO-Ahead and Behind COR1 images at 07:35~UT on the 12 December 2008. Overplotted are: the multiscale edge detections of the CME front (red); the ellipse characterisations (blue); and the resulting 3D reconstructions back-projected onto the plane-of-sky (white).}
\label{cor10735}
\end{figure}

\begin{figure}[!p]
\includegraphics[width=\linewidth]{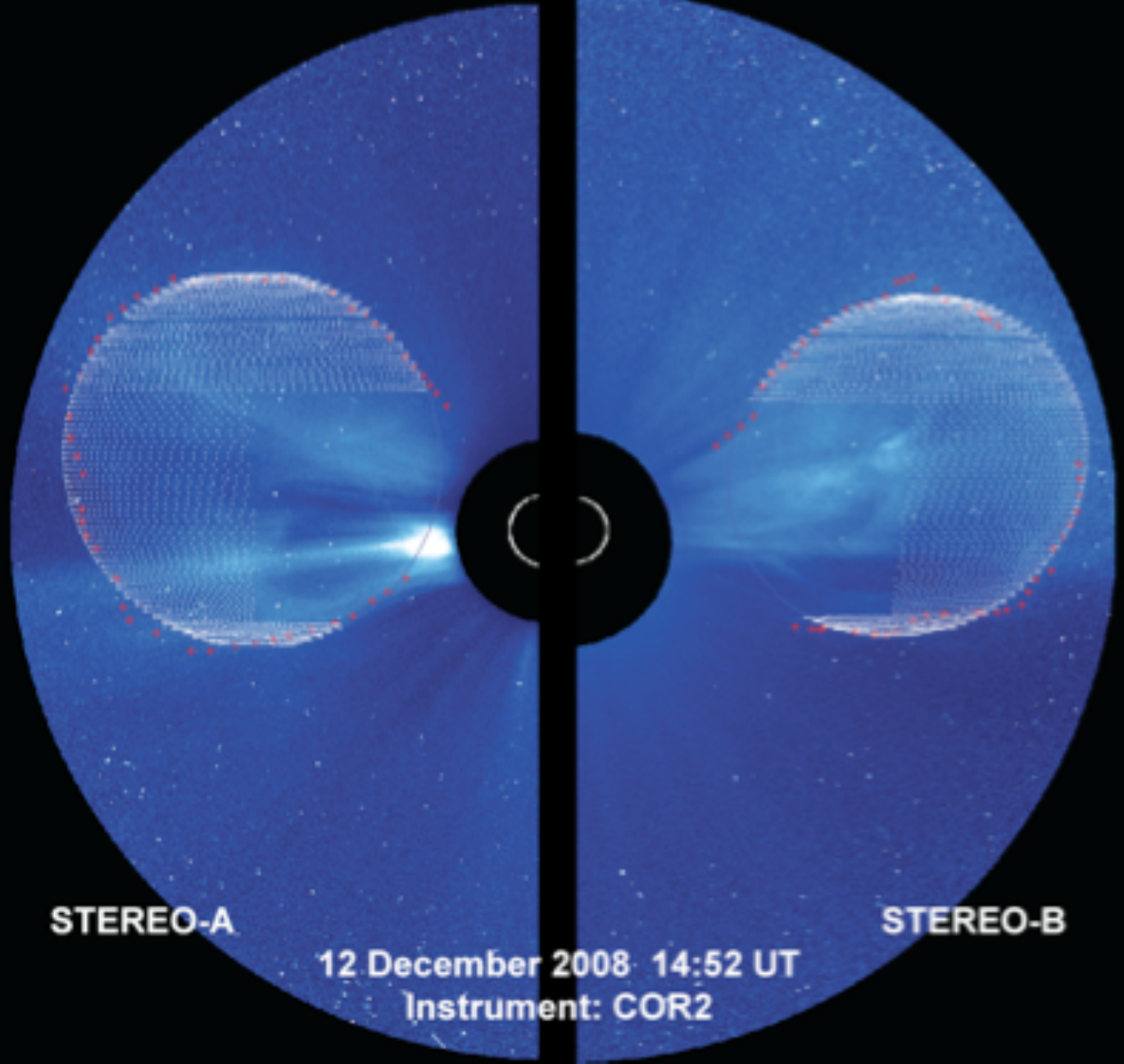}
\caption{STEREO-Ahead and Behind COR2 images at 14:52~UT on the 12 December 2008. Overplotted are: the multiscale edge detections of the CME front (red); the ellipse characterisations (blue); and the resulting 3D reconstructions back-projected onto the plane-of-sky (white).}
\label{cor21452}
\end{figure}

\begin{figure}[!p]
\includegraphics[width=\linewidth]{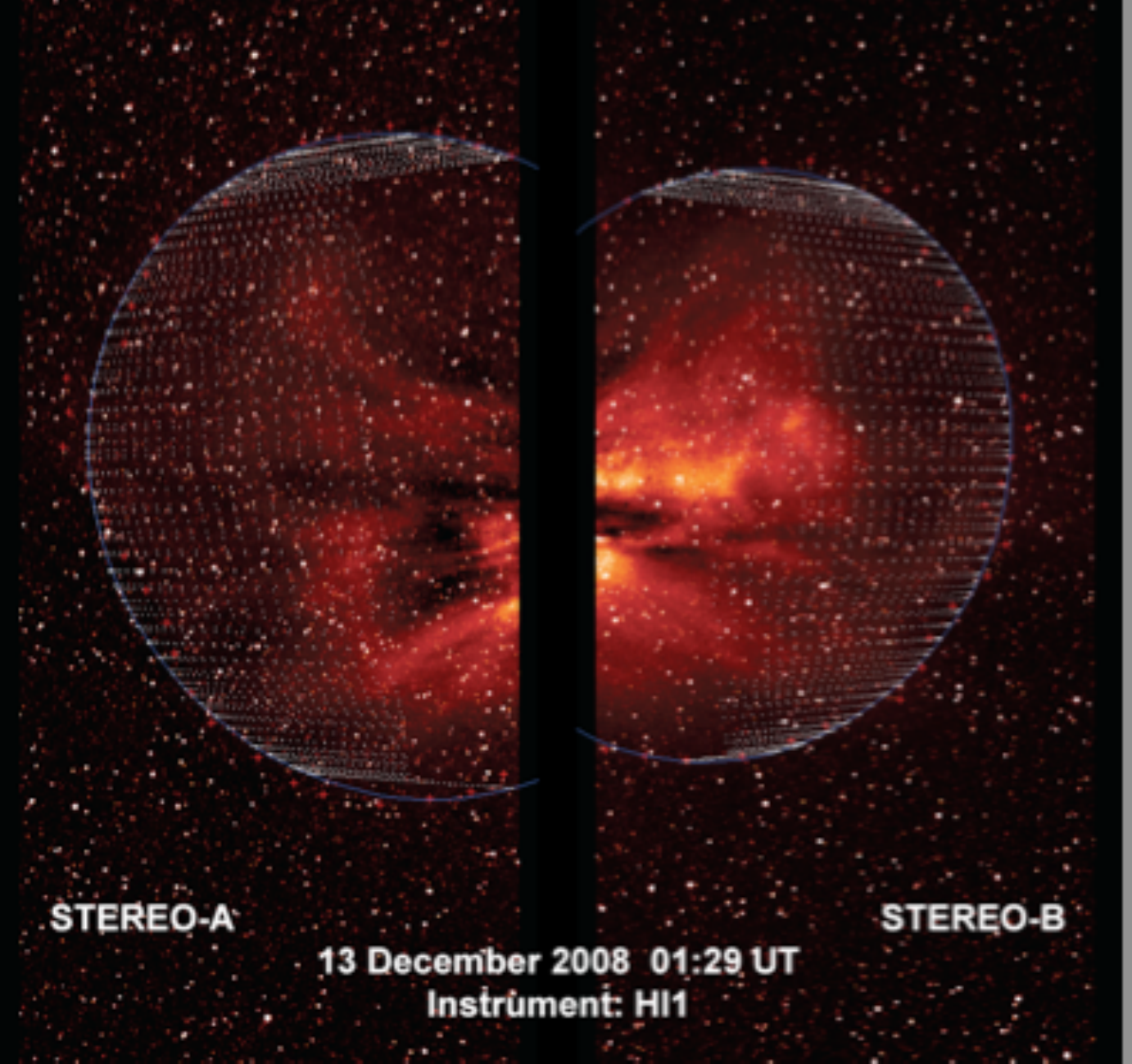}
\caption{STEREO-Ahead and Behind HI1 images at 01:29~UT on the 13 December 2008. Overplotted are: the running difference edge detections of the CME front (red); the ellipse characterisations (blue); and the resulting 3D reconstructions back-projected onto the plane-of-sky (white).}
\label{hi10129}
\end{figure}

\begin{sidewaysfigure}[!p]
\includegraphics[width=\linewidth]{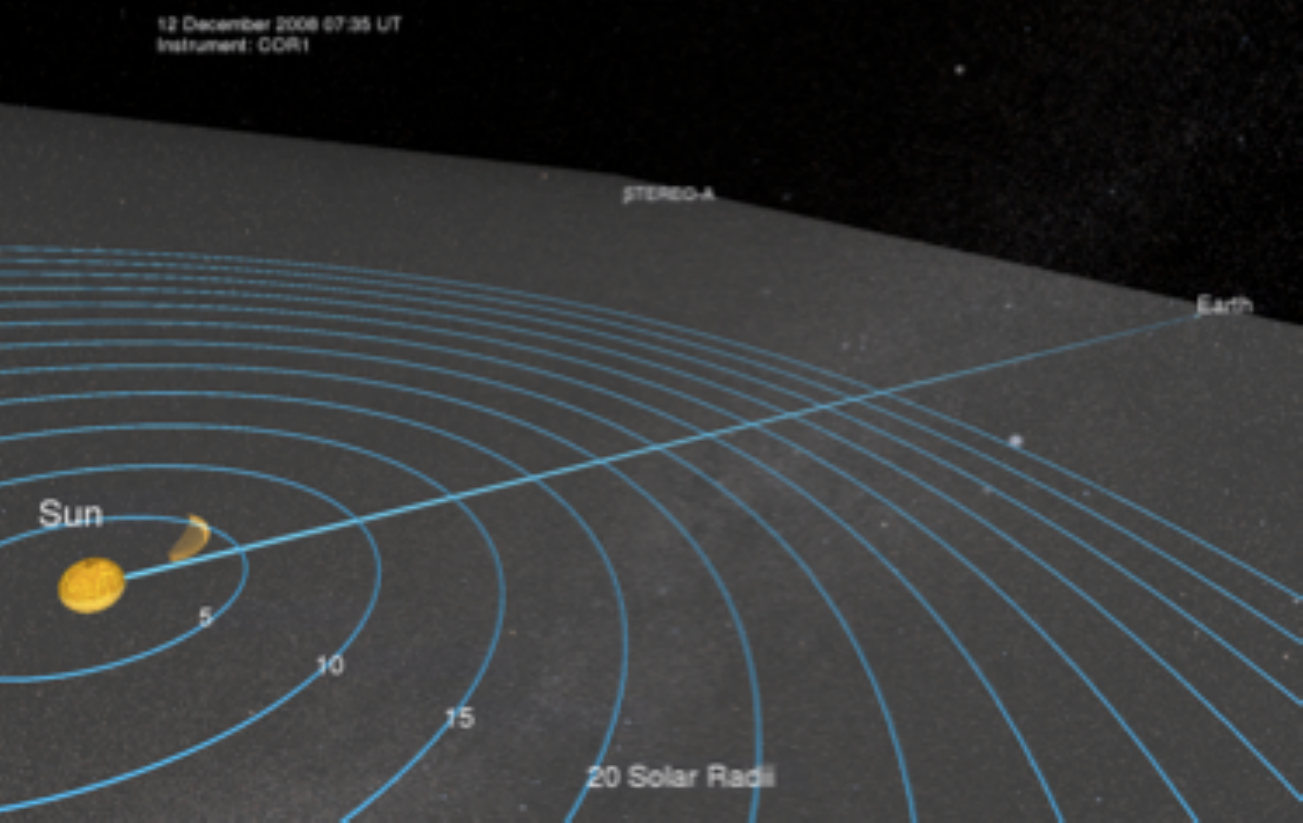}
\caption{The 3D CME front reconstruction as it propagates along the Sun-Earth line through interplanetary space. This particular frame of the reconstruction is from the observations of COR1 at 07:35~UT on 12 December 2008.}
\label{movie0735}
\end{sidewaysfigure}

\begin{sidewaysfigure}[!p]
\includegraphics[width=\linewidth]{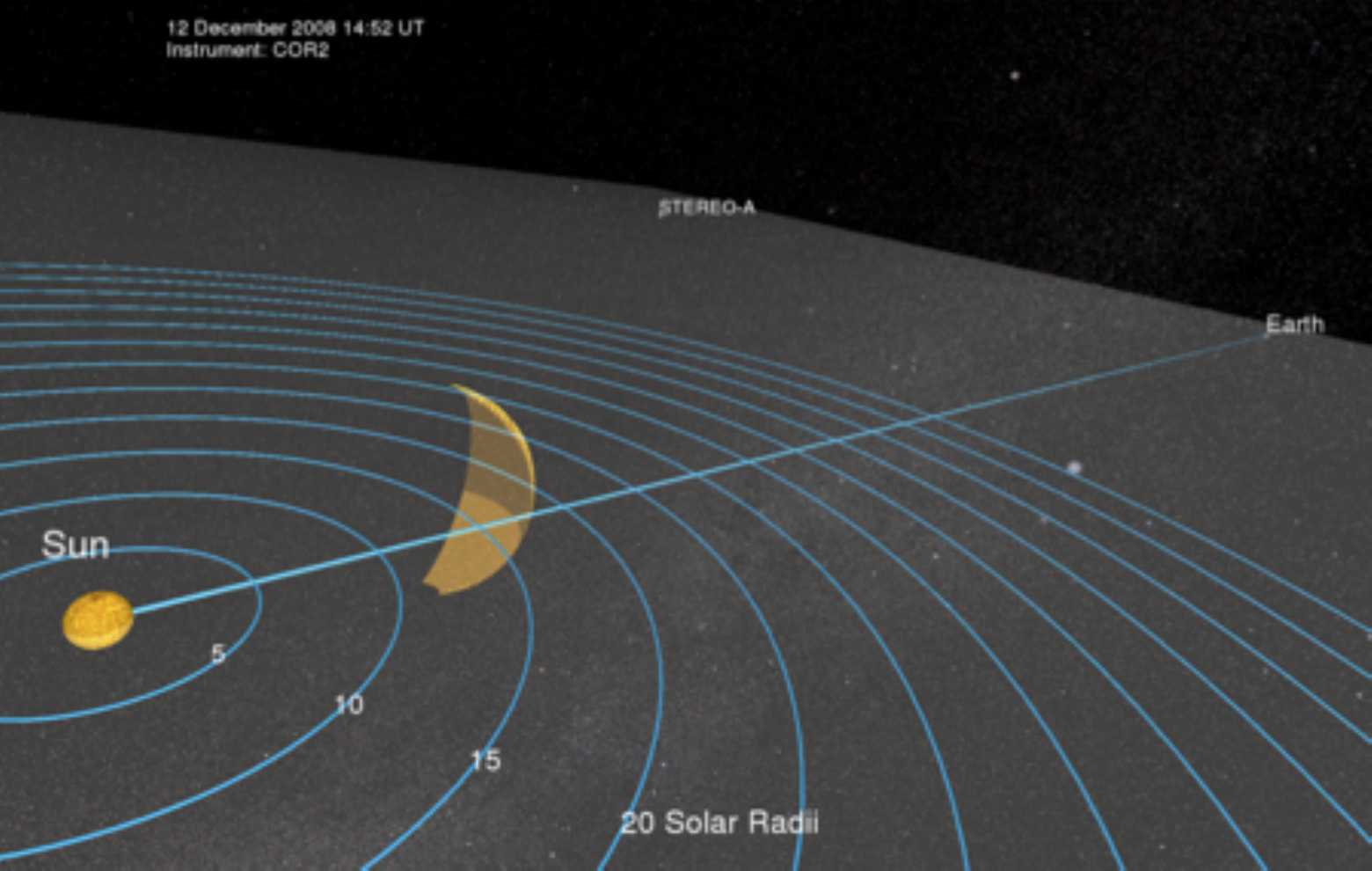}
\caption{The 3D CME front reconstruction as it propagates along the Sun-Earth line through interplanetary space. This particular frame of the reconstruction is from the observations of COR2 at 14:52~UT on 12 December 2008.}
\label{movie1452}
\end{sidewaysfigure}

\begin{sidewaysfigure}[!p]
\includegraphics[width=\linewidth]{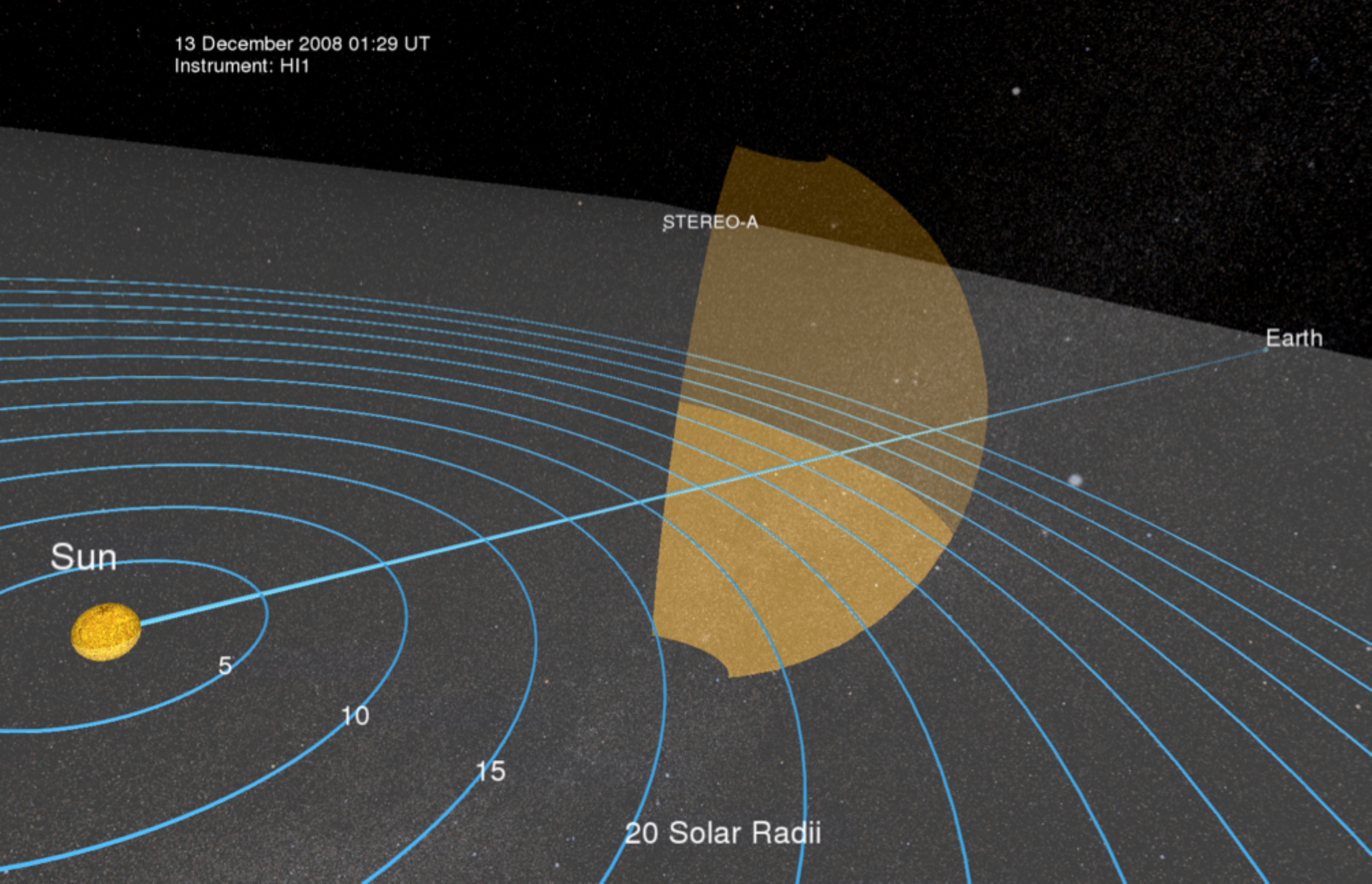}
\caption{The 3D CME front reconstruction as it propagates along the Sun-Earth line through interplanetary space. This particular frame of the reconstruction is from the observations of HI1 at 01:29~UT on 13 December 2008.}
\label{movie0129}
\end{sidewaysfigure}

\begin{sidewaysfigure}[!p]
\includegraphics[width=\linewidth]{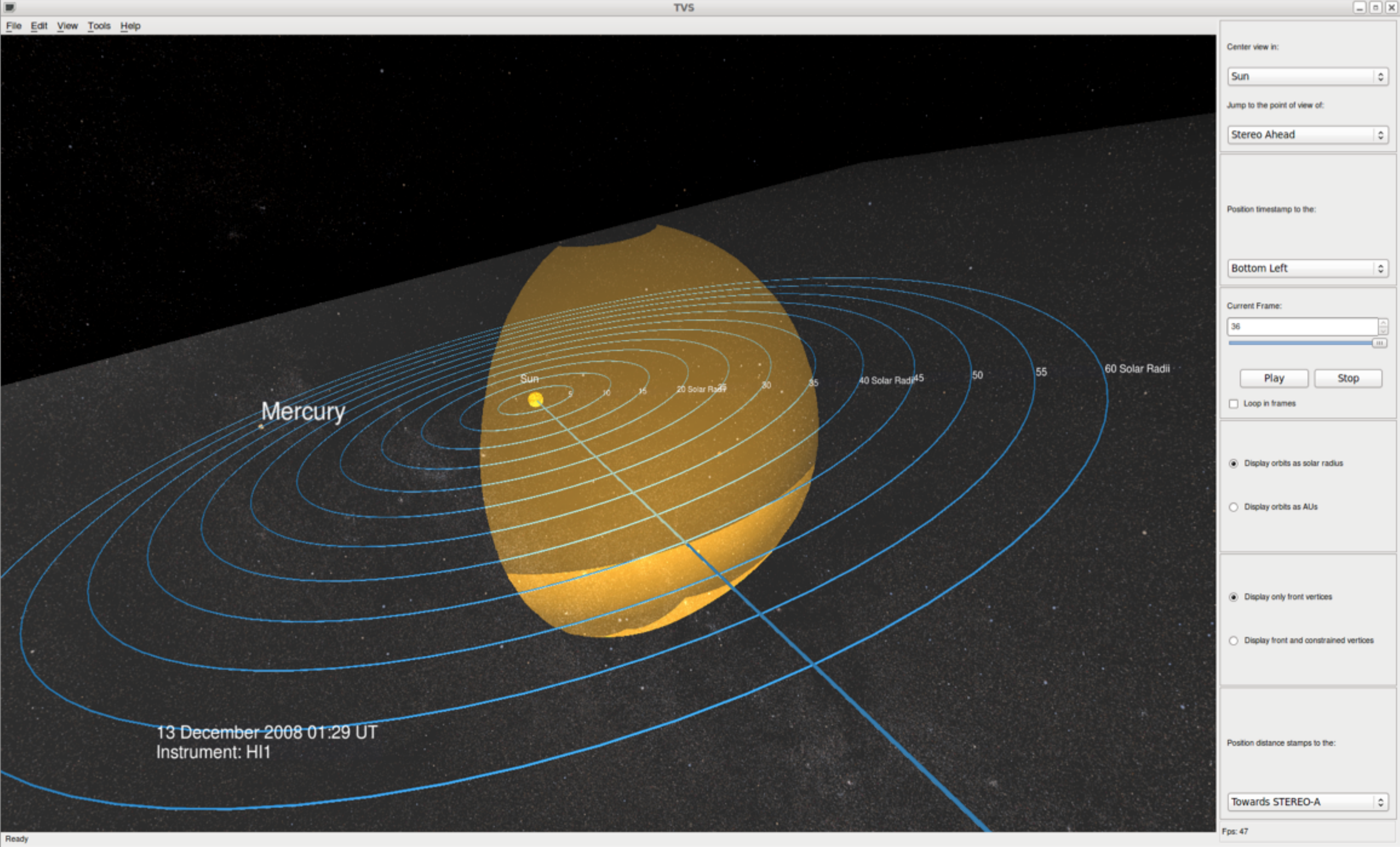}
\caption{The graphical user interface for the 3D visualisation of the CME front reconstruction, which may be used to change the orientation of the observer, or the location of the definable parameters such as time-stamps and distance measures (as displayed on the right).}
\label{hi10129rotgui}
\end{sidewaysfigure}

On 12 December 2008 an erupting prominence was observed by STEREO while the spacecraft were in near quadrature at 86.7$^{\circ}$ separation (Figure~\ref{12dec2008CME}a). The eruption is visible at 50\,--\,55$^{\circ}$ north from 03:00 UT in SECCHI/EUVI images, obtained in the 304~\AA\ passband, in the northeast from the perspective of STEREO-A and off the northwest limb from STEREO-B. The prominence is considered to be the inner material of the CME which was first observed in COR1-B at 05:35~UT (Figure~\ref{12dec2008CME}b). For our analysis, we use the two coronagraphs (COR1/2) and the inner Heliospheric Imagers (HI1) (Figure~\ref{12dec2008CME}c). In each image the front of the CME is fitted with an ellipse that characterises its propagation across the plane-of-sky \citep{2009A&A...495..325B}. This ellipse fitting is sensitive predominantly to the leading edges of the CME but equal weight is given to the CME flank edges as they enter the field-of-view of each instrument. The 3D reconstruction is then performed using a method of curvature-constrained tie-pointing within epipolar planes containing the two STEREO spacecraft (detailed in Section~\ref{sect:ellipticaltiepointing}). An example of the ellipse characterisation to the CME front and the corresponding back-projected 3D reconstruction is shown in Figure~\ref{cor10735} for COR1, Figure~\ref{cor21452} for COR2, and Figure~\ref{hi10129} for HI1. Corresponding frames from a 3D visualisation of the event are shown in Figure~\ref{movie0735} for COR1, Figure~\ref{movie1452} for COR2, and Figure~\ref{movie0129} for HI1, showing the relative locations of the Sun, Earth and STEREO spacecrafts in the inner heliosphere. Figure~\ref{hi10129rotgui} illustrates the graphical user interface developed for the 3D visualisation of the CME front reconstruction.

\section{Results}

\begin{figure}[!t]
\includegraphics[scale=0.4]{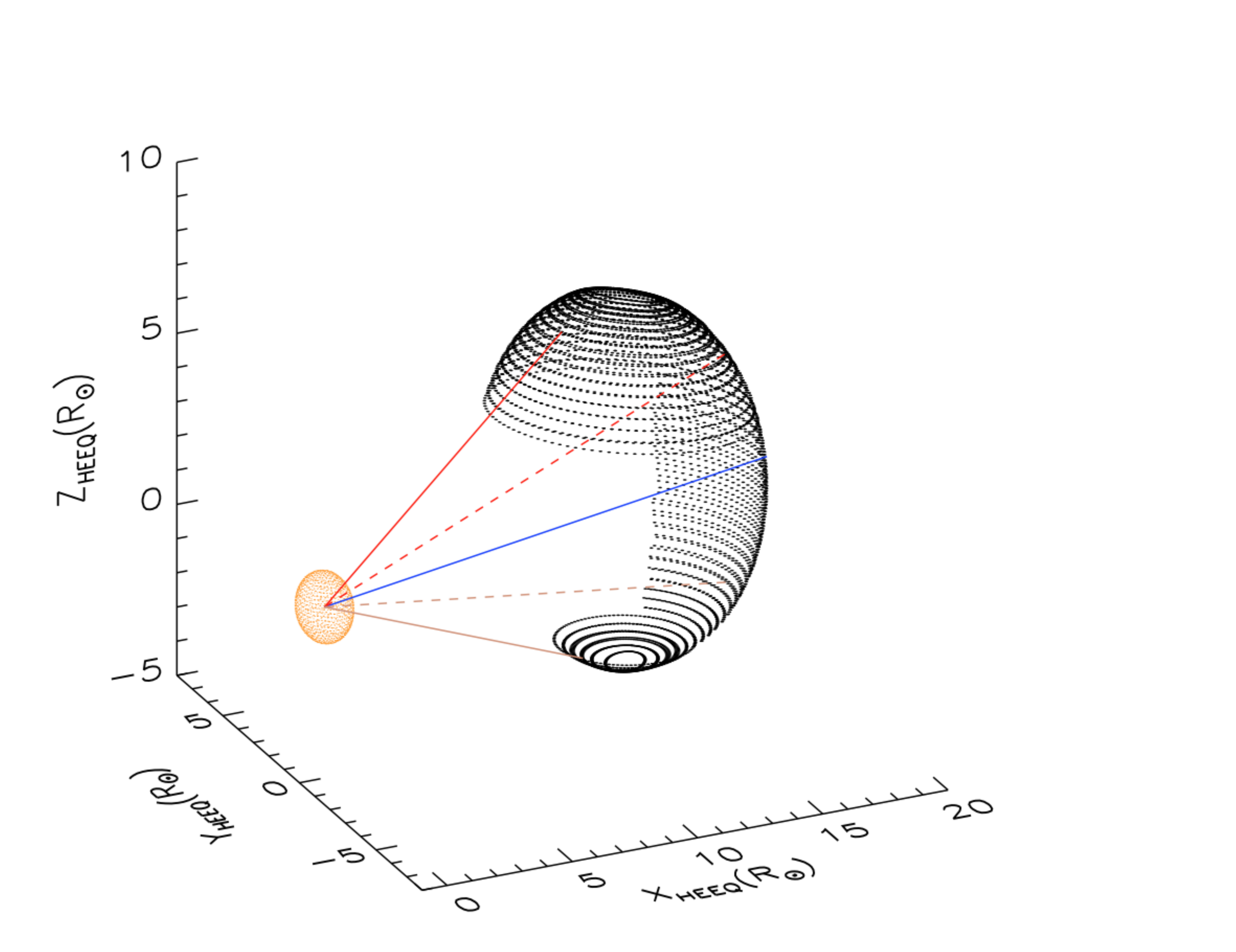}
\caption{The 3D CME front reconstruction from the COR2 Ahead and Behind frames at 14:52~UT on 12 December 2008. The lines drawn from Sun-centre indicate the `Midpoint of Front' (solid blue), the `Northern\,/\,Southern Flanks' (solid red\,/\,brown), and the `Midtop\,/\,Midbottom of Front' at angles in between (dashed red\,/\,brown). By taking these measurements across all frames we may determine the kinematics and morphology of the CME as plotted in Figures~\ref{12dec2008kins} and \ref{3Dkins}.}
\label{span1452}
\end{figure}

% It should be noted that the positions of the flanks are subject to large scatter: As the CME enters each field-of-view the location of a tangent to its flanks is prone to moving back along the reconstruction in cases where the epipolar slices completely constrain the flanks. Hence the `Midtop/Midbottom of Front' measurements better convey the southward dominated expansion.
% For each instrument the first three points of angular width measurement were removed since the CME was still predominantly obscured by each instrument's occulter

The resulting kinematics and morphology of the CME are measured along an angular span through the reconstructed CME front in the out-of-ecliptic plane along the Sun-Earth line (Figure~\ref{span1452}). These were taken by first closing tangents to the CME front (`Northern/Southern Flanks'), and then measuring the height along an angle midway between these (`Midpoint of Front'), and then similarly along the two angles midway between the midpoint and the flanks (`Midtop/Midbottom of Front'). Although these measurements are fixed along the Sun-Earth line, investigating how the CME height profile would change if taken along a trajectory slightly off the Sun-Earth line shows no significant deviation within the associated errors and thus has negligible effect on the kinematics. The same is true if the overall maximum height (of varying location) on each individual CME front is instead taken and the kinematics reanalysed.

\subsection{3D Error Propagation}

\begin{figure}[!t]
\begin{center}
\includegraphics[scale=0.25]{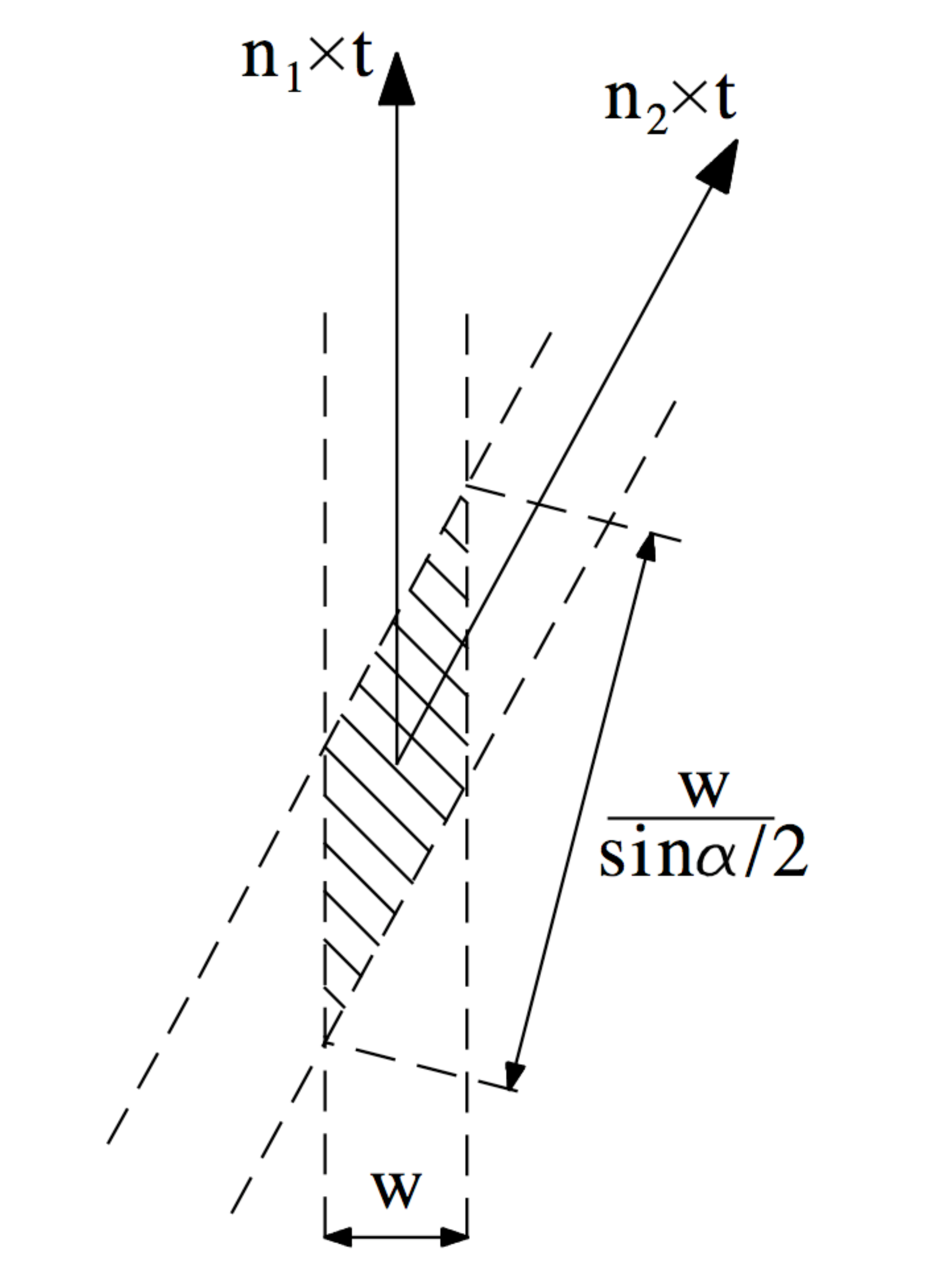}
\end{center}
\caption{The error trapezoid on the tie-pointing of two lines-of-sight in 3D space, reproduced from \citet{2006astro.ph.12649I}. The error $w$ in localising a point on each plane-of-sky results in a trapezoid with a diagonal measuring $w/\sin(\alpha/2)$ as shown, where $\alpha$ is the spacecraft separation angle.}
\label{errortrapezoid}
\end{figure}

When considering the errors that propagate from the 2D plane-of-sky of each image onto the 3D quadrilateral localising the CME, we may assume that the lines-of-sight within the error range are essentially parallel. This means the error interval on the coordinate being tie-pointed in 3D is given by a trapezoid surrounding the intersection of the lines-of-sight, illustrated in Figure~\ref{errortrapezoid}. This is done for each corner of the quadrilateral within a given epipolar plane. For a spacecraft separation of angle $\alpha$ and errorbar of magnitude $w$ on the 2D image we can define the error trapezoid as having diagonals of length $w/\cos(\alpha/2)$ and $w/\sin(\alpha/2)$.
\newline
\indent So in the case of COR1/2, the optimum filter size in the multiscale decomposition was 2$^3$ pixels wide, giving an error of $\pm$\,8 pixels, so $w=16$. Over the course of the 12 December 2008 CME the average STEREO spacecraft separation was $86.75^{\circ}$, so we calculate the error trapezoid as having diagonals of size:
\begin{eqnarray}
\left[ \frac{w}{\cos\left(\frac{\alpha}{2}\right)}, ~ \frac{w}{\sin\left(\frac{\alpha}{2}\right)} \right] \,&=\, \left[ 11.0, ~ 11.6 \right]
\end{eqnarray}
This provides a 3$\sigma$ height error of 11.6 pixels, so the corresponding 1$\sigma$ height error is given by $68(11.6)/99.7=7.9$ pixels. The time error for the multiscale edge detections is given by the exposure time of the individual frames: 1.69984 seconds for COR1 and 2.00090 seconds for COR2.
\newline
\indent In the case of HI1 a 1$\sigma$ plane-of-sky error of 3 pixels was determined \citep{2009SoPh..256..149M}, so $w=6$ and the error trapezoid is deduced to be $\left[ 4.1, ~ 4.4 \right]$. So the height error for HI1 is taken as 4.4 pixels and the time error is given by the thirty summed 60 second images to result in 1800 seconds \citep{2009SoPh..254..387E}.
\newline
\indent These errors are transformed first into arcseconds by multiplying by the plate scale of the instruments (7.5043001 arcsec/pixel for COR1, 14.7 arcsec/pixel for COR2, and 71.927554 arcsec/pixel for HI1), and then into metres, knowing the respective size of the Sun in arcseconds on the plane-of-sky observed by each instrument (given 1~R$_{\odot}=6.95508\times10^8$~m). The resultant errors are then propagated with the 3-point Lagrangian interpolation (detailed in Section~\ref{3pointlagrangian}) from the height-time curves into the velocity and acceleration profiles of Figure~\ref{3Dkins}. Due to the potentially large deviation of the end points from the general trend in 3-point Lagrangian interpolation, the endpoints of the EUVI prominence data, the COR1/2 coronagraph data, and the HI data are each removed as outliers from the velocity and acceleration plots.

\subsection{Prominence \& CME Acceleration}

\begin{figure}[!p]
\includegraphics[scale=0.7, clip=true, trim=20 100 0 0]{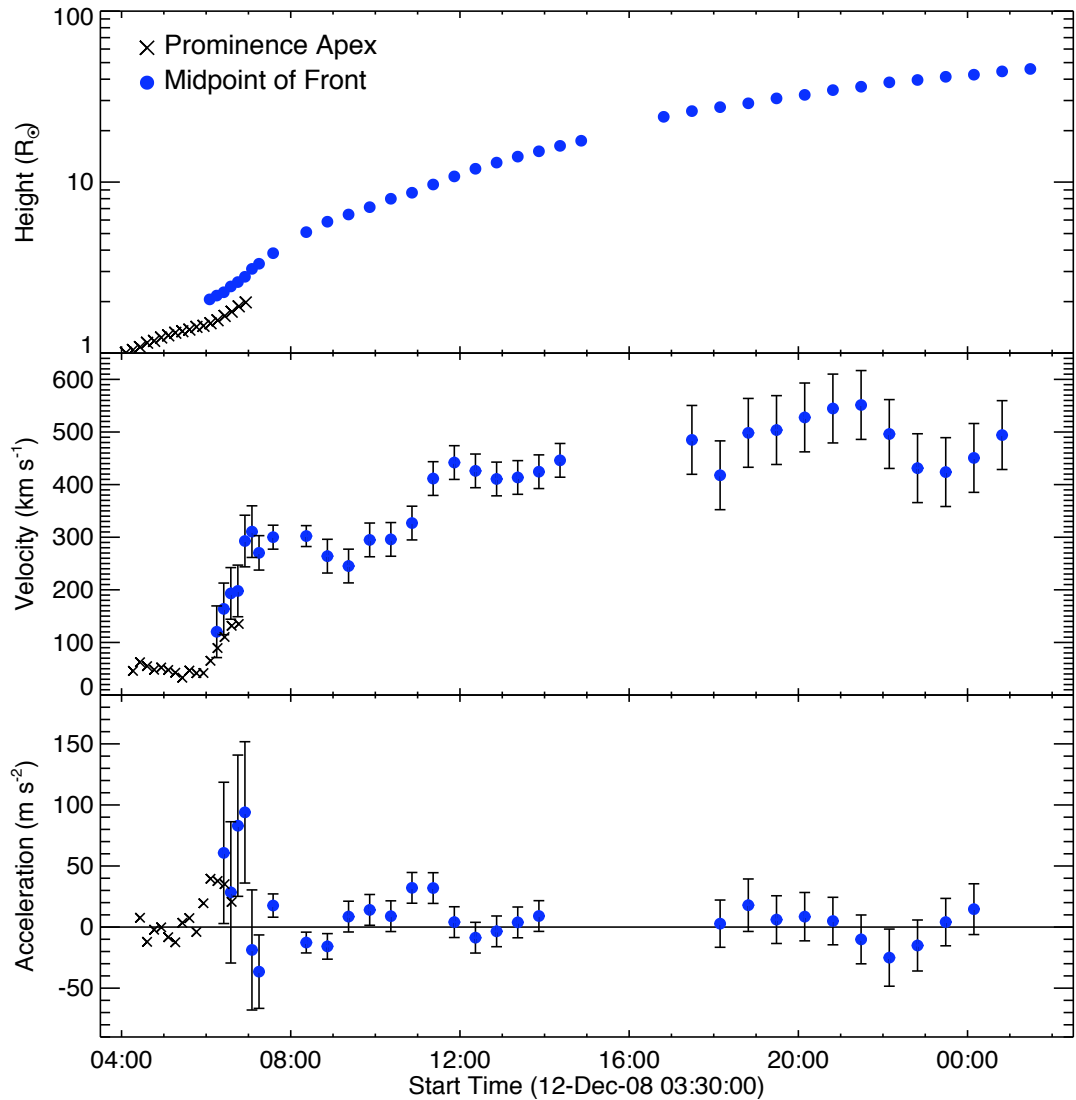}
\caption{The kinematics of the prominence and 3D reconstructed CME front of 12 December 2008. The prominence is observed as the inner material of the CME, with both undergoing acceleration from $\sim\,$06:00\,--\,07:00~UT, peaking at $\sim$\,40~m~s$^{-2}$ and $\sim$\,94~m~s$^{-2}$ respectively, before reducing to scatter about zero. Measurement uncertainties are indicated by one standard deviation error bars.}
\label{12dec2008kins}
\end{figure}

In determining the CME kinematics for the `Midpoint of Front' we find a steep increase in the velocity corresponding to an early impulsive acceleration phase of 94\,$\pm$\,58~m~s$^{-2}$ (bottom of Figure~\ref{12dec2008kins}). The prominence rises with a velocity of $\sim$\,50~km~s$^{-1}$ before the system fully erupts and the prominence undergoes an acceleration of 40\,$\pm$\,5~m~s$^{-2}$ behind the CME front. This is indicative of the onset of explosive reconnection or other loss-of-equilibrium in the system whereby the internal magnetic pressure increases sufficiently and/or the external magnetic pressure decreases sufficiently to allow the eruption to proceed in the height range $\sim$\,1.5\,--\,3~R$_{\odot}$ (top panel of Figure~\ref{3Dkins}). The velocity profile is synonymous with those produced by the 2D flux rope model as in Figure~11.5 of \citet{2000mare.book.....P}. The acceleration then reduces to scatter about zero as the explosive nature of the eruption due to the Lorentz force diminishes and the drag force due to the ambient solar wind pressure begins to dominate (see Equation~\ref{eqnmotion}).

\subsection{Non-radial Prominence \& CME Motion.}

\begin{figure}[!p]
\includegraphics[scale=0.63]{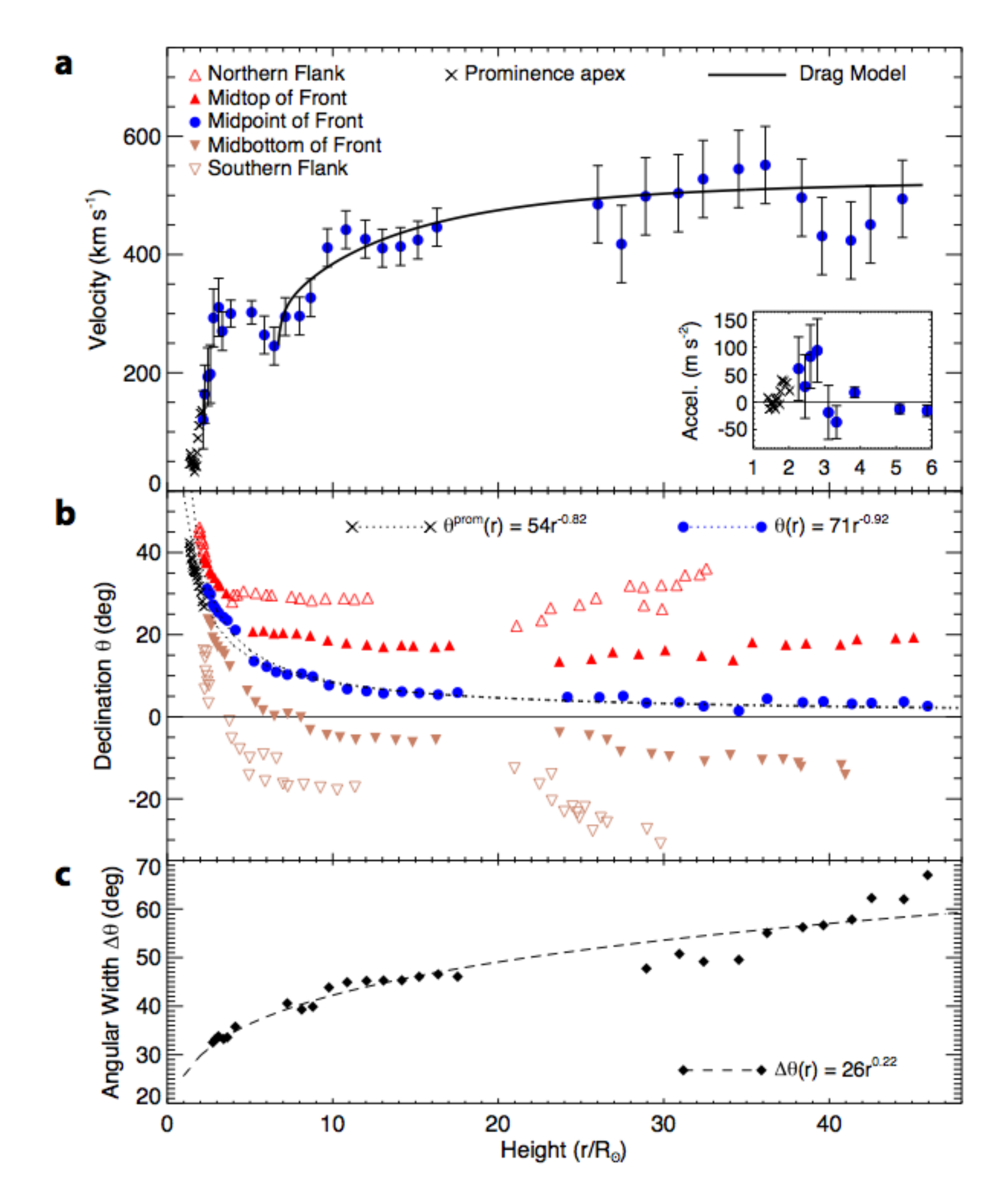}
\caption{Kinematic and morphological properties of the 3D reconstruction of the 12 December 2008 CME front. The top panel shows the velocity of the middle of the CME front with corresponding drag model and, inset, the early acceleration peak. Measurement uncertainties are indicated by one standard deviation error-bars. The middle panel shows the declinations from the ecliptic (0$^{\circ}$) of an angular spread across the front between the CME flanks with a power-law fit indicative of non-radial propagation. The bottom panel shows the angular width of the CME with a power-law expansion.}
\label{3Dkins}
\end{figure}

It is immediately evident from the reconstruction (illustrated in Figure~\ref{ellipticaltiepointing}c) that the CME propagates non-radially away from the Sun. The CME flanks change from an initial latitude span of 16\,--\,46$^{\circ}$ to finally span approximately $\pm$\,30$^{\circ}$ of the ecliptic (middle panel of Figure~\ref{3Dkins}). The mean declination, $\theta$, of the CME is well fitted by a power-law of the form $\theta(r)=\theta_{0}r^{-0.92}~(2~$R$_{\odot}<r<46~$R$_{\odot})$ as a result of this non-radial propagation. Tie-pointing the prominence apex and fitting a power-law to its declination angle results in $\theta^{prom}(r)=\theta_{0}^{prom}r^{-0.82}~(1~$R$_{\odot}<r<3~$R$_{\odot})$, implying a source latitude of $\theta_{0}^{prom}$(1~R$_{\odot}$)~$\approx$~54$^{\circ}$~N in agreement with EUVI observations. Previous statistics on CME position angles have shown that, during solar minimum, they tend to be offset closer to the equator as compared to those of the associated prominence eruption \citep{2003ApJ...586..562G}. The non-radial motion we quantify here may be evidence of the drawn-out magnetic dipole field of the Sun, an effect predicted at solar minimum due to the influence of the solar wind pressure (e.g., Figure~8 in \citet{1971SoPh...18..258P} and Figure~2 in \citet{1998A&A...337..940B}). Other possible influences include changes to the internal current of the magnetic flux rope \citep{2001SoPh..203..119F}, or the orientation of the magnetic flux rope with respect to the background field \citep{2005A&A...432..331C}, whereby magnetic pressure can act asymmetrically to deflect the flux rope pole-ward or equator-ward depending on the field configurations.

\subsection{CME Angular Width Expansion}

Over the height range 2\,--\,46~R$_{\odot}$ the CME angular width ($\Delta\theta=\theta_{max}-\theta_{min}$) increases from $\sim$\,30$^{\circ}$ to $\sim$\,60$^{\circ}$ with a power-law of the form $\Delta\theta(r)=\Delta\theta_{0}r^{0.22}$~$(2~$R$_{\odot}<r<46~$R$_{\odot})$ (bottom panel of Figure~\ref{3Dkins}). This angular expansion is evidence for an initial overpressure of the CME relative to the surrounding corona (coincident with its early acceleration inset in top panel of Figure~\ref{3Dkins}). The expansion then tends to a constant during the later drag phase of CME propagation, as it expands to maintain pressure balance with heliocentric distance. It is theorised that the expansion may be attributed to two types of kinematic evolution, namely spherical expansion due to simple convection with the ambient solar wind in a diverging geometry, and expansion due to a pressure gradient between the flux rope and solar wind \citep{2006SoPh..233..233T}. It is also noted that the southern portions of the CME manifest the bulk of this expansion below the ecliptic (best observed by comparing the relatively constant `Midtop of Front' measurements with the more consistently decreasing `Midbottom of Front' measurements in middle panel of Figure~\ref{3Dkins}). Inspection of a Wang-Sheeley-Arge (WSA) solar wind model run \citep{2000JGR...10510465A} reveals higher speed solar wind flows ($\sim$\,650~km~s$^{-1}$) emanating from open-field regions at high/low latitudes (approximately 30$^{\circ}$ north/south of the solar equator). Once the initial prominence/CME eruption occurs and is deflected into a non-radial trajectory, it undergoes asymmetric expansion in the solar wind. It is prevented from expanding upwards into the open-field high-speed stream at higher latitudes, and the high internal pressure of the CME relative to the slower solar wind near the ecliptic accounts for its expansion predominantly to the south. In addition, the northern portions of the CME attain greater distances from the Sun than the southern portions as a result of this propagation in varying solar wind speeds, an effect predicted to occur in previous hydrodynamic models \citep{1999JGR...104..493O}.

\subsection{CME Drag in the Inner Heliosphere}
\label{sect:cmedrag}

Investigating the midpoint kinematics of the CME front, we find the velocity profile increases from approximately 100\,--\,300~km~s$^{-1}$ over the first 2\,--\,5~R$_{\odot}$, before rising more gradually to a scatter between 400\,--\,550~km~s$^{-1}$ as it propagates outward (top panel of Figure~\ref{3Dkins}). The acceleration peaks at approximately 100~m~s$^{-2}$ at a height of $\sim$\,3~R$_{\odot}$, then decreases to scatter about zero. This early phase is generally attributed to the Lorentz force whereby the dominant outward magnetic pressure overcomes the internal and/or external magnetic field tension, while the subsequent increase in velocity, at heights above $\sim$\,7~R$_{\odot}$, is predicted by theory to result from the effects of drag \citep{2006SoPh..233..233T}. At large distances from the Sun, during this postulated drag-dominated epoch of CME propagation, the equation of motion can be cast in the form:
\begin{equation}
	\label{drag}
	M_{cme} \frac{d v_{cme}}{d t}=-\frac{1}{2} \rho_{sw} ( v_{cme} - v_{sw} ) |  v_{cme} - v_{sw} |  A_{cme}  C_{D}
\end{equation}
 This describes a CME of velocity $v_{cme}$, mass $M_{cme}$, and cross-sectional area $A_{cme}$ propagating through a solar wind flow of velocity $v_{sw}$ and density $\rho_{sw}$. The drag coefficient, $C_D$, is found to be of the order of unity for typical CME geometries \citep{2004SoPh..221..135C}, while the density and area are expected to vary as power-law functions of distance $R$. Thus, we parameterise the density and geometric variation of the CME and solar wind using a power-law \citep{2002JGRA..107.1019V} to obtain:
\begin{equation}
         \label{pdrag}
	\frac{d v_{cme}}{d R} = -\alpha R^{-\beta} \frac{1}{v_{cme}}\left ( v_{sw} - v_{cme} \right )^\gamma
\end{equation}
where $\gamma$ describes the drag regime, which can be either viscous ($\gamma$~=~1) or aerodynamic ($\gamma$~=~2), and $\alpha$ and $\beta$ are constants primarily related to the cross-sectional area of the CME and the density ratio of the solar wind flow to the CME ($\rho_{sw}/\rho_{cme}$). We determine a theoretical estimate of the CME velocity as a function of distance by numerically integrating this equation using a 4th order Runge-Kutta scheme and fitting the result to the observed velocities from $\sim$\,7\,--\,46~R$_{\odot}$. The initial CME height, CME velocity, asymptotic solar wind speed, and $\alpha$, $\beta$, and $\gamma$ are obtained from a bootstrapping procedure which provides a final best-fit to the observations and confidence intervals for the parameters. Best-fit values for $\alpha$ and $\beta$ were found to be (4.55$^{+2.30}_{-3.27}$)$\times$10$^{-5}$ and -2.02$^{+1.21}_{-0.95}$ which agree with values found in previous modelling work \citep{2001SoPh..202..173V}. The best-fit value for the exponent of the velocity difference between the CME and the solar wind, $\gamma$, was found to be 2.27$^{+0.23}_{-0.30}$, which is clear evidence that aerodynamic drag ($\gamma$~=~2), and not viscous drag ($\gamma$~=~1) acts during the propagation of the CME in interplanetary space.

\subsection{CME Arrival Time}
\label{sect:cmearrival}

The drag model provides an asymptotic CME velocity of 555$_{-42}^{+114}$~km~s$^{-1}$ when extrapolated to 1~AU, which predicts the CME to arrive one day before the Advanced Composition Explorer (ACE) or WIND spacecrafts detect it at the L1 point. We investigate this discrepancy by using our 3D reconstruction to simulate the continued propagation of the CME from the Alfv\'{e}n radius ($\sim$\,21.5~R$_{\odot}$) to Earth using the ENLIL with Cone Model \citep{2004JGRA..10903109X} at NASA's Community Coordinated Modeling Center. ENLIL is a time-dependent 3D MHD code that models CME propagation through interplanetary space. An ideal fluid approximation is used to describe the solar wind plasma, under time-dependent MHD processes (neglecting microscopic processes). The plasma is treated as a fully ionised hydrogen gas with equal electron and proton densities ($n=n_e=n_p$) and temperatures ($T=T_e=T_p$), and the basic equations of MHD theory applied (such as outlined in Section~\ref{section:mhdtheory}).

\begin{figure}[!p]
\includegraphics[width=\linewidth]{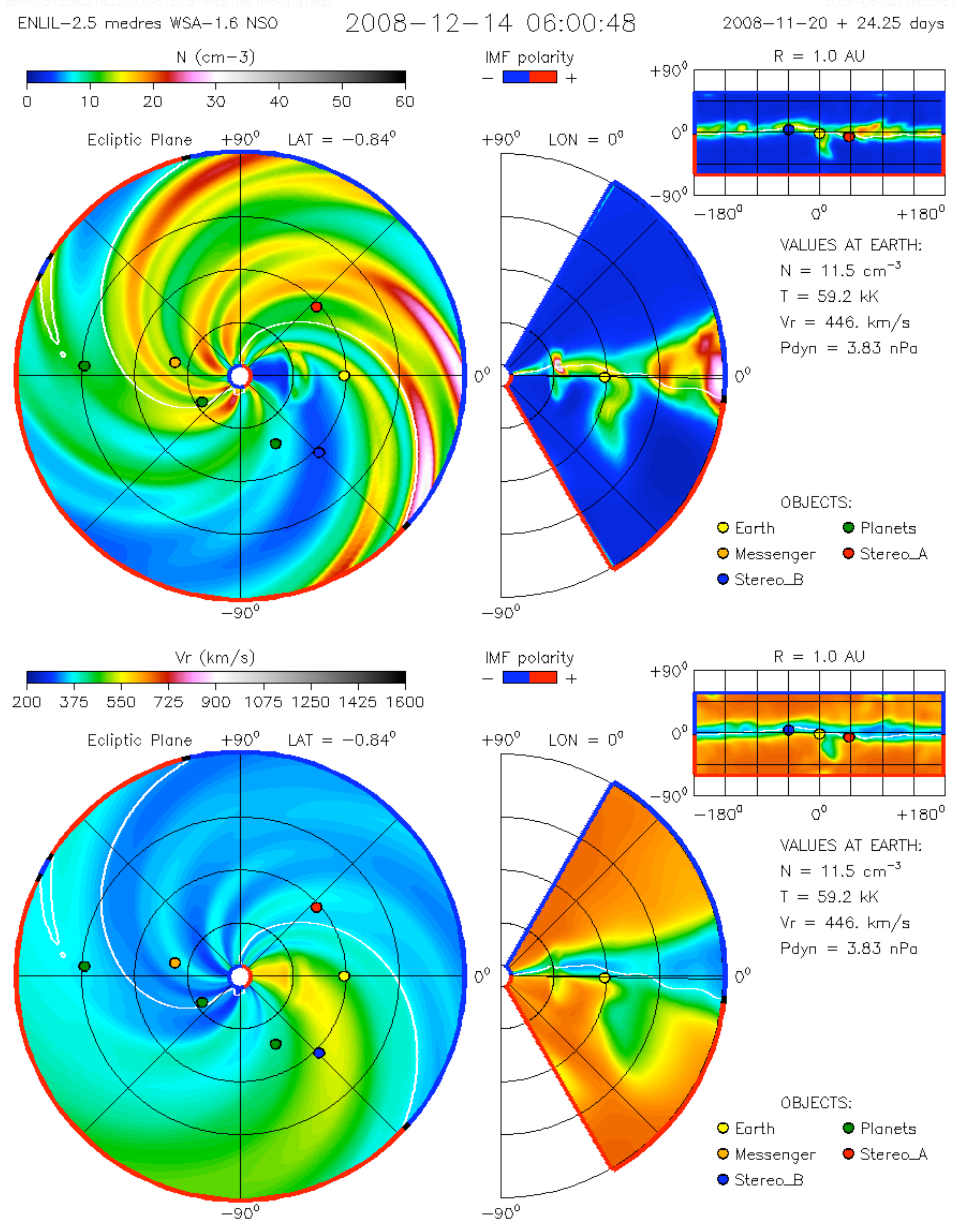}
\caption{The 3D CME front parameters are used as initial conditions for an ENLIL with Cone Model MHD simulation \citep{2004JGRA..10903109X} and the output density (top) and velocity (bottom) profiles of the inner heliosphere are illustrated here for the time-stamp of 06:00~UT on 14 December 2008. Beyond distances of $\sim$\,50~R$_{\odot}$ the CME is slowed by its interaction with the upstream, slow-speed, solar wind flow along its trajectory towards Earth, and this accounts for its arrival time as detected in-situ by the ACE and WIND spacecraft at the L1 point near Earth.}
\label{ENLIL}
\end{figure}

\begin{figure}[!p]
\includegraphics[clip=true, scale=0.7, trim=25 50 0 60]{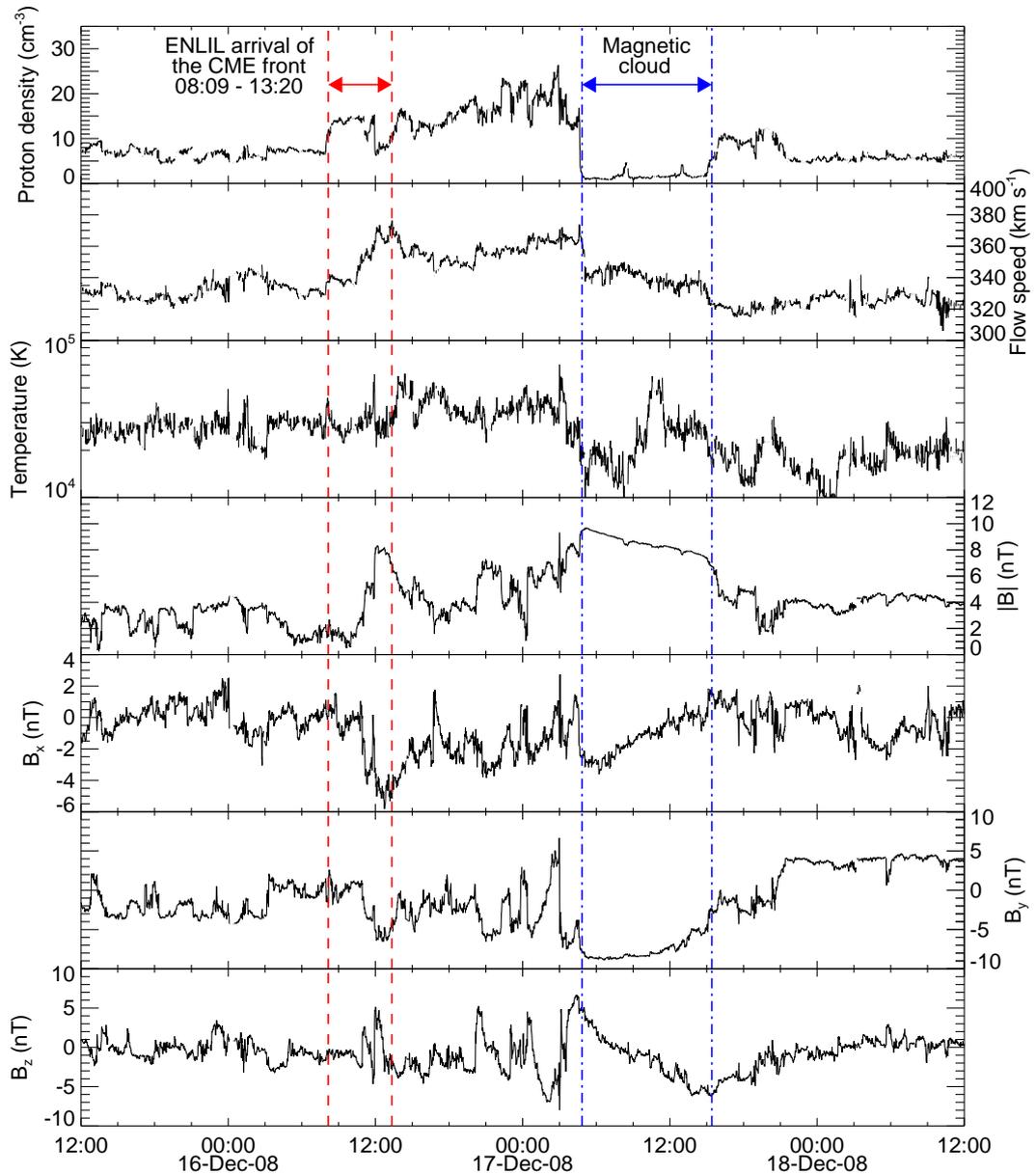}
\caption{The in-situ solar wind plasma and magnetic field data observed by the WIND spacecraft. From top to bottom the panels show proton density, bulk flow speed, proton temperature, and magnetic field strength and components. The red dashed lines indicate the arrival time of the density enhancement predicted from our ENLIL with Cone Model run providing 08:09~UT on 16 December 2008, with a potential offset error between our reconstruction and the derived model height profiles accounting for an arrival time up to 13:20~UT. We observe a magnetic cloud signature behind the front, as highlighted with blue dash-dotted lines.}
\label{wind}
\end{figure}

We use the height, velocity, and width from our 3D reconstruction as initial conditions for the simulation, and find that the CME is actually slowed to $\sim$\,342~km~s$^{-1}$ at 1~AU. This is as a result of its interaction with an upstream, slow-speed, solar wind flow at distances beyond 50~R$_{\odot}$, as seen by inspection of the solar wind profile along the trajectory of the CME in the ENLIL simulation (Figure~\ref{ENLIL}).  This CME velocity is consistent with in-situ measurements of solar wind speed ($\sim$\,330~km~s$^{-1}$) from the ACE and WIND spacecraft at L1. Tracking the peak density of the CME front from the simulation gives an arrival time at L1 of $\sim$\,08:09~UT on 16 December 2008. Accounting for the offset in CME front heights between our 3D reconstruction and ENLIL simulation at distances of $21.5~$R$_{\odot}<r<46~$R$_{\odot}$ gives an arrival time in the range 08:09\,--\,13:20~UT on 16 December 2008. This prediction interval agrees well with the earliest derived arrival times of the CME front plasma pileup ahead of the magnetic cloud flux rope from the in-situ data of both ACE and WIND (Figure~\ref{wind}) before its subsequent impact at Earth \citep{2010ApJ...710L..82L, 2009GeoRL..3608102D}.

\subsection{CME `Pancaking'}

\begin{figure}[!p]
\includegraphics[scale=0.85, clip=true, trim=40 90 0 85]{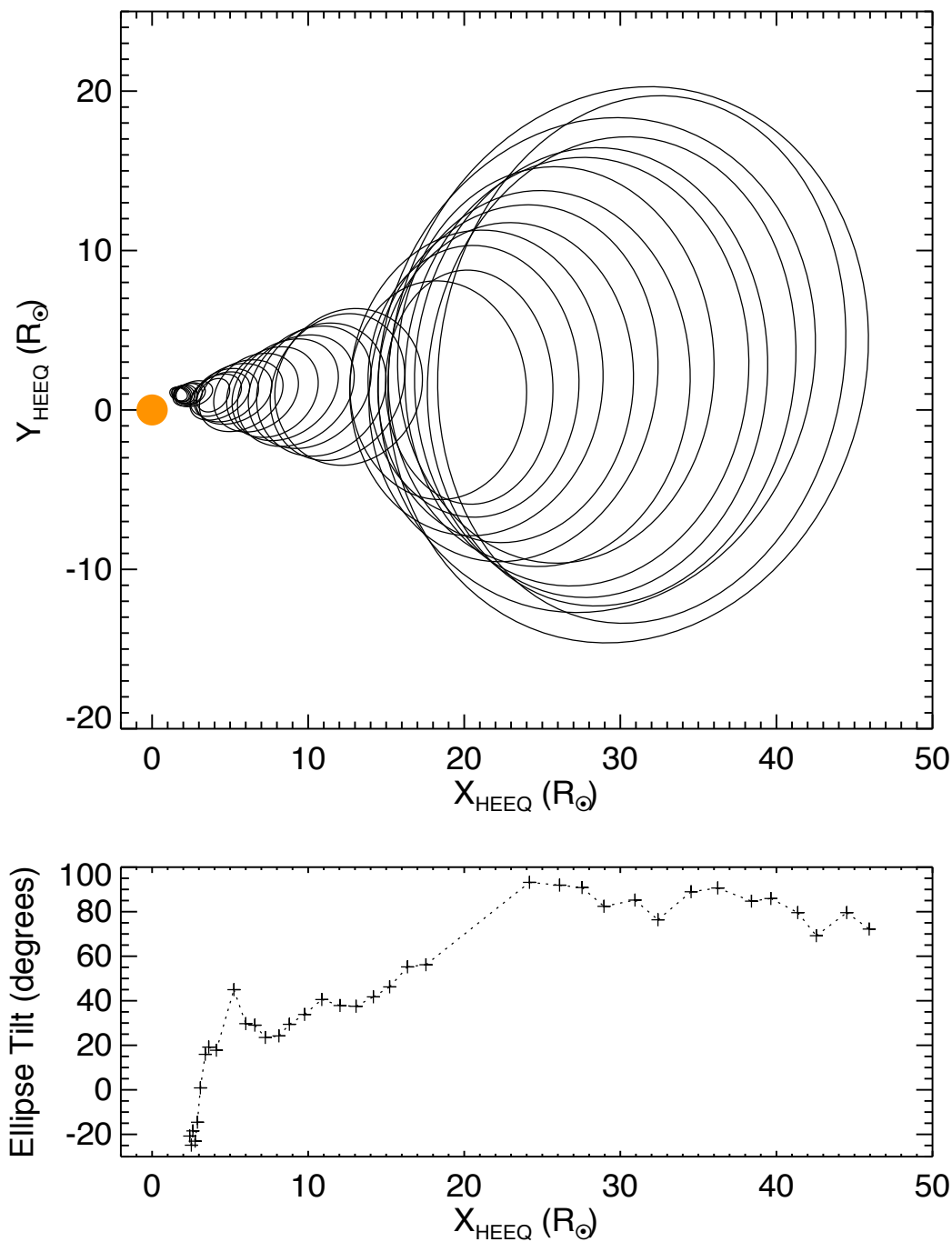}
\caption{Top: Ellipses characterising the 3D CME front reconstruction in the out-of-ecliptic plane along the Sun-Earth line. Bottom: The ellipse tilt angle is indicative of the initial effects of `pancaking', as the curvature of the CME front decreases with increasing height due to its changing morphology in the solar wind.}
\label{curvature}
\end{figure}

From the ENLIL simulation it is apparent that the CME undergoes an effect known as `pancaking' whereby the middle portion of the CME may be slowed while the flanks of the CME maintain or increase speed such that the front distorts to become concave outwards in shape. This is illustrated in the density plot for the cross-section of the CME along the Sun-Earth line in top of Figure~\ref{ENLIL}, and the effect increases with distance from the Sun. We investigate the curvature of the 3D CME front reconstruction along the Sun-Earth line in the distance 2\,--\,46~R$_{\odot}$ by fitting the front in this plane with an ellipse and inspecting the changing morphology with distance. A plot of this characterisation against height is shown in Figure~\ref{curvature} where it can be seen that the curvature of the front decreases as the CME propagates, with the front initially optimised by high-curvature horizontal ellipses, then becoming better optimised by more spherical ellipses, before finally being optimised by smoother vertical ellipses. The observations through the latter half of the HI1 into the HI2 fields-of-view also show this pancaking effect on the CME (though geometrical and instrumental effects must be considered when interpreting these images).

\section{Discussion \& Conclusions}

Since its launch, the dynamic twin-viewpoints of STEREO have enabled studies of the true propagation of CMEs in 3D space. Our new elliptical tie-pointing technique uses the curvature of the CME front as a necessary third constraint on the two viewpoints to build an optimum 3D reconstruction of the front. Here the technique is applied to an Earth-directed CME, to reveal numerous forces at play throughout its propagation.
\newline
\indent The early acceleration phase results from the rapid release of energy when the CME dynamics are dominated by outward magnetic and gas pressure forces. Different models can reproduce the early acceleration profiles of CME observations though it is difficult to distinguish between them with absolute certainty \citep{2008ApJ...674..586S, 2010A&A...516A..44L}. For this event the acceleration phase coincides with a strong angular expansion of the CME in the low corona, which tends toward a constant in the later observed propagation in the solar wind. While, statistically, expansion of CMEs is a common occurrence \citep{1994SSRv...70..215B}, it is difficult to accurately determine the magnitude and rate of expansion across the 2D plane-of-sky images for individual events. Some studies of these single-viewpoint images of CMEs use characterisations such as the cone model \citep{2002JGRA..107.1223Z, 2004JGRA..10903109X, 2005JGRA..11008103X} but assume the angular width to be constant (rigid cone) which is not always true early in the events \citep{2009CEAB...33..115G, 2009A&A...495..325B}. Our 3D front reconstruction overcomes the difficulties in distinguishing expansion from image projection effects, and we show that early in this event there is a non-constant, power-law, angular expansion of the CME. Theoretical models of CME expansion generally reproduce constant radial expansion, based on the suspected magnetic and gas pressure gradients between the erupting flux rope and the ambient corona and solar wind \citep{1999JGR...104..493O, 2003PhRvE..67c6405B, 2000JGR...105.7509C}. To account for the angular expansion of the CME, a combination of internal overpressure relative to external gas and magnetic pressure drop-offs, along with convective evolution of the CME in the diverging solar wind geometry, must be considered \citep{2004ApJ...600.1035R}.
\newline
\indent During this early phase evolution the CME is deflected from a high-latitude source region into a non-radial trajectory as indicated by the changing inclination angle (middle panel of Figure~\ref{3Dkins}). While projection effects again hinder interpretations of CME position angles in single images, statistical studies show that, relative to their source region locations, CMEs have a tendency to deflect toward lower latitudes during solar minimum \citep{2003ApJ...586..562G, 2004JGRA..10907105Y}. It has been suggested that this results from the guiding of CMEs towards the equator by either the magnetic fields emanating from polar coronal holes \citep{2009SoPh..259..143X, angeo-27-4491-2009}, or the flow pattern of the background coronal magnetic field and solar wind/streamer influences \citep{1986JGR....91...31M, 2004A&A...422..307C, 2009SoPh..259..143X}. Other models show that the internal configuration of the erupting flux rope can have an important effect on its propagation through the corona. The orientation of the flux rope, either normal or inverse polarity, will determine where magnetic reconnection is more likely to occur, and therefore change the magnetic configuration of the system to guide the CME either equator- or pole-ward \citep{2005A&A...432..331C}. Alternatively, modelling the filament as a toroidal flux rope located above a mid-latitude polarity inversion line results in non-radial motion and acceleration of the filament, due to the guiding action of the coronal magnetic field on the current motion \citep{2001SoPh..203..119F}. Both of these models have a dependence on the chosen background magnetic field configuration, and so the suspected drawn-out magnetic dipole field of the Sun by the solar wind \citep{1971SoPh...18..258P, 1998A&A...337..940B} may be the dominant factor in deflecting the prominence/CME eruption into this observed non-radial trajectory.
\newline
\indent At larger distances from the Sun ($>$\,7~R$_{\odot}$) the effects of drag become important as the CME velocity approaches that of the solar wind. The interaction between the moving magnetic flux rope and the ambient solar wind has been suggested to play a key role in CME propagation at large distances where the Lorentz driving force and the effects of gravity become negligible \citep{1996JGR...10127499C}. Comparisons of initial CME speeds and in-situ detections of arrival times have shown that velocities converge on the solar wind speed \citep{2009SoPh..256..149M, 2003JGRA..108.1039G}. For this event we find that the drag force is indeed sufficient to accelerate the CME to the solar wind speed, and quantify that the kinematics are consistent with the quadratic regime of aerodynamic drag (turbulent, as opposed to viscous, effects dominate). The importance of drag becomes further apparent through the CME interaction with a slow-speed solar wind stream ahead of it, slowing it to a speed that accounts for the observed arrival time at L1 near Earth. This agrees with the conjecture that Sun-Earth transit time is more closely related to the solar wind speed than the initial CME speed \citep{2009IAUS..257..271V}. Other kinematic studies of this CME through the HI fields-of-view quote velocities of 411\,$\pm$\,23~km~s$^{-1}$ (Ahead) and 417\,$\pm$\,15~km~s$^{-1}$ (Behind) when assumed to have zero acceleration during this late phase of propagation  \citep{2009GeoRL..3608102D}, or an average of 363\,$\pm$\,43~km~s$^{-1}$ when triangulated in time-elongation J-maps \citep{2010ApJ...710L..82L}. These speeds through the HI fields-of-view, lower than those quantified through the COR1/2 fields-of-view, agree somewhat with the deceleration of the CME to match the slow-speed solar wind ahead of it in our MHD simulation. Ultimately we are able to predict a more accurate arrival time of the CME front at L1.
\newline
\indent A cohesive physical picture for how the CME erupts, propagates, and expands in the solar atmosphere remains to be fully developed and understood from a theoretical perspective. Realistic MHD models of the Sun's global magnetic field and solar wind are required to explain all processes at play, along with a need for adequate models of the complex flux rope geometries within CMEs. Additionally, spectral observations of CME onset signatures using such instrumentation as the Coronal Diagnostic Spectrometer \citep[CDS;][]{1995SoPh..162..233H} on SOHO and EUV Imaging Spectrometer \citep[EIS;][]{2007SoPh..243...19C} on Hinode, along with future space exploration missions such as Solar Orbiter \citep[ESA;][]{2009SolarOrbiter} and Solar Probe$+$ \citep[NASA;][]{2008SolarProbePlus}, will be required to give us a better understanding of the fundamental plasma processes responsible for driving CMEs and determining their adverse effects at Earth. 

%\subsection{Description of Power Law Polynomial Fits?!?!}

%The fits to the derived kinematics and morphological profiles of Figure~\ref{3Dkins} are presented as Marquardt-Levenberg least squares fits with the power law form $f(x)=ax^b$. In the simplest model the values of $x_i$ and $y_i$ would be expected to be normally distributed about $x$ and $y$, so our investigation begins with fitting the profiles by linear, then quadratic, and finally a variable power polynomial. It was the power law which resulted in the minimum chi-squared value and deemed optimum for representing the observed trends. But $\log(x_i)$ and $\log(y_i)$ are not necessarily normally distributed about $x$ and $y$, so the log space was investigated further to observe how the data points behave (Figure

%\begin{figure}[!t]
%\includegraphics[clip=true, scale=0.75, trim=0 0 0 345]{kins_powerlaws_occulter_prom.pdf}
%\caption{}
%\label{kins_powerlaws}
%\end{figure}

\chapter{Conclusions \& Future Work}
\label{chapter:futurework}

This thesis has sought to increase our understanding of solar activity and its effects on Earth through a study of the phenomenon known as CMEs. This was undertaken by studying CME propagation with new data and techniques. This chapter presents the main results and conclusions, and outlines possible future directions for this work.

\section{Principal Results}

The primary objective of this study was to further our understanding of the kinematic and morphological evolution of CMEs as they propagate from the Sun into the heliosphere. This was done by applying new methods of multiscale image processing to CME observations in order to identify and track the CME front and characterise its propagation through coronagraph data. Following this, the development of a new elliptical tie-pointing technique applied to STEREO observations allowed a reconstruction of a CME front in 3D, overcoming projection effects and revealing its true 3D motion. The principal results arising from these studies may be summarised as follows:

\begin{enumerate}

\item The multiscale nature of CMEs was revealed through the application of a high- and low-pass image filtering technique \citep{2008SoPh..248..457Y}. The multiscale filtering was shown to effectively suppress noise and small-scale features in order to reveal CME structure on a scale that best identifies the CME front. The specific implementation of multiscale filtering introduced by \citet{2008SoPh..248..457Y} also allowed the chaining of pixels along edges in the decomposed images to reveal the CME front for tracking through time. Such an algorithm provides a robustness in CME front detection that alleviates issues of subjective user biases and unreproducibility of results. The technique is also more accurate than running-difference techniques which are widely used for determining CME heights, both spatially since it requires no arbitrary scaling and/or thresholding to find the edges, and temporally as it operates on individual images without a need for subtracting antecedent frames. The technique was also extended for use as a potentially automated CME front detection algorithm, discussed further in Section~\ref{sect:automation} below.

\item An ellipse characterisation of the CME front in coronagraph data was implemented and shown to be an effective method for retrieving information on the changing CME morphology through sequences of observations. An ellipse was chosen for the innate freedom in its parameters, having the ability to best fit the varying curvature of CME fronts across different image sequences, while still being constrained to close upon itself. The characterisation provided a robust method for obtaining CME heights since it is not affected by deviations along the often ill-defined and/or kinked CME front. The opening cone angle to the ellipse also provided a measure of angular width for testing CME expansion, and the eccentricity of the ellipse provided information on the changing CME front curvature as it was observed to propagate across the plane-of-sky.

\item The degree of accuracy provided by the multiscale methods and characterisation of the CME front allowed a test of the constant acceleration model upon a variety of CMEs, and it was found not to be true of all cases. This has implications for how CMEs may be modelled theoretically since the forces acting must have different regimes of dominance to cause non-constant acceleration in certain events. This warrants further investigation, especially if projection effects can be corrected for using STEREO data, and the kinematics determined with the best available accuracy (see the discussion of Sections~\ref{sect:3dcmecandidates} and \ref{sect:derivingkins} below).

\item The accuracy of these methods further revealed an early acceleration phase for some CMEs, and those with high speeds tended to reach greater angular widths, indicative of an initially dominant outward driving force ramping up the CME speed and expansion within the first few solar radii. Testing for this outward force that drives the CME acceleration and causes it to expand, is vitally important for comparing with theory in an effort to understand the interplay of forces acting on the CME as it propagates through the corona and heliosphere.

\item The newly developed elliptical tie-pointing technique was shown to overcome the plane-of-sky projection effects of previous single vantage-point observations in a manner better suited to studying the CME of 12 December 2008 than previous stereoscopic efforts. By using the characterised curvature of the CME front in the observations from the twin-viewpoints of STEREO it was possible to constrain a reconstruction of the true CME front between the two planes-of-sky. This revealed the CME's true 3D motion, thus greatly increasing the accuracy with which we can interpret its kinematics and morphology.

\item The acceleration phase of the prominence eruption and CME was determined in 3D to occur within the first 3~R$_{\odot}$, with the speed of the outer CME front being higher than the speed of the prominence that forms the inner core material of the CME. Their different speeds are indicative of the CME expansion, and indeed the early phase acceleration corresponds to the event's strongest characterised (out-of-ecliptic) angular width increase. This is again indicative of an initially dominant outward driving force that causes the CME speed and width to increase dramatically (generally considered to be the Lorentz magnetic force of CME initiation and propagation).

\item The deflected motion of the prominence eruption and CME was quantified in 3D; originating at high solar latitudes of $\sim$\,55$^{\circ}$ but determined to move on a trajectory along the ecliptic at $\sim$\,0$^{\circ}$. This is indicative of the drawn-out background magnetic field of the quiet Sun. The pressure of the solar wind acts to drag out field lines in the $\beta>1$ coronal regime, influencing the overall magnetic field configuration right down to the surface of the quiet Sun. At these low heights the magnetic pressure of the drawn-out field guides the motion of the gradual eruption into a non-radial trajectory, placing it on a course towards Earth. This highlights the importance of the overall solar magnetic field with respect to the CME and its source region, and reveals how necessary it is to understand their interdependence, both in the context of CME physics and also space weather monitoring.

\item The effects of drag on the 3D CME motion were investigated and it was found that, subsequent to the early acceleration phase of the CME, its speed continued to increase towards the speed of the ambient solar wind. A comparison of the CME velocity with the in-situ detection of its arrival at L1 implied the CME had to be further slowed along its trajectory, revealed to be true when investigated with the 3D MHD `ENLIL with Cone' Model. This highlights the importance of drag on the CME throughout its propagation as it was found to undergo further deceleration by the slow-speed solar wind stream ahead of it as it propagated through interplanetary space. Understanding these drag effects is thus of great importance for predicting CME arrival times at Earth and improving space weather forecasts.%, as shown for the CME of 12 December 2008. (Sections~\ref{sect:cmedrag} and \ref{sect:cmearrival}).

\end{enumerate}

\section{Future Work}

The methods developed and implemented in this thesis are a first use of multiscale analysis and characterisation in obtaining CME kinematics and morphology with better constrained errors than previous efforts, of benefit in testing theoretical CME models and in forecasting space weather at Earth or elsewhere in the heliosphere. The possibility for automation has been demonstrated and could provide a new catalogue of CMEs working in realtime with SolarMonitor.org for example. Extending these multiscale techniques to curvelets or ridgelets may better suit the detection of the typically curved nature of CMEs in coronagraph data. Furthermore, the data from the twin viewpoints of the STEREO mission have allowed us to overcome plane-of-sky projection effects in studying CMEs with our newly developed elliptical tie-pointing technique, resulting in a very cohesive description of how the CME of 12 December 2008 propagates from the Sun to the Earth. There are numerous candidate events which should also be studied in this manner to test the conclusions drawn from the results of the CME studied here and gain further insight into CME dynamics. Other methods for deriving the velocity and acceleration profiles of CMEs also warrant investigation, since the 3-point Lagrangian interpolation is sensitive to scatter in the data (though less so than the standard forward/reverse difference) and bootstrapping or inversion techniques, for example, may help overcome this.

\subsection{Automation}
\label{sect:automation}

\begin{figure}[!p]
\centerline{\includegraphics[width=\linewidth]{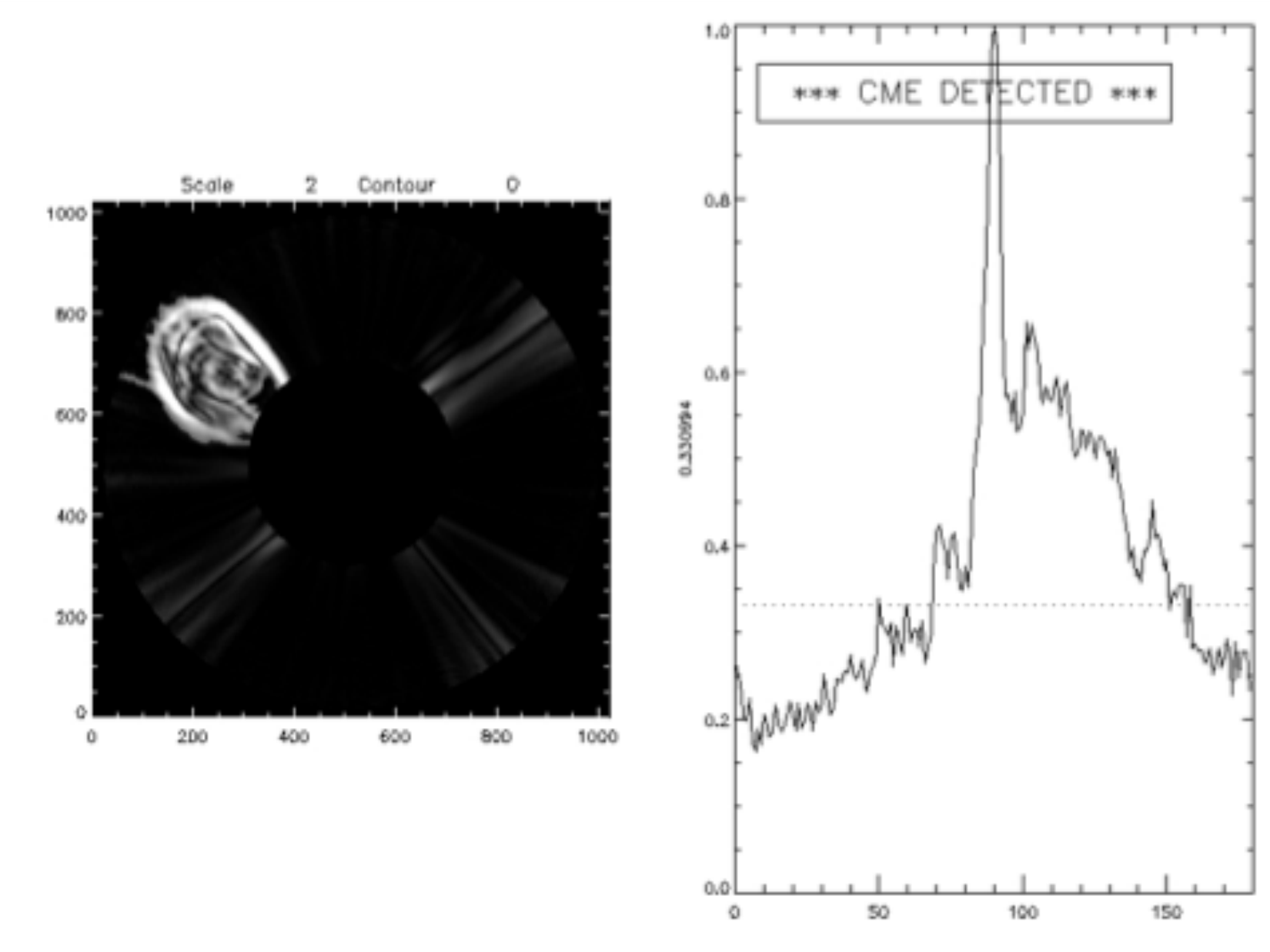}}
\caption{An example interface of the automated multiscale detection algorithm based upon thresholding the magnitude and angular information, discussed in Section~\ref{automatedmultiscale}. {\it Left:} The algorithm can be made to cycle through different scales, and threshold any number of contoured regions of magnitude (edge strength). {\it Right:} Contours that contain a wide distribution of angles signal a CME detection. The angular information is normalised to 1 and folded into a 0\,--\,180$^{\circ}$ range due to symmetry of the edge normals. A threshold on the normalised angular distribution is specified to flag regions as CMEs or otherwise (e.g., $>$\,20\%). A detection mask is then built through multiple scales. The limitations to be overcome for a robust automation of this technique include developing dynamic thresholds such that multiple contours of CME edges are not fragmented, and increased angular distributions due to streamer deflections are not mistakenly labelled as CMEs (a non-trivial task). Moreover, an automated pixel chaining of the CME edges must be included in order to produce an ellipse characterisation of the CME front in the image.}
\label{automated}
\end{figure}

\begin{figure}[!p]
\centerline{\includegraphics[width=\linewidth]{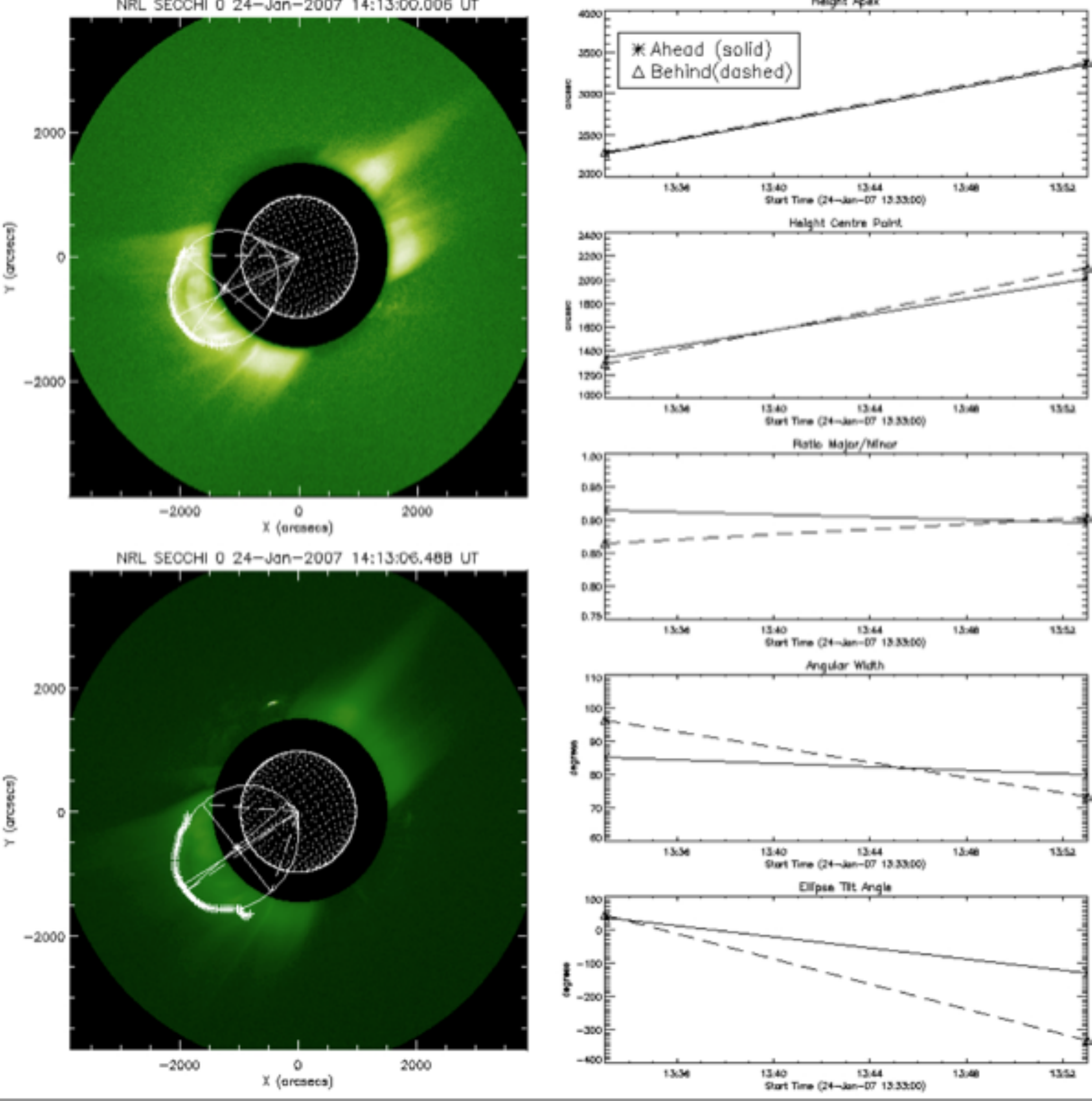}}
\caption{An example interface which could be developed for the potentially automated multiscale filtering and ellipse tracking of CMEs observed by STEREO-A and B. Images from the respective instruments appear on the left, and the resulting parameters from the characterisation of the CME front appear on the right. Any manner of information may be chosen for display, and clocked through time as the CME progresses.}
\label{ellipse_graphs}
\end{figure}

As discussed in Chapter~\ref{chapter:multiscale} there is great benefit in implementing an automated CME detection and tracking algorithm for cataloguing their kinematics and similar properties of interest in large data sets. The algorithms (specifically CACTus, SEEDS and ARTEMIS) remove the need for a manual inspection of the images (as performed in the CDAW catalogue), which can be both laborious and suffer from user-specific biases. The automated techniques also produce a robust output of parameters for large statistical analyses of CME properties over long periods of solar activity, since the thresholds for detection and height/width measurements are hard-coded into the algorithm. However, the ultimate aim is to determine the kinematics and morphology of CMEs with as great an accuracy as possible with the available data, thus operations such as image rebinning, smoothing, and differencing are not ideal. Also, accurately measuring the CME height (and subsequently deriving the velocity and acceleration) can be difficult since the CME front is often diffuse and ill-defined.
\begin{figure}[!p]
\centerline{\includegraphics[scale=0.5]{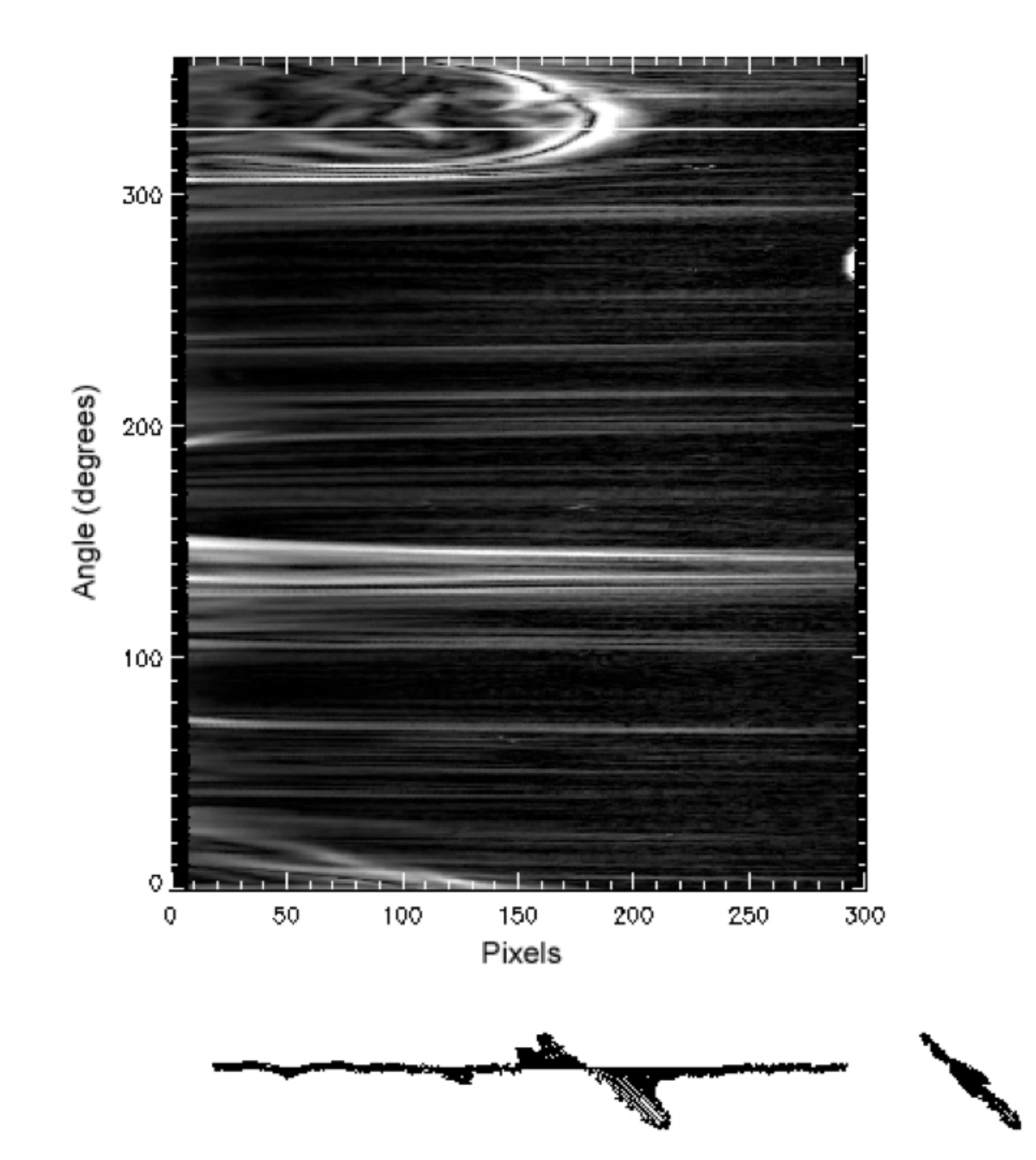}}
\caption{The magnitude (edge strength) from the multiscale filtering of a CME observation, unwrapped into polar coordinates $(r,\theta)$ with axis units of pixels and degrees, respectively. A scanning-line runs over the image to produce the scatter of edge normals at the bottom of the image, where each normal is plotted with a vector magnitude and associated angular information from the multiscale filter (cf. the edge normals of Figure~\ref{Vectors}). An end-projection of the normals along the scanning-line is plotted in the bottom right of the image. The scanning-line is located at angle 328$^{\circ}$, passing over a CME in the image. The resulting edge normals show a slice through the angular distribution of the CME, and as the scanning-line moves along the image, the end-projection shows a continuous rotation of angles due to the curvature of the CME front. Detecting this continuous rotation of angles may be used to distinguish CMEs from streamers which show an abrupt, stepwise change in angle as the scanning-line crosses over them. This may alleviate a need for the intensity thresholding and contouring which can fail to detect faint CMEs or portions thereof.}
\label{unwrap_scan}
\end{figure}
\newline
\indent These issues motivated the application of multiscale filtering to enhance the CME front in coronagraph images without reducing the quality of the image since noise and small-scale features can be removed in the multiscale decomposition for optimum CME detection. The multiscale technique outlined in Chapter~\ref{chapter:multiscale} also results in an edge detection on the image akin to the Canny edge detector as a result of the horizontal and vertical directions in which the multiscale filter is applied. Regions with intensities above a specific threshold in the image are contoured as possible CME candidates, though this often includes streamers since they appear on the same scales as CMEs. CME detection is then shown to be effective through the thresholding of angular spreads across the contoured regions of the image, since CMEs will have a large distribution of edge normals compared with the generally radial nature of coronal structures such as streamers (Figure~\ref{automated}). Ideally the thresholds should be made dynamic so as to better detect the varying CME intensities and minimise the fragmentation of intensity contoured regions of interest. Too strong a hard threshold will neglect faint CMEs or parts thereof. The algorithm also currently lacks a satisfactory technique for discarding the regions of an image that do not correspond to the detected CME and thus still requires a user to perform an inspection of the edges to be maintained/discarded in the images. Including multiple scales of the decomposition, rather than just the one with the best signal-to-noise ratio for the CME, allows a scoring system to refine the CME detection mask, although cases of streamer interaction or deflection are prone to skew the resulting detection. The algorithm needs to be improved with a form of image segmentation included to make the CME detection masks more robust, and minimise the effects of streamers. Then the ellipse characterisation can be automatically applied for studying the CME front kinematics and morphology with greater accuracy than traditional image processing techniques, and produce a multiscale-methods based catalogue of events (e.g., Figure~\ref{ellipse_graphs}).
\newline
\indent In an effort to overcome the hard-thresholding biases of the intensity contouring of CMEs, a detection based solely on the angular distribution may be considered in future work. First a multiscale filtered coronagraph image, like that of Figure~\ref{modalpmapcontour}, is unwrapped into polar coordinates $(x,\,y) \rightarrow (r,\, \theta)$. Then the image is scanned with regard to the combined magnitude and angular information of the multiscale filtering process, resulting in a dynamic change of vector arrows (edge normals), as shown in Figure~\ref{unwrap_scan}. When scanning over the quiet corona or streamers, their radial nature means the angles along the scanning-line will all predominantly point in one direction and undergo abrupt, stepwise changes. As the scanning-line moves across a CME the angles will change in a more linear, continuous fashion. When such a continuous angular change is measured, the region under the scanning-line can be flagged as a CME detection.

\subsection{Ridgelets/Curvelets}

The implementation of multiscale analysis has been demonstrated for its efficacy in highlighting CMEs and coronal structure against noise and small-scale features in coronagraph images \citep{2003A&A...398.1185S, 2008ApJ...674.1201S, 2008SoPh..248..457Y, 2009A&A...495..325B}. However, wavelets are better suited to identifying point-like features in images, but their extension to ridgelets and curvelets has been shown to better resolve the visibility of the curved form of a typical CME front \citep{2010gallagher}. Thus their development may provide a more reliable CME detection algorithm than the multiscale technique investigated in this work.

\begin{figure}[!t]
\centerline{\includegraphics[scale=0.3]{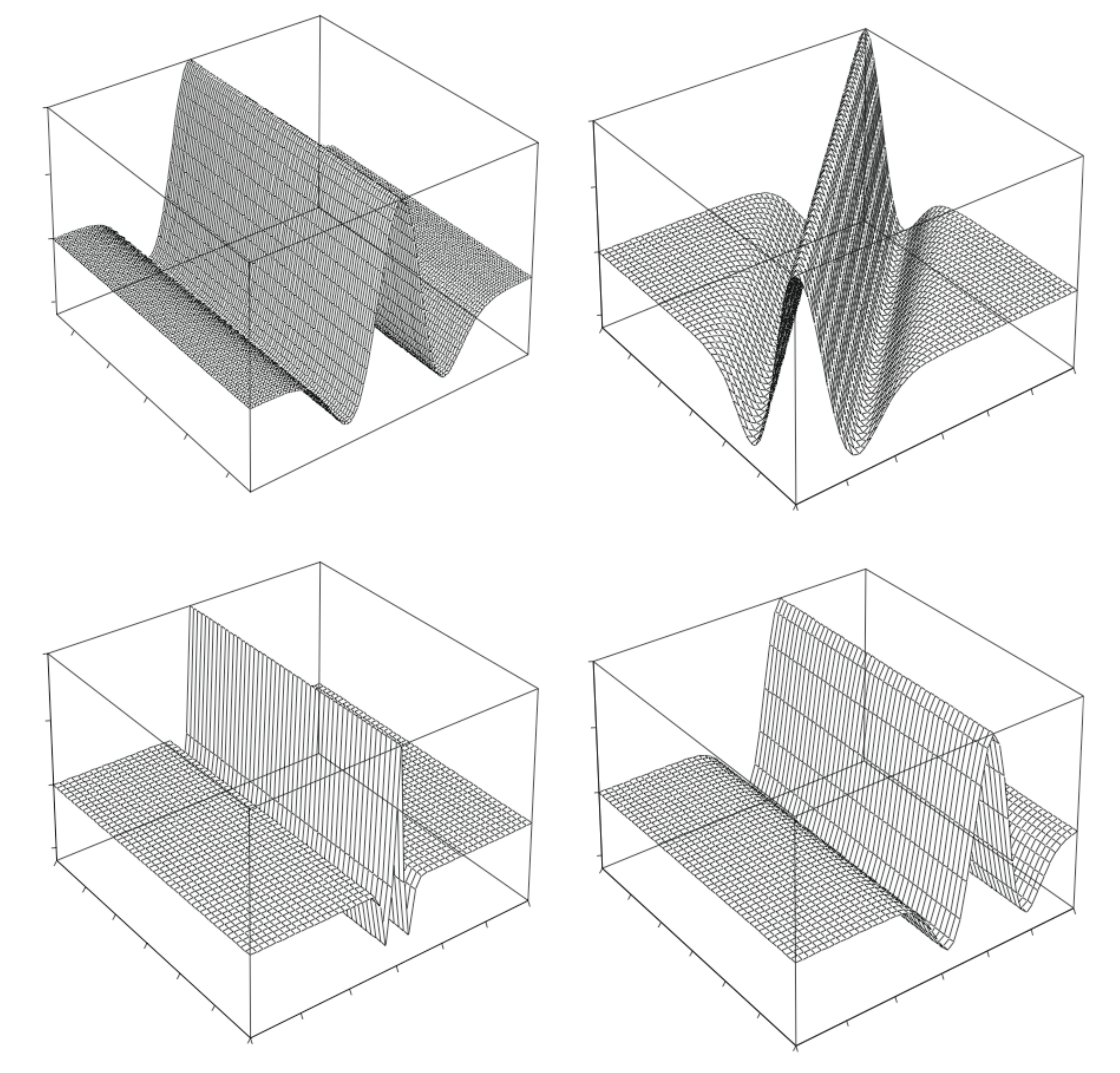}}
\caption{An example ridgelet, reproduced from \citet{2010gallagher}. The first graph shows a typical ridgelet, and the second to fourth graphs are obtained from simple geometrical manipulations, namely rotation, rescaling and shifting.}
\label{ridgelets}
\end{figure}
\begin{figure}[!t]
\centerline{\includegraphics[width=\linewidth]{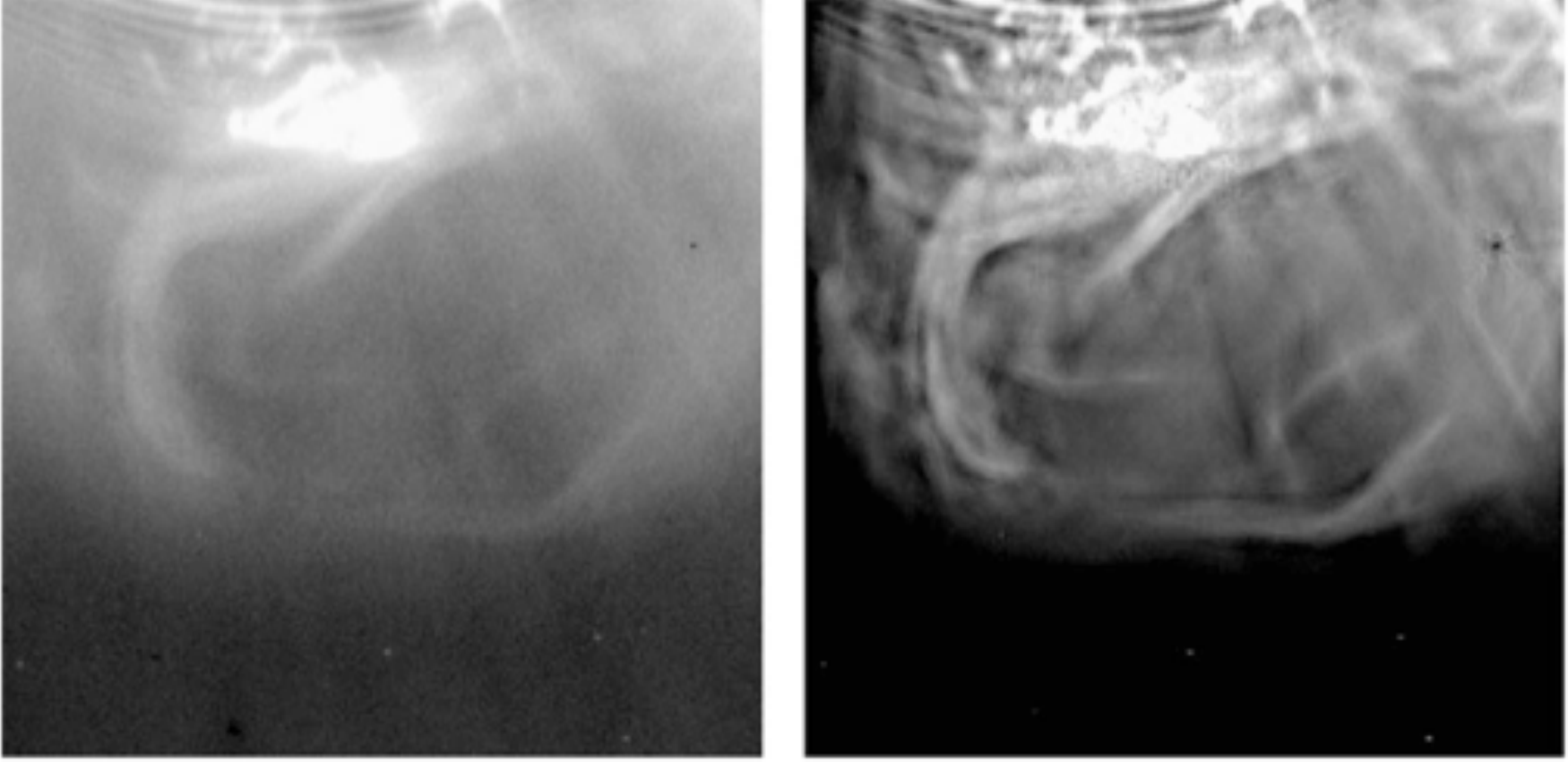}}
\caption{The curvelet filtered image of a CME observed by LASCO/C2 on 18 April 2000, reproduced from \citet{2010gallagher}. The detail along the curved CME front is enhanced as a result of the curvelet technique following the removal of certain coefficients probably due to noise.}
\label{curvelet_cme}
\end{figure}
The continuous wavelet transform of an image may be defined as:
\begin{equation}
w \left( s, a, b \right) \; = \; \int{\int{ I \left(x, y \right) \psi_{s, a, b} \left( x, y \right) dx}\, dy}
\end{equation}
where $w(s,a,b)$ are the wavelet coefficients of the image $I(x,y)$, $\psi_{s}(x,y)$ is the mother wavelet, and $s$ is the term describing scale at a position $(a, b)$. The mother wavelet can take many forms, e.g., the Morlet wavelet or the Mexican hat wavelet. The ridgelet transform takes a similar mathematical form to the wavelet transform, i.e., it is a convolution of an image with a predefined basis function, but they are anisotropic and thus more directionally sensitive (Figure~\ref{ridgelets}). The ridgelet uses a radon transform that transforms lines into points, upon which a wavelet transform may be applied to provide a sparse representation of the points. The basis function of the ridgelet takes the form:
\begin{equation}
\psi_{s,b,\theta}\left(x,y\right)\;=\;s^{-1/2}\psi\left(\frac{x\cos\theta+y\sin\theta-b}{s}\right)
\end{equation}
The ridgelet is constant along lines $x\cos\theta+y\sin\theta=const.$ and the ridgelet coefficients are given by the convolution:
\begin{equation}
R_{I}\left(s,b,\theta\right)\;=\; \int{ \int{ I\left(x,y\right)\psi_{s,b,\theta}\left(x,y\right)dx}\,dy}
\end{equation}
Curvelets generalise the idea of ridgelets to multiscale curves in images. The detailed maths on the application of curvelets to CME images is discussed in \citet{2010gallagher} based on the developments by \citet{1999candes}. % are defined at scale 2$^{-j}$, orientation $l$ and position $x_k^{j,l}=R_{\theta_l}^{-1}(2^{-j}k_1,2^{-j/2}k_2)$ by translation and rotation of a mother curvelet $\phi_j$ as:
%\begin{equation}
%\phi_{j,l,k}(x) \;=\; \phi_j \left( R_{\theta_l} \left( x-x_k^{j,l} \right) \right)
%\end{equation}
%where $R_{\theta_l}$ is the rotation by $\theta_l$ radians. $\theta_l$ is the equi-spaced sequence of rotation angles $\theta_l=2\pi2^{-j/2}l$, with integer $l$ such that $0 \leq \theta_l \leq 2\pi$ (the number of orientations varies as $1/\sqrt{scale}$. ${\bf k}=(k_1,k_2)$ is the sequence of translation parameters. The waveform $\phi_j$ is defined by means of its Fourier transform $\hat{\phi}_j(\nu)$ written in polar coordinates in the Fourier domain:
%\begin{equation}
%\hat{\phi}_j(r,\theta) \;=\; 2^{-3j/4}\hat{w}\left(2^{-j}r\right)\hat{\nu}\left(\frac{2^{j/2}\theta}{2\pi}\right)
%\end{equation}
%In continuous frequency the curvelet coefficients of $I(x)$ are defined as the inner product:
%\begin{equation}
%c(j,l,k)\;=\; \int{ \int{ \hat{I}(\nu)\hat{\phi}_j\left(R_{\theta_l}\nu\right)e^{ix_k^{j,l}\cdot \nu} d\nu}}
%\end{equation}
Figure~\ref{curvelet_cme} shows how well the curvelet filter performs on an image of a CME. It enhances the structure within the curved CME front while reducing the noise and small-scale intensity features elsewhere in the image. This may provide a more accurate method of determining the CME front location in individual images, rather than relying on differencing techniques which can be prone to scaling errors and spatio-temporal cross-talk (discussed in Section~\ref{multiscalekinsintro}). 

\subsection{3D CME Reconstruction}
\label{sect:3dcmecandidates}

\begin{figure}[!t]
\centerline{\includegraphics[scale=0.8]{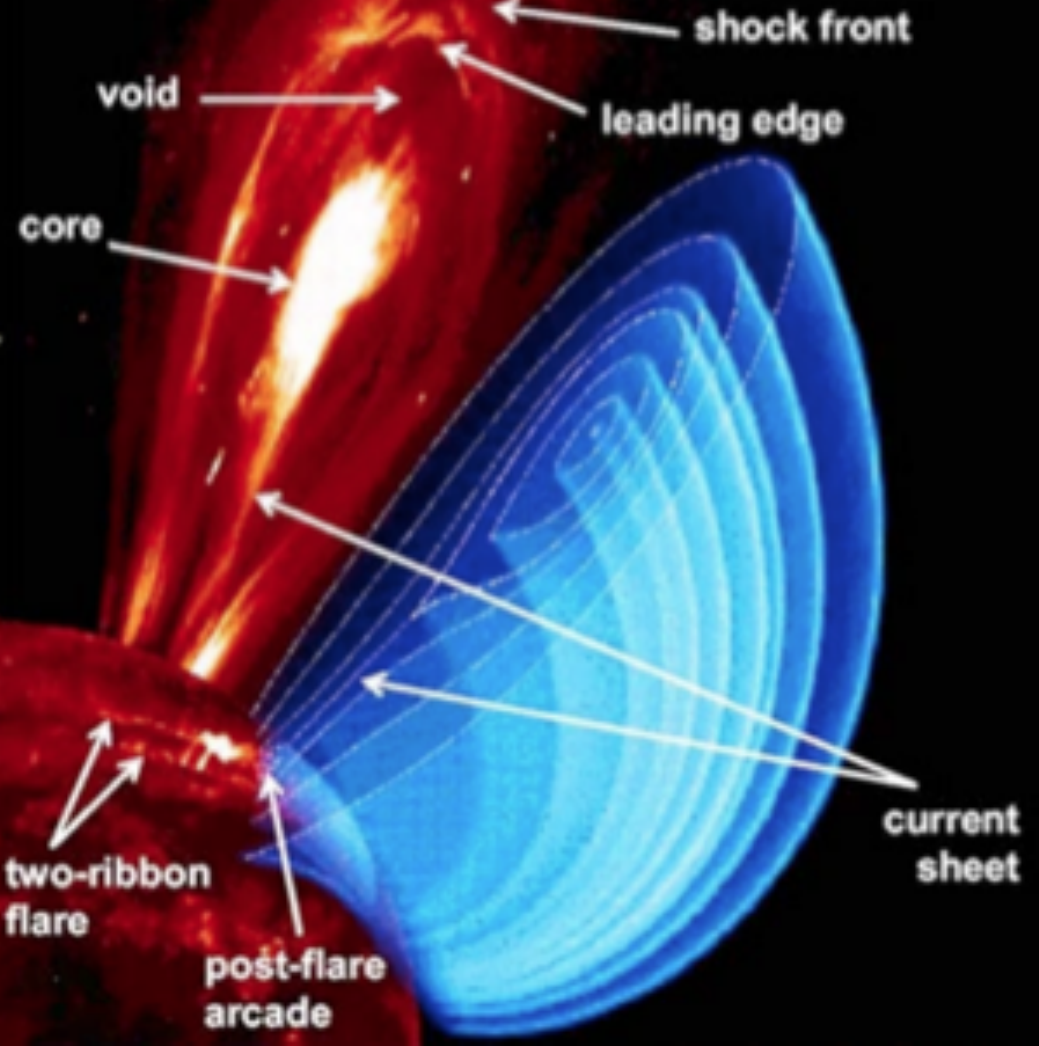}}
\caption{A schematic of the possible 3D magnetic field topology of a CME, as inferred from typical observations of a leading edge, dark void and bright core. There may also be an associated on-disk post-flare arcade at the base of the postulated current sheet. 3D reconstructions of CMEs are important for testing the validity of theoretical models.
\newline
{\it Image credit: http://www.nrl.navy.mil}}
\label{cme_3dcartoon}
\end{figure}

The development of the elliptical tie-pointing technique has proven effective at reconstructing the CME front of the 12 December 2008 event, discussed in detail in Chapter~\ref{chapter:3D}. Such studies are necessary for obtaining the true kinematics and morphology of CMEs and minimising the uncertainties in their 3D quantification, of benefit in comparing observations with theoretical models (Figure~\ref{cme_3dcartoon}). This event was an ideal case for study since the STEREO spacecraft were at an angular separation of almost 90$^{\circ}$, so the lines-of-sight intersected to form optimum quadrilaterals localising the CME front. Very small, or very large, angles of separation would not be as effective since the quadrilaterals will be skewed and so too will the corresponding inscribed ellipses. So while a reconstruction would match the observations of both spacecraft it may not perfectly represent the true curvature of the CME front (although it will still offer a better approximation than tie-pointing alone).
\newline
\indent The fact that the 12 December 2008 CME propagated along the Sun-Earth line midway between the two spacecraft meant the reconstruction could be performed out to heights of almost 50~R$_{\odot}$ before the `pancaking' of the CME (along with concerns on the scattering geometry and instrumental effects) meant the ellipse fit was no longer appropriate. If we can correct for these effects, or include them in the uncertainty, then it is possible to use more than one ellipse fit to characterise the different portions of the CME front and perform elliptical tie-pointing on these individually. This could provide an insight into the 3D propagation through the rest of the HI1 field-of-view. It would also be interesting to test how the method fares with the observations of HI2.
\newline
\indent A number of other CMEs will warrant studying with the elliptical tie-pointing technique throughout the lifetime of the STEREO mission. Many events have been observed through the COR1 and COR2 fields of view and been studied by several authors through a variety of stereoscopic methods \citep{2009SoPh..256...57L, 2009SoPh..259..213S, 2009SoPh..259..163W, 2008SoPh..252..385M, 2009SoPh..256..131B, 2009SoPh..256..167D, 2009SoPh..256..183T}. Direct comparisons may be made with the results of these studies and future stereoscopic CME analyses as the spacecraft move into orbits on the far side of the Sun, approaching quadrature again as the Sun becomes more active through the rise of solar cycle 24. In conjunction with further 3D CME reconstruction analyses, it will also be useful to perform a simulation of different model CME structures, with varying signal-to-noise ratios, cadences, image compressions, etc., and at different spacecraft separation angles. Such simulations will provide a method for testing the inherent uncertainties associated with the different events being reconstructed at different stages of the STEREO mission. Furthermore, as discussed in Section~\ref{sohothirdeye}, the observations from SOHO/LASCO can also be used as a third perspective on the events.
\begin{figure}[!t]
\centerline{\includegraphics[scale=0.4]{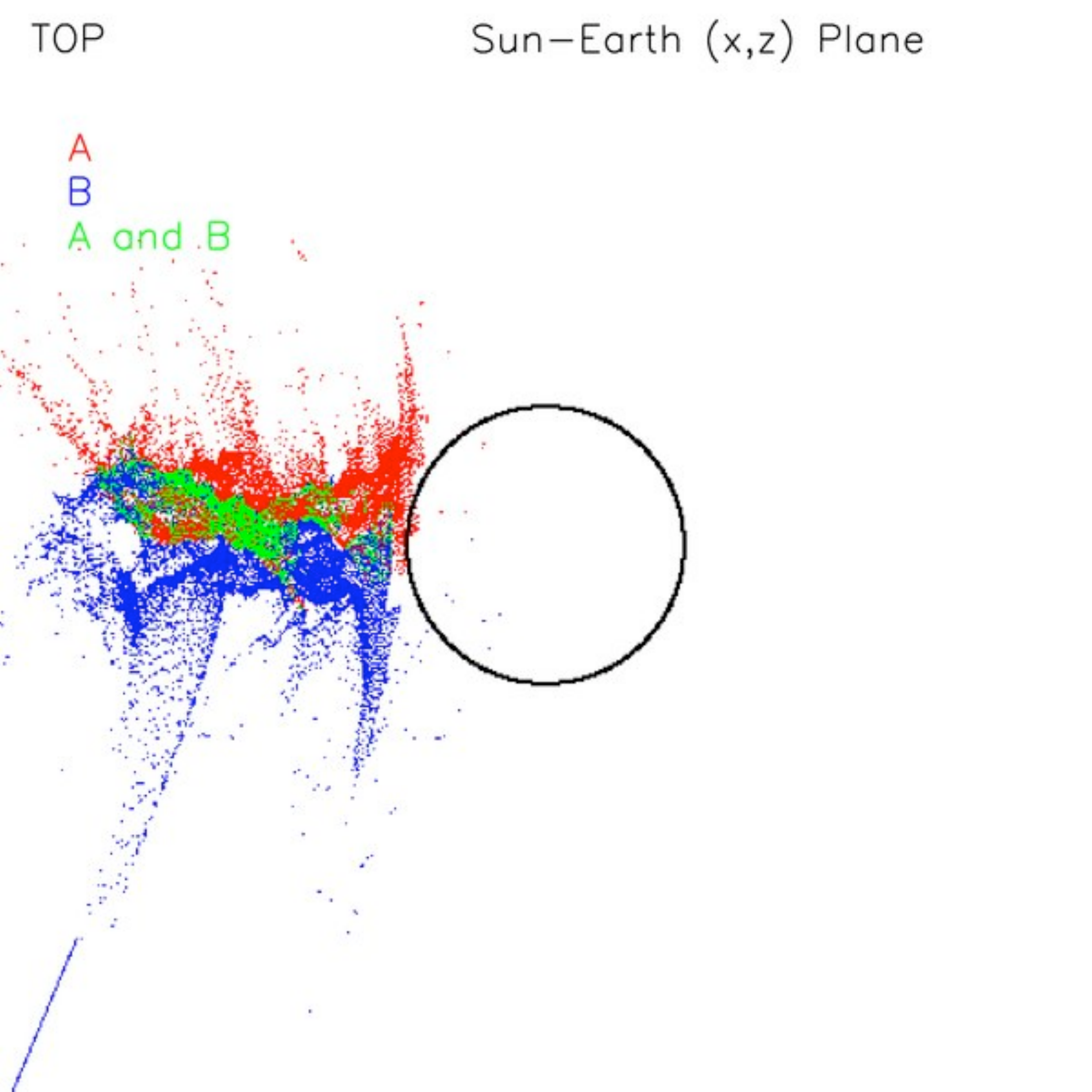}}
\caption{`Top' view of the polarimetric reconstruction method applied to both STEREO-A and B observations of a CME at 01:30~UT on the 31 December 2007 (rotated to the Sun-Earth frame), reproduced from \citet{2010ApJ...712..453M}. The red points are the reconstruction from STEREO-A, and the blue points from STEREO-B, with the green points showing the regions of overlap.}
\label{moran_3d}
\end{figure}
\newline
\indent In particular, the polarisation technique of \citet{2004Sci...305...66M} could be used in conjunction with the elliptical tie-pointing reconstruction, to potentially reveal detail of the structure of the CME behind the front. The technique relies upon the geometric dependence of the polarisation of Thomson-scattered light, whereby the polarisation fraction in CME emission provides a line-of-sight averaged `mean distance to the plane of the sky' for selected, or all, CME pixels. The validity of the method was tested and proven using STEREO observations of two CMEs and shown to be in good agreement with other triangulation methods \citep{2010ApJ...712..453M}. A combined use of the technique with the geometric localisation on COR2 beacon data has been explored by \citet{2009SPD....40.1609D}, although the polarisation technique becomes unreliable at heights $\gtrsim$\,5~R$_{\odot}$ \citep{1966gtsc.book.....B}. An example of how the technique applies to STEREO data is shown in Figure~\ref{moran_3d} for a CME on 31 December 2007. It can be seen how the final polarimetric reconstruction could be combined with the 3D CME front reconstruction for a more cohesive rendering of the overall CME structure.
\newline
\indent Due to the nature of the STEREO mission, telemetry limitations are an important point to note as the spacecraft separate, specifically regarding image degradation and the consequences for CME studies. The overall mission is split into three phases: early operations, prime science phase, and extended mission phase. During the early operations phase, when the spacecraft were still close to the Earth, telemetry rates were restricted compared to the rates of the prime science mission, being 30~kbps in the first two weeks. Once the spacecraft were in heliocentric orbits the prime science phase began and nominal telemetry rates of 720~kbps were employed. As the spacecraft separate, the data rate is diminished, going down to 160~kbps in early 2011, and 120~kbps by 2013. In the case of the SECCHI suite, progressive telemetry reductions are introduced through increased image binning and compression, reduced image cadences, and prioritisation of total brightness coronagraph images, as per the STEREO Science Operations Plan\footnote{http://stereo-ssc.nascom.nasa.gov/publications.shtml}. With regards the elliptical tie-pointing technique, the uncertainties involved in identifying and tracking CMEs will increase due to the image binning and reduced cadences. When the spacecraft are in near-quadrature (the optimum positioning for the technique) on the far side of the Sun again in late 2012 to early 2013 the low-rate telemetry stream will limit the precision with which the true kinematics of future events may be derived.

%\begin{description}
%\item EUVI: \newline Increase compression for majority of images in highest-cadence wavelength. \newline Reduce image cadence to 2 hours in other wavelengths ($\times$2.7 combined telemetry reduction).
%\item COR1: \newline Bin images to 512$\times$512 pixels, and increase compression ($\times$3 telemetry reduction). \newline Sum images onboard and send down a single 512$\times$512 total B image, instead of three polarized images ($\times$3 telemetry reduction). \newline Reduce image cadence.
%\item COR2: \newline Bin images to 1024$\times$1024 pixels, and increase compression ($\times$3 telemetry reduction). \newline Change all polarisation sequences to total B images (currently interleave pB and B) ($\times$2.25 telemetry reduction). \newline Reduce image cadence.
%\item HI: \newline Reduce cadence of high resolution calibration images from once per day to once per week. \newline Change HI1 image cadence from 40 minutes to 60 minutes. \newline Use subfield masks to send down only a portion of each image. \newline Change HI2 image cadence from 2 hours to 3 hours. \newline Bin images to 512$\times$512 pixels.
%\end{description}

\subsection{Deriving CME Kinematics}
\label{sect:derivingkins}

\begin{figure}[!t]
\centering
\subfigure{\includegraphics[scale=0.43, trim=110 50 70 70]{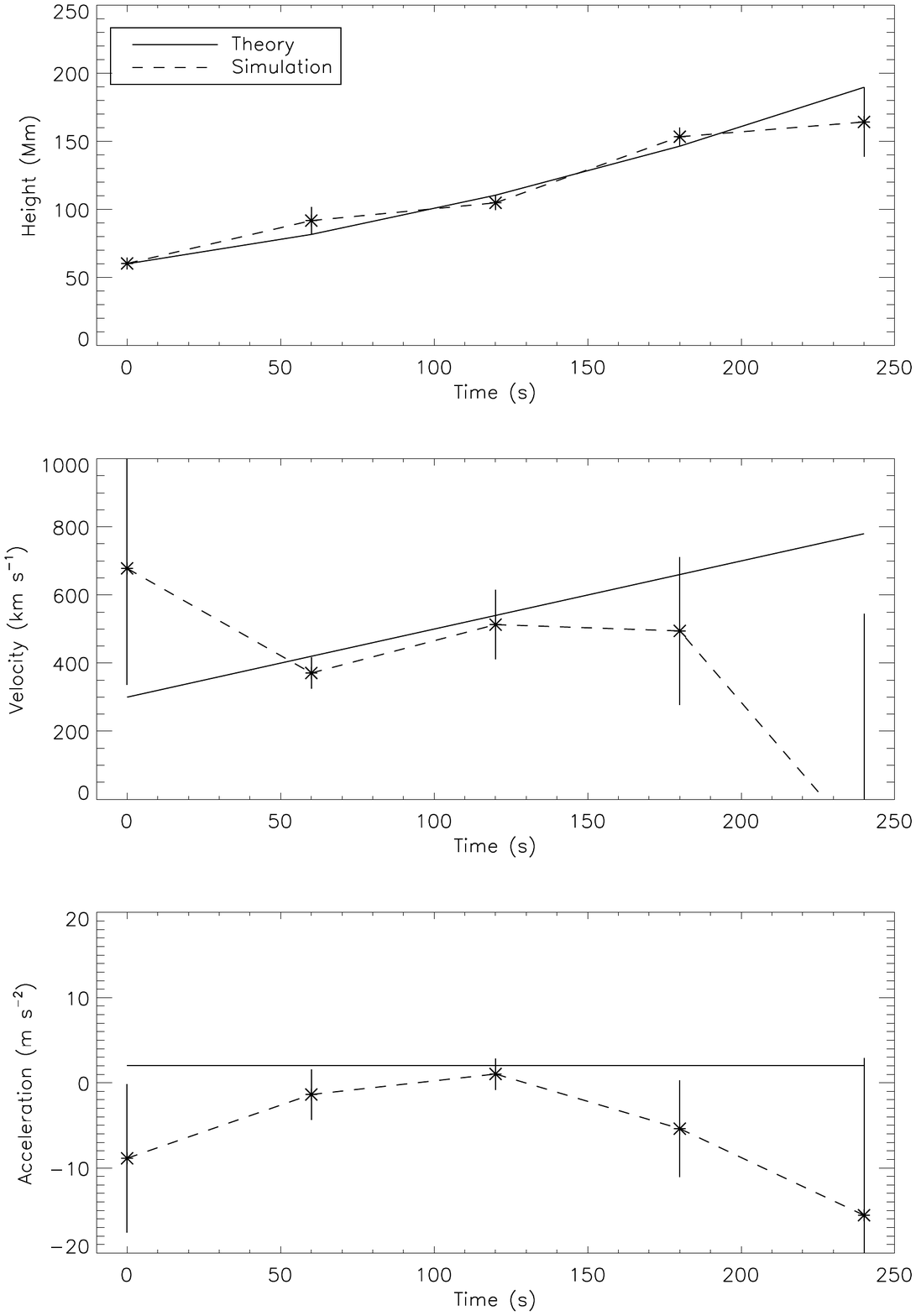}}
\subfigure{\includegraphics[scale=0.43, trim=70 50 110 70]{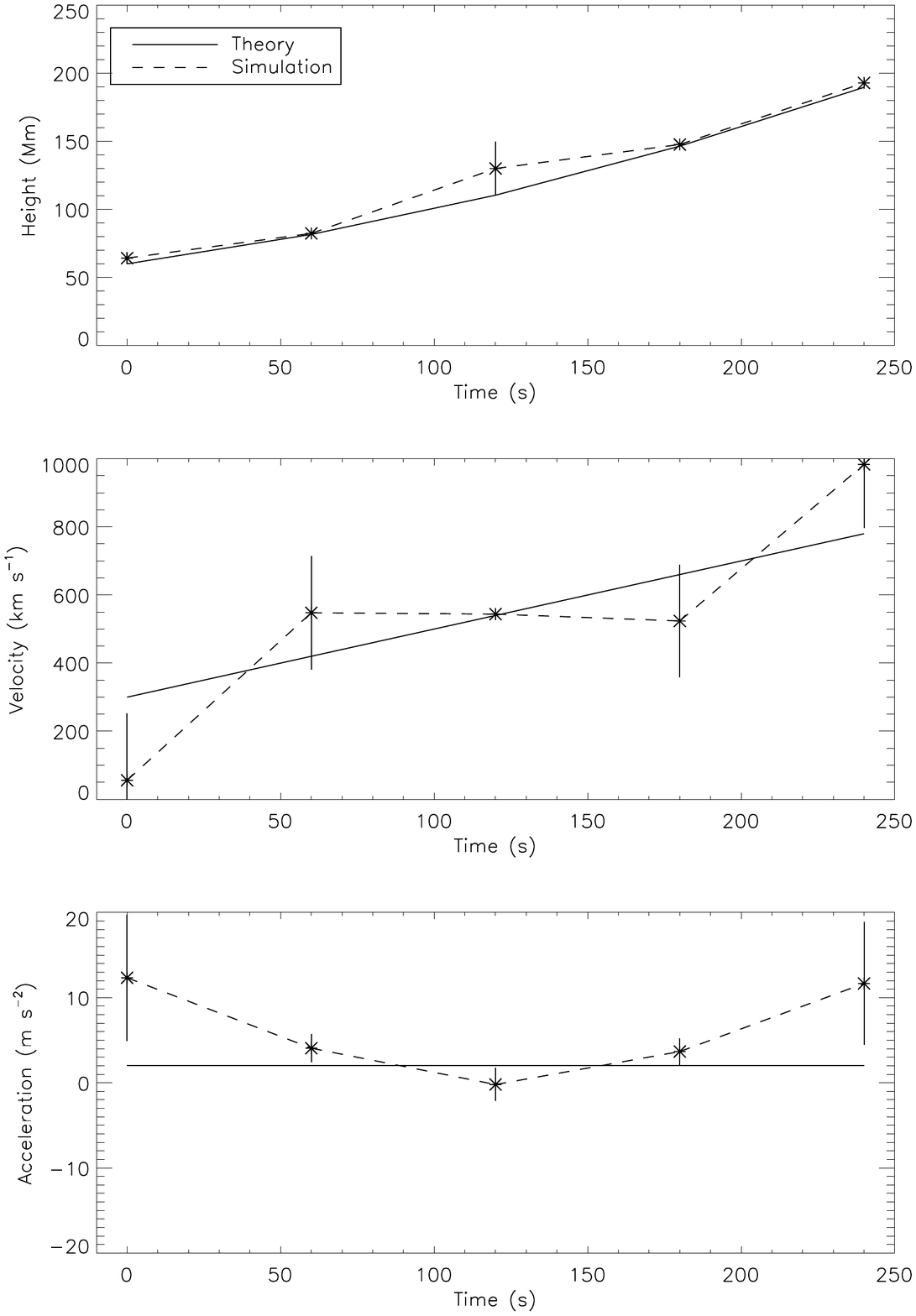}}
\caption{A theoretical model for a CME with constant acceleration 2~m~s$^{-2}$ and initial velocity 300~km~s$^{-1}$, and two simulations of how the resulting profiles for a noisy sample of datapoints behave using 3-point Lagrangian interpolation.}
\label{sim_vels_thesis}
\end{figure}

The standard method for determining the kinematics of a CME is to obtain a sequence of height-time measurements and perform a 3-point Lagrangian interpolation to derive the velocity and acceleration of the event (detailed in Section~\ref{3pointlagrangian}). This method alone is somewhat dated since the advent of strong numerical computing power and development of bootstrapping and spline-fitting techniques, for example. As a first approximation the 3-point Lagrangian provides a good estimate of how the velocity and acceleration corresponding to a height-time curve are likely to behave, by revealing the trends in the profiles that indicate increasing/decreasing velocity and/or acceleration. However, this is only true if the scatter and errors of the height measurements are not unreasonably large, since a large scatter would be enhanced by the derivatives, and a large error will increase the uncertainty on the derivative points, such that trends may become untestable. A quick simulation of how the 3-point Lagrangian fares for a theoretical model CME undergoing a constant acceleration of 2~m~s$^{-2}$ and initial velocity of 300~km~s$^{-1}$ reveals the unreliability of the resulting kinematic profiles. A scatter of height-time datapoints is chosen with varying levels of noise up to $\sim$\,20\%. An errorbar on each datapoint is determined by its distance from the theoretical height-time profile. Various instances of datapoint scatters result in erroneous trends in the velocity and acceleration profiles - even with the proper error treatment. Figure~\ref{sim_vels_thesis} shows two examples of how different the derived kinematics can be, in comparison with the theoretical model and each other. They show how different scatters of the datapoints can result in what appear to be completely opposing acceleration trends, meaning the nature of the scatter is not satisfactorily reflected in the derived errorbars of the resulting profiles. This warrants further investigation for future CME, and other, kinematic studies, since the implications for how we interpret observations could be profound.
\begin{figure}[!t]
\centerline{\includegraphics[scale=0.71, trim=50 320 120 40]{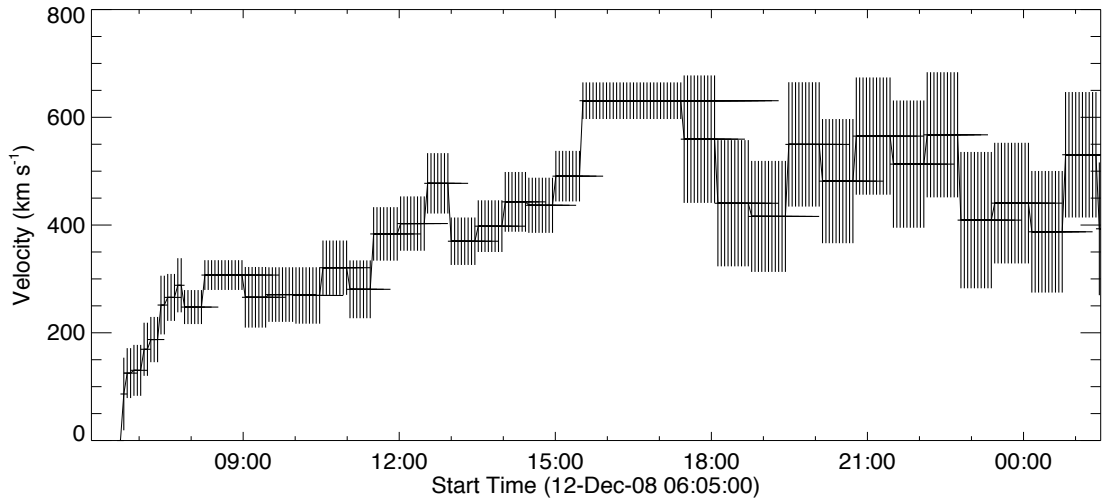}}
\caption{The resulting velocity profile for the 3D reconstructed CME front of the 12 December 2008 using the inversion technique of \citet{2005SoPh..227..299K}.}
\label{20081212_inversion}
\end{figure}
\newline
\indent It should also be noted that the 3-point Lagrangian counter-intuitively increases the errorbars on the resulting velocity and acceleration profiles when the number of height-time measurements are increased (i.e., for smaller cadences which new missions continue to provide). This is due to the algorithm's inverse dependence on the spacing between points. It is therefore worth investigating how other techniques might be applied to more confidently derive the kinematics of CMEs. Bootstrapping of a presumed model fit to the CME height-time profile would result in the best match of parameters to the data, but questions would remain on the appropriateness of the chosen model itself. Alternatively a simulation of data could be bootstrapped regarding the resulting derived kinematics and an estimate of the errors involved could be deduced to apply to true observations, though this again may be model dependant. An inversion technique could also be investigated for obtaining derivatives \citep{2005SoPh..227..299K}, an example of which is shown in deriving the velocity of the 12 December 2008 CME (Figure~\ref{20081212_inversion}). It works by essentially solving for the smoothest spline fit that minimises the distance between the end-points while still being bound by any constraints on the data. If a quantification of the height-time errors is provided, then inversion techniques may result in a more robust determination of the kinematic profiles with statistically sound uncertainties.%, in comparison with the 3-point Lagrangian which tends to result in large deviations at the end-points and is sensitive to the data scatter and cadence.
%\citep{Example}
%\citep{Example_book}

%\include{ch7} byrne_thesis_corrections_report

% --------------------------------------------------------------
%:                  BACK MATTER: appendices, refs,..
% --------------------------------------------------------------

% the back matter: appendix and references close the thesis

%: ----------------------- bibliography ------------------------

% The section below defines how references are listed and formatted
% The default below is 2 columns, small font, complete author names.
% Entries are also linked back to the page number in the text and to external URL if provided in the BibTex file.

% PhDbiblio-url2 = names small caps, title bold & hyperlinked, link to page 
%\begin{multicols}{2} % \begin{multicols}{ # columns}[ header text][ space]
%\begin{tiny} % tiny(5) < scriptsize(7) < footnotesize(8) < small (9)

%\bibliographystyle{Latex/Classes/CUEDbiblio-url2} % Title is link if provided
\bibliographystyle{jmb} % Title is link if provided
\renewcommand{\bibname}{References} % changes the header; default: Bibliography

\bibliography{references.bib} % adjust this to fit your BibTex file

%\end{tiny}
%\end{multicols}

% --------------------------------------------------------------
% Various bibliography styles exit. Replace above style as desired.

% in-text refs: (1) (1; 2)
% ref list: alphabetical; author(s) in small caps; initials last name; page(s)
%\bibliographystyle{Latex/Classes/PhDbiblio-case} % title forced lower case
%\bibliographystyle{Latex/Classes/PhDbiblio-bold} % title as in bibtex but bold
%\bibliographystyle{Latex/Classes/PhDbiblio-url} % bold + www link if provided

%\bibliographystyle{Latex/Classes/jmb} % calls style file jmb.bst
% in-text refs: author (year) without brackets
% ref list: alphabetical; author(s) in normal font; last name, initials; page(s)

%\bibliographystyle{plainnat} % calls style file plainnat.bst
% in-text refs: author (year) without brackets
% (this works with package natbib)

% --------------------------------------------------------------

% according to Dresden med fac summary has to be at the end
%\include{0_frontmatter/abstract}

\end{document}